\documentclass[]{msu-thesis}
\usepackage[T1]{fontenc}
\usepackage{newtxtext,newtxmath} 
\usepackage{siunitx}  

\usepackage{booktabs}
\usepackage{graphicx}
\usepackage{pgf-pie}

\usepackage{amssymb}
\usepackage[colorlinks=false, linkcolor=black, citecolor=blue, urlcolor=blue, hidelinks]{hyperref}
\usepackage[backend=biber, style=numeric, backref=true, sorting=none]{biblatex} 
\DeclareFieldFormat{url}{\normalfont\url{#1}}
\DeclareFieldFormat{doi}{\normalfont\url{#1}}

\usepackage{lineno}
\usepackage{placeins}
\usepackage{etoolbox}

\renewcommand{\msuappendixdelim}{:}
\newcommand{\br}{$b_{R}$~}
\newcommand{\btr}{$b$ from $t_R$~}
\newcommand{\bta}{$b$ from $t_A$~}

\title{Search for a Heavy-philic $W'$ Boson using Proton-Proton Collisions at $\sqrt{s}=13~\textrm{TeV}$ using the ATLAS Detector}
\author{Jason Peter Gombas}
\dualmajor{Physics}{Computational Mathematics, Science and Engineering} 

\date{2025}

\begin{document}

\frontmatter
\maketitlepage

\begin{abstract}
This thesis presents a search for a new, hypothetical particle predicted by theories extending the Standard Model of particle physics. This heavy $W'$ boson interacts only with the heaviest known quarks, top and bottom (heavy-philic). Such a particle could provide insight into the fundamental forces of nature and be the first hint at extra dimensions or a composite Higgs. The $W'$ boson is produced in high-energy proton-proton collisions, mainly through gluon fusion. Its decay leads to a distinctive final state: $tbW'\rightarrow tbtb$. This search uses data collected by the ATLAS detector during Run 2 at the Large Hadron Collider, focusing on events with a single charged lepton, at least five jets, and at least three jets identified as originating from bottom quarks. To improve sensitivity to this rare process, advanced machine learning techniques are applied. A profile likelihood fit to the machine learning output is used to evaluate the data. No significant excess above the Standard Model background is observed, and exclusion limits are set at the 95\% confidence level on the production cross-section of the heavy-philic $W'$ boson.
\end{abstract}

\clearpage


%
%
%
\clearpage
%
%

\SingleSpacing
\tableofcontents* 
%
%
%
%
\mainmatter
%

\chapter{Introduction}
Ever since the discovery of the Higgs boson at the LHC by the ATLAS collaboration \cite{HIGG-2012-27} and the CMS collaboration \cite{CMS:2012gua}, the Standard Model has been subjected to increasingly rigorous experimental scrutiny. It has been repeatedly confirmed, each time in novel and more precise ways. During Run 2, the ATLAS experiment achieved remarkable milestones: measuring the properties of the Higgs boson \cite{atlas_run2_higgs}, advancing our understanding of electroweak and QCD processes and flavor physics \cite{atlas_run2_electroweak}, exploring the top quark \cite{atlas_run2_top}, probing for additional scalars and exotic Higgs decays \cite{atlas_run2_scalars}, continuing the search for supersymmetry \cite{atlas_run2_ss}, and conducting a wide range of searches for exotic particles \cite{atlas24highenergyfront}.

While the Standard Model remains extraordinarily successful at describing known fundamental particles and interactions, compelling evidence suggests it is incomplete. Observations of galactic rotation curves, along with other evidence, point to the presence of non-luminous, non-baryonic matter (dark matter) that cannot be explained by the Standard Model alone \cite{Rubin1970, Rubin1980}. This, along with other open questions in particle physics, motivates the ongoing search for new physics beyond the Standard Model.

It is quite remarkable that, despite living in a universe filled with chaos and destruction, scientists are able to find moments of peace and venture towards discovery and understanding. It should be surprising that the universe is not only governed by order and natural laws, but also it can be understood through natural laws and expressed through mathematical equations. As Einstein famously remarked, ``The most incomprehensible thing about the universe is that it is comprehensible" \cite{einstein1936}. This sentiment is echoed by Eugene Wigner, who reflected on the deep harmony between mathematics and physics, calling it ``the unreasonable effectiveness of mathematics in the natural sciences" \cite{wigner1960}.

The natural sciences has also sought to uncover the fundamental nature of matter for centuries. At first, scientist observed the macroscopic world through the lens of visible light, uncovering the beauty and complexity of cells, bacteria, and surface patterns with microscopes \cite{hooke1665}. With the discovery of the electron \cite{thomson1897}, scientist pushed further, developing electron microscopes \cite{knoll1932} that revealed atoms, molecules, and the intricate structures they form.

However, there are limits to how far visible light and electrons can take us. To look at smaller things, scientists turned to nuclear science, propelling atoms together with particle accelerators to unlock the quantum world \cite{griffiths2008}. This exploration revealed protons, neutrons, and eventually the quarks \cite{gellmann1964} and leptons that form part of the Standard Model. Today, we probe even further, seeking to understand not only these fundamental particles, but whether they themselves are composed of something even more elementary or if there are even more particles that we have missed all along.

\section{Exotic Particle Searches}
At the writing of this dissertation, no particle beyond the Standard Model has been discovered. There have been extensive exotic particle searches conducted by the ATLAS \cite{atlas24highenergyfront} and CMS \cite{lyon24reviews} collaborations. The Standard Model has been remarkably successful in describing the known particles and their interactions. However, it is well established that the Standard Model does not provide a complete description of the fundamental forces and particles in the universe. This motivates ongoing searches for new and exotic physics beyond its framework.

\begin{figure}[ht]
    \centering
    \includegraphics[width=0.8\linewidth]{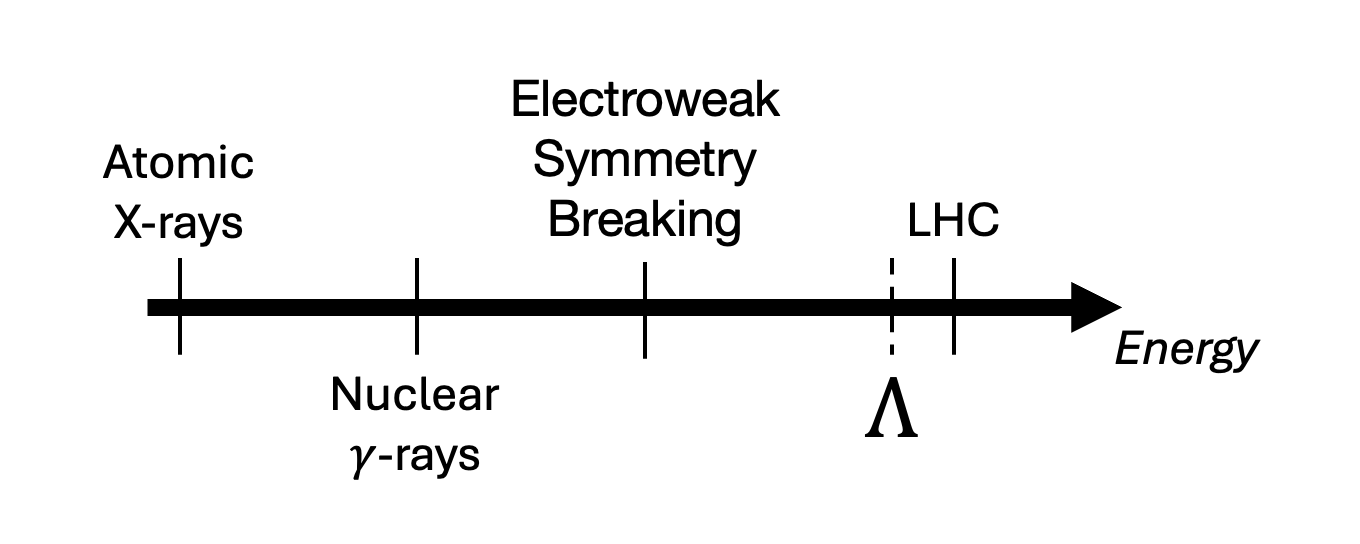}
    \caption{A diagram showing the possible separation of energy scales. Starting from the atomic scale with X-rays and increasing through $\gamma$-rays, electroweak symmetry breaking where the Higgs and top quarks are probed, to the energy scale of the LHC. At a currently unknown scale, new physics is predicted which would have couplings to the third generation of quarks. }
    \label{fig:non-universal_int}
\end{figure}

Oftentimes, theories which predict new physics Beyond the Standard Model (BSM theories) involve symmetry breaking at the energy scale reached by the LHC. There are some theories which suggest that there are certain scales which are flavor non-universal \cite{Muller_1996, PhysRevD.74.075008, MALKAWI1996304, PatiSalam1974}. This means the couplings do not adhere to all flavors of quarks and leptons. If this is true, then the first hint of new physics would come from couplings to the third generation of quarks first, then to the second, and then to the first generation of quarks. Each new flavor would be included once one reaches a higher and higher energy scale. Fig.~\ref{fig:non-universal_int} shows a diagram representing this schematically.

Almost all BSM theories predict the existence of a new gauge boson, usually denoted as $W'$ or $Z'$ boson, arising from mechanisms such as extra dimensions \cite{Dienes_1999}, strong dynamics \cite{PhysRevD.13.974, EICHTEN1980125}, or a composite Higgs \cite{Muller_1996}. 

Previous searches for the $W'$ boson have focused on Standard Model-like couplings to all quarks \cite{JHEP2023-073}. A recent effort examined a top-philic  $Z'$, where couplings to lighter quarks are suppressed, resulting in different kinematic signatures \cite{atlas:24zprime}. Building on this idea, this dissertation explores a heavy-philic $W'$ boson, hypothesized to couple exclusively to the third-generation quarks: top and bottom. By turning off the couplings to the first and second generations, the production dynamics shift significantly, relying primarily on gluon-gluon fusion through gluon pair-production rather than quark-gluon interactions. Also, an additional top and bottom quark are created in the initial state, making the heavy-philic $W'$ boson model significantly different from previous $W'$ boson models. 

While there have been numerous studies on the hypothetical $W'$ boson \cite{JHEP2023-073, ATLASexot109.112008, cms:wprimetb}, there have been no studies so far that have attempted to turn off the couplings to the first and second generation of quarks. If new physics beyond the Standard Model exists but has remained hidden due to suppression in third-generation production channels, it would represent a significant missed opportunity for discovery. This study aims to address a gap that remains despite the substantial progress made in $W'$ boson and other BSM searches.

\section{Detecting Particles}
Particles are often conceptualized as tiny specks moving through space with well-defined positions. However, Quantum Field Theory (QFT) has a more complex description of nature. In QFT, particles are not discrete objects like billiard balls; rather, they are excitations of underlying quantum fields that permeate space. Particles are described by probability distributions—the modulus squared of their wave functions—representing the likelihood of finding a particle at a given location.

The framework used to model physical phenomena depends on the relevant energy and time scales. At the smallest scales—characterized by the highest energies and shortest time intervals—Quantum Field Theory (QFT) is essential for describing the creation and interactions of fundamental particles within the Standard Model. However, as particles travel away from the collision site, the relevant time and energy scales shift. At this scale, a more simplified model of particle interactions becomes applicable, eliminating the need for the full complexity of QFT while still providing an accurate description of the observed behavior of nature.

\section{Natural Units}
Beam energies at the LHC are in the TeV range, corresponding to extremely high energies. An interesting aspect of using natural units in theoretical calculations is the ability to highlight key insights. In the case of the LHC, the natural units are defined by setting $\hbar = 1$, which simplifies the treatment of quantum phenomena, and $c = 1$, which corresponds to extremely high speeds. By setting \(c=1\), mass, energy, and momentum have the same units (MeV, GeV).

Because $\hbar$ (with units of energy $\times$ time) is set to 1, energy and time become directly related, i.e., energy is inversely proportional to time. Additionally, since $c$ (with units of distance/time) is set to 1, distance becomes proportional to time. Therefore higher energies probe smaller distances.

This brief exercise highlights why particles are collided at such high energies: to study the fundamental particles of the universe, which exist at the smallest scales.

\section{Thesis Structure}

This dissertation focuses on a novel search for a particle Beyond the Standard Model (BSM). Chapter 1 provides an introduction to the field of high-energy physics. Chapter 2 contains a brief summary of the Standard Model and Quantum Field Theory (QFT) as the foundation for this work. Chapter 3 describes the Large Hadron Collider (LHC) facility and the ATLAS detector, essential tools for probing the frontiers of particle physics. Chapter 4 outlines the modeling of proton-proton collisions and the reconstruction of events from the data measured by the ATLAS detector. Chapter 5 introduces the heavy-philic $W'$ boson, a theoretical BSM particle and the focus of this dissertation, and discusses the Standard Model background relevant to the phase space of the search. Chapter 6 presents the results of the analysis, including fits and statistical interpretations. Finally, Chapter 7 concludes with a discussion of the findings and potential directions for future research.
\chapter{Theory}
\section{Quantum Field Theory}
In Lagrangian Field Theory, the action \(S\) is defined as the time integral of the Lagrangian \(L\). In local field theory, one typically can write the Lagrangian as an integral over the Lagrangian density which is a function of a field \(\phi_X(x)\), or fields, and their derivatives \(\partial_{\mu}(x)\). This can be written as the following

\begin{equation}
    S=\int Ldt = \int \mathcal{L}(\phi_X, \partial_{\mu}\phi_X)d^4x
    \label{eq:action}
\end{equation}

\noindent according to \cite{Peskin:1995ev}. The principal of least action states that configurations will proceed along a path in such a way that minimizes \(S\) as time progresses from \(t_1\) to \(t_2\). This statement can be summed up as $0=\delta S$ since the extremum is normally the minimum. The Euler-Lagrange formula can then be derived by taking the derivative of \(S\) as defined in Eq.~\ref{eq:action}, and finding where it vanishes. To do this one must assume that \(\delta S\) is zero at the temporal beginning and end of the integration region. The condition that satisfies these requirements independent of \(\partial \phi_X\) is the Euler-Lagrange formula

\begin{equation}
    \partial\left( \frac{\partial \mathcal{L}}{\partial( \partial_{\mu}\phi_X)} \right) - \frac{\partial \mathcal{L}}{\partial \phi_X} = 0
    \label{eq:euler_lagrange}
\end{equation}

\noindent this process is independent of the field which means that there exists an Euler-Lagrange equation for each and every field illustrated with the subscript \(X\). 

The Lagrangian is then used to derive the quantities of interest. In high-energy physics, the Lorentz-invariant phase space form of Fermi's Golden Rule takes the form of

\begin{equation}
    d\Gamma = \frac{(2\pi)^4 \delta^4(p_i - p_f)}{2E_1 2E_2} |\mathcal{M}_{fi}|^2 \prod_f \frac{d^3p_f}{(2\pi)^3 2E_f}
\end{equation}

\noindent $d\Gamma$ is the differential decay rate, $(2\pi)^4 \delta^4(p_i - p_f)$ is a four-dimensional Dirac delta function ensuring that four-momentum is conserved, $E_1$ and $E_2$ are the energies of the initial state particles, $|\mathcal{M}_{fi}|^2$ is the squared matrix element that contains all the dynamics of the interaction, $\prod_f \frac{d^3p_f}{(2\pi)^3 2E_f}$ is the phase space volume element for each final-state particle $f$. Integrating this yields the total decay rate over the allowed final momenta. 

Cross section is a quantity that describes the probability of a given interaction occurring between particles. It is the effective area that creates a particular interaction. Convention has been established to use units of barns (\(1 \text{ barn} = 10^{-28} \text{ m}^2\)). The cross section is central to experimental particle physics since it directly determines the number of events observed in a detector. For the collision of two particles, it can be written as
\begin{equation}
\sigma (X_1X_2\rightarrow Y) = \frac{1}{4 E_1 E_1 |\vec{v}_1 - \vec{v}_2|} \int |\mathcal{M}|^2 \, d\Phi
\end{equation}

\noindent The differential cross section is also important, as it tells what kinematic region to expect a flux of particles. 

The total number of expected events with final state \( Y \) also depends on the luminosity of the colliding beams. This is why the instantaneous luminosity, \( I(t) \), is measured. The integrated luminosity, \( \mathcal{I} \), is then the total number of incident particles per unit area, and is given by \( \mathcal{I} = \int I(t) dt \). The number of observed events \( N \) for a particular process that has cross section $\sigma$ is therefore given by \(N = \mathcal{I} \cdot \sigma\). The total integrated luminosity measured for the ATLAS detector during Run 2 was $139~\text{fb}^{-1}$ \cite{DAPR-2021-01}.

To compute the cross section along with the kinematic differential distributions, the integral of the matrix element must be taken. Since this integral results in IR divergences due to soft and co-linear gluon emission, the direct calculation is impossible to calculate by hand. Hence, this study uses the Monte Carlo method to produce these distributions. 

At small distances, the strong coupling constant is small enough where the modeled process in QCD can be approximated by a finite amount of interactions \cite{PhysRevLett.30.1343}. In perturbative QCD, the calculation of hadronic cross sections requires the separation of short- and long-distance physics. This is achieved through the factorization theorem, which introduces the factorization scale $\mu_F$.

In addition, the renormalization scale $\mu_R$ appears through the running of the strong coupling constant $\alpha_S(\mu_R)$, as a consequence of renormalizing ultraviolet divergences in loop calculations. Although physical observables are, in principle, independent of both $\mu_F$ and $\mu_R$, fixed-order calculations retain some residual dependence on these scales. In perturbative QCD, the desired process can be calculated up to a certain order in the strong coupling constant. This is called a fixed order calculation in QCD. 

The 4-flavor scheme (4FS) is used, in which top and bottom quarks are excluded from the parton distribution functions (PDFs). The advantages and disadvantages of using the 4-flavor versus the 5-flavor scheme (5FS) are discussed in detail in \cite{Maltoni_2012}. While each scheme has its own domain of validity and theoretical motivations, they generally yield consistent results within uncertainties for inclusive observables.

\begin{figure}
    \centering
    \includegraphics[width=0.65\linewidth]{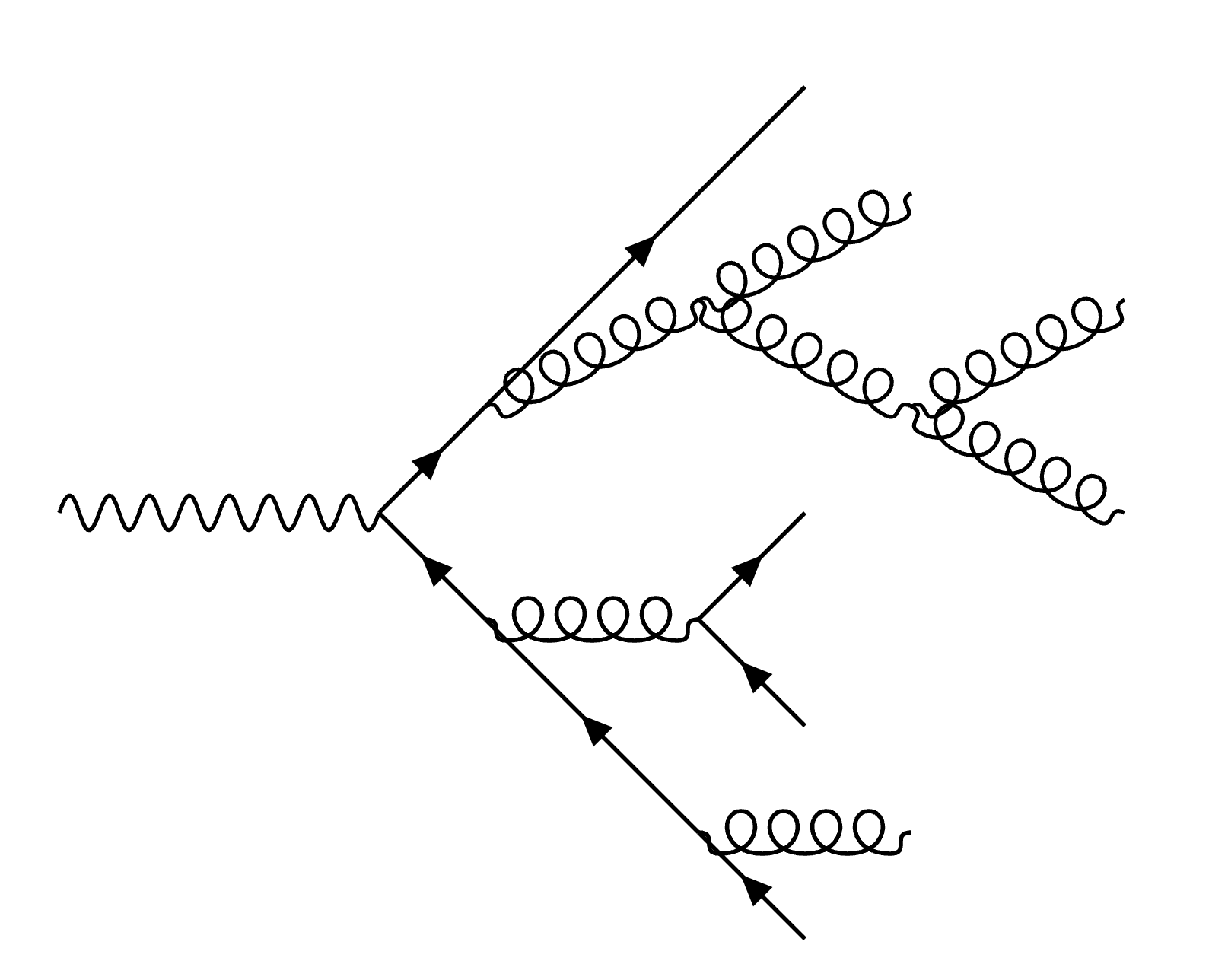}
    \caption{A Feynman diagram of a boson decaying into a quark-antiquark pair which produces parton showering where the curly lines represent gluons and the solid lines represent quarks. The horizontal direction is the temporal dimension and the vertical direction is the spatial dimension.}
    \label{fig:parton_shower}
\end{figure}

When highly energetic quarks or gluons are produced in a collision, they will undergo a phenomenon known as parton showering. This occurs as quarks or gluons travel through space and emit additional gluons or split into quark-antiquark pairs. The emitted gluons can themselves undergo further branching, creating a cascade of partons. This process is a result of the self-interacting nature of gluons, governed by the QCD coupling constant at high energies, such as those at the LHC. As the gluons continue to propagate, they can undergo pair production and split into additional gluons or quark pairs, leading to the creation of thousands of particles in a rapid cascade. This process can be visualized through a Feynman diagram, as shown in Fig.~\ref{fig:parton_shower}.

After parton showering, hadronization of each unstable particles occurs. Due to quantum confinement, quarks cannot exist on their own. They must be paired with at least 1 other quark in a period of time. Therefore, quarks will immediately hadronize into mesons and hadrons after the parton shower.

\section{The Standard Model}

The Standard Model is a model of all the fundamental particles that are known about and their interactions. Gluons are the force carriers for the strong force, $W$ and $Z$ bosons are the force carriers for the weak force, and photons are the force carriers for the electromagnetic force. Mathematically, the Standard Model is a non-abelian gauge quantum field theory that has the symmetries of SU(3)$\times$ SU(2)$\times$ U(1) group. 

There are 6 quarks in the Standard Model, arranged in 3 generations with 2 quarks in each generation. Each generation has increasing mass and each flavor within the generation share similar properties. The first generation of quarks are the up and down quarks ($u$,$d$), the second generation of quarks are charm and strange quarks ($c$,$s$), and the third generation of quarks is are the top and bottom quarks ($t$,$b$). Each quark has a pair anti-quark that is denoted with an overhead bar (ie $\bar{u}$, $\bar{c}$, $\bar{t}$). 

\begin{figure}
    \centering
    \includegraphics[width=0.6\linewidth]{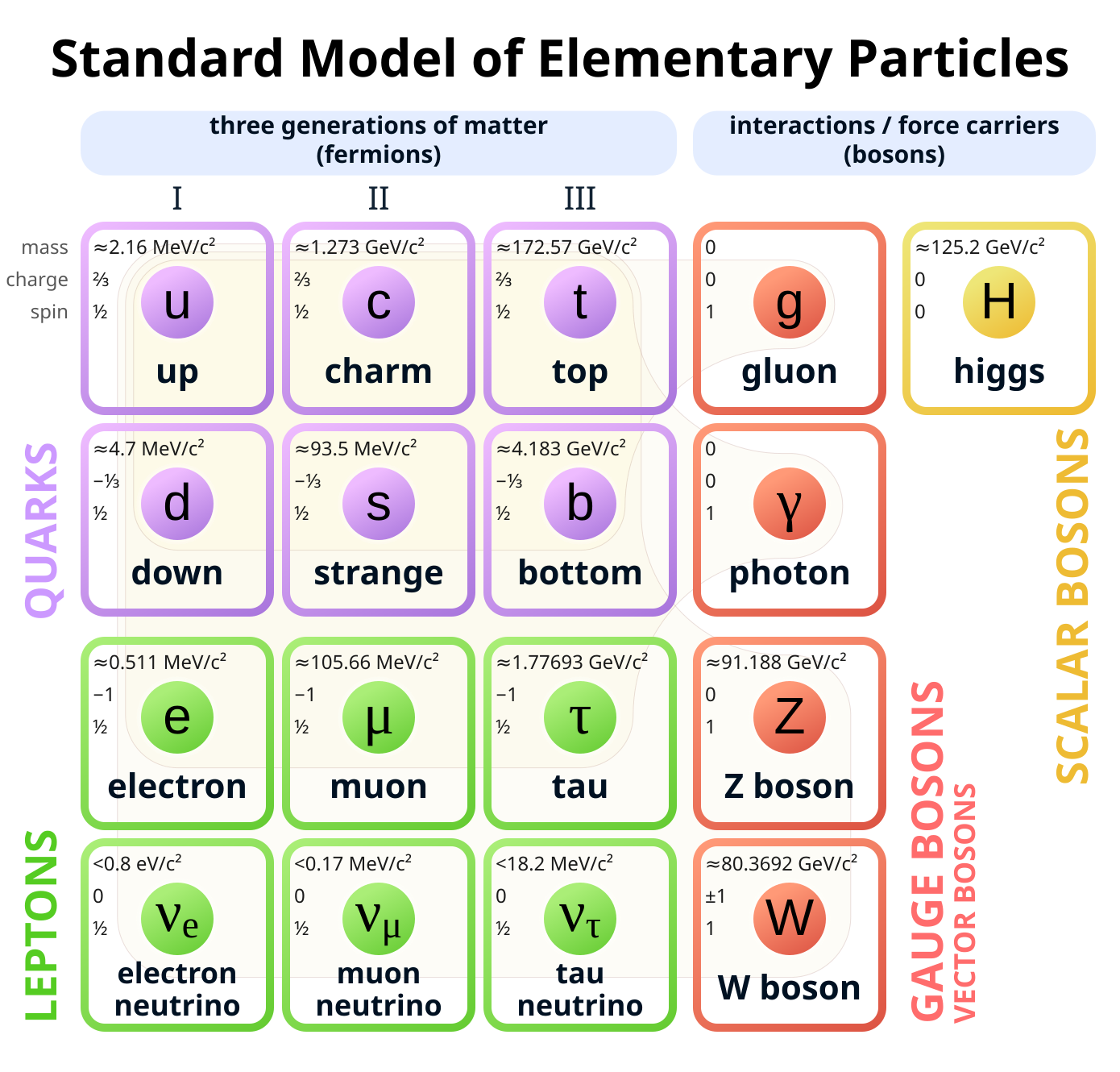}
    \caption{Standard Model diagram showing approximate values for masses, and grouping particles according to their symmetries. The Standard Model particle couplings are also shown in the background.}
    \label{fig:standardmodel}
\end{figure}

There are 3 leptons, electrons, muons, and tauons (\(e^-,\mu^-,\tau^-\)), and each lepton has a matching neutrino (\(\nu_{e},\nu_{\mu},\nu_{\tau}\)), and matching anti-matter (\(e^+,\mu^+,\tau^+,\bar{\nu_{e}},\bar{\nu_{\mu}},\bar{\nu_{\tau}}\)). See Fig.~\ref{fig:standardmodel} for a diagram of all these particles. 

In high energy proton-proton collision experiments, most of the energy in the quarks are in the momentum of the particles. Therefore, particles with less mass than the $b$ quark are approximated to be massless in most event simulations.

The charged current interaction is included in the overall Lagrangian with the following expression:

\begin{equation}
    \mathcal{L}^{\text{eff}} = \frac{g}{2\sqrt{2}}\bar{f}_i\gamma^\mu (1-\gamma^5) (T^+W^+_{\mu} +T^-W^-_{\mu})f_j + \text{h.c.}
\end{equation}

\noindent Here \(g\) is the weak coupling constant, \(\gamma^{\mu}\) are the Dirac \(\gamma\)-matrices, \(T^+\) and \(T^-\) are the weak isospin raising and lowering operators, and \(W^{\pm}\) are the charged weak boson fields. The $W'$ boson Lagrangian will have a structure similar to the charged current Lagrangian. The full SM Lagrangian after electroweak symmetry breaking that is complete with other interactions can be viewed in \cite{PhysRevD.110.030001}.

\section{Top Quark Physics}

The top quark is by far the most massive quark in the Standard model with a nominal mass of $172.5$ GeV \cite{ATLASCMS2024TopMass}. This large mass gives rise to its extremely short lifetime at about $10^{-25}$ s \cite{atlas_2018_topwidth}, which is an order of magnitude shorter than the process of hadronization in QCD which is about $10^{-24}$ s \cite{Song_2016}. This leads to the unusual phenomenon where the top quark decays before it has time to hadronize. Top quark bound states (toponium) have been confirmed, and there is ongoing discussion of how the inclusion of toponium in QCD calculations might improve the description of an excess observed in data near the $t\bar{t}$ threshold region~\cite{ATLAS-CONF-2025-008,cms_toponium}. 

Production of top quarks at the LHC proceeds mainly via gluon-gluon fusion where two gluons collide to produce a top-antitop quark pair. Quark-antiquark fusion also contributes to top quark pair production, but at a much lower rate. 

The LHC produces many top quarks as the current cross-section measurement for inclusive $t\bar{t}$ is $829~\text{pb}$ corresponding to $N = 829~\text{pb}\times139~\text{fb}^{-1} = $ 116 million events within ATLAS during Run 2 \cite{atlas_2023_ttbar_inclu}.

The top quark decays almost exclusively into a $W$ boson and a bottom quark (99\%). The bottom quark undergoes decay and hadronization forming a conical spray of particles (jet) which will be discussed in Chapter 4. The $W$ boson undergoes decay into either a lepton neutrino pair (32\%), or a quark-antiquark pair (68\%) \cite{PhysRevD.98.030001}. 

In the case of a quark-antiquark pair decay of the $W$ boson, two additional jets will be produced. In the case of a lepton neutrino pair, a lepton and missing transverse energy will be observed in the detector. 

\begin{figure}[h]
    \centering
    \begin{tikzpicture}
        \pie{15/$\tau$ + jets, 46/All Hadronic, 15/$\mu$ + jets, 15/$e$ + jets,9/Dilepton}
    \end{tikzpicture}   
    \caption{Breakdown on $t\bar{t}$ decay channels. $\mu$ + jets and $e$ + jets are considered to be the single lepton channel. $\tau+\text{jets}$ is a very small channel usually considered on its own. }
    \label{fig:ttbar_channels}
\end{figure}
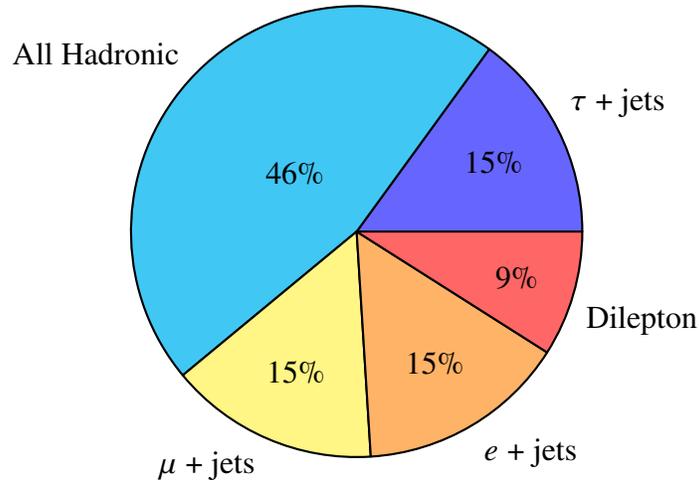

In top quark pair production, then, there are a number of possible final states that can be observed in the detector. If neither $W$ bosons from the top pairs decay into a lepton, the event will be fully hadronic. If one $W$ boson decays leptonically, then one lepton will be observed which is called the single lepton channel. If both $W$ bosons decay leptonically, then two leptons will be observed which is called the dilepton channel in $t\bar{t}$ production. The branching fraction of the top quark decaying leptonically is \(\approx0.326\) \cite{PhysRevD.110.030001}. A pie chart describing the probability of decaying into each channel is shown in Fig.~\ref{fig:ttbar_channels}.

The all-hadronic channel is dominated by QCD multijet which has high modeling uncertainties. The single lepton channel has the amount of recorded events due to the efficient ATLAS lepton triggers. Single lepton channel has single top t-channel, $W$+jets and some multi-jet backgrounds. The dilepton channel composes the smallest branching ratio compared to the other 2 channels and has smaller systematic uncertainties due to the presence of fewer jets. The analysis in this work focuses on the single lepton plus jets channel which excludes the all hadronic channel and the dilepton channel.

\section{Proton-Proton Collisions and Parton Distribution Functions}
To study the fundamental particles of nature, proton-proton collisions at the Large Hadron Collider (LHC) at CERN are observed. At the LHC, the interacting constituents of protons are typically gluons, which, along with quarks, form the proton’s internal structure. The probability of finding a specific parton (quark or gluon) carrying a fraction x of the proton’s momentum at a given energy scale Q is described by the parton distribution functions (PDFs). These functions are extracted from experimental data, primarily from deep inelastic scattering, such as those performed at HERA~\cite{Brinkmann:366965}. PDFs also evolve with scale according to the DGLAP equations \cite{thomson2013modern}.

\begin{figure}
    \centering
    \includegraphics[width=0.7\linewidth]{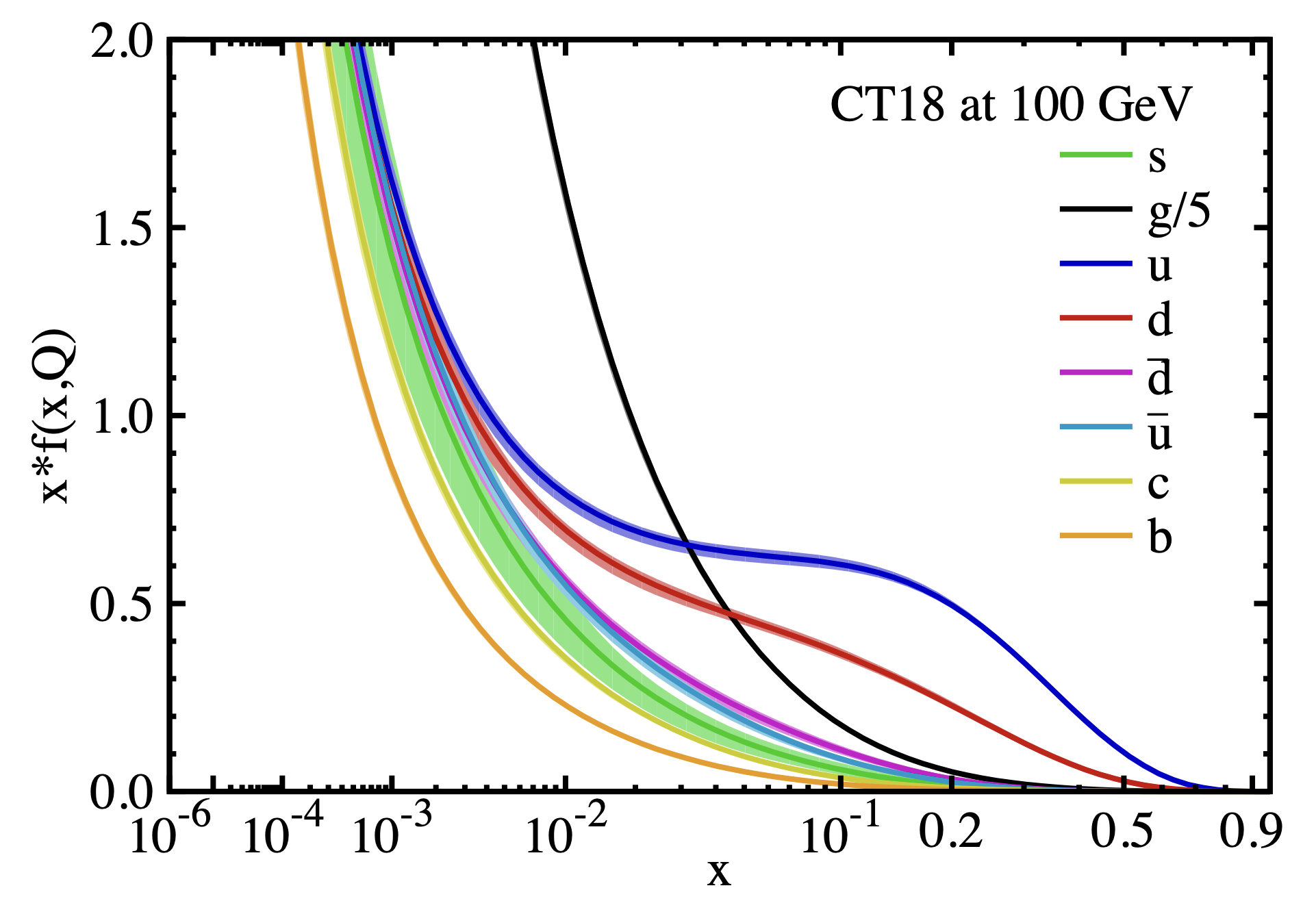}
    \caption{CT18 PDF set \cite{Hou:2019efy} with \(Q^2=100\) GeV plotted as the PDF multiplied by the momentum fraction \(x\). }
    \label{fig:pdf}
\end{figure}

PDF distributions are typically plotted as the product of the probability density and momentum fraction to account for steep slopes at small  \(x \). The CT18 PDFs are shown in Fig.~\ref{fig:pdf}. As seen in the figure, it is most likely to find that the up quark carries most of the momentum fraction of the proton due to it being a valence quark. The down quark carries the next highest momentum fraction, followed by the heavier flavor quarks which tend to carry a small momentum fraction of the proton \cite{Hou:2019efy}. At energy scales relevant for top-quark pair production (around 500 GeV and above), the typical values of the momentum fraction $x$ are approximately 0.1 or lower, making it much more likely for gluons inside the proton to participate in the interaction.

At $\mu_F$, non-perturbative physics is absorbed into the PDFs, while the remaining hard scattering process can be computed perturbatively. PDFs are thus evaluated at $\mu_F$, and their scale dependence is governed by the DGLAP evolution equations.

The leading order QCD prediction cross section with a final state \( Y\) of a collision between two protons therefore takes the form

\noindent
\begin{align}
        \sigma(p(P_1)+p(P_2) \rightarrow Y + X)  =\int_0^1dx_1\int_0^1dx_2 \sum_f f_f(x_1,\mu_F) f_{\bar{f}}(x_2,\mu_F) \notag \\
        \quad \times \sigma(q_f(x_1P)+q_{\bar{f}}(x_2P) \rightarrow Y, \mu_F,\mu_R),
\end{align}

Here the sum runs over every quark flavor and \( X\) denotes any hadronic final state. $f_f(x_1,\mu_F)$ and $f_{\bar{f}}(x_2,\mu_F)$ denote the parton distribution functions with quark flavor \( f \) and \( \bar{f} \). The cross section is now calculated as a function of momentum fractions. In this computation the cross section is calculated with quark $q_f$ with momentum $x_1P$ and quark $q_{\bar{f}}$ with momentum $x_2P$. The final step of the calculation is then to compute the integral over all momentum fractions \( x_1\) and \( x_2\). More details on PDFs can be found in \cite{griffiths2008}.

\section{Heavy $W'$ Boson Signal Model}
This study defines a specific Beyond the Standard Model (BSM) theory, following the philosophy that, while model-agnostic approaches are valuable for confirming consistency with the Standard Model, they are limited in specifying new physics, and thus this work focuses on a particular BSM scenario involving a heavy-philic $W'$ boson. This particle has a Lagrangian set up to ensure that the $W'$ boson only couples to the third generation of quarks, and doesn't couple to leptons. The mass is set to an unknown parameter that can be adjusted. When calculating the fixed order QCD calculation of the heavy-philic $W'$ boson, orders up to next-to-leading order (NLO) are performed. This includes up to 3 vertices in the considered interaction diagrams. 

There have been several studies that have searched for a $W'$ boson such as \cite{JHEP2023-073} where the search is performed with a SM like $W'$ boson which decays into a top and bottom and looking in the 0 lepton and single lepton channel. In \cite{ATLASexot109.112008}, the search is conducted on a SM like $W'$ boson that decays into a tau-neutrino pair. However, both assume a SM-like $W'$ boson where the $W'$ boson is produced from light quarks like the SM $W$ boson.

Since the heavy-philic $W'$ boson does not couple to the first two generations of quarks, the production mode of the $W'$ boson changes drastically. Production of the $W'$ boson can no longer occur directly from the proton partons. Primary production comes from \(gg\) with small amplitudes coming from a gluon interacting with a quark from the other proton. 

\begin{figure}
    \centering
    \includegraphics[width=0.5\textwidth]{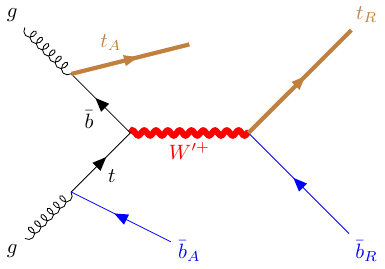}
    \caption{Leading order Feynman diagram for heavy-philic $W'$ boson  production. This study uses the convention where the spatial dimension is vertical. To produce the heavy-philic $W'$ boson, gluons from the incoming protons undergo pair production of a top pair and bottom pair which can then produce the resonance particle. }
    \label{fig:wp_lo_diagram}
\end{figure}

In Fig.~\ref{fig:wp_lo_diagram}, a leading order Feynman diagram of the heavy-philic $W'$ boson in the 4-flavor parton scheme is shown. This diagram includes the effects of two gluons coming together, each splitting into a quark pair where one of them splits into a bottom pair, and the other a top pair. One of the top quarks and one of the bottom quarks then form the $W'$ boson. 

This Feynman diagram represents a contribution to the matrix element calculation for the $W'$ process, which determines the probability of producing a bottom quark, a top quark, and a $W'$ boson when two gluons interact. 

Interaction of the heavy-philic $W'$ boson with the quarks is described by the Lagrangian term:

\begin{equation}
    \mathcal{L}^{\text{eff}} = \frac{V_{f_if_j}}{2\sqrt{2}}g_{W'}\bar{f}_i\gamma_\mu \left[ \alpha_R^{f_if_j}(1+\gamma^5)+\alpha_L^{f_if_j}(1-\gamma^5) \right] W^{'\mu}f_j + \text{h.c.},
\end{equation}

\noindent where \( V_{f_if_j}\) is the analogue of the Cabibbo-Kobayashi-Maskawa (CKM) matrix if \(f_i\) and \( f_j\) represent quarks. Specifically, in this heavy-philic $W'$ boson model, the matrix \( V_{f_if_j}\) has values given by

\[
V_{f_if_j} = 
\begin{bmatrix}
V_{ud} & V_{us} & V_{ub} \\
V_{cd} & V_{cs} & V_{cb} \\
V_{td} & V_{ts} & V_{tb}
\end{bmatrix}
= \begin{bmatrix}
0 & 0 & 0 \\
0 & 0 & 0 \\
0 & 0 & 1
\end{bmatrix}
\]

\begin{figure}
\centering
\includegraphics[width=0.8\linewidth]{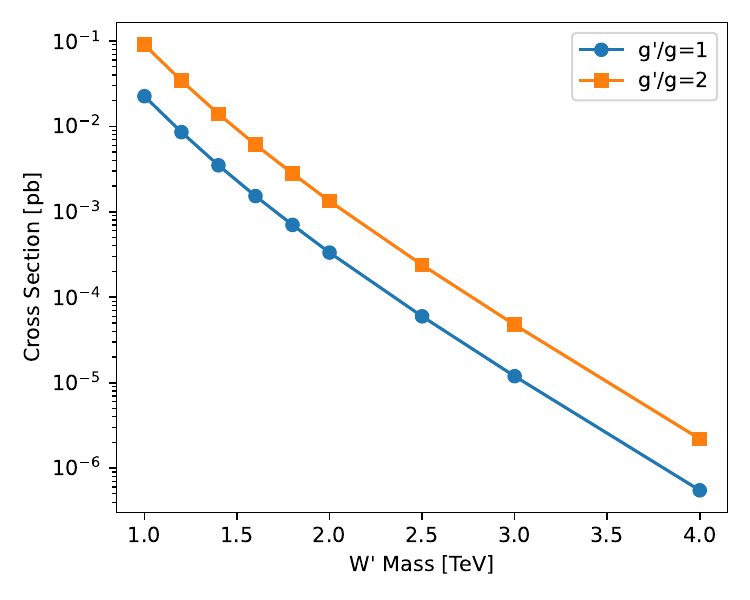}
\caption{Theoretical production cross section of a left-handed, heavy-philic $W'$ boson as a function of its mass. The blue line corresponds to a coupling strength equal to that of the Standard Model $W$ boson, while the orange line represents a coupling twice as strong. Lines interpolate between calculated mass points.}
\label{fig:wp_crosssec}
\end{figure}

While for leptons \( V_{f_if_j}\) is the null matrix. The coupling strength of the $W'$ boson is \(g_{W'}\) (aka \(g'\)). Two parameters, \(\alpha_R^{f_if_j}\) and \(\alpha_L^{f_if_j}\), are inserted into the modified Lagrangian to be able to regulate the chirality fraction of the $W'$ boson. For samples that are generated for a pure right-handed (RH) $W'$ boson, the coupling parameters are defined as \(\alpha_R^{f_if_j} = 1\) and \(\alpha_L^{f_if_j} = 0\). For samples generated with a purely left-handed (LH) $W'$ boson, the coupling parameters are defined as \(\alpha_L^{f_if_j} = 1\) and \(\alpha_R^{f_if_j} = 0\). The cross section for the heavy-philic $W'$ boson production can be seen in Fig.~\ref{fig:wp_crosssec}.

No significant interference is expected between the heavy-philic $W'$ boson and the SM $W$ boson. This is due to the distinct coupling structure of the $W'$, which couples exclusively to third-generation quarks, and the fact that the relevant initial state, composed of a top and bottom quark, is highly suppressed in the SM. As a result, the SM contribution to this process is small, and any potential interference is negligible.


\chapter{The LHC and the ATLAS Detector}

\section{The LHC Facility}
The Large Hadron Collider (LHC) \cite{Evans:2008zzb} is a particle accelerator located on the French-Swiss border near Geneva, Switzerland \cite{Cowan:2010js}. It was built by the European Organization for Nuclear Research (CERN). This accelerator was originally designed to be a Higgs boson factory, which had previously been theorized, and then experimentally observed after several years of operation and analysis. The LHC has a synchrotron design which collides two beams of protons that travel in opposite directions at near the speed of light. There are 4 collision points around the LHC ring. The LHC's 27 km circular tunnel is buried underground to prevent background radiation from interfering with data collection, and to protect personnel from exposure to the high amount of radiation. 

The energy of the LHC started at 7~TeV during the first run, which occurred between 2010 and 2011. The energy was then increased to 8~TeV during the next period of data-taking during 2012. Subsequent shutdown of the LHC lasted for 2 years where some equipment was replaced and upgraded. The next period of data taking was called Run-2 which lasted between 2015 and 2018. The center of mass energy was increased to 13~TeV for this period. A total of about 139~$\text{fb}^{-1}$ of data was collected at this time \cite{DAPR-2021-01}.

The next period of data taking then began in 2023 after a long shutdown. This period of data is called Run-3 and is ongoing and currently colliding protons at an center of mass energy of 13.6~TeV. 

In order to get to the unprecedented collision energy, the proton beam goes through several steps that subsequently increase the proton energy. The protons begin by being accelerated through the LINAC 2 linear accelerator. They then enter into the Proton Synchrotron Booster (PSB). At this stage, they have 1.4~GeV\@. The protons then enter into the Proton Synchrotron (PS), and are accelerated to 450~GeV\@. Then finally the protons enter into the LHC in bunches composed of about $10^{11}$ protons. These bunches are separated by 25~ns which gives the possibility to store up to 2500 bunches for each beam. 

The LHC continues to circulate protons in its ring until the beam is dumped. This occurs if the beam becomes unstable or loses significant intensity. Over time, protons gradually fall out of the ring despite the magnetic fields designed to keep them confined, and many are also consumed in collisions. Beam dumps typically occur after about 10 hours.

Four collision sites exist at the LHC, and at each of these locations, a unique particle detector measures collisions at its site. The two large detectors are located on opposite sides of the LHC: ATLAS(A Toroidal LHC ApparatuS) and CMS(Compact Muon Solenoid). The two other detectors are ALICE (A Large Ion Collider Experiment) \cite{alice_detector} and LHCb (LHC-beauty) \cite{lhcb_detector}. ALICE specifically studies the properties of the quark-gluon plasma with proton-lead and lead-lead collisions whereas LHCb studies $b$-hadron physics. This thesis will focus on experiments within the ATLAS detector.

\section{The ATLAS Detector}
\begin{figure}
    \centering
    \includegraphics[width=0.93\linewidth]{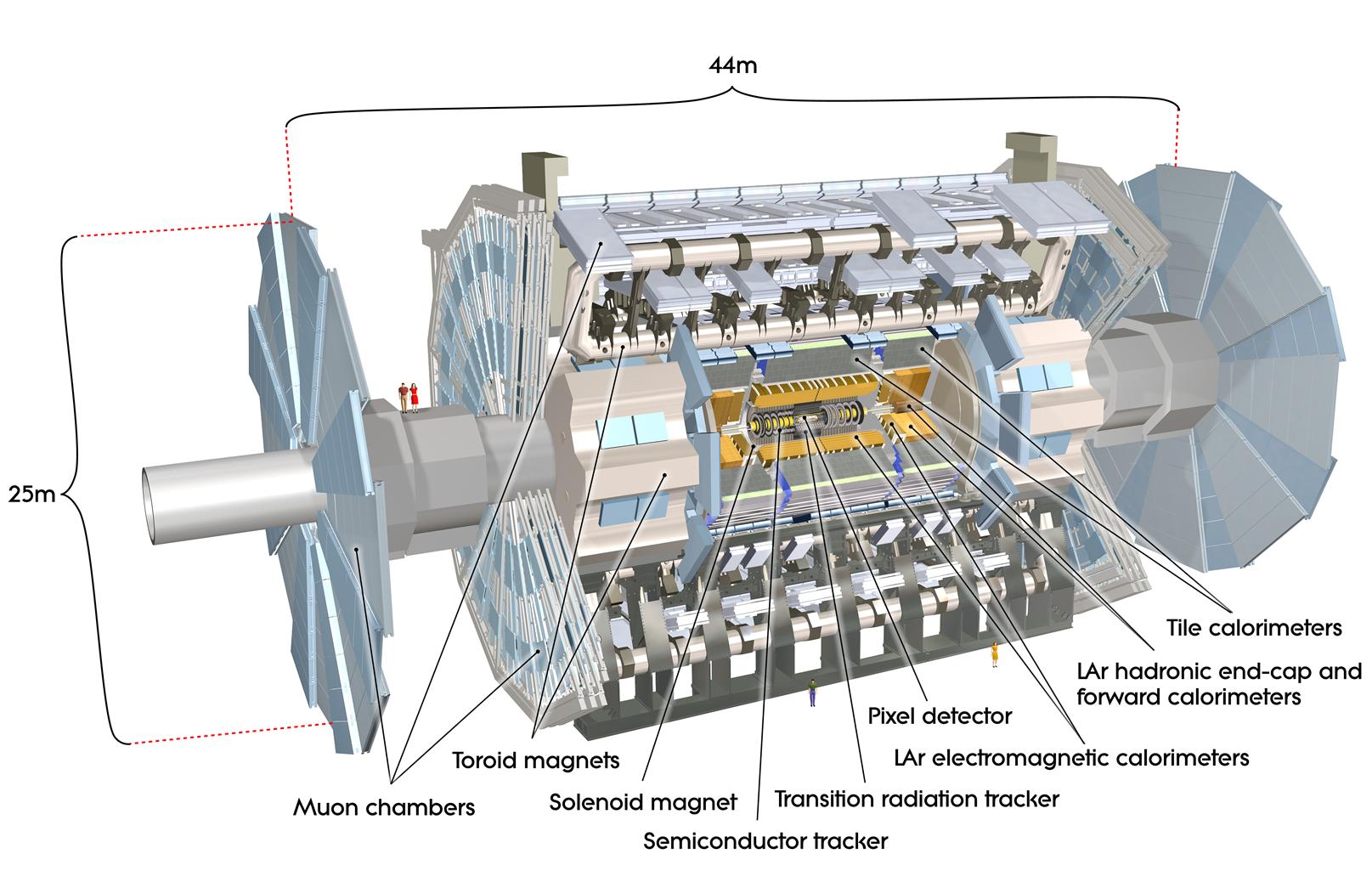}
    \caption{A cutout of the ATLAS detector labeling all the significant detector components \cite{ATLASOpenDataFigure}.}
    \label{fig:atlas_sys}
\end{figure}

The ATLAS experiment at the LHC is a multipurpose particle detector with a cylindrically symmetric geometry and a near $4\pi$ solid angle coverage \cite{ATLAS:2008xda}. See Fig.~\ref{fig:atlas_sys} to reference the locations of all the detector components.

It is approximately 46 meters in length, 25 meters in diameter, and weighs 700 tons. It consists of an inner detector for tracking surrounded by a thin superconducting solenoid which provides a 2~T axial magnetic field, electromagnetic and hadronic calorimeters, and a muon spectrometer. ATLAS uses a right-handed coordinate system with its origin at the nominal interaction point in the center of the detector and its $z$-axis along the beam pipe. The $x$-axis points from the interaction point towards the center of the LHC ring, and the $y$-axis points upwards. Cylindrical coordinates $\left(r, \phi\right)$ are used in the transverse plane, with $\phi$ being the azimuthal angle around the $z$-axis. The pseudo-rapidity is defined in terms of the polar angle $\theta$ as $\eta = -\ln\tan\left(\theta/2\right)$. Angular distance is measured in units of $\Delta R \equiv \sqrt{\left(\Delta\eta\right)^2 + \left(\Delta\phi\right)^2}$.

Particles created from the proton collisions begin in the center of the cylinder where they are tracked by the inner detector which consists of silicon pixel, silicon microstrip, and transition-radiation tracking detectors. An innermost pixel layer is inserted at a radius of 3.3 cm. Next particles traverse through the Lead/liquid-argon (LAr) sampling calorimeters provide electromagnetic (EM) energy measurements with high granularity. Finally, hadrons traverse through a hadronic calorimeter which covers the central pseudo-rapidity range of $|\eta| < 1.7$. All hadrons are stopped here.

Each particle will travel through the detector in different paths. Particles with charge will traverse a curved trajectory due to the strong magnetic field. Photons, and electrons are stopped in the electromagnetic calorimeter. Hadrons are stopped in the hadronic calorimeter. Muons are measured in the muon spectrometer but are likely to escape the detector volume. Neutrinos almost always escape without any interaction within the detector volume. 

\begin{figure}
    \centering
    \includegraphics[width=0.7\linewidth]{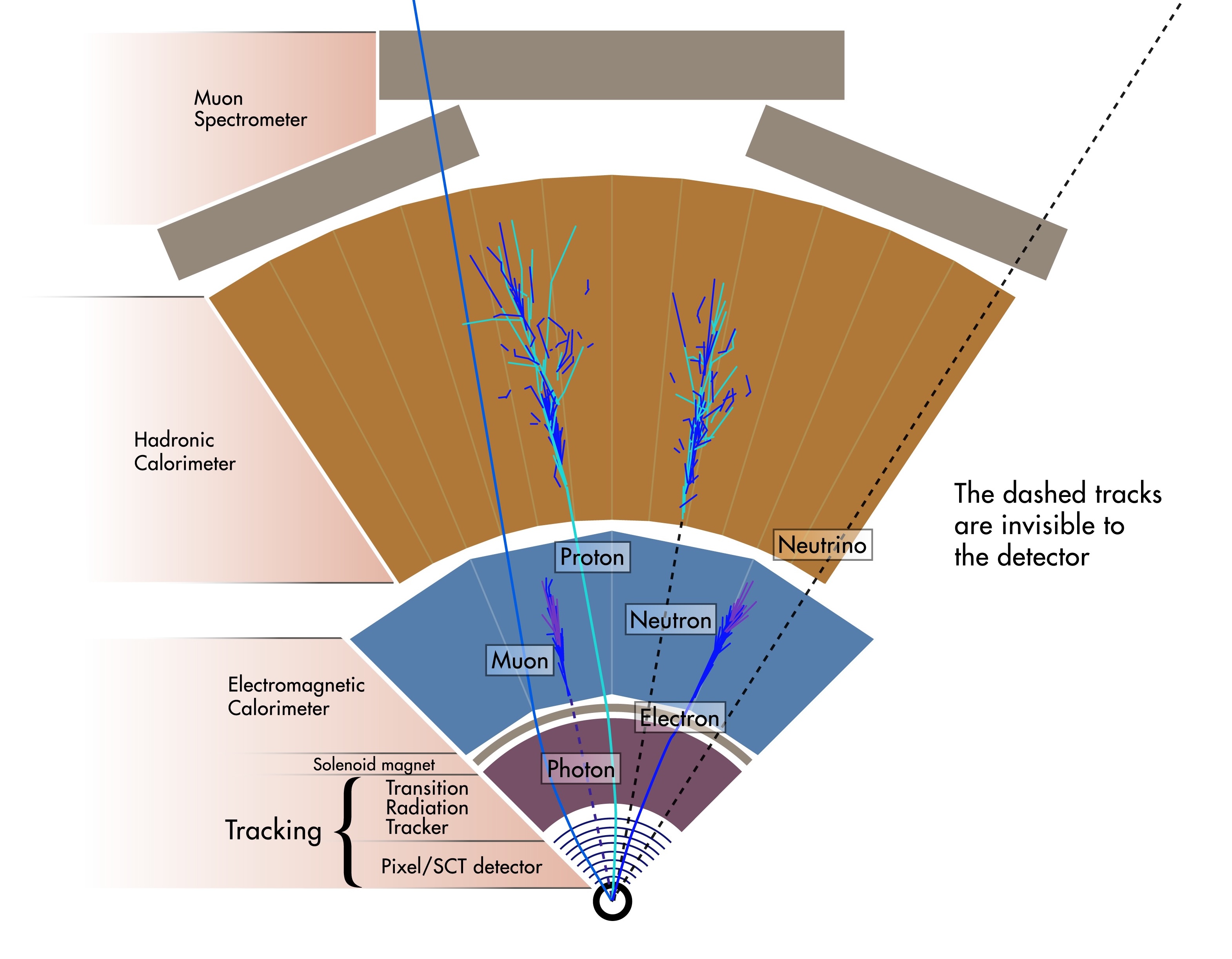}
    \caption{A depiction on the different paths particles produced at the collision site within the ATLAS detector traverse. Image from \cite{Pequenao:1505342}}
    \label{fig:particle_paths}
\end{figure}


\subsection{Calorimetry}

The ATLAS calorimeter system covers the range $|\eta| < 4.9$ and is designed to meet the specific needs of obtaining measurements for different physics objects while adapting to its radiation environment. It consists of electromagnetic (EM) and hadronic calorimeters, optimized for measuring electrons, photons, jets, and missing transverse energy ($E_{T}^{\text{miss}}$).

The EM calorimeter is a liquid argon (LAr) sampling detector with lead absorber plates, divided into a barrel ($|\eta| < 1.475$) and two end-cap components ($1.375 < |\eta| < 3.2$), housed in separate cryostats. Its accordion geometry ensures uniform energy resolution, while the total thickness exceeds 22 radiation lengths ($X_0$) in the barrel and 24 $X_0$ in the end-caps. $X_0$ is the average distance over which the energy of an electron is reduced by the factor 1/$e$ by bremsstrahlung \cite{Gupta:1279627}. The EM calorimeter plays a crucial role in electron and photon energy reconstruction.

The hadronic calorimeter comprises three sections: the tile calorimeter \cite{TCAL-2021-01}, LAr hadronic end-caps, and the LAr forward calorimeter \cite{ATLAS-TDR-02, ATLAS-TDR-22}. The tile calorimeter ($|\eta| < 1.7$) consists of steel absorber plates and plastic scintillators, segmented into depth layers for precise energy measurements. The LAr hadronic end-caps ($1.5 < |\eta| < 3.2$) share cryostats with the EM end-caps, ensuring seamless energy measurement transitions. The forward calorimeter, integrated into the end-cap cryostats, is designed for high radiation tolerance, with copper and tungsten modules measuring electromagnetic and hadronic interactions, respectively.

The entire system is structured to provide strong containment of electromagnetic and hadronic showers while maintaining high energy resolution, ensuring accurate particle energy measurements crucial for physics analyses at ATLAS.

\subsection{The Muon Spectrometer}
Muons, due to their higher mass and relatively long lifetimes, can traverse the inner detector (ID) and calorimeters without depositing significant energy. To detect and measure muons, the muon spectrometer is employed \cite{PERF-2014-05, ATLAS-TDR-26}. This spectrometer, the largest and outermost component of the detector, is surrounded by a system of three toroidal magnetic fields. As muons pass through these fields, they experience the Lorentz force, causing their trajectories to curve. By analyzing this curvature, the momentum of the muons can be determined.

The muon system is divided into two main regions:
\begin{itemize}
    \item \textbf{Barrel Region:} Consists of three cylindrical layers concentric with the beam axis.
    \item \textbf{End-Cap Region:} Features three layers of chambers arranged perpendicular to the beam axis.
\end{itemize}

The system is composed of both precision tracking and triggering components designed for high-resolution measurements and efficient particle detection across various pseudorapidity regions. The precision tracking subsystem includes Monitored Drift Tubes (MDTs), which provide a fine z-axis spatial resolution of $35 \mu m$ and cover the central to forward region of the detector up to $|\eta| < 2.7$. Complementing the MDTs in the forward region are the Cathode Strip Chambers (CSCs), which operate in the range $2.0 < |\eta| < 2.7$ and offer an $R \times \phi$ resolution of $40 \, \mu m \times 5 \, \text{mm}$. For triggering, the system utilizes Resistive Plate Chambers (RPCs) in the barrel region ($|\eta| < 1.05$), capable of providing $10 \text{mm}$ resolution in both the z and $\phi$ directions. In the end-cap region ($1.05 < |\eta| < 2.7$), Thin Gap Chambers (TGCs) are employed, delivering resolutions between $2-6 \, \text{mm}$ in R and $3-7 \, \text{mm}$ in $\phi$. Together, these components ensure both precise tracking and fast, reliable triggering across the detector.

In addition to tracking, the trigger components contribute independent tracking information in the orthogonal direction of the inner tracking detector. It also provides timing resolution critical for bunch-crossing identification.

Recent upgrades to the ATLAS detector have replaced the MDT with small Muon Drift Tubes (sMDT) that significantly increases the position resolution. This work is for the HL-LHC upgrade, the chambers won’t be installed until 2028, during the long shutdown.

\section{Luminosity and the Trigger Systems}
The integrated luminosity of all run-2 data is 139 fb$^{-1}$. This corresponds to a staggering $10^{13}$ inelastic proton-proton scattering events within the ATLAS detector. Due to limitations in extracting and recording events, it is impossible to record every proton-proton collision event. In order of overcome this problem, a 2-tier trigger system is implemented. One is implemented at the hardware level which cuts the bandwidth information coming from the ATLAS detector. The second is implemented on the software side where cuts are made on the reconstructed objects (more on reconstruction in the next chapter). 

The ATLAS trigger system rapidly identifies LHC collision events that meet predefined criteria. It consists of three progressively more selective stages: Level-1 (L1), Level-2 (L2), and the Event Filter (EF). The L1 trigger utilizes muon trigger chambers and coarse-grained calorimeter data to select events containing high-$p_T$ muons, electrons, photons, jets, $\tau$-leptons, and those with significant missing transverse energy ($E_T^{miss}$). This initial selection reduces the event rate from approximately 40 MHz to 100 kHz \cite{atlas_op:2020}. Additionally, the L1 trigger designates regions of interest in $\eta$-$\phi$ space around objects that surpass certain criteria, such as energy thresholds.

The L2 and EF triggers apply further filtering using additional detector information, including data from the tracking detector, within the designated regions of interest. These stages employ stricter selection criteria, such as higher energy thresholds and more precise object definitions, further refining the event selection process. Together, the L2 and EF triggers reduce the event rate from approximately 100 kHz to 1 kHz \cite{atlas_op:2020}.

Careful measurement of the luminosity is crucial for any measurement because information about the total number of proton-proton collisions per unit area before the trigger system is necessary. Luminosity is measured in several ways to ensure redundancy, consistency and accuracy. The primary luminosity measurement comes from the Luminosity Detector (LUCID) which  consists of Cherenkov tubes positioned near the beamline that detect particles from the proton-proton collisions. The inner detector and hadronic calorimeter provide additional cross-checks on these measurements. To calibrate these luminosity measurements, ATLAS periodically performs Van der Meer scans \cite{DAPR-2021-01}. This involves moving the LHC beams across each other while recording collision rates. 

\section{Data Flow}
Due to the large amount of data that is collected from the ATLAS detector, data is transmitted to the ``Tier 0" which has the job of processing the raw data, achieving the data to tape, and registering data with the relevant catalogs \cite{Markus:Elsing_2010}. The data are then distributed to ``Tier 1" clusters for offline processing. Much of the offline processing happens in the 13 Tier 1 computer clusters which are made available 24/7 with support from the CERN Computing GRID. Each is responsible for storing a proportional share of the raw and reconstructed data, performing large-scale reprocessing, and storing the resulting output. ``Tier 2" systems are smaller computing sites that offer universities and other scientific institutions sufficient data storage and computing power to further process data and Monte Carlo simulations from the Tier 1 sites. There are about 155 Tier 2 sites around the world.

After the data and simulation have been processed on the cloud computing clusters, further analysis is done locally. Datasets on the order of terabytes are downloaded locally to train machine learning models on local GPUs, perform the final analysis, and the final statistical fit. 
\chapter{Event Simulation and Reconstruction}

The central aim of physics—and the natural sciences more broadly—is to predict the outcome of a given initial condition. In the context of high-energy physics, outcomes follow probability distributions, and many of these distributions cannot be calculated analytically. Therefore, the Monte Carlo method is employed to approximate the theoretical probability distributions. By randomly sampling millions of events, this method approximates the theoretical distributions, with greater accuracy achieved through larger samples.

\section{Levels of Computation}

Several stages of computation have been standardized in the ATLAS collaboration, and similarly in the CMS collaboration. The first level is what is commonly called ``truth" level. This is the earliest time frame where Standard Model particles are produced immediately after the proton-proton collision. This is the domain of QFT and perturbative QCD. This level is where the heavy-philic $W'$ boson is produced, the associated top and bottom quarks, and the resulting resonant top and bottom quarks from the $W'$ boson decay. This level is unobservable because the truth level particles decay into more stable particles before they can reach the detector volume. 

The generation of truth level events comes out from the high energy tool called Madgraph \cite{Alwall:2014hca} where the matrix element is integrated over through the use of the Monte Carlo method. This tool allows one to generate events for a given process of interest. Fig.~\ref{fig:wp_lo_diagram} shows a Feynman diagram for $W'$ boson production, which is one of many diagrams Madgraph accounts for during event generation. The output consists of a list of events with features for each particle. There are typically tens of particles at this level and tens of features for each particle such as the 4-momentum, particle ID, decay parents and decay children. There are also other particle-specific features like spin, electromagnetic charge, and QCD color charge. In comparison to the other levels of computation, this step is relatively fast. 

After particle generation, particles undergo parton showering and subsequent hadronization. The tool that is used here is Pythia \cite{Sjostrand:2006za} (or Herwig \cite{Corcella:2000bw}). This level of computation is commonly called the ``particle level". At this level of computation, the output consists of a list of events with particles and features. The number of particles at this stage is typically in the thousands, with several features such as 4-momentum, particle ID, and charge. Each particle at this stage is a on-shell particle such as a lepton, neutrino, or hadron. This level of computation obtains the relatively stable particles right before it enters into the detector volume. The particle level is what is used to match the truth level particles to the reconstructed level jets. This increases the accuracy of the matched reconstructed jets to the truth level particles. 

The particles that are simulated from the parton showing and hadronization tool are then sent through a detector simulator. ATLAS uses GEANT4 \cite{Agostinelli:2002hh} for this. GEANT4 uses the current understanding of how particles interact with matter in each detector module in ATLAS. A virtual ATLAS detector is carefully created to simulate how particles will interact with each sub component. Fast simulation is also common which replaces the full detector simulation with an approximate detector response to improve computation time. Fast simulation samples are not used in this work.

The results from detector simulation produce signals that can be compared to ATLAS data. However, this is very difficult to do as looking at detector responses is difficult to interpret. For example, a reading from sector 2 of the A detector side, index 5, for the odd sector on the middle barrel MDT is unwieldy. Instead, higher-level objects are created which carry much more interesting physical properties that can be understood. This process is called reconstruction where ATLAS detector responses are used as input, and the output are these higher-level objects which are discussed in the following section.




\section{Reconstructed Objects and Event Selection}

The reconstruction of physics objects in this analysis includes electrons, muons, hadronically decaying taus, jets, b-jets, and missing transverse momentum.

Electrons are reconstructed from energy clusters in the electromagnetic calorimeter matched to tracks in the inner detector (ID) \cite{EGAM-2021-01,EGAM-2021-02}. They are required to have  $p_T > 27$  GeV and  $|\eta| < 2.47$, with candidates in the calorimeter barrel–endcap transition region ( $1.37 < |\eta| < 1.52$ ) excluded. Identification is based on a likelihood discriminant, and electrons must satisfy impact parameter constraints ($|z_0| < 0.5$ mm and $|d_0|/\sigma_{d_0} < 5$), as well as the gradient isolation criteria.

Muons are reconstructed from either track segments or full tracks in the muon spectrometer, matched to ID tracks, and re-fitted using information from both detector systems \cite{MUON-2022-01}. They must satisfy  $p_T > 27$  GeV,  $|\eta| < 2.5$ , impact parameter constraints ($|z_0| < 0.5$ mm and $|d_0|/\sigma_{d_0} < 3$), and the tight identification and FixedCutTightTrackOnly isolation criteria.

Hadronically decaying tau leptons ($\tau_{\text{had}}$) are identified using track multiplicity and a boosted decision tree incorporating track collimation, jet substructure, and kinematic properties \cite{ATLAS:2015xbi}. Candidates must satisfy  $p_T > 25$  GeV,  $|\eta| < 2.5$ , and pass the Medium $\tau$-identification working point. While taus are not directly used in the analysis, they are considered in the overlap removal procedure and event selection \cite{JHEP2023-073}.

Jets are reconstructed from topological energy clusters in the calorimeter using the anti-$k_t$ algorithm \cite{Cacciari:2008gp} with a radius parameter of 0.4. Each cluster is first calibrated to the electromagnetic scale response, with additional energy corrections applied using simulation and in situ data. Jets must have  $p_T > 25$  GeV,  $|\eta| < 2.5$ , and pass quality criteria to remove those originating from non-collision sources or detector noise. To mitigate pileup effects, jets with  $p_T < 120$  GeV and  $|\eta| < 2.4$  are further required to be consistent with the primary vertex using the jet vertex tagger (JVT) \cite{Tomiwa_2017}.

Jets are b-tagged if their decay products come from a b-quark. The b-tagging is done by using multivariate techniques that incorporate impact parameter information and secondary/tertiary vertex properties. Jet b-tagging in this analysis uses the GN2v01 algorithm \cite{atlas_gn2_btag} at the 77\% efficiency working point, trained on simulated  $t\bar{t}$  events to distinguish b-tagged jets from other flavored jets.

The missing transverse momentum ( $E_T^{\text{miss}}$ ) is computed as the negative sum of the  $p_T$  of all physics objects in the event, with an additional correction for soft energy unassociated with hard objects. This correction is based on ID tracks matched to the primary vertex to ensure resilience against pileup contamination. Only events with $E_T^\text{miss}>20$ GeV are considered in this analysis.

To prevent multiple detector responses from being counted as separate objects, an overlap removal procedure is applied. Jets within  $\Delta R_y = 0.2$  of a selected electron are removed to prevent double-counting of energy deposits. If a jet remains within  $\Delta R_y = 0.4$  of an electron, the electron is discarded. Muons are removed if they are within  $\Delta R_y < 0.4$  of a jet, unless the jet has fewer than three associated tracks, in which case the muon is retained and the jet is removed. Tau candidates are removed if within  $\Delta R_y < 0.2$  of selected electrons or muons. This follows the ATLAS standard overlap removal procedure, except for tau-jet overlap removal, which is not applied to preserve analysis integrity.

\chapter{Search for the Heavy-philic $W'$ Boson}

\section{Truth Level Kinematic Properties}

This section describes the truth-level kinematic properties of particles involved in the production and decay of the heavy-philic $W'$ boson, focusing on their relevance to signal selection and background rejection.

\begin{figure}
    \centering
    \includegraphics[width=0.9\linewidth]{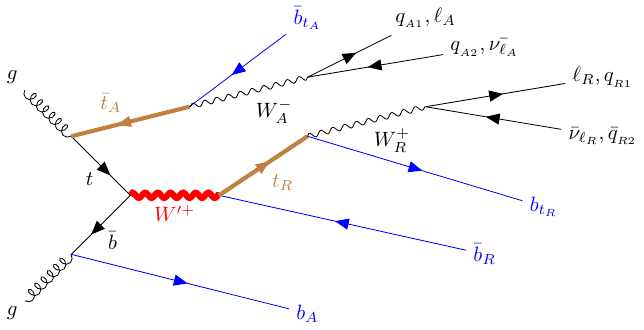}
    \caption{A representative Feynman diagram of the full heavy-philic $W'$ boson production and decay into the one-lepton final state. The $W'$ boson is rendered in bold red, the top quarks are rendered in brown, and the bottom quarks are rendered in blue for clarity. Two channels are represented in this diagram as the $W$ bosons from the top quarks can decay leptonically or hadronically. Since this search is conducted in the single lepton channel, the lepton can come from either the \(t_A\) or the \(t_R\).}
    \label{fig:full_diagram}
\end{figure}

During heavy-philic $W'$ boson production, two quarks are produced alongside the $W'$ boson. These are referred to as the associated particles: one associated top quark and one associated bottom quark. The heavy-philic  $W'$ boson subsequently decays, and since its couplings to all generations other than the third are set to zero, it decays exclusively into a top quark and a bottom quark. These decay products are called the resonant particles. A Feynman diagram of the full process can be seen in Fig.~\ref{fig:full_diagram}. Thus, in total, the final state contains a resonant top quark ($t_R$), a resonant bottom quark ($b_R$), a associated top quark ($t_A$), and a associated bottom quark ($b_A$) to form a final state of $tbtb$.

Since there are two top quarks, there are three main channels; the all-hadronic, single-lepton, and dilepton channel. The branching fractions are identical to $t\bar{t}$ production and the pie chart showing the branching fractions into the different channels can be seen in Fig.~\ref{fig:ttbar_channels}

Since this is the first ever search for a heavy-philic $W'$ boson with couplings exclusively to the third generation of quarks, the 1-lepton channel is selected as it minimizes the background contributions while maintaining a significant number of signal events.
  
\begin{figure}
\centering
\includegraphics[width=0.49\linewidth]{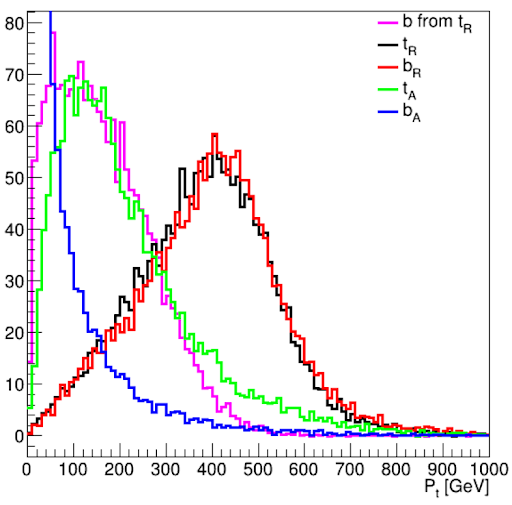}
\includegraphics[width=0.49\linewidth]{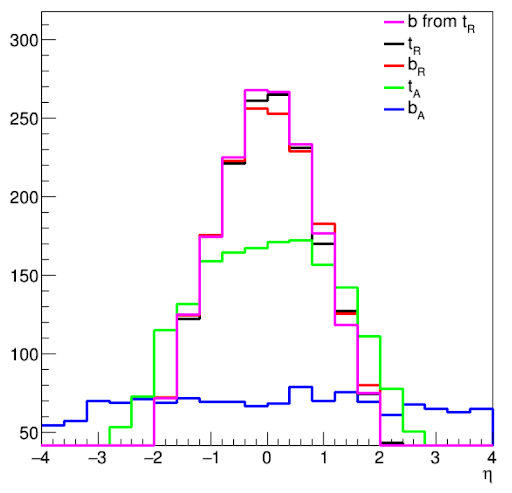}
\caption{Kinematic properties of the truth LH 1 TeV $W'$ production process. The left panel shows the transverse momentum ($p_T$) distribution for the resonant and associated top and bottom quarks, while the right panel shows their rapidity ($\eta$) distributions.}
\label{fig:wp_kinematics}
\end{figure}

Understanding the kinematic properties of the resonant and associated particles is critical for optimizing the signal region selection and for identifying the dominant Standard Model (SM) backgrounds to reject. Figure~\ref{fig:wp_kinematics} shows the transverse momentum ($p_T$) and rapidity ($\eta$) distributions of the key particles in this analysis. These distributions reveal that \br and $t_R$ have high $p_T$, with the other particles having lower $p_T$. The $b_A$ have the lowest $p_T$, and are emitted in a direction closest to the beam line.  

\section{Reconstruction Level Selection}

Most of the events measured in the ATLAS detector are irrelevant to this analysis, either lacking a reconstructed electron or muon or containing only a small number of reconstructed jets. To focus on an enriched signal region and suppress complex backgrounds, a series of selection criteria are applied. These cuts are designed to minimize contributions from complicated SM backgrounds like QCD multijet background while maintaining reasonable signal efficiency.

The signal region is defined by the following criteria:
\begin{itemize}
    \item At least five jets
    \item At least two b-tagged jets
    \item The presence of exactly one electron or one muon
\end{itemize}

The requirement of one electron or muon helps reject the dominant QCD multijet background, which is both challenging to model and associated with large systematic uncertainties.


Theoretical cross-section predictions for heavy-philic  $W'$  boson production as a function of the  $W'$  mass are shown in Figure~\ref{fig:wp_crosssec}. These predictions indicate that the cross section for this process is very small. This low cross section highlights the need for a precise analysis and stringent event selection to maximize sensitivity. 

After pre-selection, events are categorized into different kinematic regions based on the number of b-tagged jets. Each region has 5 or more jets, but separated based on the number of b-tags. A $t\bar{t}$ control region is defined by requiring exactly 2 b-tagged jets with $p_T>50~\text{GeV}$. Signal region 1 (SR1) is defined by requiring exactly 3 jets being b-tagged with $p_T>50~\text{GeV}$. Signal region 2 (SR2) is defined by requiring 4 or more b-tagged jets with $p_T>50~\text{GeV}$. 

Another cut designed to further reduce multijet background is implemented. The $p_T$ of the MET must be greater than 20 GeV and MET+$M^W_T$ must be greater than 60 GeV \cite{Faraj:2024qrn}. $M^W_T$ is defined as $M_T^W = \sqrt{2 p_T^\ell E_T^{\text{miss}} \left(1 - \cos \Delta\phi\right)}$. QCD multijet events typically don’t have large MET, and typically populate low $M^W_T$ regions.

\section{Backgrounds}

Top quark pair production ($t\bar{t}$) is by far the dominant background in the phase space region relevant to this search. Single top quark production is the second most significant Standard Model process contributing to the signal region. However, it yields fewer events due to the production of fewer jets, often with only one originating from a bottom quark. The next most relevant Standard Model background is $t\bar{t}+\mathrm{boson}$ production. Although the acceptance of these events in the signal region is high, their cross sections are very low, resulting in only a small contribution to the overall background. 

\subsection{\texorpdfstring{$t\bar{t}~+$ jets}{ttbar + jets}}

\begin{figure}
    \centering
    \includegraphics[width=0.5\linewidth]{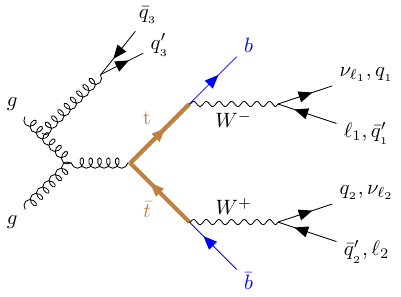}
    \caption{Feynman diagram of $t\bar{t}$ production via the more common gluon-gluon production mechanism. The $W$ bosons either decay hadronicaly or leptonically. Depending on these decays, the event will fall into the three main categories (all hadronic, single lepton, and dilepton). }
    \label{fig:ttbar_feynman}
\end{figure}

The production of $t\bar{t}$ events can easily fall within the signal region as can be seen in the Feynman diagram in Fig.~\ref{fig:ttbar_feynman}. To fall within the signal region one of them needs to decay hadronically and the other leptonically. Then another jet could come from QCD correction, final state radiation, or additional radiation from the parton showering in order to be put in the selection region for this analysis. 

The $t\bar{t}$ cross section is relatively large compared to other Standard Model processes in the analysis region, and its higher acceptance further contributes to its dominance as a background in this phase space.

The production of $t\bar{t}$ events in this analysis is modeled using the \textsc{PowhegBox}~v2 generator \cite{Nason:2004rx,Frixione:2007vw,Alioli:2010xd}, which provides the matrix element (ME) at NLO in the strong coupling constant ($\alpha_S$) with the NNPDF3.0NLO PDF set\cite{Ball:2014uwa} and the $h_{damp}$ parameter set to 1.5$m_{top}$. The functional form of $\mu_f$ and $\mu_r$ is set to the default scale $\sqrt{m_t^2+p_{T,t}^2}$. The events are showered with Pythia 8.230 \cite{Mrenna:2016sih}.

The uncertainty due to initial-state-radiaton (ISR) is estimated using weights in the ME and in the parton shower (PS). To simulate higher parton radiation $\mu_f$ and $\mu_r$ are varied by a factor of 0.5 in the ME while using the $Var3c$ upward variation from the A14 tune. For lower parton radiation, $\mu_f$ and $\mu_r$ varied by a factor of 2.0 while using the $Var3c$ downward variation in the PS. The impact of final-state-radiation (FSR) is evaluated using PS weights which vary $\mu_r$, for QCD emission in the FSR by a factor of 0.5 and 2.0, respectively. The impact of the PS and hadronisation model is evaluated by changing the showering of the nominal \textsc{PowhegBox} events from Pythia to Herwig 7.04 \cite{Bellm:2015jjp}.

To assess the uncertainty due to the choice of the matching scheme, the Powheg sample is compared to a sample of events generated with MG5.aMC v2.6.0 \cite{Alwall:2011uj} and the NNPDF3.0NLO PDF set showered with Pythia 8.230. The shower starting scale has the functional form $\mu_q=H_T/2$, where $H_T$ is defined as the scalar sum of all outgoing partons. Choice of $\mu_f$ and $\mu_r$ is the same as that for the Powheg setup. 

\subsection{\texorpdfstring{$t\bar{t}$ HF Classification}{ttbar HF Classification}}
The $t\bar{t}$ + jets background is categorized according to the flavor of additional jets in the event, using the same procedure as described in Ref.~\cite{2021_resolvehp}. Generator-level particle jets are reconstructed from stable particles (mean lifetime $\tau>3\times10^{-11}$ seconds) using the anti-$k_t$ algorithm with a radius parameter $R=0.4$, and are required to have $p_{T}>15$ GeV and $|\eta |<2.5$. The flavour of a jet is determined by counting the number of B- or C-hadrons within $\Delta R<0.4$ of the jet axis. Jets matched to at least one B-hadron with $p_T > 5$ GeV are labeled as B-jets. Similarly, jets matched to at least one C-hadron (and not already identified as a B-jet) are labeled as C-jets. Events that have at least one B-jet, not counting heavy-flavour jets from top-quark or $W$-boson decays, are labeled as $t\bar{t}+\geq1b$; those with no B-jets but at least one C-jet are labeled as $t\bar{t}+\geq1c$. Finally, events not containing any heavy-flavour jets aside from those from top-quark or $W$-boson decays are labeled as $t\bar{t}$ + light. This classification is used to define the background categories in the likelihood fit.

\subsection{\texorpdfstring{$t\bar{t}$ Reweighting Technique}{ttbar Reweighting Technique}}

The $t\bar{t}$ simulation using the Powheg+Pythia generator does not accurately reproduce data at high jet multiplicities. Standard Model $t\bar{t}$ mismodeling is a well known problem with high jet multiplicities \cite{Spalla:2776941}. To improve the agreement between data and Monte Carlo predictions, data-driven corrections are applied to the MC samples, to aid the convergence of the fit. 

Reweighting factors are derived by comparing the data and MC predictions within the $2b$-tagged control region. Since the primary source of mismodelling is assumed to be additional radiation from the parton shower these reweighting factors are also applied in the $3b$ and $\geq4b$ regions. This is possible because additional radiation from the parton shower is largely independent of the jet flavor. This helps improve data/MC agreement, with any remaining discrepancies expected to be covered by the systematic uncertainties.

The reweighting factors are defined as:
\begin{equation}
    R(x) = \frac{\text{Data}(x) - \text{MC}^{\text{non-}t\bar{t}}(x)}{\text{MC}^{t\bar{t}}(x)}
\end{equation}
where \(x\) represents the variable misrepresented by the simulation. In this context, $t\bar{t}$ includes $t\bar{t}+>1b$, $t\bar{t}+>1c$, and $t\bar{t}+\text{light}$-flavor jets.

\begin{figure}
    \centering
    \includegraphics[width=0.49\linewidth]{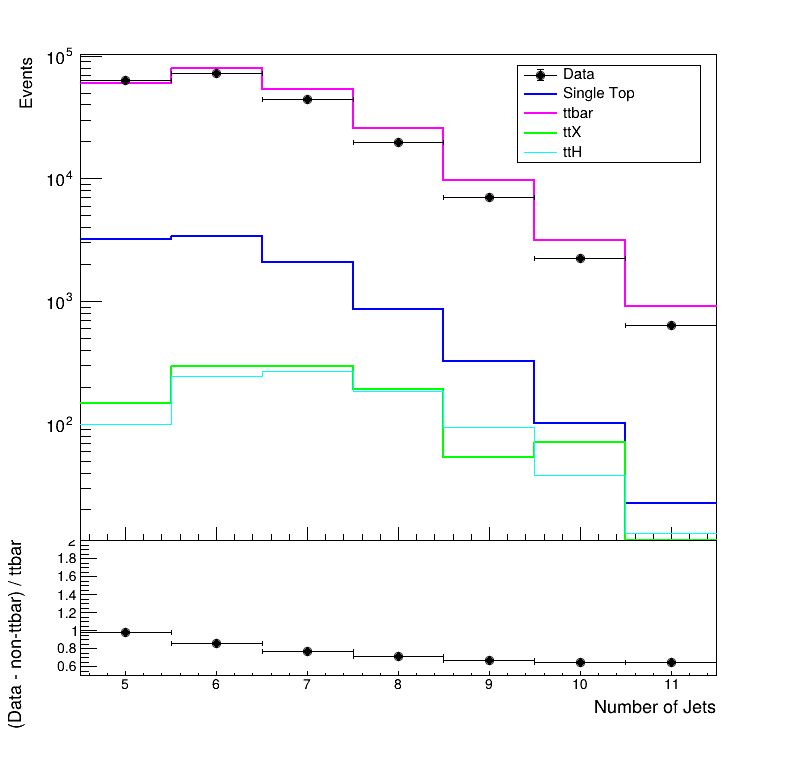}
    \includegraphics[width=0.49\linewidth]{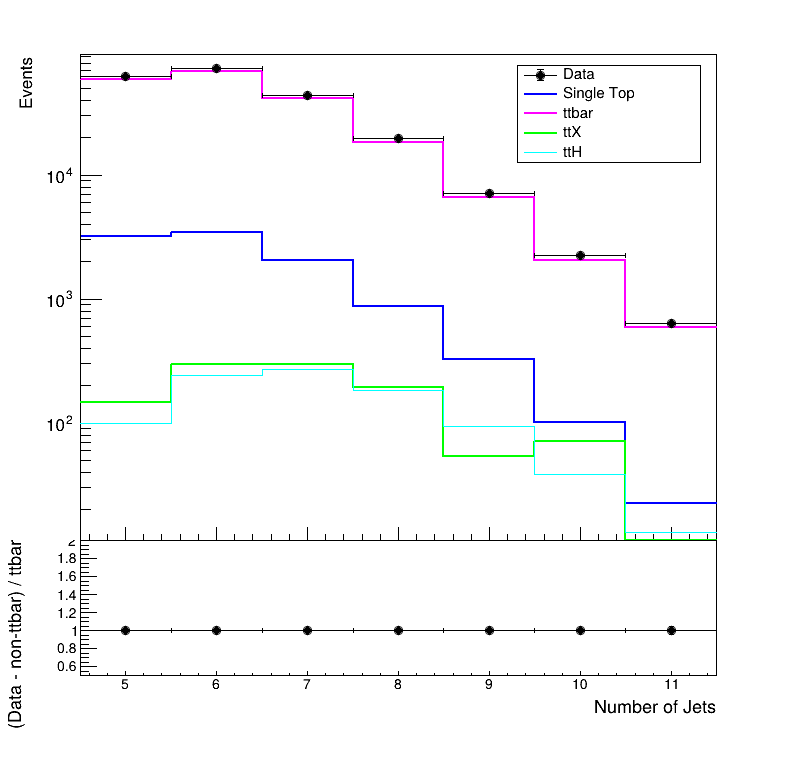}
    \caption{The jet multiplicity distribution before any weights are applied is shown on the left. The resulting distribution after weights based on the number of jets is applied is shown on the right. These plots compare the background sum to the data and do not include overflow in the last bin. }
    \label{fig:njets_ttbar_reweight}
\end{figure}

\begin{figure}
    \centering
    \includegraphics[width=0.49\linewidth]{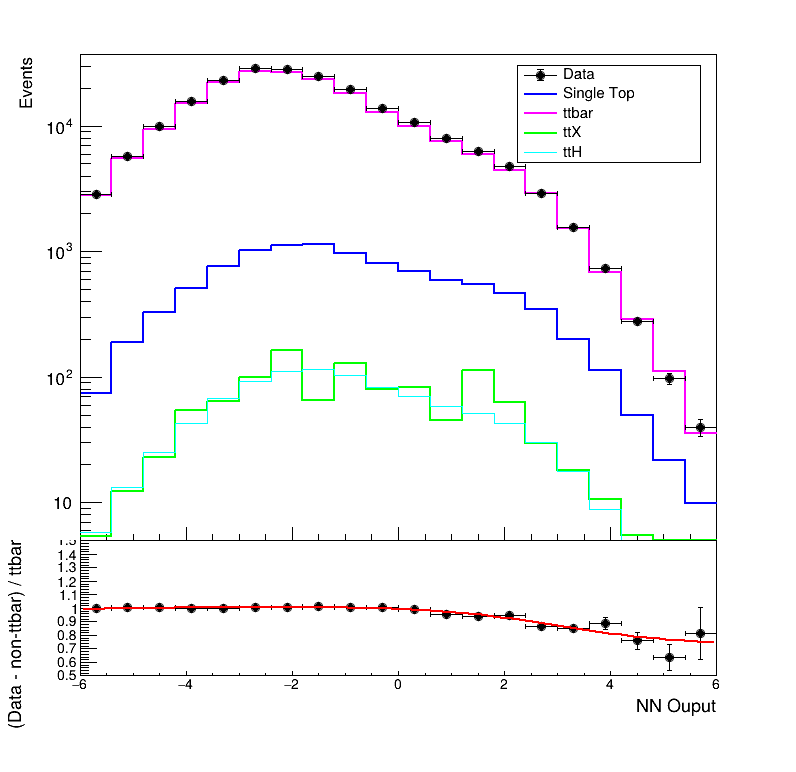}
    \includegraphics[width=0.49\linewidth]{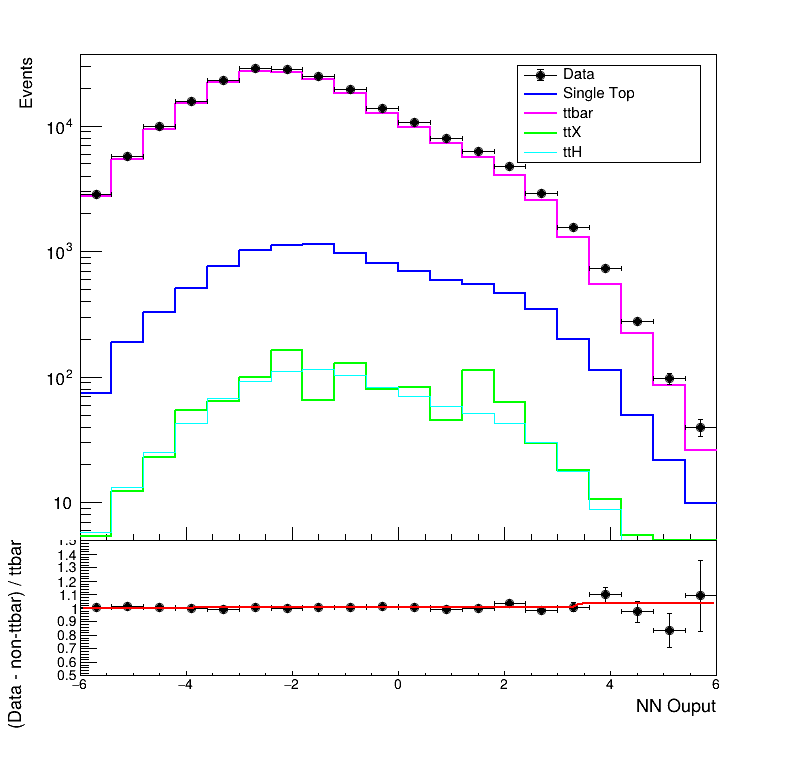}
    \caption{The neural network output distribution before applying the reweighting function is shown on the left, and the distribution after reweighting is shown on the right. These plots compare the sum of background contributions to the data and exclude underflow and overflow in the first and last bins, respectively.}
    \label{fig:mlp_ttbar_reweight}
\end{figure}

Reweighting is performed sequentially using the number of jets, followed by the neural network (NN) output distribution within the $t\bar{t}$ control region. The neural network architecture and training procedure are described in detail in Section5.4. Figure\ref{fig:njets_ttbar_reweight} shows the original jet multiplicity distribution and the corresponding weight distribution after the first reweighting step. Figure~\ref{fig:mlp_ttbar_reweight} displays the NN output distributions before and after the successive reweighting procedures. These two figures only show the $t\bar{t}$ CR.

The $t\bar{t}$ weights are applied to all $t\bar{t}$ events, independent of their heavy flavor classification. Their event weights are based on their jet multiplicity and neural network output values. To perform the neural network reweighting, a functional form combining a hyperbola and a sigmoid, given by:
\[
\omega = a + \frac{b}{(x^{\text{NN}}+10)^c} - \frac{d}{1 + \exp(e - f \cdot (x^{\text{NN}}+10)}
\]
is fitted to the already weighted distributions based on jet multiplicity in $t\bar{t}$ control region (see Figure~\ref{fig:mlp_ttbar_reweight}). Table~\ref{tab:function_fit_parameters} provides the fitted parameters. 

\begin{table}[h]
    \centering
    \caption{The set of fit parameters used to perform $t\bar{t}$ reweighting based on the neural network output score. }
    \begin{tabular}{c|cccccc}
         Region&  a & b& c&d &e &f\\ \hline
         $\geq 5j2b$& 
     1.205& -0.2628& 0.15926& -0.30738& 10.683&0.81727\end{tabular}
    \label{tab:function_fit_parameters}
\end{table}

\begin{figure}
    \centering
    \includegraphics[width=0.49\linewidth]{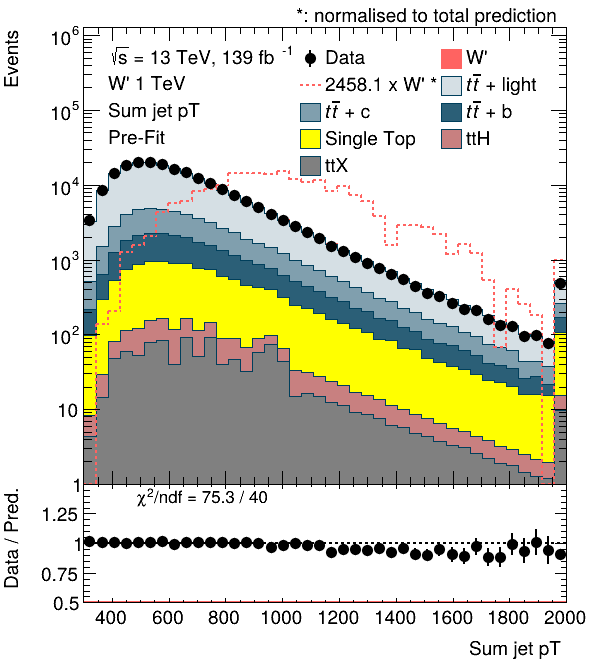}
    \includegraphics[width=0.49\linewidth]{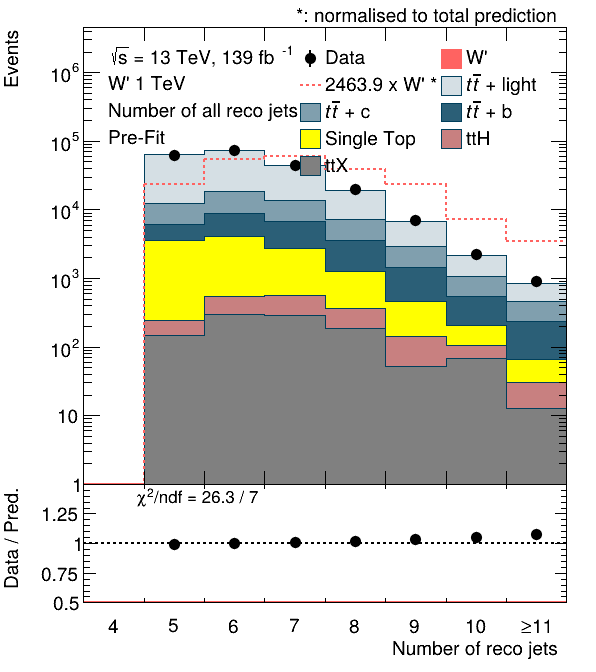} \\
    \includegraphics[width=0.49\linewidth]{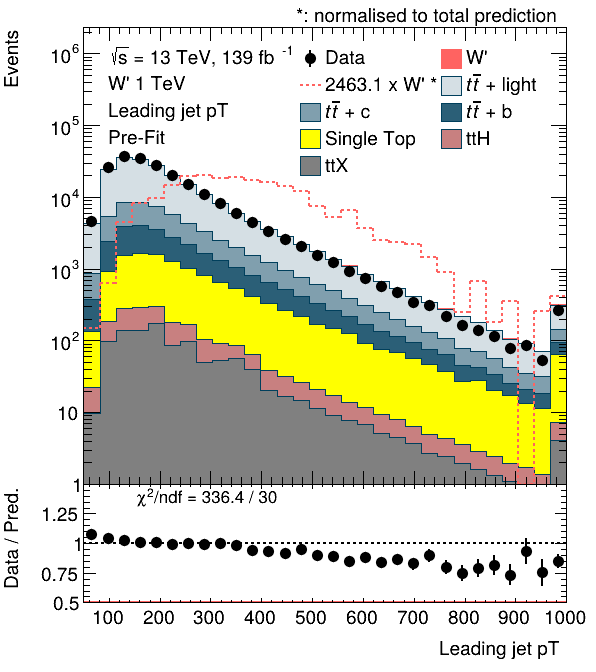}
    \caption{Kinematic distributions showing the full Standard Model background in the $t\bar{t}$ control region. The top left plot shows the scalar sum of all jets' transverse momenta. The top right plot shows the number of reconstructed jets. The bottom plot shows the transverse momentum of the leading jet. Underflow and overflow are added to the first and last bin in these histograms.}
    \label{fig:cr_kinematics}
\end{figure}

Other kinematic distributions within the $t\bar{t}$ control region are shown in Fig.~\ref{fig:cr_kinematics}. The plots included within this figure show the HT, number of jets, and the leading jet $p_T$. Perfect agreement between data and MC samples is not expected for HT or leading jet $p_T$, as these variables are not used in the reweighting. Additionally, discrepancies in the jet multiplicity distribution are also anticipated, since applying weights based on the neural network output slightly alters this distribution. While not shown, all distributions agree within the approximately 25\% systematic uncertainty in this region. The $\chi^2$ of each distribution is shown in Fig.~\ref{fig:cr_kinematics} and later plots. It is calculated as
\[
\chi^2 =  \sum_{i=1}^{N_\text{bins}} \frac{(N_i^\text{data} - N_i^\text{MC})^2}{\sigma_i^2},
\]
where $N_i^\text{data}$ and $N_i^\text{MC}$ are the data and Monte Carlo yields in bin $i$, and $\sigma_i$ is the statistical uncertainty in that bin.

Separate histograms are created that include these $t\bar{t}$ weights and another one that doesn't include the weights. This is done to model the systematic uncertainty involved with the $t\bar{t}$ reweighting uncertainty.

\FloatBarrier

\subsection{Single Top}
Single top quark production is modeled separately for the $t$-channel, $s$-channel, and $tW$ associated production modes. For each process, the nominal Monte Carlo (MC) sample is generated at next-to-leading order (NLO) in QCD using the \textsc{PowhegBox}~v2 generator, interfaced to \textsc{Pythia}~8.230 for parton showering and hadronisation.

\textbf{\boldmath$t$-channel:} The ME is calculated at NLO in the four-flavor scheme (4FS) using the NNPDF3.0NLOnf4 PDF set. The renormalisation and factorisation scales are set to $\sqrt{m_b^2 + p_{T,b}^2}$, as recommended in Ref.~\cite{Frederix:2012dh}.

\textbf{\boldmath$s$-channel:} The ME is calculated at NLO in the five-flavor scheme (5FS) using the NNPDF3.0NLO PDF set. The renormalisation and factorisation scales are set to the top quark mass.

\textbf{\boldmath$tW$-channel:} The ME is calculated at NLO in the 5FS using the NNPDF3.0NLO PDF set, with renormalisation and factorisation scales set to the top quark mass. The \textsc{PowhegBox}~v2 generator uses the diagram removal (DR) scheme \cite{Frixione:2008yi} to handle interference with $t\bar{t}$ production \cite{ATL-PHYS-PUB-2016-020}.

A 5\% uncertainty on the cross section of each of the 3 production modes of single top is assumed in these samples.

\subsection{\texorpdfstring{$t\bar{t}V$ and $t\bar{t}H$}{ttV and ttH}}
The production of $t\bar{t}V$ events is modeled using the MG5\_aMC v2.3.3 generator \cite{Alwall:2014hca}, which provides the ME at NLO in $\alpha_S$ with the NNPDF3.0NLO PDF set \cite{Ball:2014uwa}. The functional form of $\mu_f$ and $\mu_r$ is set to the default scale $0.5 \times \Sigma_i\sqrt{m_i^2+p_{T_i}^2}$, where the sum runs over all the particles generated from the ME calculation. The events are showered with Pythia 8.210. A 15\% uncertainty in the total cross section for $t\bar{t}V$ is assumed. In this work, the notations $t\bar{t}V$ and $t\bar{t}X$ are used interchangeably.

The production of \(t\bar{t}H\) events is modeled in the 5F scheme using the \textsc{PowhegBox} generator \cite{Hartanto_2015} at NLO in $\alpha_S$ with the NNPDF3.0NLO PDF set. The $h_{damp}$ parameter is set to 3/4 $\dot (m_t + m_{tbar} + m_H)=352.5$ GeV. The events are showered with Pythia 8.230. A 10\% uncertainty in the total cross section for $t\bar{t}H$ is assumed.

\subsection{Other Rare Backgrounds}

The rare top processes that are considered in this analysis are $tZq$, $tZW$, 4-top, and diboson events. Once these samples were run through the pre-selection requirements, there were determined to be insignificant ($<1\%$) and are excluded from the analysis. 

\subsection{Background Composition and Kinematic Properties}
In this analysis, we look at the region that includes 1 electron or 1 muon exclusively. At least 5 jets are required, and 2 of them must be b-tagged. In this region, there is still abundant SM background. The biggest contribution is $t\bar{t} + \text{jets}$ , but several other processes still exist in this region and the summary of the backgrounds in these regions can be found in Table~\ref{tab:back_comp}. The background sum agrees with data within uncertainties. There also exists normalization and shape uncertainties which are not included here, but will be discussed in Chapter 6.

\begin{table}[h]
\begin{center}
\caption{Yields of the analysis. Statistical uncertainties are shown in parenthesis. } 
\label{tab:back_comp}
\begin{tabular}{|l|S|S|S|}
\hline 
 & {CR: \text{2b}} & {\text{SR1}: \text{3b}} & {\text{SR2}: $\geq$ \text{4b}}\\
\hline 
  $t\bar{t}$ + light   & 152000 \pm 34000 & 3400 \pm 1000 & 19 \pm 11 \\ 
  $t\bar{t}+\geq 1c$   & 29100 \pm 6700 & 2090 \pm 860 & 37.6 \pm 12 \\ 
  $t\bar{t}+\geq 1b$  & 15000 \pm 3500 & 8400 \pm 2300 & 1120 \pm 520 \\ 
  Single Top   & 10000 \pm 1300 & 703 \pm 100 & 51.2 \pm 13 \\ 
  ttH   & 945 \pm 96 & 507 \pm 51 & 145 \pm 15 \\ 
  ttV   & 1070 \pm 160 & 48.0 \pm 7.3 & 2.79 \pm 0.44 \\ 
  1 TeV LH $W'$   & 89.8 \pm 4.6 & 85.8 \pm 4.6 & 26.1 \pm 1.5 \\ 
\hline 
  Total  & 208000 \pm 40000 & 15200 \pm 3000 & 1400 \pm 520 \\ 
\hline 
  Data   & 209345 & 18865 & 1836 \\ 
\hline 
\end{tabular} 
\end{center} 
\end{table}   

 \begin{figure}[h]
     \centering
     \includegraphics[width=0.6\linewidth]{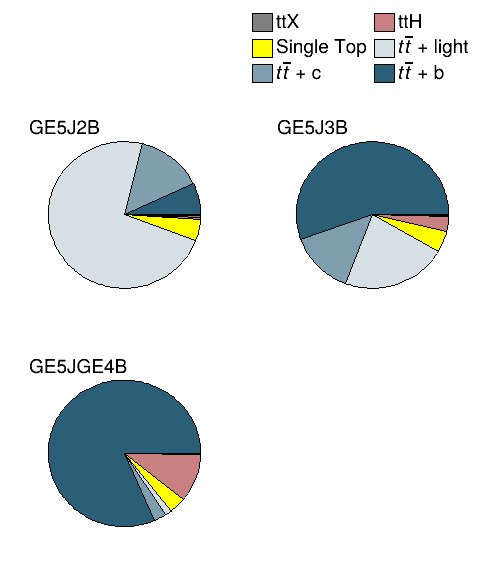}
     \caption{A pie chart representing the dominant background contributions in the region for this search.}
     \label{fig:pie_chart_backgrounds}
 \end{figure}

\begin{figure}[h]
     \centering
    \includegraphics[width=0.49\linewidth]{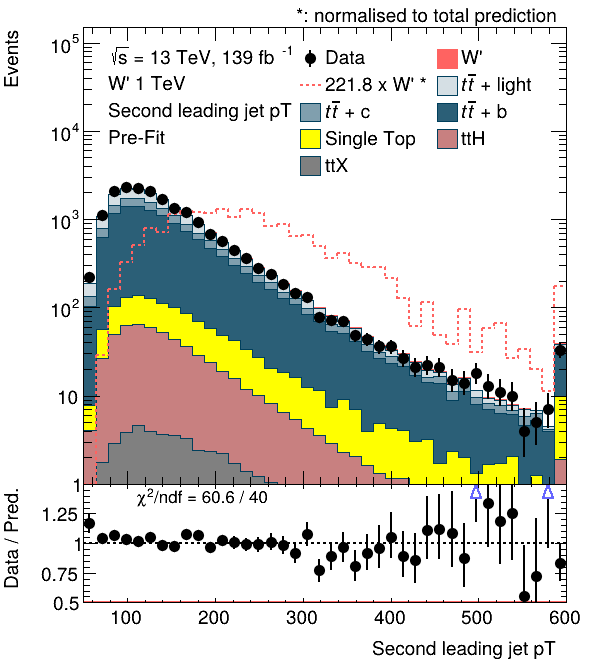}
    \includegraphics[width=0.49\linewidth]{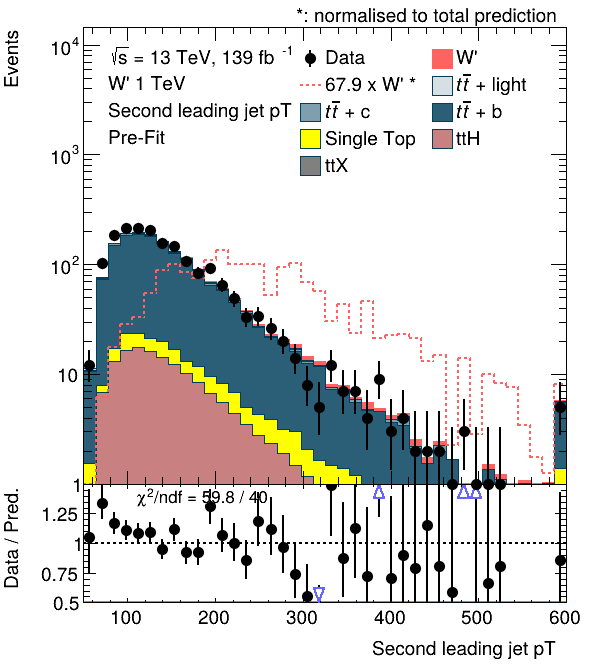}\\
    \includegraphics[width=0.49\linewidth]{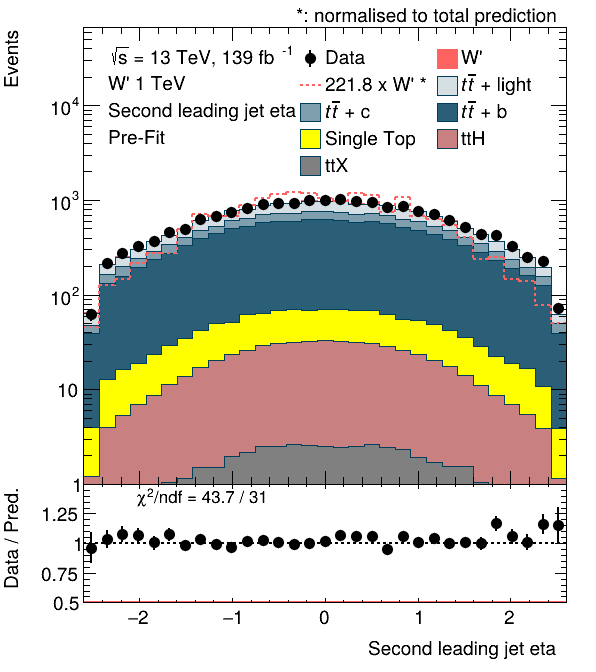}
    \includegraphics[width=0.49\linewidth]{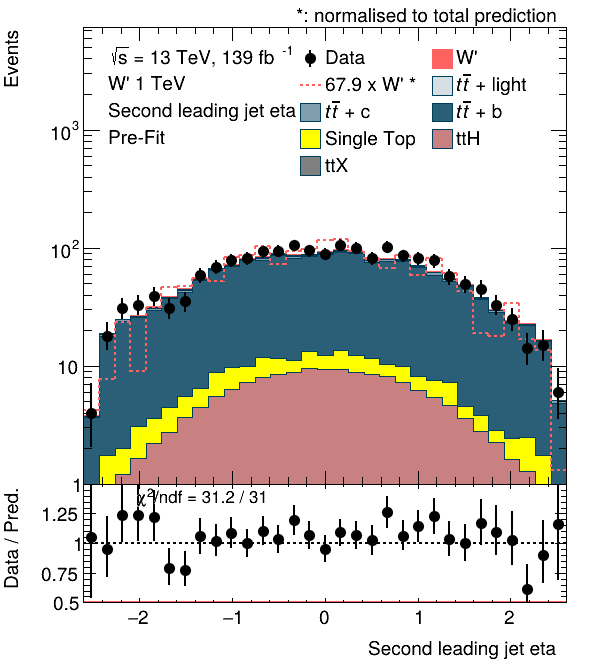}
     \caption{Kinematic distributions of background and signal in SR1 (left) and SR2 (right). Each bin in SR1 carries about 15\% and in SR2 about 25\% systematic uncertainty.}
     \label{fig:kinematics-2}
 \end{figure}

\begin{figure}[h]
     \centering
    \includegraphics[width=0.49\linewidth]{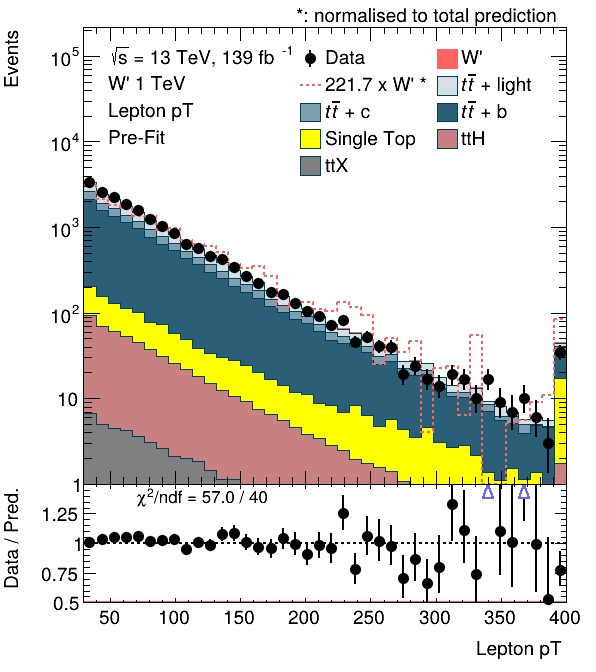}
    \includegraphics[width=0.49\linewidth]{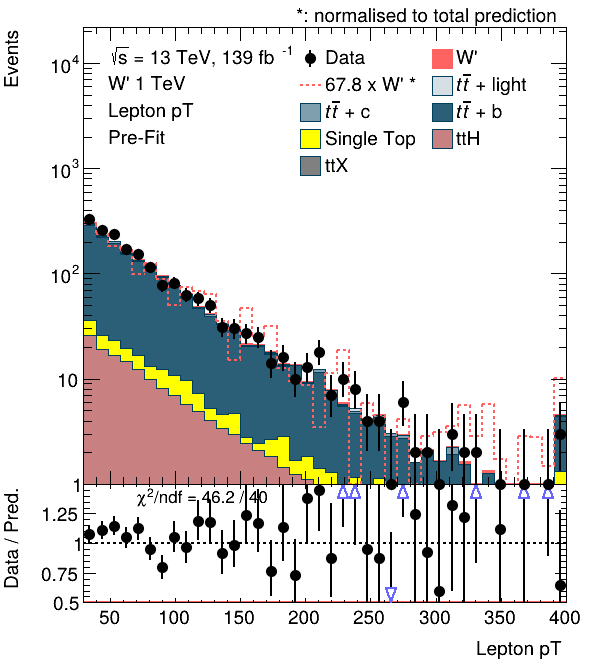}\\
    \includegraphics[width=0.49\linewidth]{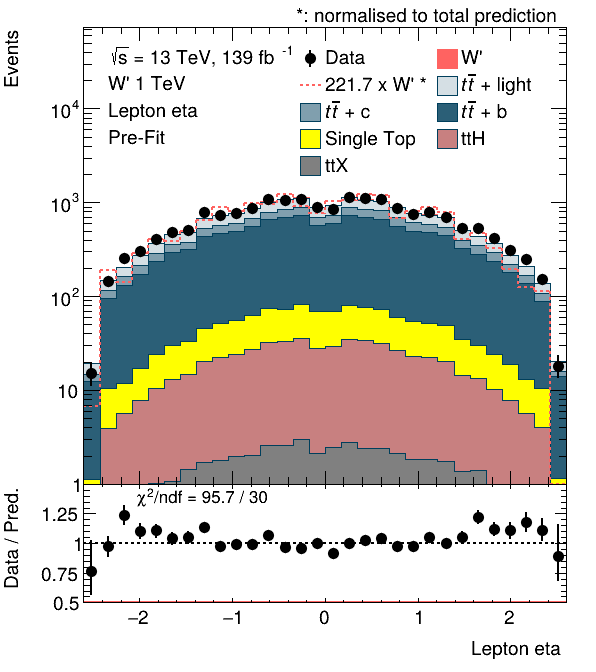}
    \includegraphics[width=0.49\linewidth]{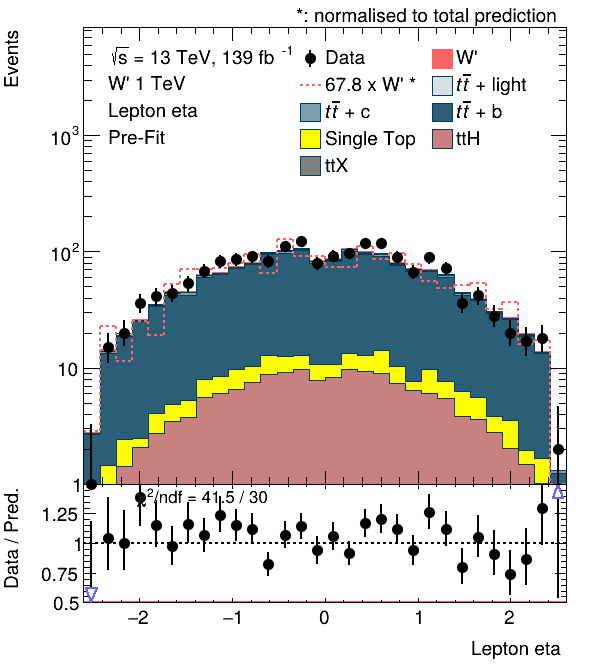}
     \caption{Kinematic distributions of background and signal in SR1 (left) and SR2 (right). Each bin in SR1 carries about 15\% and in SR2 about 25\% systematic uncertainty.}
     \label{fig:kinematics-3}
 \end{figure}

\begin{figure}[h]
     \centering
    \includegraphics[width=0.49\linewidth]{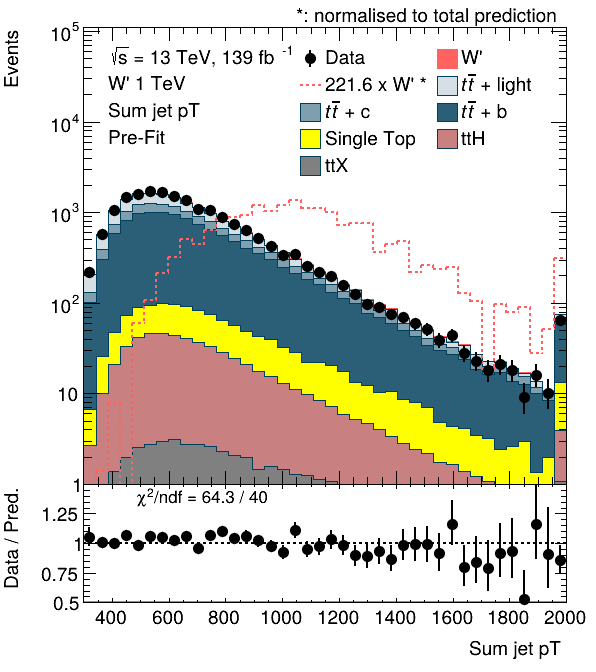}
    \includegraphics[width=0.49\linewidth]{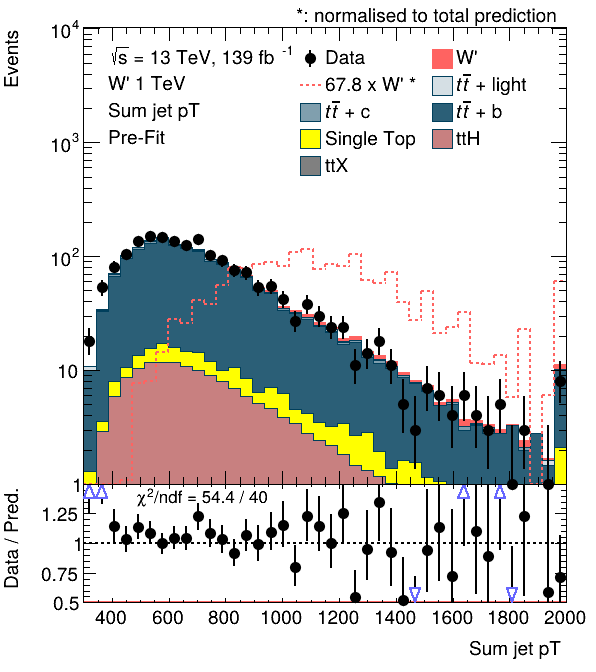}\\
    \includegraphics[width=0.49\linewidth]{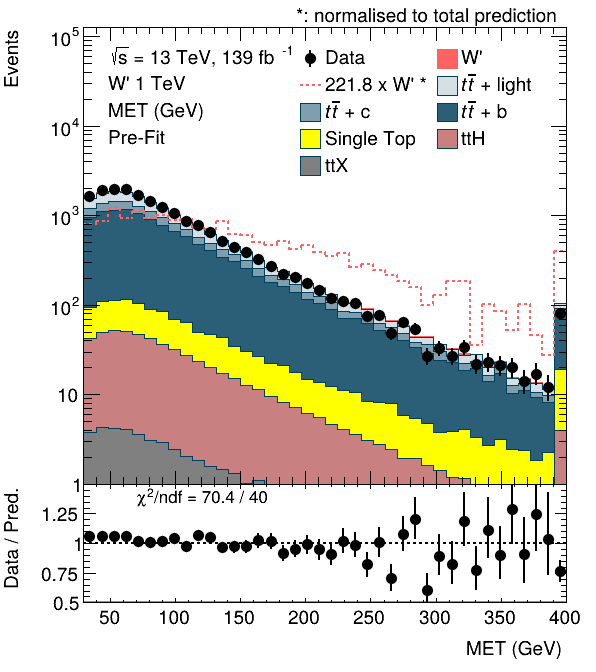}
    \includegraphics[width=0.49\linewidth]{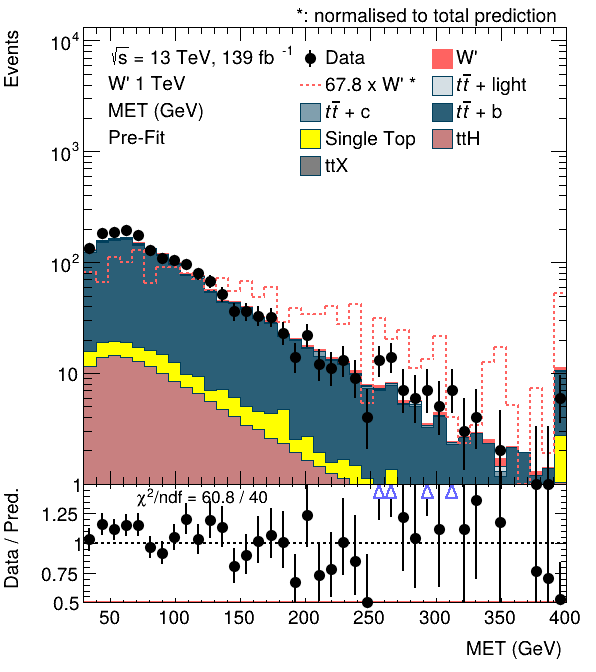}
     \caption{Kinematic distributions of background and signal in SR1 (left) and SR2 (right). Each bin in SR1 carries about 15\% and in SR2 about 25\% systematic uncertainty.}
     \label{fig:kinematics-4}
 \end{figure}

\begin{figure}[h]
     \centering
    \includegraphics[width=0.49\linewidth]{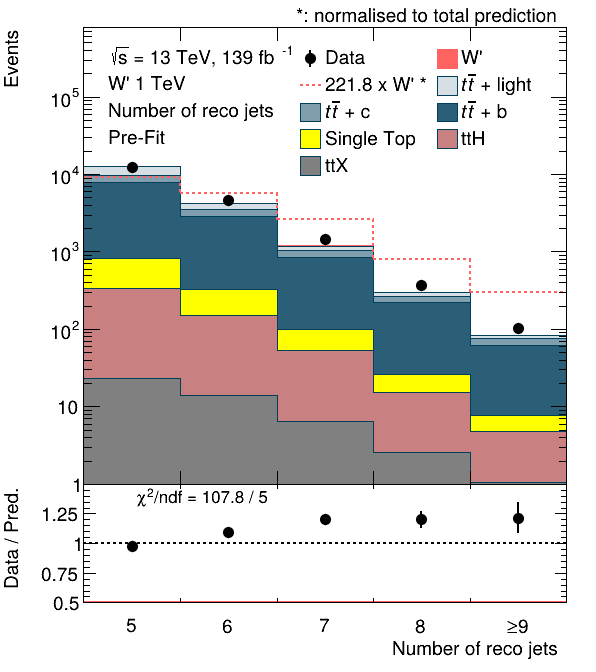}
    \includegraphics[width=0.49\linewidth]{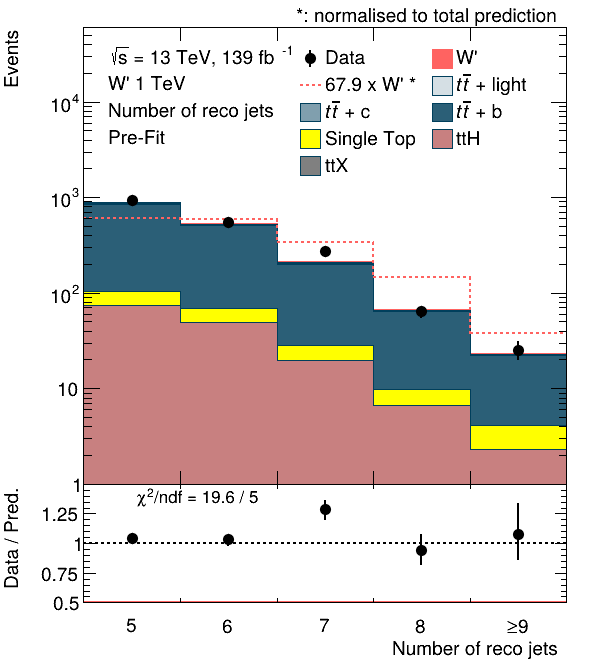} \\
    \includegraphics[width=0.49\linewidth]{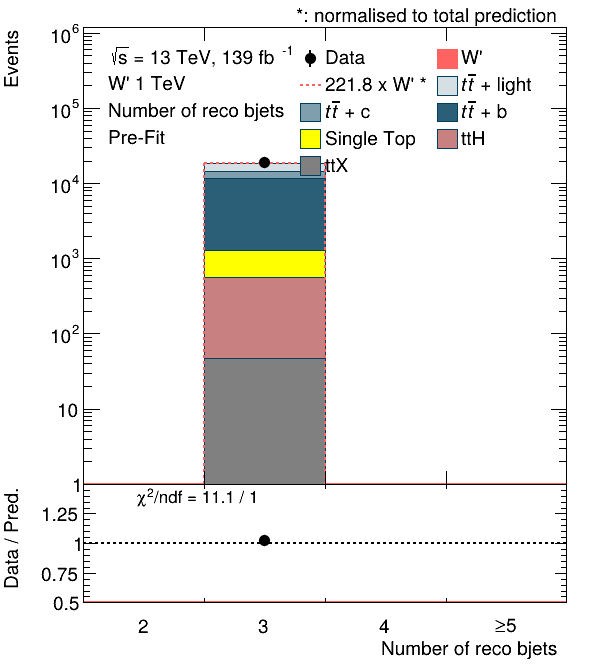}
    \includegraphics[width=0.49\linewidth]{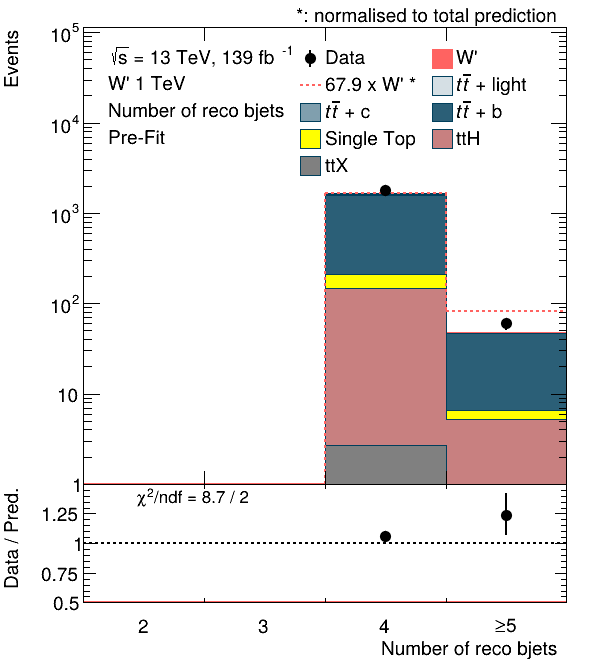}
     \caption{Kinematic distributions of background and signal in SR1 (left) and SR2 (right). Each bin in SR1 carries about 15\% and in SR2 about 25\% systematic uncertainty.}
     \label{fig:kinematics-5}
 \end{figure}

Several kinematic distributions of the reconstructed events in SR1 and SR2 are shown in the figures below. Figure~\ref{fig:kinematics-1} presents the leading jet $p_T$ and $\eta$, both of which show good agreement between data and the SM prediction. The leading jet $p_T$ also provides some discrimination between SM background and $W'$ boson signal. Figure~\ref{fig:kinematics-2} shows the second-leading jet $p_T$ and $\eta$, again demonstrating good agreement with SM expectations, with the $p_T$ distribution offering modest discriminating power. Figure~\ref{fig:kinematics-3} displays the reconstructed lepton $p_T$ and $\eta$, where both variables agree well with the SM background and show no significant separation from the $W'$ signal. Figure~\ref{fig:kinematics-4} shows $H_T$ and MET, both of which exhibit clear differences between the $W'$ signal and SM background while maintaining good agreement with the data. Lastly, Figure~\ref{fig:kinematics-5} shows the jet multiplicity, which indicates some tension between data and SM prediction. However, the level of systematic uncertainty ($\approx 20\%$) is sufficient to cover the observed differences between the SM and $W'$ signal.

 \begin{figure}[h]
     \centering
    \includegraphics[width=0.49\linewidth]{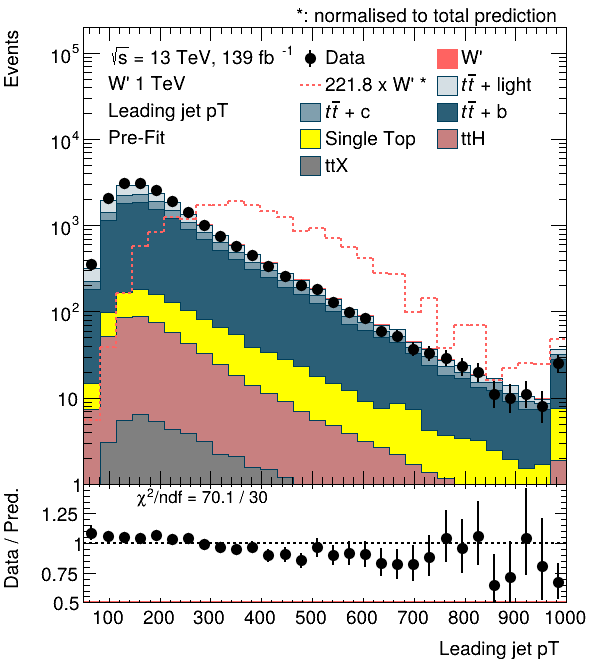}
    \includegraphics[width=0.49\linewidth]{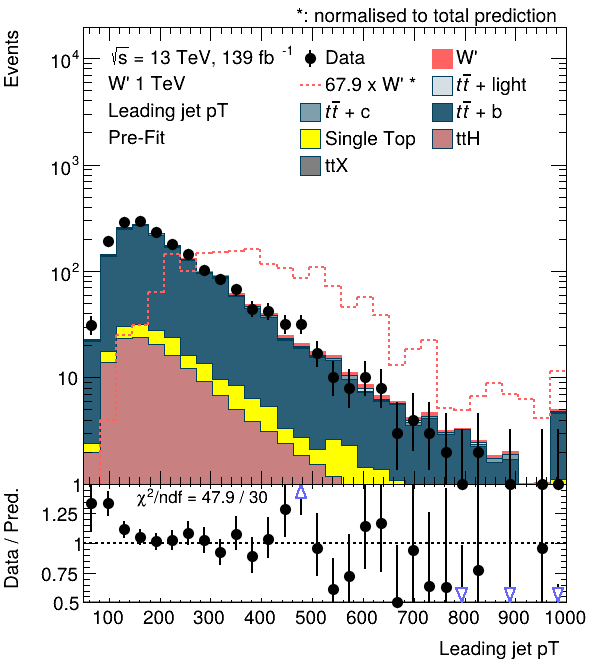}\\
    \includegraphics[width=0.49\linewidth]{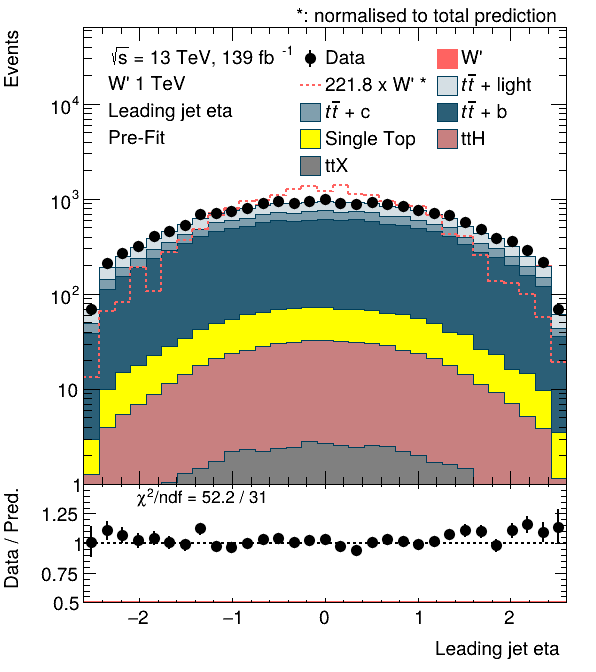}
    \includegraphics[width=0.49\linewidth]{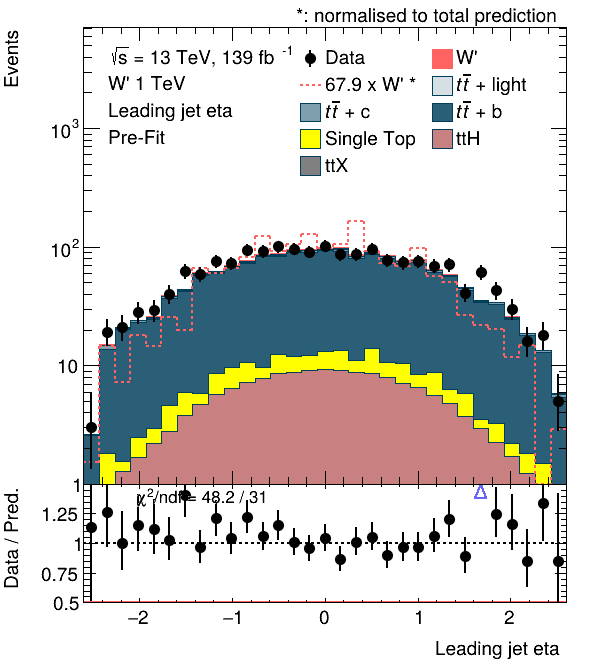}
     \caption{Kinematic distributions of background and signal in SR1 (left) and SR2 (right). Each bin in SR1 carries about 15\% and in SR2 about 25\% systematic uncertainty.}
     \label{fig:kinematics-1}
 \end{figure}

\FloatBarrier

\section{Neural Network Strategy}


In the production of SM-like $W'$ boson with no associated particles, the $t_R$ and $b_R$ can to be reconstructed as in ~\cite{JHEP2023-073}. This method struggles when searching for a heavy-philic $W'$ boson. The heavy-philic $W'$ boson is produced in association with two more particles, a top and bottom quark. Several assumptions, such as the $b_R$ having the highest $p_T$ is no longer a good assumption. Furthermore, figuring out the jet that comes from the decayed $t_R$ is much more complicated. 

As seen in the previous section, no kinematic variable has a high enough sensitivity to discriminate between the $W'$ boson signal and SM background. As in many heavy resonance searches, the sum of jet $p_T$ in Fig.~\ref{fig:kinematics-3} comes close, but is still not sensitive enough. To perform this search, a more complex analysis needs to be developed. 

Two neural networks are employed in this analysis to improve sensitivity to the heavy-philic $W'$ boson. The first is developed within the \textsc{SPANet} framework~\cite{10.21468/SciPostPhys.12.5.178}, and plays a central role in the reconstruction of the $W'$ boson decay chain. It uses key event-level information to identify jets originating from the $b_R$ and the $b$ from $t_R$, classify the decay as resonance or associated production, and predict the reconstructed $W'$ boson mass. \textsc{SPANet} thus enables the extraction of physics-motivated variables that are essential for characterizing the heavy-philic $W'$ boson, making it a crucial component of the analysis strategy.

The second neural network is designed to separate the heavy-philic $W'$ boson signal from all the relevant SM backgrounds. This network is a multilayer perceptron (MLP) that predicts a final signal output score (SB discriminator). This score is then used for the final profile likelihood fit. 

\begin{figure}
    \centering
    \includegraphics[width=\linewidth]{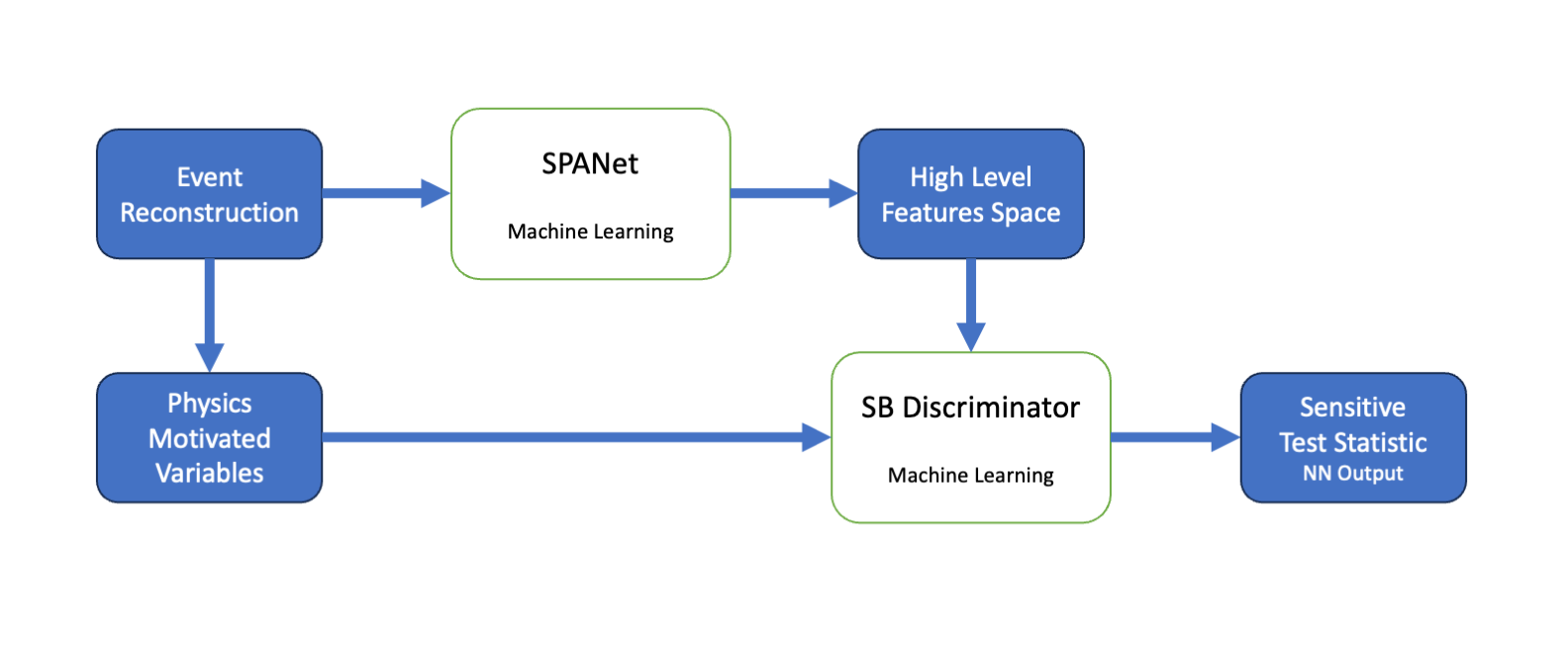}
    \caption{This flow chart shows how information is transformed throughout the analysis. The two neural networks are depicted as green blocks, and transformed states of the data and MC samples are depicted in solid blue boxes. This analysis structure combines the power of machine learning and directed by physics-motivated approach.}
    \label{fig:nn_flowchart}
\end{figure}

Figure~\ref{fig:nn_flowchart} provides a high-level overview of the analysis structure, highlighting where and how the two neural networks are employed. The analysis strategy emphasizes modularity by separating distinct components into interpretable, task-specific blocks rather than treating the workflow as a black box. This design improves the interpretability of each stage and allowing the neural networks to be specialized for their respective objectives.

\subsection{\textsc{SPANet} and Transformers on $W'$ Boson Signal}

\begin{figure}
    \centering
    \includegraphics[width=0.8\linewidth]{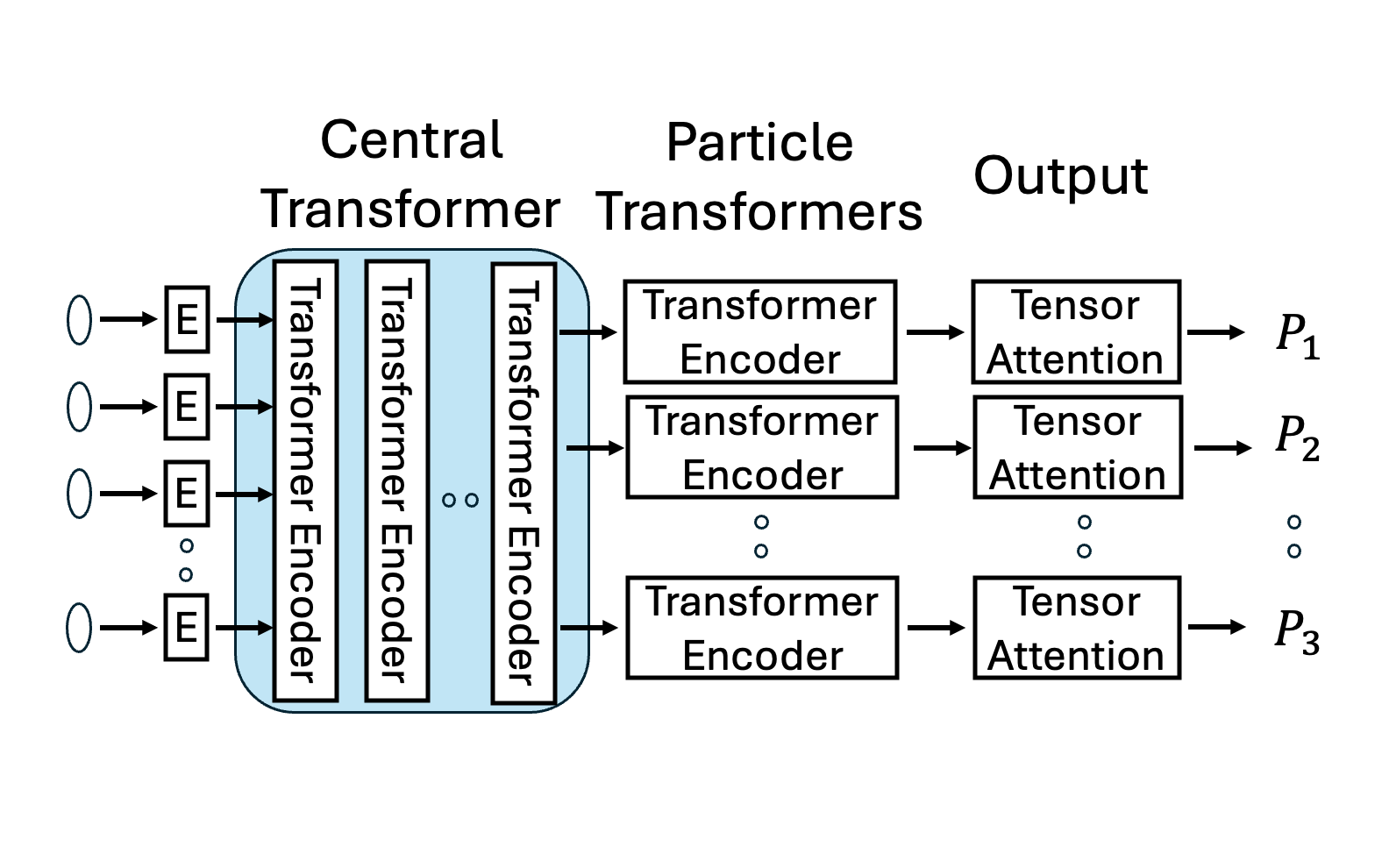}
    \caption{A high level diagram of the \textsc{SPANet} architecture \cite{10.21468/SciPostPhys.12.5.178}.}
    \label{fig:spanet_arch}
\end{figure}

The number of jets produced from a proton-proton collision is variable which means that a simple MLP will struggle in this domain. To overcome the problem of variable jet multiplicity in the feature space, transformers \cite{vaswani2023attentionneed} are used in a package called \textsc{SPANet} \cite{10.21468/SciPostPhys.12.5.178}. The high-level architecture of \textsc{SPANet} can be seen in Fig.~\ref{fig:spanet_arch}. 

Each jet which is represented as a vector of fixed length is fed through a position independent embedding which transforms the input feature vector into a more useful embedded latent space representation. Then, each embedding is fed into several transformer encoder layers which process the vectors within the context of the entire event. Transformer encoders follow the central transformer. After the particle transformers, tensor attention layers process the symmetries of the process which then output the final prediction. The final prediction is a vector of probabilities that are assigned to each jet which are used to match the reconstructed jet to the predicted truth level particle. These are the predictions for assigning each given jet of matching the original \br, \btr, or \bta. 

A global maximum jet multiplicity of 20 is used in this architecture. Jets are ordered by $p_T$ and low $p_T$ jets are discarded to obtain the maximum of 20 jets per event. When fewer than 20 jets are present, a mask is used to identify how many jets are present in the event. 

The output of \textsc{SPANet} consists of probability vectors for each jet, indicating the likelihood of assignment to specific truth particles. To perform the assignment, the algorithm iteratively selects the jet with the highest remaining probability, removes it from the pool, and repeats this process until all jets are uniquely assigned. This method ensures one-to-one jet assignments while maximizing the likelihood at each step. 

\subsection{Jet Truth Matching}

\begin{figure}
    \centering
    \includegraphics[width=0.49\linewidth]{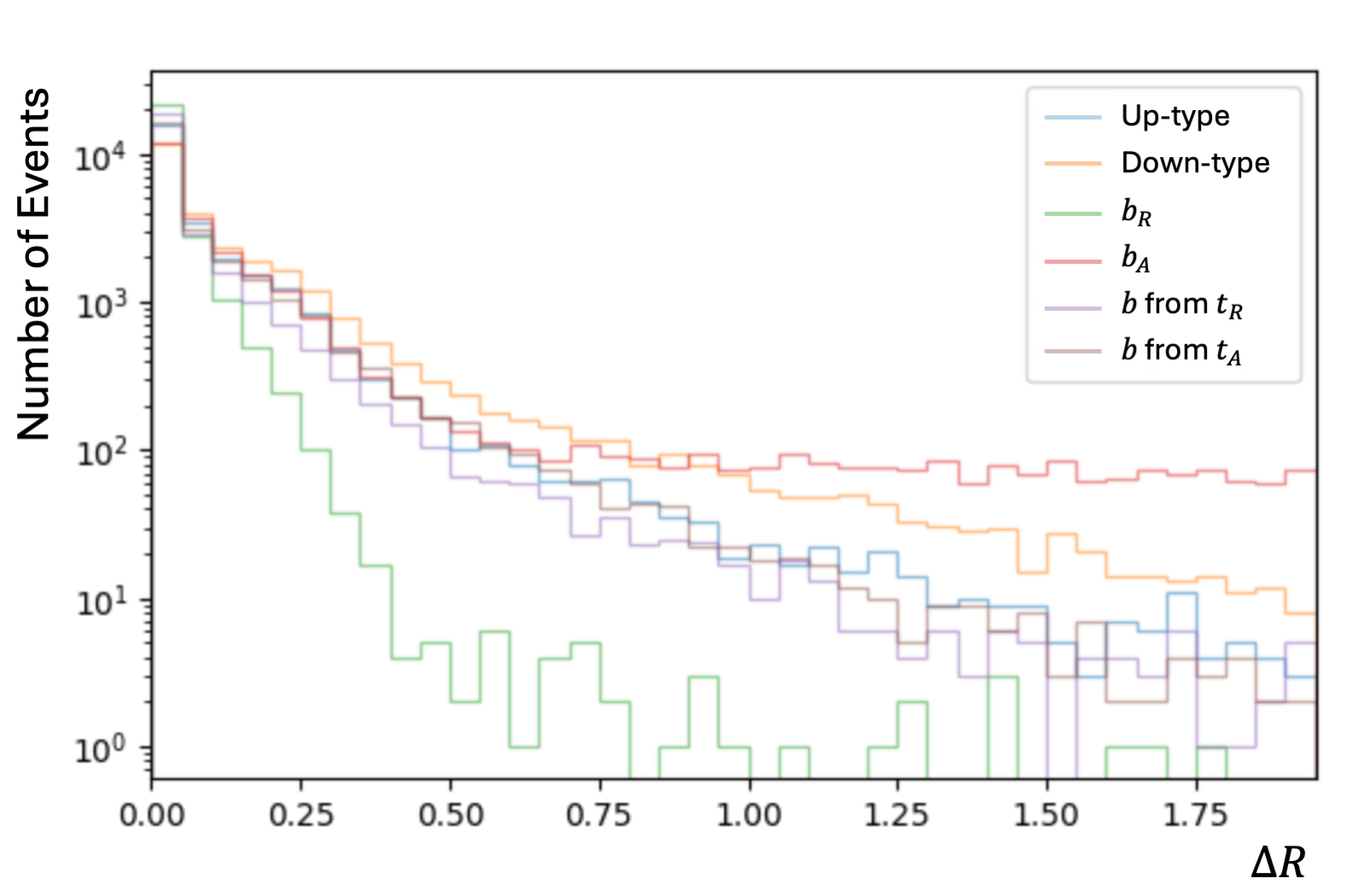}
    \includegraphics[width=0.49\linewidth]{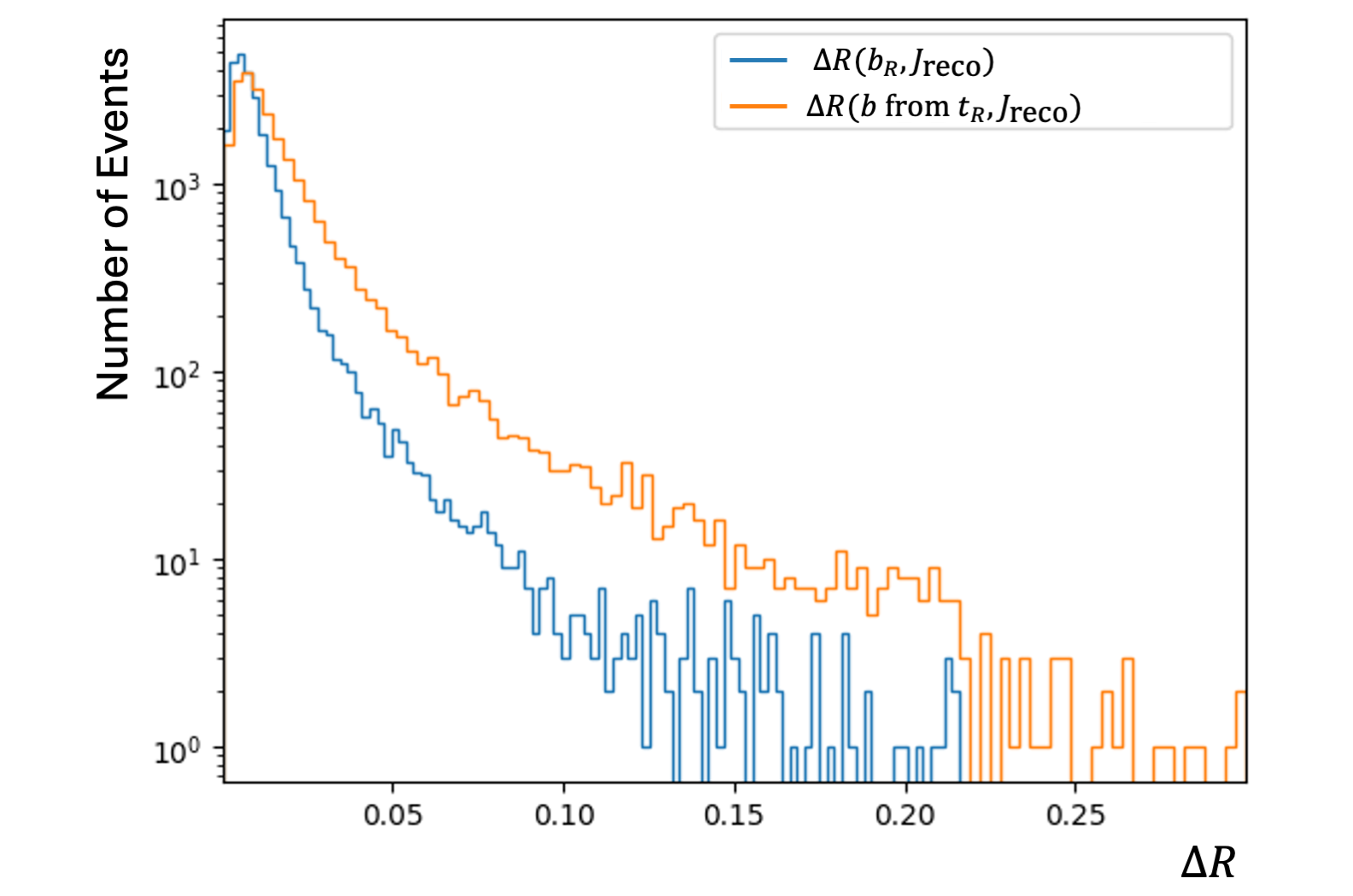}
    \caption{This figure shows the $\Delta R$ distributions for the 1 TeV left-handed $W'$ boson. On the left, the minimum $\Delta R$ between truth-level particles and their matched particle-level jets is shown for the up- and down-type quarks from the hadronic $W$ decay, the $W'$ decay products, and associated particles. On the right, the minimum $\Delta R$ between each particle-level jet and its matched reconstructed jet is shown for the two resonance decay products.}
    \label{fig:jet_matching}
\end{figure}

The neural network is trained using supervised machine learning with a labeled dataset. In order to acquire the labeled dataset, reconstructed jets needed to be matched to the original truth particles. This is done in two steps. First the truth particles are matched to their particle level jet (AntiKt4TruthDressedWZJets) by matching the geometrical trajectory of the truth particle to the nearest particle level jet. The second step is to match each particle level jet to a reconstructed jet geometrically again. Truth particles are allowed to be matched to the same jet, and if a match doesn't fall within \(\Delta R<0.3\), then it isn't matched. The distance between truth particles and particle level jets can been seen in Fig.~\ref{fig:jet_matching} as well as the distance between corresponding particle level jets and reconstructed jets.

This matching scheme ensures accurate particle matching. Furthermore, by allowing the truth particles to be unmatched gives rise naturally to a detection probability that \textsc{SPANet} works out. This adds a weight in the final signal-background discriminator to add an element of confidence for a specific jet to be matched to $b_R$, \btr, or \bta. 

\subsection{\textsc{SPANet} Training}

During training, a two-fold cross validation method is then used. Even numbered events are used as the training set while the odd numbered events are used for testing and validation. Then, the network is trained again by switching the roles of the odd and even events. This process ensures that the neural network is learning something meaningful, and isn't being over trained. The full dataset is used for the signal and background discriminator using the properly trained network over the odd and even number Monte Carlo events. Training is performed over 32 epochs using the full $W'$ boson signal dataset, which includes both left-handed and right-handed samples across all mass points. Approximately 4 million events are used for training, with an additional 4 million reserved for validation.

The loss function in \textsc{SPANet} is designed to balance regression of the $W'$ boson mass and $W_R$ kinematic variables, decay channel classification, and truth particle and reconstructed jet matching prediction. Each prediction from \textsc{SPANet} is then utilized by the signal and background discriminator.

\FloatBarrier

\subsection{Results from \textsc{SPANet}}
\begin{figure}[h]
    \centering
    \includegraphics[width=0.39\linewidth]{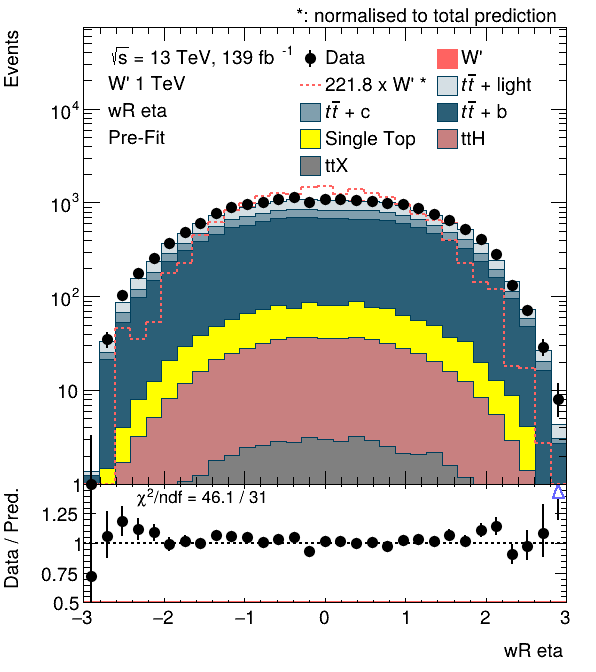}
    \includegraphics[width=0.39\linewidth]{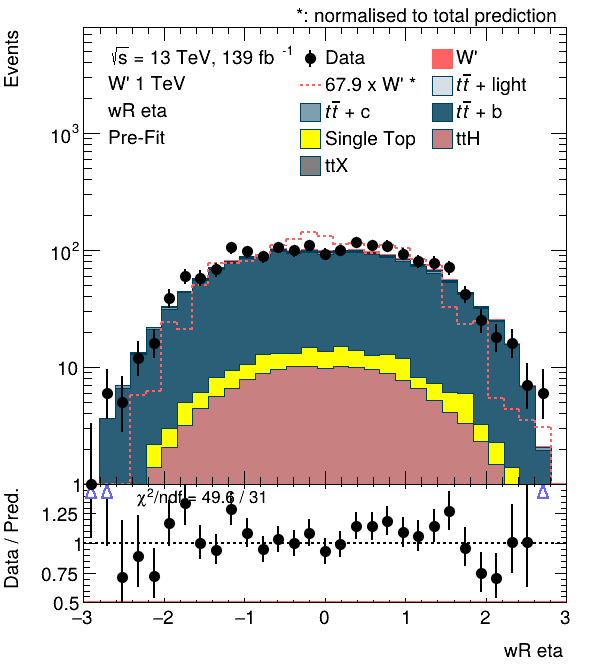}\\
    \includegraphics[width=0.39\linewidth]{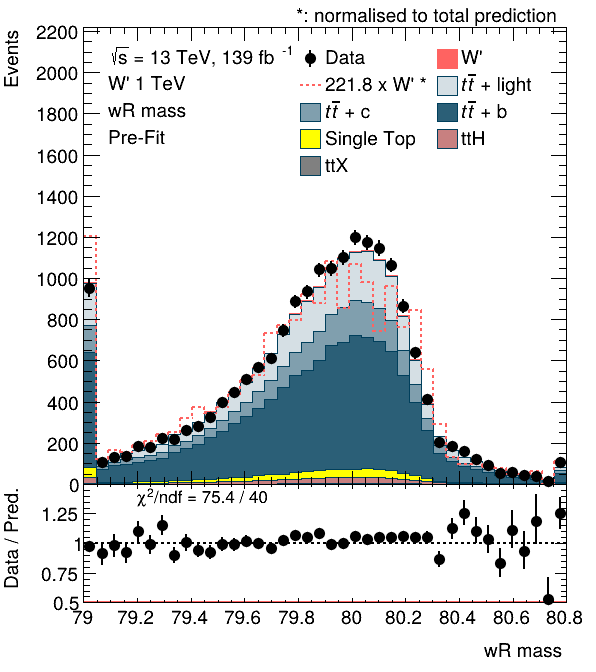}
    \includegraphics[width=0.39\linewidth]{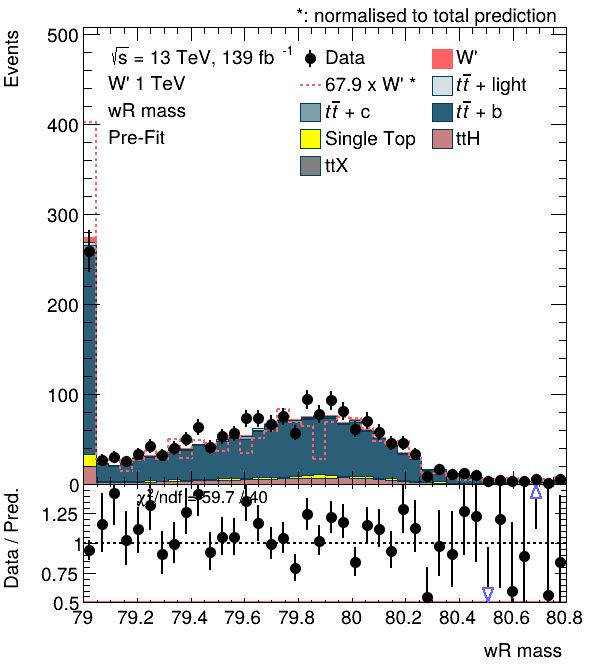}\\
    \includegraphics[width=0.39\linewidth]{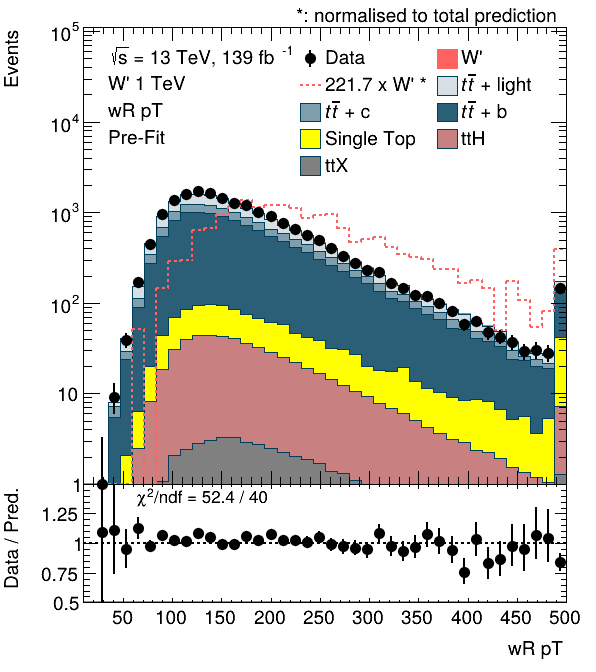}
    \includegraphics[width=0.39\linewidth]{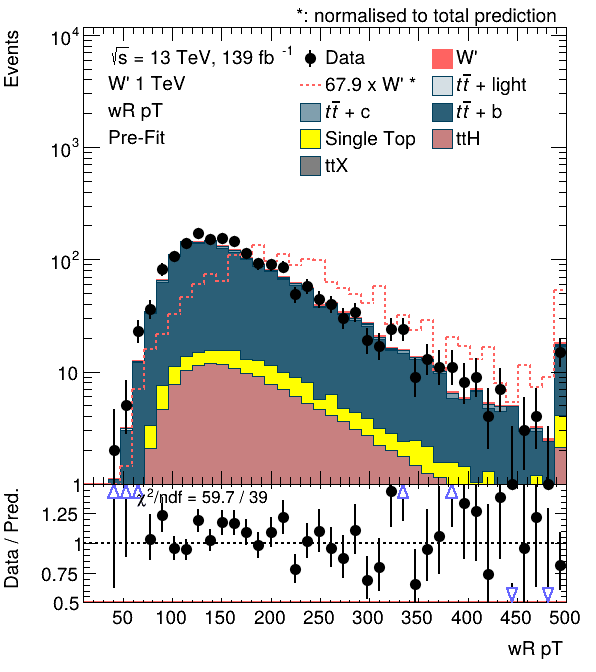}
    \caption{\textsc{SPANet}-predicted regression variables in SR1 (left) and SR2 (right), with about 15\% and about 25\% per-bin systematic uncertainties, respectively.}
    \label{fig:WR_regression}
\end{figure}

There are several variables that are simultaneously regressed from reconstruction level objects from \textsc{SPANet}. One important physics variable is the kinematic 4-vector of the resonance $W$ boson that comes from the $t_R$. The regression results for these variables are shown in Fig.~\ref{fig:WR_regression}.

\begin{figure}
    \centering
    \includegraphics[width=0.39\linewidth]{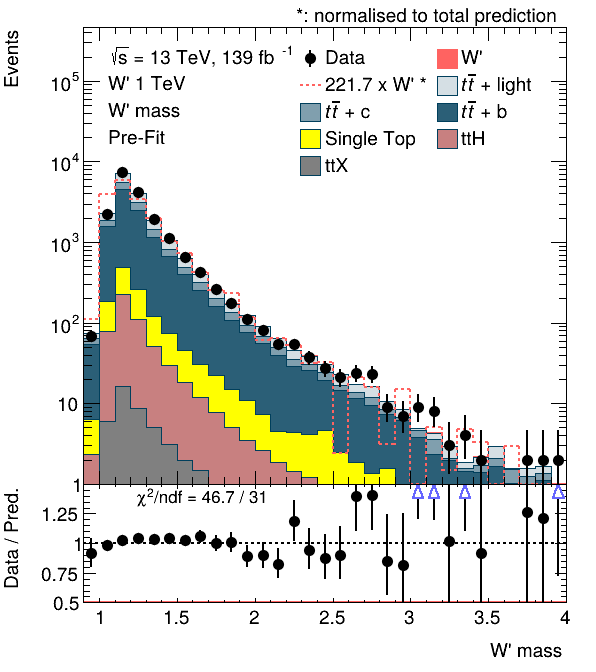}
    \includegraphics[width=0.39\linewidth]{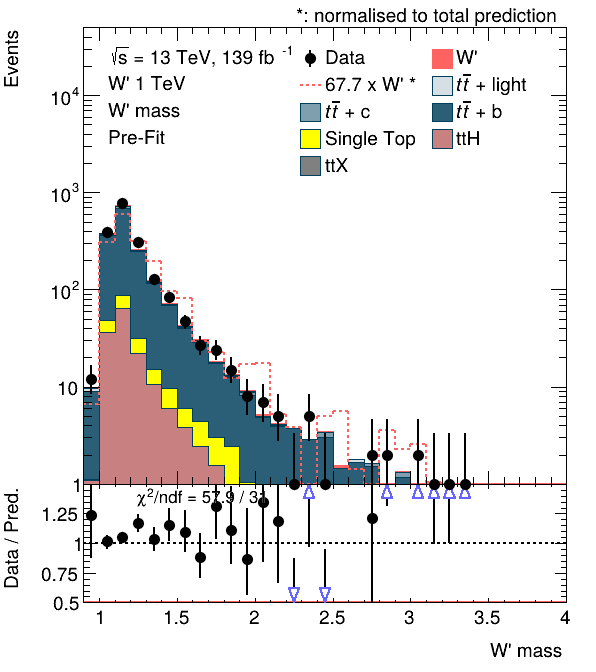}\\
    \includegraphics[width=0.39\linewidth]{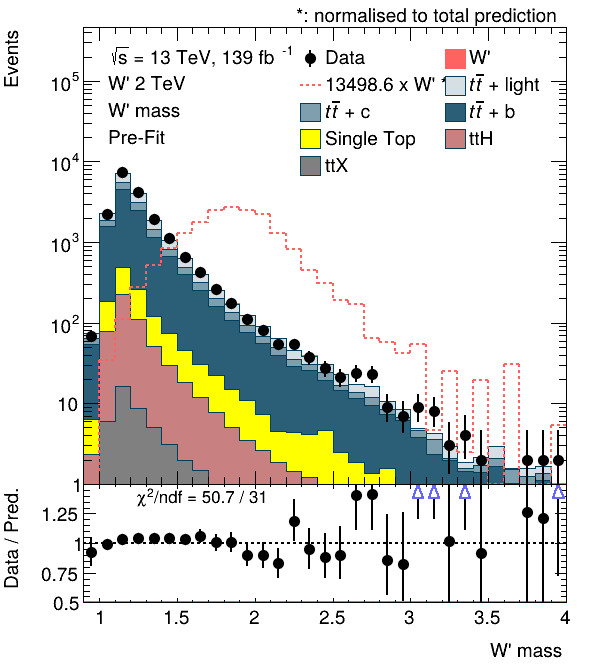}
    \includegraphics[width=0.39\linewidth]{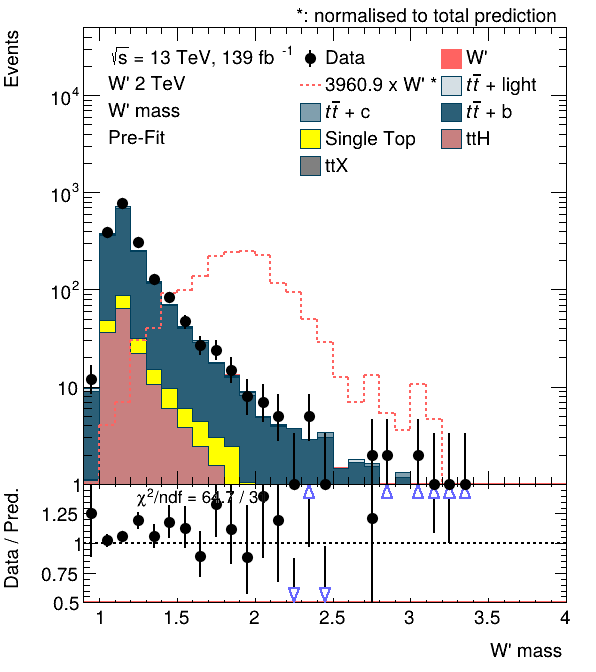}
    \caption{$W'$ regression variable that are predicted by the trained \textsc{SPANet} neural network. SR1 variables are on the left and SR2 variables are on the right. The top two depict the regression results for the 1 TeV mass point, and the bottom two depict the regression results for the 2 TeV mass point. The distributions in SR1 have about a 15\% systematic uncertainty in each bin while the distributions in SR2 have about a 35\% systematic uncertainty in each bin.}
    \label{fig:wp_regression}
\end{figure}

\begin{figure}
    \centering
    \includegraphics[width=0.79\linewidth]{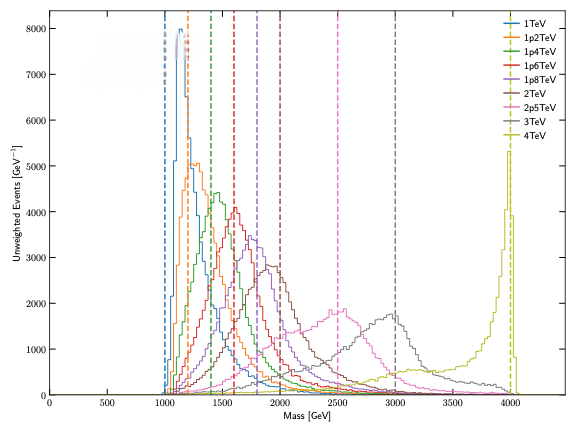}
    \caption{The distributions of the predicted $W'$ masses for each sample set for the heavy-philic LH $W'$.}
    \label{fig:wp_spanet_reg_masses}
\end{figure}

In addition to the $W_R$ kinematic variables, the mass of the $W'$ boson is also estimated within the \textsc{SPANet} neural network. The distributions of the LH $W'$ boson masses can be seen in Fig.~\ref{fig:wp_spanet_reg_masses}. This figure demonstrates that \textsc{SPANet} successfully reconstructs each input mass with good mass resolution. The distributions are asymmetric and exhibit sharp cutoffs between 1~TeV and 4~TeV, a characteristic often seen in neural networks whose predictions are constrained by the range of their training data.

Table ~\ref{tab:matching_alg} shows how \textsc{SPANet} compares to different matching algorithms that could be used for selecting matching truth partons to reconstructed jets. The random choice algorithm shows the worst case scenario where the truth particles are matched randomly to the reconstructed jets. The matching algorithm in \cite{JHEP2023-073} describes how previous searches for the SM-like $W'$ boson performs when trying to match the heavy-philic $W'$ boson resonance truth particles. This shows that the previous algorithm is far from being optimized for matching the truth particles of the heavy-philic $W'$ boson. The third algorithm is selecting the hightest $p_T$ jet for the $b_R$ and the second highest $p_T$ jet for \btr. The neural network approach matches the two jets the best. 

\begin{table}[]
    \centering
    \caption{Different matching efficiencies showing improvement of the new neural network approach over simple algorithms and previous algorithms. The percent correct is calculated based on the number of times the algorithms matches the truth particle to the geometrically matched reconstructed jet.}
    \begin{tabular}{lcc}
        Algorithm Name & $b_R$ & $b$ from $t_R$ \\ \hline
        Random Choice & 15\% & 15\% \\
        Algorithm in \cite{JHEP2023-073}  & 29\% & 24\% \\
        Leading Jet & 69\% & 37\% \\
        \textsc{SPANet} & 65\% & 50\% \\
    \end{tabular}
    \label{tab:matching_alg}
\end{table}

\FloatBarrier

\subsection{Signal and Background Discriminator}

\begin{figure}
    \centering
    \includegraphics[width=0.5\linewidth]{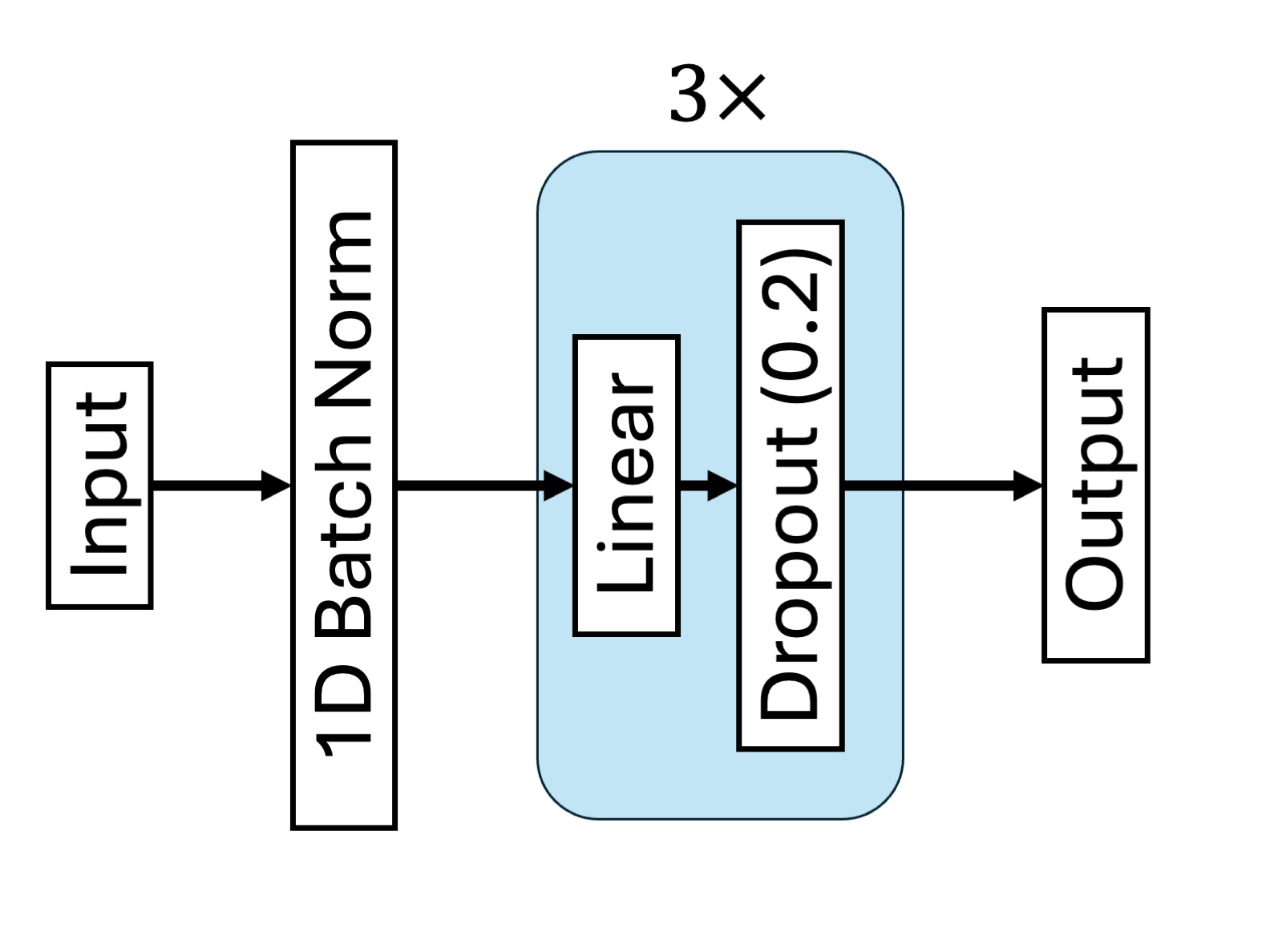}
    \caption{A diagram representing the simple MLP design of the signal-background discriminator.}
    \label{fig:sb_discriminator}
\end{figure}

The next phase of the analysis is to discriminate between signal and background. To do this, a multiplayer perceptron (MLP) is used with 5 hidden layers. A dropout layer is added after each layer to ensure that the model wouldn’t be over-trained. There are 38 inputs to the model, and each hidden layer has 512 nodes. The activation function that is chosen for these layers are simple rectified linear unit functions. The final layer is a simple linear layer that outputs a float that scores each event as either signal or background. A diagram of the signal-background discriminator can be seen in Fig.~\ref{fig:sb_discriminator}.

One of the key features of this neural network is that it is a mass-parameterized neural network (MP-NN). In order to efficiently train a network that works for each $W'$ boson mass point, one dimension of the neural network is set to the true $W'$ boson mass. This means that in training, the true $W'$ boson mass is input for signal events, and a random number is drawn from the probability density distribution of the $W'$ boson mass distribution of all the samples for the background events. This pools relevant information about a specific mass point into centralized regions of the NN phase space. An alternative would be to train a separate neural network for each mass point. However, this makes inferences between mass points more discrete. With the MP-NN, during evaluation and prediction, this mass parameter is taken to be a random number following the truth level $W'$ boson mass distribution.

To discriminate the dominate backgrounds in this region from the $W'$ boson signal, several high level variables along with some traditional kinematic variables are used as inputs in the multilayer perception. A summary of these inputs can be found in Table~\ref{tab:sb_discriminator_table}.

\begin{table}[h!]
\centering
\caption{Summary of input features used in the signal vs. background discriminator neural network. A total of 36 input variables are used in the first layer of the MLP.}
\renewcommand{\arraystretch}{1.3}
\begin{tabular}{@{}lll@{}}
\toprule
\textbf{Category} & \textbf{Feature Variables} & \textbf{\# Variables} \\
\midrule
\textbf{\textsc{SPANet} Variables} & & \\
\quad Assignment probabilities & $P(b_R)$, $P(b_{t_A})$, $P(b_{t_R})$ & 3 \\
\quad Predicted Jet 4-vectors + btag & $b_R$, $b$ from $t_A$, $b$ from $t_R$ & $3 \times 5 = 15$ \\
\quad Classification & Resonant channel classification & 1 \\
\quad Regression & $W'$ ($m$),  $W_R$ ($\eta$, $m$, $\phi$, $p_T$) & 5 \\
\midrule
\textbf{Jet kinematic properties} & & \\
\quad Next leading jet & btag, $\eta$, $\phi$, $p_T$ & 4 \\
\quad Next-next leading jet & btag, $\eta$, $\phi$, $p_T$ & 4 \\
\quad Jet multiplicity & $N_\text{jets}$, $N_\text{btag}$ & 2 \\
\quad Jet $p_T$ sum & Sum of jet $p_T$ & 1 \\

\midrule
\textbf{$W'$ Boson Truth Mass Parameter} & & 1 \\
\bottomrule
\end{tabular}
\label{tab:sb_discriminator_table}
\end{table}

\begin{figure}
    \centering
    \includegraphics[width=0.49\linewidth]{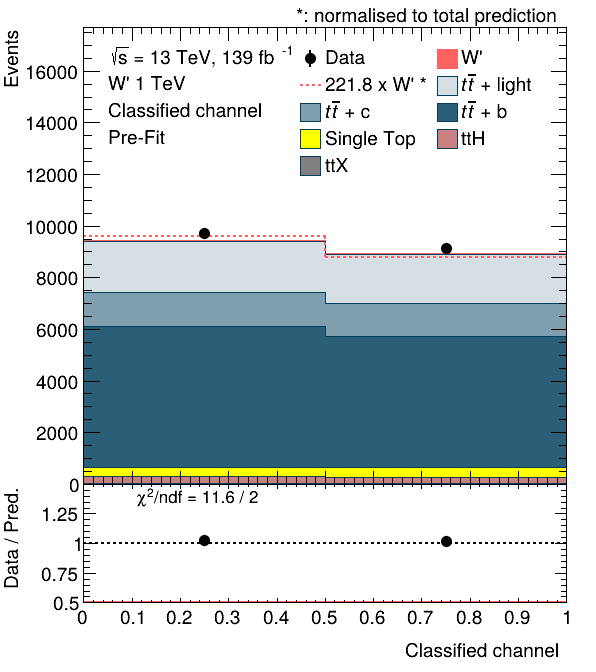}
    \includegraphics[width=0.49\linewidth]{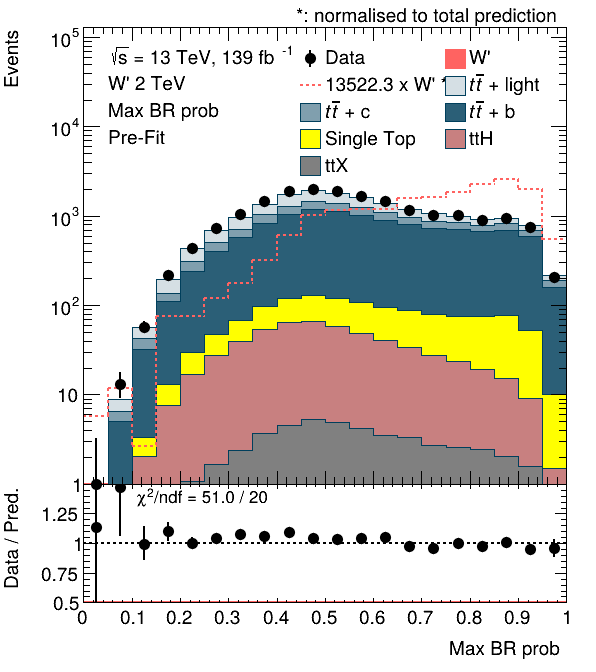}
    \caption{The resonance channel classification which comes from the output of \textsc{SPANet} is shown on the left. A hadronically decaying $W'$ is classified as 0, and a leptonically decaying $W'$ boson is classified as 1. The right shows the assignment probability of $b_R$.}
    \label{fig:spanet_output_example}
\end{figure}

Examples of additional output variables from \textsc{SPANet} which is used within the SB discriminator is shown in Fig.~\ref{fig:spanet_output_example}. By themselves, they are not useful to search for signal excess, but the SB discriminator benefits from the addition of these variables as it provides critical information on the kinematic properties of the $W'$ boson, and the certainty on the assignment of key particles such as the $b_R$.

\subsection{SB Discriminator Training}

Training of the signal and background discriminator is done through the use of a two-fold cross validation method similar to the way \textsc{SPANet} is trained. Training occurs over 10 epochs, and is done with unbalanced datasets. There is a significantly larger number of $t\bar{t}$ events compared to other backgrounds and $W'$ boson signal. To utilize the full phase space available for the $t\bar{t}$ dataset, but balance the training, weights are used during the training phase. The cross-section weights for each background event are used, and then the average background weight are used for the $W'$ boson signal. This ensures the network is balanced between signal and background events equally. It also ensures that the different backgrounds are considered with the appropriate weight. Both LH and RH $W'$ boson samples are used in the training of the SB discriminator. 

\subsection{Results from SB Discriminator}

\begin{figure}
    \centering
    \includegraphics[width=0.75\linewidth]{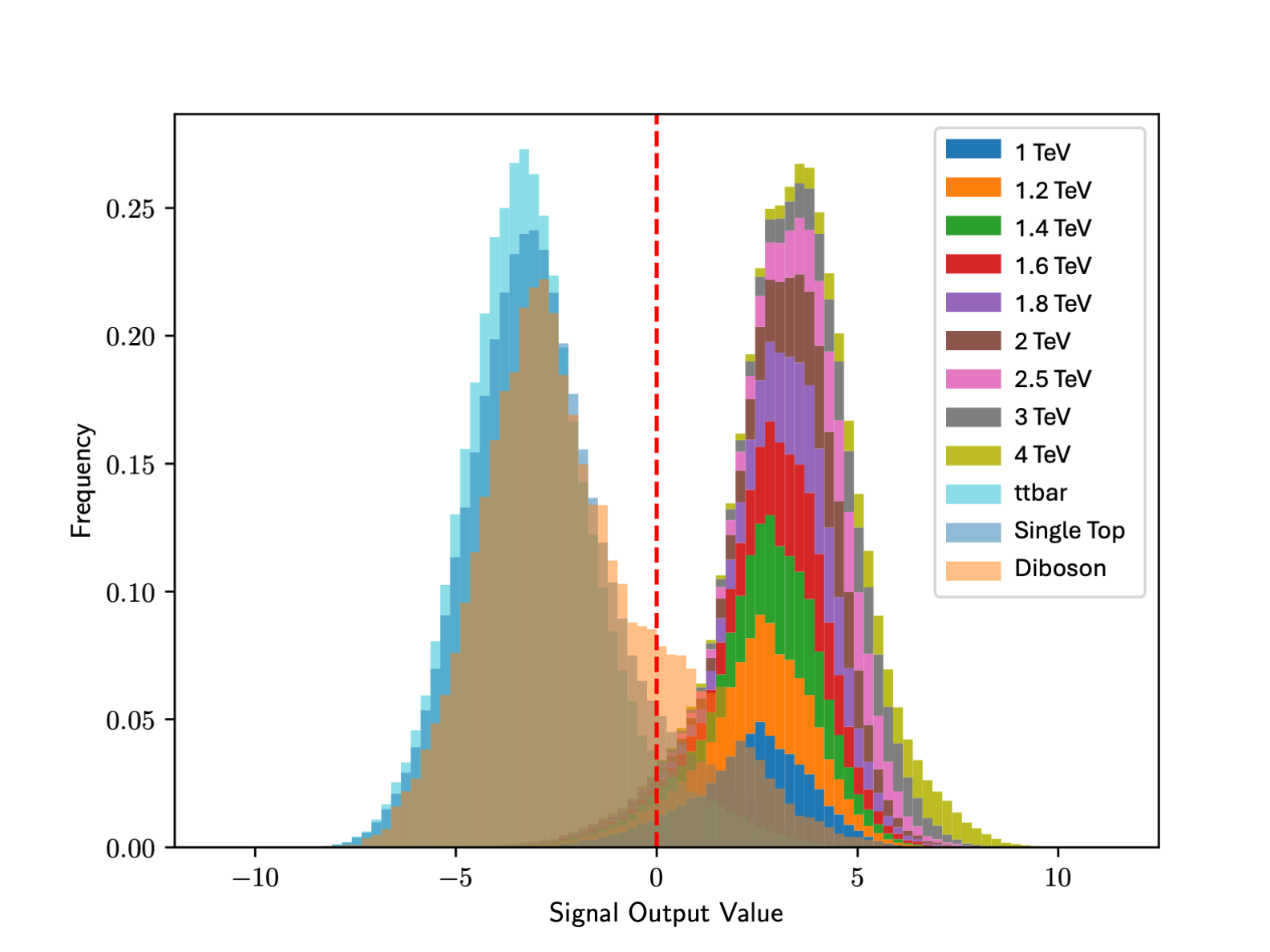}
    \caption{Results from the trained neural network showing the output score for each mass stacked on top of each other. The SM backgrounds are normalized, and the sum of the $W'$ boson signal samples are normalized for comparison.}
    \label{fig:sb_output}
\end{figure}

\begin{figure}
    \centering
    \includegraphics[width=0.75\linewidth]{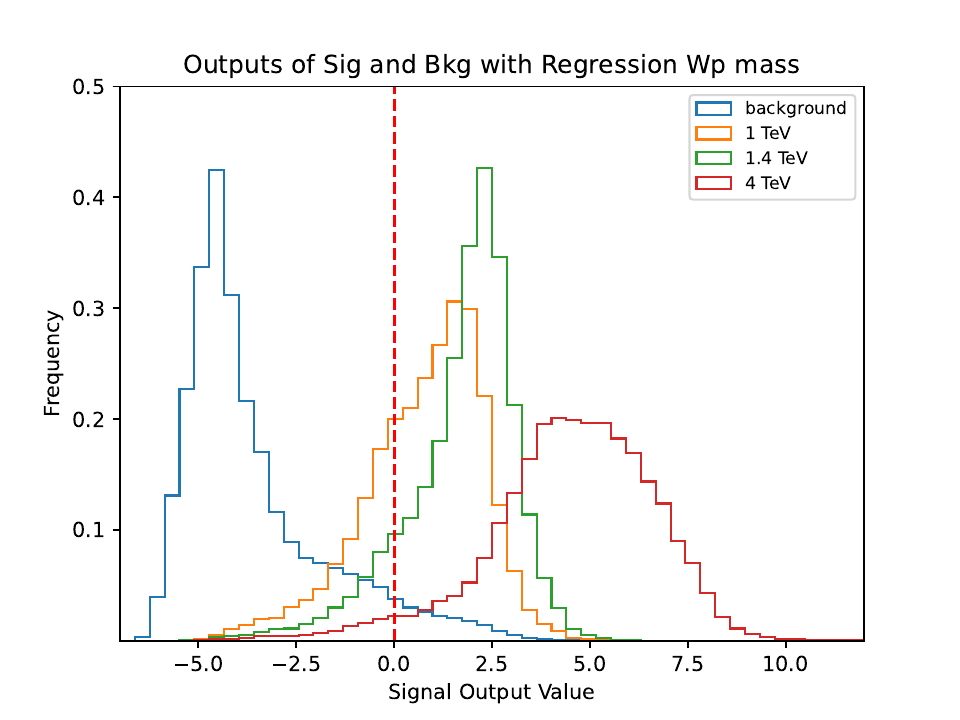}
    \caption{The neural network output for the total background is show in comparison to $W'$ signal samples. The dashed red line is drawn for visibility and isn't a cut on the samples that are made. This output score is binned and a profile likelihood fit is done on this distribution.}
    \label{fig:sig_back}
\end{figure}

\begin{figure}
    \centering
    \includegraphics[width=0.75\linewidth]{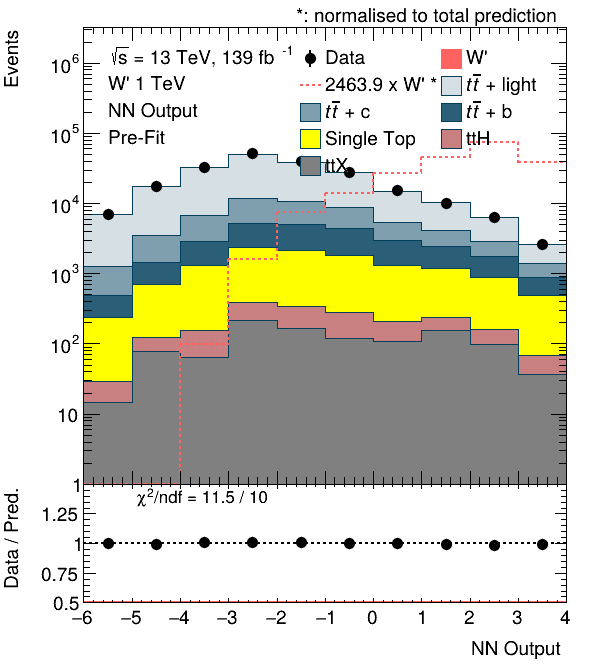}
    \caption{The neural network output with all backgrounds in the $t\bar{t}$ CR.}
    \label{fig:nn_ttbar_cr}
\end{figure}

The results from training this neural network are shown in Fig.~\ref{fig:sb_output} and Fig.~\ref{fig:sig_back}. Each $W'$ boson mass point has similar sensitivity, with higher masses having slightly improved discriminating power. This is because the higher the $W'$ boson mass, the easier it is to distinguish from SM $t\bar{t}$ events. The kinematic properties begin to look very different because jets from the resonance particles begin to have even greater transverse momentum. Fig.~\ref{fig:nn_ttbar_cr} also shows the NN output score in the $t\bar{t}$ CR for each background and compared with the $W'$ boson signal.

\begin{figure}
    \centering
    \includegraphics[width=0.79\linewidth]{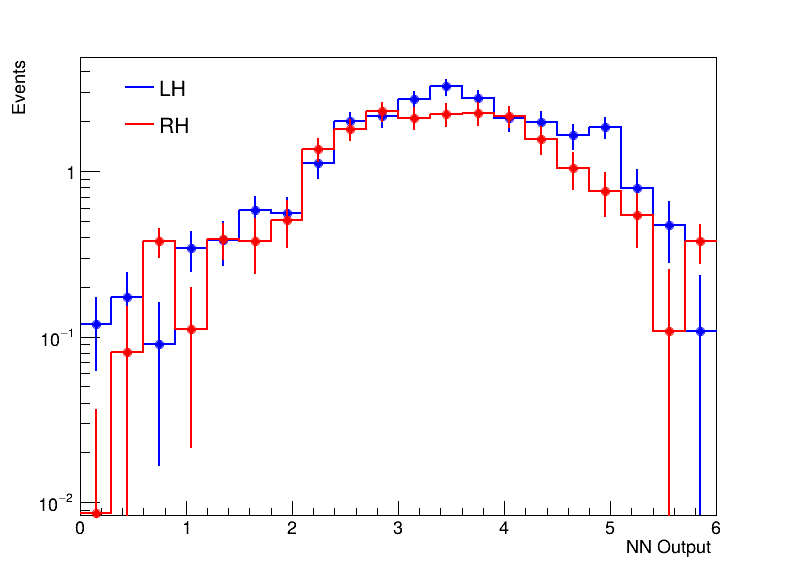} \\
    \includegraphics[width=0.79\linewidth]{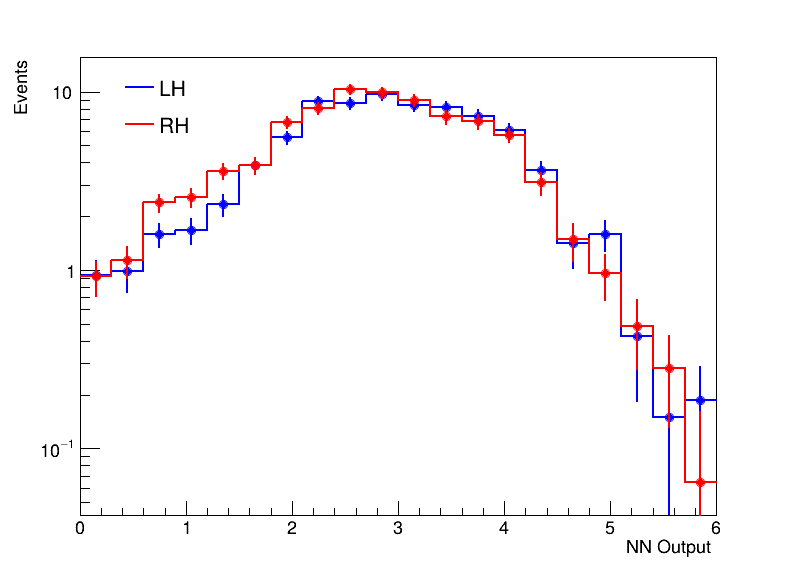}
    \caption{Shown is the comparison between purely LH and RH $W'$ boson signal events. The NN output score in SR1 is shown on the top while SR2 is shown on the bottom. This distributions are similar but not identical. The distributions are mostly consistent to within statistical uncertainty.}
    \label{fig:LH_RH_nn}
\end{figure}

The resulting distributions can be found in Fig.~\ref{fig:LH_RH_nn} which compares the samples of the two chiralities in the SR1 and SR2. 

\chapter{Statistical Analysis}
The statistical interpretation of the observed data is initiated by evaluating whether there exists a statistically significant excess of events relative to the predicted SM background. In the absence of such an excess, an upper limit is set on the production cross section of the signal process, $\sigma(pp \rightarrow tbW')$, using a model-dependent approach based on the modified frequentist method, $CL_s$~\cite{Read:2002hq}. Although exclusion limits are often the final results presented, they are only meaningful once it has been demonstrated that the observed data is consistent with the SM-only hypothesis, i.e., that no significant deviation is observed.

A basic strategy for identifying BSM signals involves a cut-based counting experiment, where events are selected using optimized criteria designed to maximize the signal-to-background ratio. The number of selected events observed in data is then compared to the SM background expectation, which includes both statistical and systematic uncertainties. A significant excess in the observed yield relative to the expected background may indicate potential evidence for new physics. In the absence of such an excess, exclusion limits are placed on the parameter space of the signal model under consideration~\cite{Cowan:2010js} at the 95\% confidence level (CL).

However, in the current precision-driven era of particle physics, more sophisticated statistical tools are required to extract the full sensitivity of the data. Rather than relying solely on event counts, analyses typically exploit the shape information of discriminating variables by using binned distributions spanning multiple kinematic regions. In this work, the statistical inference is performed via a binned profile likelihood fit~\cite{Cowan:2010js} over the signal regions SR1 and SR2, as defined in Chapter~5. This approach enhances the sensitivity of the analysis by incorporating the full distributional information, thereby improving the robustness and reach of the resulting constraints. \textsc{TRExFitter} \cite{Esch:2018ccs} is the software package that is used to implement the binned, maximum-likelihood fit. The binning algorithm used for binning for SR1 and SR2 is called TransfoD \cite{Calvet:2017thesis}. With the selected options for this analysis, this algorithm merges bins together, but enforces no more than 25\% of signal and no more than 20\% of total SM background to be in any bin. The algorithm assists in selecting an optimal binning scheme that avoids the loss of sensitivity associated with overly coarse binning, while also mitigating issues arising from overly fine binning, such as low or negative expected event counts in signal or background distributions.

\section{Blinding Procedure}
A blinding procedure is used in this analysis to prevent biased results while searching for new particles or rare phenomena. The purpose of blinding is to obscure the signal-sensitive results until all data selection, calibration, and analysis techniques have been finalized. This ensures that methods are not subconsciously tuned to produce a desired outcome.

In this analysis, the neural network output for data in the two signal regions (SR1 and SR2), which are sensitive to the heavy-philic $W'$ boson, is blinded. All optimization steps, including event selection, classifier architecture, and hyperparameters, are performed using only simulation and an Asimov pseudo-dataset. Once the full analysis is fixed, an Asimov fit is performed to validate the statistical model and ensure that the fit behaves as expected. Only after these steps are completed is the final fit to data in SR1 and SR2 performed.

To verify that the background modeling is well understood, the $t\bar{t}$ control region is unblinded throughout the analysis. This region has negligible signal contribution and offers insight into the dominant $t\bar{t}$ background. Blinding is crucial in analyses like this because the statistical nature of the experiment makes results susceptible to small but impactful biases.

\section{Profile Likelihood Fit and Nuisance Parameters}

To test for the presence of a heavy-philic \( W' \) in Run 2 data collected from the ATLAS detector, a binned maximum likelihood fit \cite{Cowan:2010js} is performed across all analysis regions simultaneously. The fit is done on the binned distribution of the neural network (NN) output, and conducted separately for each mass hypothesis. Two unconstrained normalization factors are included to estimate the normalizations of the \( t\bar{t}+\geq1b \) and \( t\bar{t}+\geq1c \) backgrounds, and another to estimate the normalization of the \( t\bar{t}+\text{light}\) background. The parameter of interest is the signal significance, defined as the production cross section \( \sigma(pp\rightarrow tb W') \).  

To estimate the signal strength, the likelihood function \( \mathcal{L}(\mu, \theta) \) is defined as a product of Poisson probability terms, with one term per bin of the neural network (NN) output distribution in each analysis region. The expected event yield in each bin depends on the signal strength \( \mu \) and a set of nuisance parameters \( \theta = (\theta_1, \theta_2, \dots, \theta_N) \), which encode systematic effects and per-bin statistical uncertainties in the simulated samples. Shape uncertainties are implemented as correlated bin-by-bin distortions of the NN distributions. These uncertainties account for detector calibration, theoretical modeling, and generator-level variations All nuisance parameters are modeled using Gaussian or log-normal probability density functions. The fit includes approximately 150 such parameters, with minor variations across signal hypotheses. Their impact on the signal strength is propagated through the profiling procedure, and their relative influence is assessed via nuisance parameter ranking, defined by the shift in the best-fit signal strength when each parameter is individually fixed to its nominal value.

The negative log-likelihood function is then defined as  

\begin{equation}
    -\log\mathcal{L}(\hat{\mu}, \hat{\theta}_1, \hat{\theta}_2, \dots, \hat{\theta}_N) = \min \{-\log\mathcal{L}(\mu; \vec{\theta}) : \mu, \theta_1, \theta_2, \dots, \theta_N \in \mathbb{R} \}.
\end{equation}

\section{Exclusion Limit Calculation Using a Test Statistic}

To extract the 95\% confidence level upper exclusion limit on the signal strength \( \mu \), a likelihood-based test statistic is employed \cite{Cowan:2010js}. The signal strength serves as the parameter of interest (POI) in this analysis and is the free-floating normalization factor (NF) for the $W'$ boson signal samples. The test statistic is defined as  

\begin{equation}
    \tilde{t}_\mu =
    \begin{cases} 
        -2\ln\frac{\mathcal{L}\left(\mu,\hat{\hat{\theta}}\right)}{\mathcal{L}\left(0,\hat{\hat{\theta}}(0)\right)}, & \text{if } \hat{\mu} < 0 \\
        -2\ln\frac{\mathcal{L}\left(\mu,\hat{\hat{\theta}}\right)}{\mathcal{L}(\hat{\mu},\hat{\theta})}, & \text{if } \hat{\mu} \geq 0 
    \end{cases}
\end{equation}

where:  
\begin{itemize}
    \item \( \mathcal{L}(\mu, \theta) \) is the likelihood function for a given signal strength \( \mu \) and a set of nuisance parameters \( \theta \).  
    \item \( \hat{\mu} \) and \( \hat{\theta} \) are the values of the signal strength and nuisance parameters that maximize the likelihood function.  
    \item \( \hat{\theta}(\mu) \) represents the values of the nuisance parameters that maximize the likelihood function for a fixed value of \( \mu \).
    \item   Here \(\hat{\hat{\theta}}(0)\) and \(\hat{\hat{\theta}}\) refer to the conditional ML estimators of \(\theta\)  given a strength parameter of 0 or \(\mu\), respectively. Thus \(\hat{\hat{\theta}}(0)\) represents the set of nuisance parameters that maximizes the likelihood function for the background only hypothesis, and \(\hat{\hat{\theta}}\) represents the set of nuisance parameters that maximizes the likelihood function for the background plus signal hypothesis given a signal strength \(\mu\). By profiling out the nuisance parameters, the method accurately incorporates uncertainties from background normalizations, detector effects, and other systematic sources.

\end{itemize}

The test statistic \( \tilde{t}_\mu \) follows the profile likelihood ratio approach, which quantifies how well a hypothesized signal strength \( \mu \) agrees with the observed data compared to the best-fit value \( \hat{\mu} \). Larger values of \(\tilde{t}_{\mu}\) indicates increasing incompatibility between data and the hypothesized value of \(\mu\). The two cases in the definition ensures the proper treatment in the case where \(\hat{\mu}<0\) in models where this is unphysical.  If the best-fit signal strength \( \hat{\mu} \) is negative, the likelihood ratio is computed relative to the background-only hypothesis (\( \mu = 0 \)).  If \( \hat{\mu} \) is non-negative, the likelihood ratio is taken relative to the best-fit signal strength, ensuring the most optimal constraint on \( \mu \).  This approach maximizes over all nuisance parameters for each assumed signal strength, allowing for a comprehensive treatment of systematic uncertainties. 

The observed value of \( \tilde{t}_\mu \) is compared against the distribution \(P(\tilde{t}_{\mu})\) to determine the probability, also known as the \( p \)-value, of obtaining a test statistic at least as \( \tilde{t}_\mu \) can be found in Ref.~\cite{Cowan_2011}.

The upper limit on \( \mu \) at 95\% CL is determined by finding the largest signal strength \( \mu_{\text{up}} \) such that the probability of obtaining a test statistic more extreme than the observed one in the background-only pseudo-experiments is at most 5\%. In other words, \( \mu_{\text{up}} \) satisfies  

\begin{equation}
    \int_{\tilde{t}_\mu}^{\infty} P(\tilde{t}_\mu | 0) d\tilde{t}_\mu = 0.05.
\end{equation}

This ensures that if the true signal strength is equal to \( \mu_{\text{up}} \), the data would yield a stronger exclusion (larger \( \tilde{t}_\mu \)) in only 5\% of cases, thus setting a limit at the 95\% confidence level. This statistical approach provides a well-defined criterion for determining the exclusion limit based on the observed data and expected background fluctuations.

However, there are situations where the p-value may fall below the exclusion threshold, even when neither the signal-plus-background (\( s+b \)) nor the background-only (\( b \)) hypothesis provides a satisfactory fit to the data. A simple example of this occurs when background events are systematically underestimated in a counting experiment, leading to misleading exclusions. 

To mitigate this issue, a common refinement is the \( \text{CL}_s \) method \cite{Read:2002hq} is used which modifies the p-value definition to ensure a more conservative rejection of the \( s+b \) hypothesis:

\begin{equation}
    p_{s+b} = \frac{p_s}{1 - p_b} < 0.05,
\end{equation}

where $p_s$ represents the p-value for the signal-plus-background hypothesis ($\mu > 0$), and $p_b$ corresponds to the background-only hypothesis ($\mu = 0$). The denominator, $1 - p_b$, quantifies the sensitivity of the statistical test.

\section{Detector Systematic Uncertainties}

Systematic uncertainties exist for the reconstruction of the physics objects used in this analysis which include jets, leptons, and \(E_T^{\text{miss}}\) . To account for these uncertainties, calibrations in the Monte Carlo samples are adjusted according to the prescribed uncertainty. The variations that follow in the physics objects are then carried throughout the analysis which includes new event selections, rerunning of the neural networks, and resulting in new final NN distributions. It is required to rerun the entire framework. For example, if the rapidity calibration altered then some events which previous passed event selection may not pass with the altered rapidity calibration or vice versa. 

The two main components of systematic uncertainties associated with jets are jet energy scale (JES) and jet energy resolution (JER). The jet energy scale uncertainty reflects the degree to which the calorimeter’s response to a particle at a given energy is understood. This uncertainty varies with transverse momentum and pseudorapidity. There are many different sources that comprise of the JES uncertainty such as pseudorapidity calibrations, and flavor response \cite{PERF-2016-04}. To reduce the total number of uncertainties, the covariance matrix of these uncertainties are diagonalized and the most important eigenvectors are kept. 

The jet energy resolution uncertainty is how precisely energy of the jet is measured. Parts of this uncertainty come from things like electronic noise and pile-up \cite{JETM-2023-08}. In much the same way as JES, effective NPs are constructed to model the systematic uncertainties that come from the resolution of each jet.  

Systematic uncertainties related to leptons ($e$ and $\mu$) arise from imperfect knowledge of the detector performance and corrections applied to simulation. These uncertainties are primarily associated with the trigger selection, the object reconstruction, identification and isolation criteria, and the lepton momentum scale and resolution, and track-to-vertex association (TTVA). Efficiency scale factors, typically derived from tag-and-probe techniques, are applied to simulated events to match data, and the uncertainties on these scale factors are propagated through the analysis \cite{2016_muonreco_perform,EGAM-2018-01}. They account for statistical limitations in control samples, differences between generators, background modeling, and potential mismodeling in specific kinematic regions such as low $p_T$ or high $\eta$. In addition, variations related to the combination of inner detector and muon spectrometer information are included. All electron and muon related systematic uncertainties are incorporated as nuisance parameters in the statistical fit, affecting both event yields and shapes, and are treated coherently across control, validation, and signal regions to ensure proper estimation of their impact on the final result.

The \(E_T^{\text{miss}}\) object is constructed from jets, leptons, as well as the energy not associated with any reconstructed object (soft terms). Therefore, the uncertainty on this object can be divided into two components. One that stems from the physics objects detailed above (leptons and jets), while the other stems from soft terms. These soft terms have uncertainty on their resolution and scale which is included in this analysis \cite{2018_met_perf,ATLAS-CONF-2018-023}.

A total of 108 detector-related systematic uncertainties are considered in the final fit. Additional details are provided in Section 6.6, and a summary of the systematic uncertainties that are pruned prior to inclusion in the fit is shown in Fig.~\ref{tab:detect_summary}.

\section{$t\bar{t}$ Systematic Uncertainties}
To account for the various sources of uncertainty on the dominant $t\bar{t}$ background different strategies are employed. A summary of this can be found in Table~\ref{tab:ttbar_sys_sum}.

Systematic uncertainties are extracted by comparing the nominal results to the results of different MC samples with different settings. The nominal Powheg+Pythia sample is compared to the Powheg+Herwig sample to assess the effect of the PS and Hadronization models. To account for the effects of varying $h_{\text{damp}}$ parameters on the $t\bar{t}$ samples, an additional sample with a modified $h_{\text{damp}}$ value is run through the analysis. To account for the modeling uncertainty associated with large jet multiplicity phase space of the $t\bar{t}$ samples, a $t\bar{t}$ reweighting technique similar to \cite{2021_resolvehp} is used, and two alternative distributions are used to create a nuisance parameter. One distribution is the nominal $t\bar{t}$ sample using the weights derived from the reweighting procedure. The other distribution is the nominal $t\bar{t}$ sample without these weights.

\begin{table}[]
    \centering
    \caption{Summary of the sources of systematic uncertainty for $t\bar{t} + \text{jets}$ modeling.}
    \label{tab:ttbar_sys_sum}
    \begin{tabular}{c|l|c}
        \hline \hline 
        Uncertainty Source & Description & Components \\ \hline
        $t\bar{t}$ Reweighting & Distributions with and without weights & All \\
        PS \& Hadronization & Powheg+Herwig & All \\
        $h_{\text{damp}}$ & Varying $h_{\text{damp}}$ parameter & All \\ \hline
        $t\bar{t}+\geq 1b/c$ Normalization & Free-floating &  $t\bar{t}+\geq 1b$,$\geq 1c$ \\
        $t\bar{t}+\text{light}$ Normalization  & Free-floating & $t\bar{t}$ light \\
        \hline \hline
    \end{tabular}
\end{table}

\section{Pruning and smoothing of systematic uncertainties}
In the fits, pruning is applied at the 1\% level, meaning that if the effect of a nuisance parameter is smaller than 1\% (separately for shape and normalization) it does not enter into the fit. This pruning procedure reduces the CPU time and helps the fit to converge. Since the pruning threshold is very small, this has no impact to the final fit results.

 A table which summarizes the pruning of the detector systematic uncertainties is shown in Fig.~\ref{tab:detect_summary}. This table only shows the pruning for the majority SM background, and the $W'$ boson signal. Pruning is done separately for each sample in each of the two signal regions.

\begin{table}[ht]
\centering
\caption{Number of systematic uncertainties that get pruned before entering into the fit. Only the majority background samples and signal sample are shown. All detector systematic uncertainties for $t\bar{t}H$ and $t\bar{t}V$ are excluded in both regions.}
\begin{tabular}{lcccc}
\hline \hline
 & $W'$ & $t\bar{t}+>1b$ & $t\bar{t}+>1c$ & $t\bar{t}+\text{light}$ \\ \hline
Kept in both regions          & 9  & 51 & 56 & 57 \\
Shape removed in 1 region    & 13 & 52 & 2  & 1  \\
Shapes removed in both regions & 37 & 5  & 50 & 41 \\
Excluded in 1 region          & 1  & 0  & 0  & 9  \\
Excluded in both regions      & 48 & 0  & 0  & 0  \\
\hline \hline
\end{tabular}
\label{tab:detect_summary}
\end{table}

\begin{figure}
    \centering
    \includegraphics[width=0.49\linewidth]{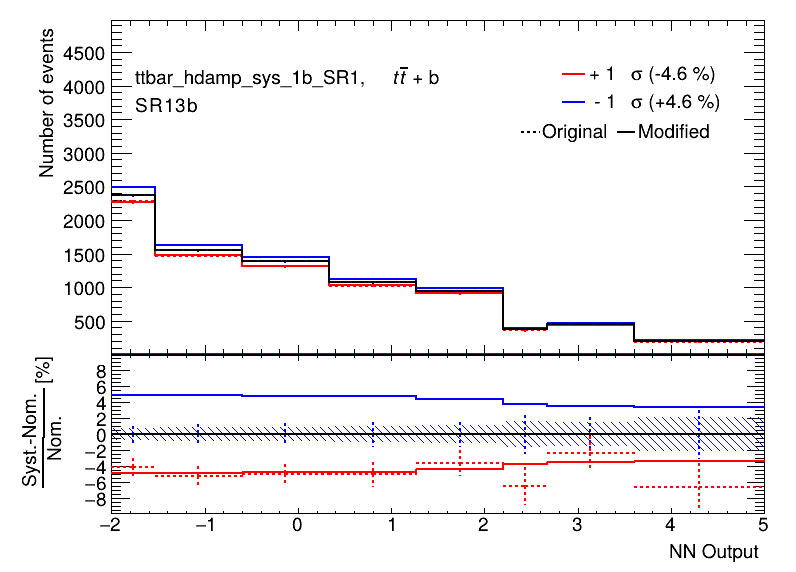}
    \includegraphics[width=0.49\linewidth]{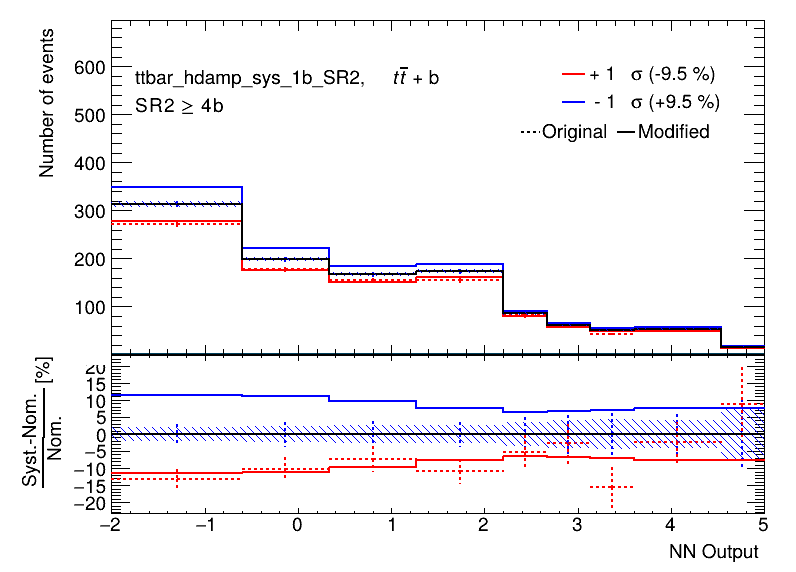}
    \caption{Example of systematic smoothing for the $t\bar{t}$ uncertainty associated with the $h_{\text{damp}}$ parameter. The left plot shows the impact in SR1, and the right plot shows the impact in SR2. This uncertainty applies only to the $t\bar{t}+>1b$ sample.}
    \label{fig:smoothing}
\end{figure}

Smoothing is applied for systematic uncertainties on $t\bar{t}$ modeling \cite{ttbar_2025}. No smoothing is applied for modeling systematic uncertainties on small backgrounds or for experimental systematic uncertainties. An example of smoothing is shown in Fig.~\ref{fig:smoothing} which shows $h_{\text{damp}}$ parameter $t\bar{t}$ systematic uncertainty.

\section{Asimov Fit Results}
Below are the results of the fit using Asimov data where pseudo-data is artificially created from only the nominal SM background. Asimov fits are used to get the expected behavior by fitting to pseudo-data \cite{Cowan:2010js}. This section only shows the results for the fit to the 1 TeV LH $W'$ boson signal. All the other Asimov fits can be seen in Appendix B. 

Fig.~\ref{fig:exp_normfactor_g2} shows the normalization factors that are extracted from the Asimov fit. These normalization factors show that the analysis and statistical fit is sensitive enough to test for the presence of the $W'$ boson signal. The POI, $\mu$, is fitted to 0 as expected, with a low uncertainty and rejects the nominal value of 1.

\begin{figure}[h]
    \centering
    \includegraphics[width=0.8\linewidth]{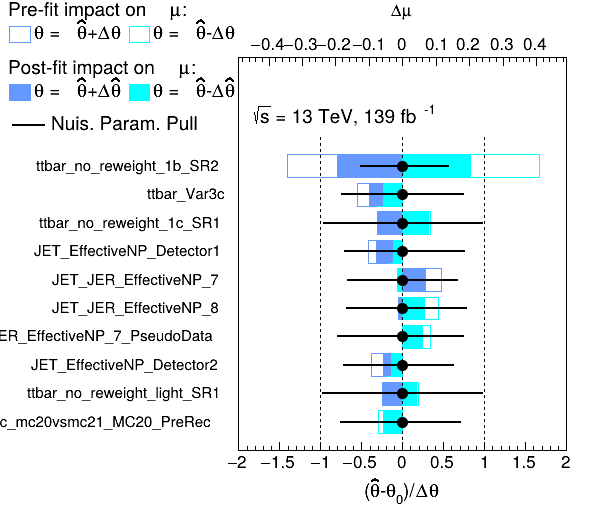}
    \caption{Ranking plot of the NPs for $\mu$ for the Asimov fit. The first systematic uncertainty is the $t\bar{t}$ reweighting systematic uncertainty in SR2 which incorporate the uncertainty of the $t\bar{t}$ reweighting correction. The second systematic uncertainty is the $t\bar{t}$ uncertainty that uses the Var3c downward variation in the PS. The next 5 systematic uncertainties involve the uncertainties of the jet variables. The ninth systematic uncertainty accounts for $t\bar{t}$ reweighting correction in SR1. The final systematic uncertainty in this figure is another uncertainty accounting for JER. }
    \label{fig:ranking_expected}
\end{figure}

Figure~\ref{fig:ranking_expected} shows the ranking of the most impactful NPs on \( \mu \) from the Asimov fit. The dominant contributions come from systematic uncertainties associated with the \( t\bar{t} \) background, with the most significant NP being the reweighting uncertainty for the \( t\bar{t}+>1b \) sample. Following the \( t\bar{t} \)-related uncertainties, the next most impactful NPs are associated with the JES and JER. This is expected as jet-based kinematic variables are central to variables in this analysis (HT for example). The output of the SB discriminator, which relies on these variables is therefore impacted by these uncertainties which degrade the separation between signal and background, thereby impacting the fit's sensitivity to \( \mu \). One-sided pre- and post-fit impact on $\mu$ means that the systematic variation ($\pm 1\sigma$) pull $\mu$ in the same direction.

\begin{figure}[h]
    \centering
    \includegraphics[width=0.49\linewidth]{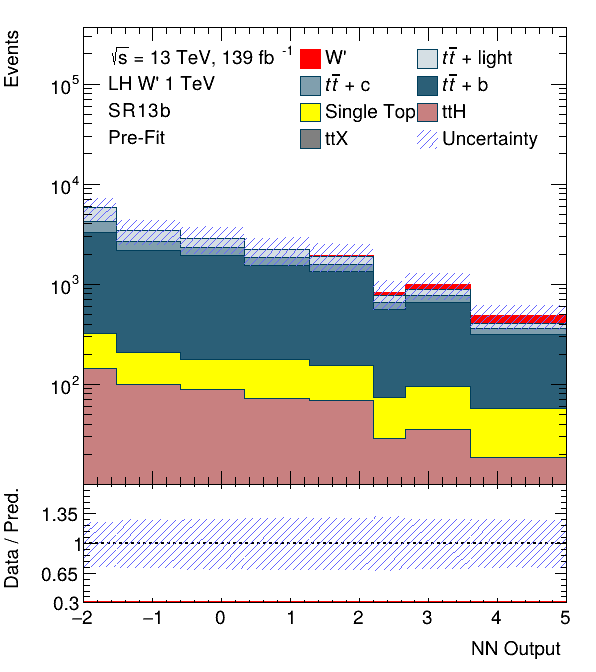}
    \includegraphics[width=0.49\linewidth]{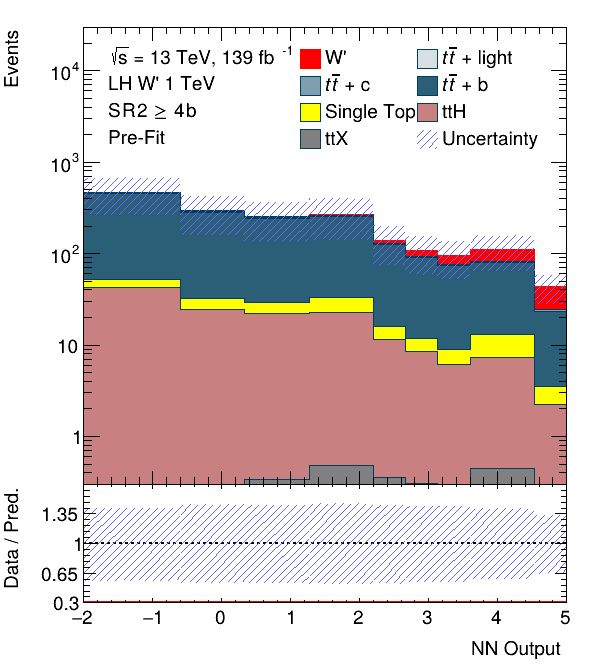}
    \caption{Asimov pre-fit plots for LH 1 TeV $W'$ boson signal sample where g'/g=2.}
    \label{fig:as_pre_1tev_wp_g2}
\end{figure}

\begin{figure}[h]
    \centering
    \includegraphics[width=0.49\linewidth]{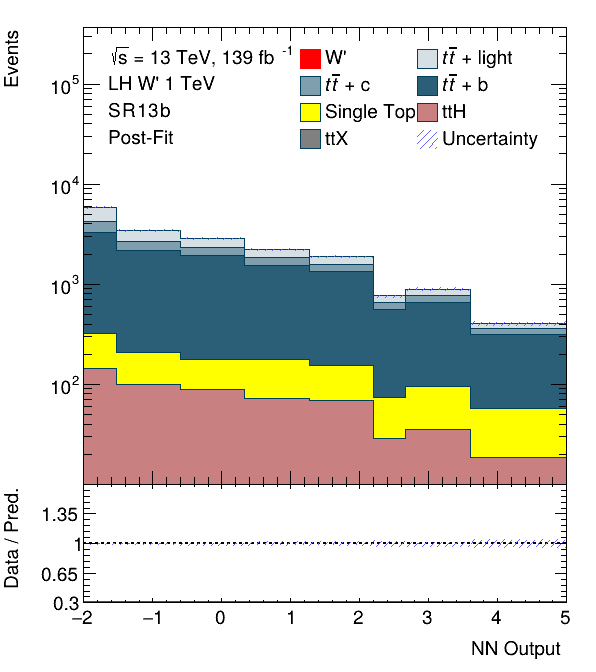}
    \includegraphics[width=0.49\linewidth]{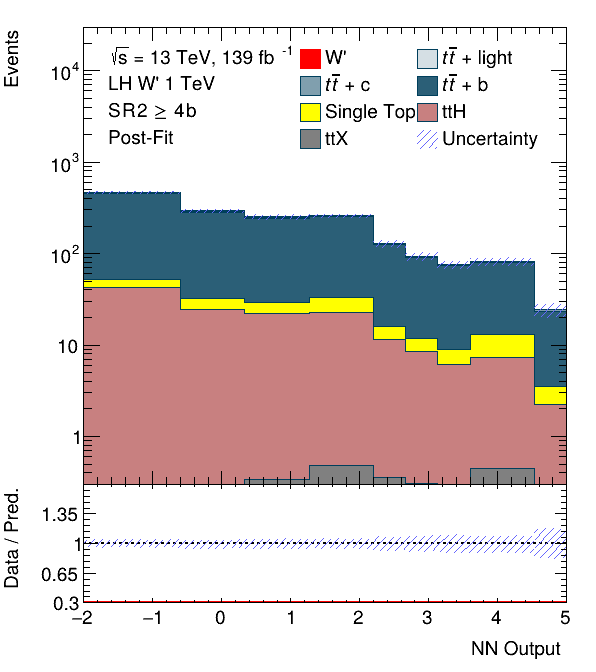}
    \caption{Asimov post-fit plots for LH 1 TeV $W'$ boson signal sample where g'/g=2.}
    \label{fig:as_post_1tev_wp_g2}
\end{figure}

\begin{figure}[h]
    \centering
    \includegraphics[width=0.49\linewidth]{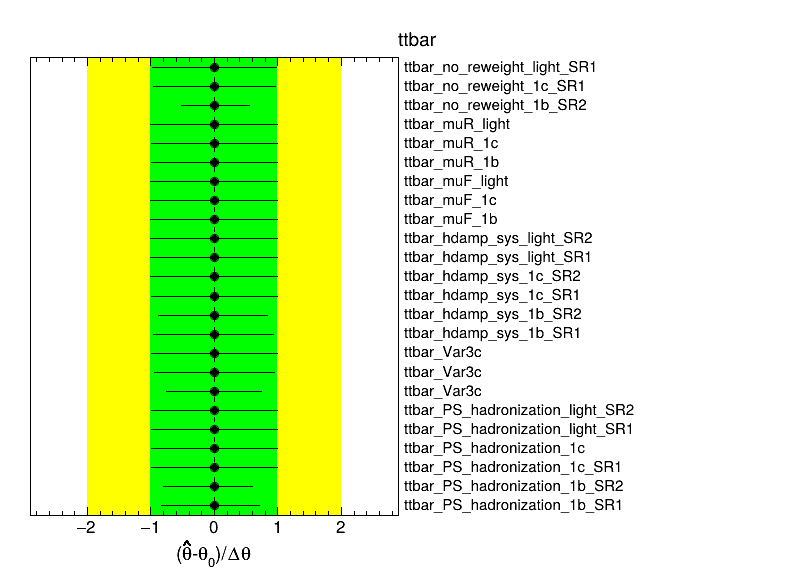}
    \includegraphics[width=0.49\linewidth]{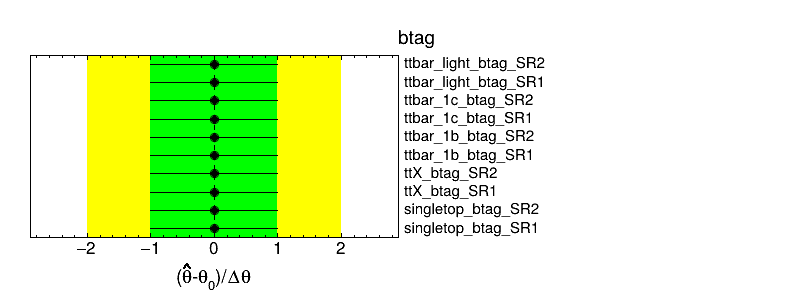}\\
    \includegraphics[width=0.49\linewidth]{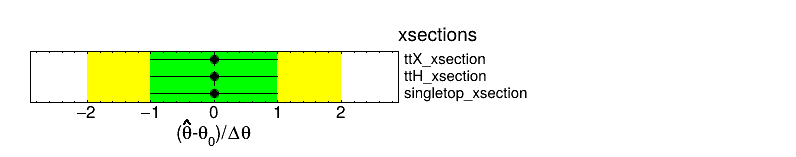}
    \caption{Systematic uncertainty pull plot for Asimov fit on LH 1 TeV $W'$ boson sample where g'/g=2. Top left is the $t\bar{t}$ NPs, top right is the btagging NPs, and bottom are the cross-section NPs.}
    \label{fig:ttbar_sys_pull_expected_g2}
\end{figure}

Fig.~\ref{fig:as_pre_1tev_wp_g2} shows the pre-fit plots that are determined for this Asimov fit. The uncertainty bands flat across the NN output distribution and as expected, there are significant signal event counts in the last few bins. Fig.~\ref{fig:as_post_1tev_wp_g2} shows the post-fit plots which show the updated uncertainty bands after the fit. Fig.~\ref{fig:ttbar_sys_pull_expected_g2} shows the systematic uncertainty pull plots from the NPs. A few of the $t\bar{t}$ systematic uncertainties are constrained, but the cross section and b-tagging NPs are not modified. 

\begin{figure}[h]
    \centering
    \includegraphics[width=0.7\linewidth]{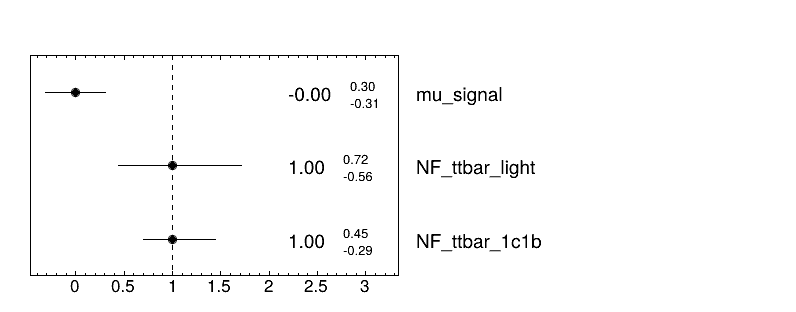}
    \caption{Normalization factors determined by the binned maximum likelihood fit for the LH 1 TeV $W'$ boson signal where g'/g=2.}
    \label{fig:exp_normfactor_g2}
\end{figure}

Several of the \( t\bar{t} \) NPs are constrained in the Asimov fit. This is primarily because the statistical uncertainties are relatively small, allowing the fit to effectively constrain those parameters. Additionally, some NPs, such as the \( t\bar{t} \) reweighting uncertainties, are conservative in their pre-fit estimates. As a result, the fit reduces their uncertainties post-fit, reflecting the fact that the data prefer a smaller variation than initially assumed.

\FloatBarrier

\section{Fit Results on Data}
This section only shows results for the LH 1 TeV $W'$ boson. The other fits to the RH $W'$ boson and all other mass points can be found in Appendix C. Fig.~\ref{fig:obs_normfactor} shows the normalization factors that are extracted from the fit to data. The fitted signal strength is very close to the results from the Asimov fit, and it is consistent with zero which shows that the data in this fit is not compatible with the presence of the 1 TeV LH $W'$. The $t\bar{t}$ normalization factors are adjusted, but as seen in the Asimov fit, the uncertainties on these normalizations are quite high. Both normalization factors are consistent within $1\sigma$ from their nominal value. 

\begin{figure}[h]
    \centering
    \includegraphics[width=0.8\linewidth]{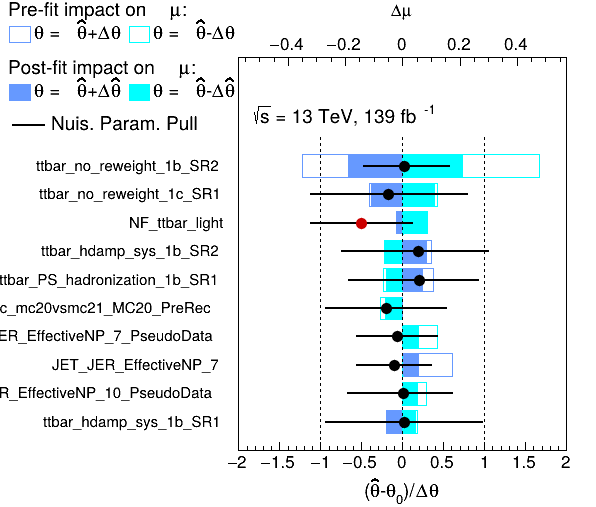}
    \caption{Ranking plot of the NPs for $\mu$ for the fit to data. The first two systematic uncertainties are the $t\bar{t}$ reweighting systematic uncertainties in the two signal regions which incorporate the uncertainty of the $t\bar{t}$ mismodeling corrections. The third systematic uncertainty is the normalization factor for $t\bar{t}+\text{light}$. The fourth and fifth systematic uncertainties are $t\bar{t}$ modeling uncertainties associated with the $h_{\text{damp}}$ parameter and the PS and hadronization calculations. The sixth and seventh systematic uncertainties are associated with JER. The final systematic uncertainty is another uncertainty associated with the $h_{\text{damp}}$ parameter.}
    \label{fig:ranking_observed}
\end{figure}

Fig.~\ref{fig:ranking_observed} shows a ranking plot for the most impactful NPs on $\mu$ for the fit to data. As expected, the NPs that have the highest effect on $\mu$ are the $t\bar{t}$ systematic uncertainties. More specifically, the $t\bar{t}$ reweighting NPs are the top two most important NPs. All NPs are consistent with their nominal value to within $1\sigma$. It can also be seen in the Asimov fit, that JER is an important NP in this analysis. This figure also shows that the fit to data was not able to constrain the systematics as much as the fit to Asimov data. 

\begin{figure}[b]
    \centering
    \includegraphics[width=0.49\linewidth]{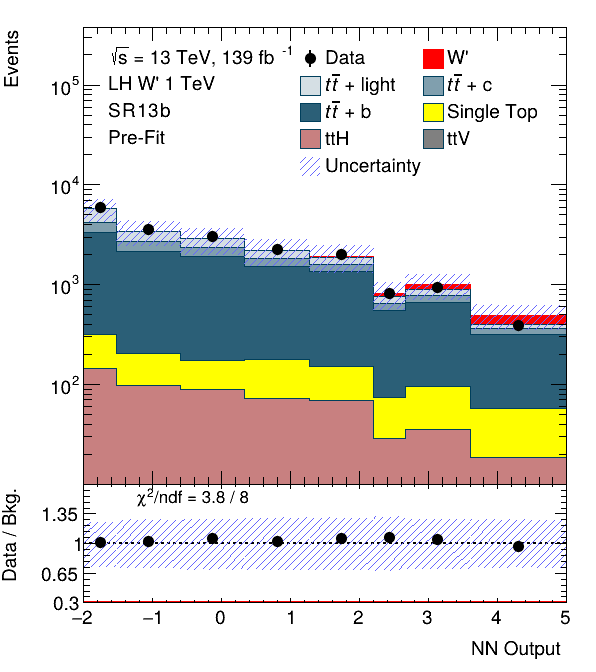}
    \includegraphics[width=0.49\linewidth]{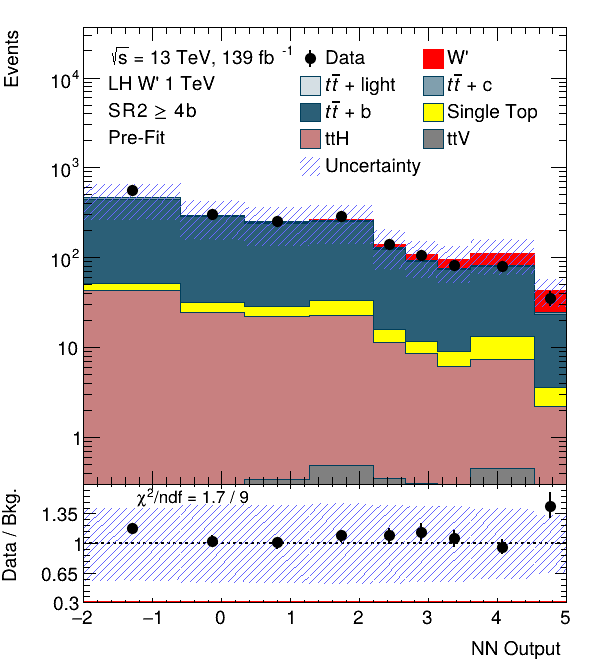}
    \caption{Pre-fit plots to data for LH 1 TeV $W'$ boson signal sample where g'/g=2.}
    \label{fig:obs_pre_1tev_wp}
\end{figure}

\begin{figure}[h]
    \centering
    \includegraphics[width=0.49\linewidth]{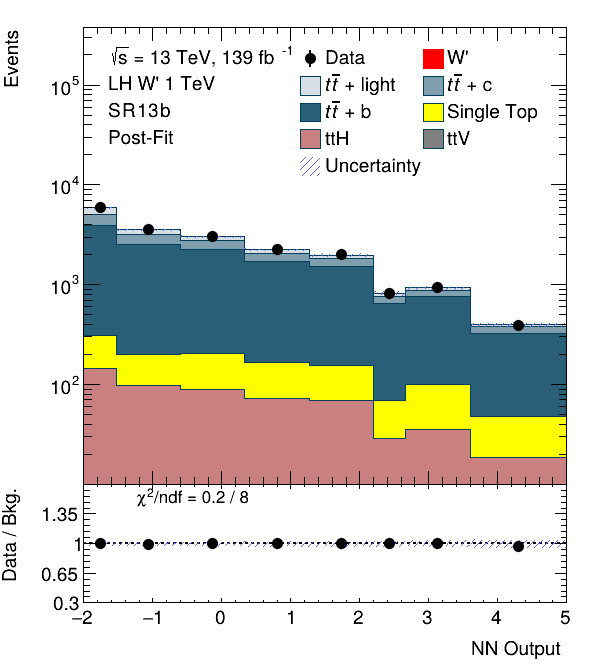}
    \includegraphics[width=0.49\linewidth]{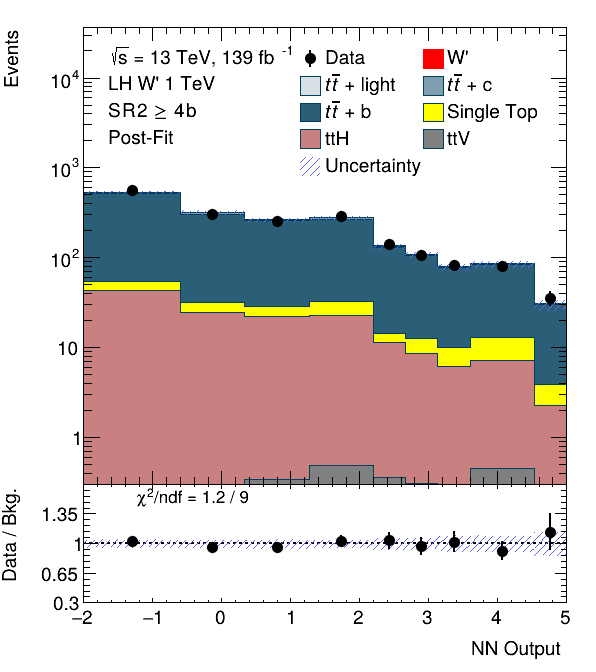}
    \caption{Post-fit plots to data for LH 1 TeV $W'$ boson signal sample where g'/g=2.}
    \label{fig:obs_post_1tev_wp}
\end{figure}

\begin{figure}[h]
    \centering
    \includegraphics[width=0.49\linewidth]{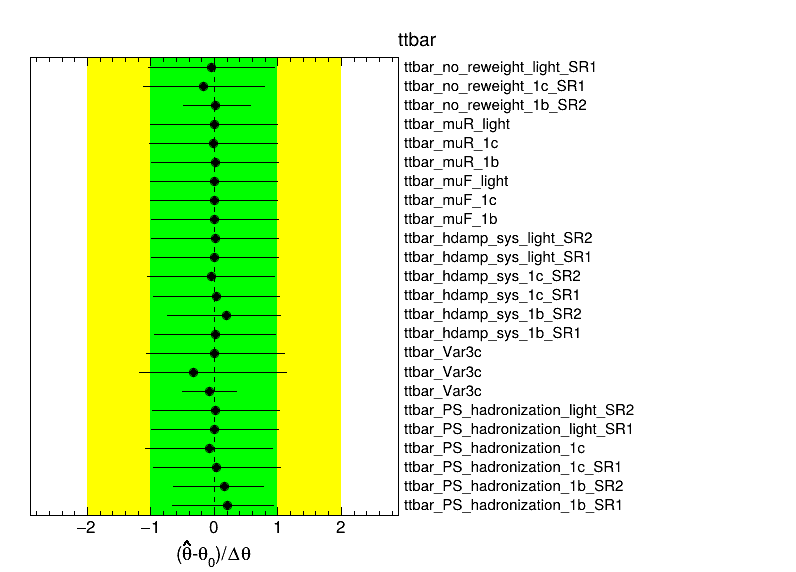}
    \includegraphics[width=0.49\linewidth]{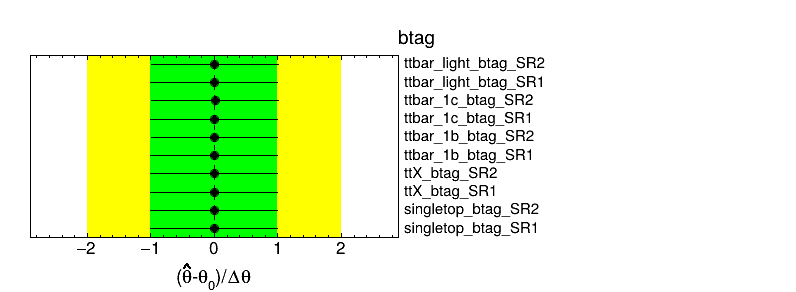}\\
    \includegraphics[width=0.49\linewidth]{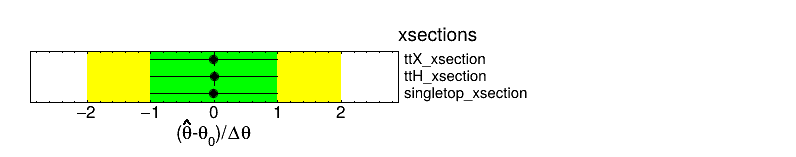}
    \caption{$t\bar{t}$ systematic uncertainty pull plot for Asimov fit on LH 1 TeV $W'$ boson sample where g'/g=2.}
    \label{fig:ttbar_sys_pull_obs}
\end{figure}

Fig.~\ref{fig:obs_pre_1tev_wp} shows the resulting pre-fit plots, which demonstrate good agreement between data and background in both signal regions. Fig.~\ref{fig:obs_post_1tev_wp} shows the post-fit plots, which show a strong reduction of uncertainty after the fit. The post-fit results also indicate good convergence of the fit. Fig.~\ref{fig:ttbar_sys_pull_obs} shows the systematic uncertainty pull plots from the $t\bar{t}$ systematic uncertainties. The $t\bar{t}$ systematic uncertainties are pulled slightly and constrained in the fit while the b-tagging and cross-section systematic uncertainties remain unchanged by the fit. 

\begin{figure}[b]
    \centering
    \includegraphics[width=0.8\linewidth]{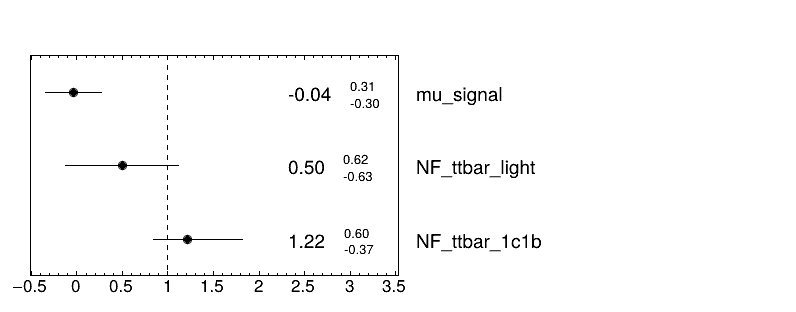}
    \caption{Normalization factors determined by the binned maximum likelihood fit for the LH 1 TeV $W'$ boson signal where g'/g=2.}
    \label{fig:obs_normfactor}
\end{figure}

The results of the the fit to data reveal some tension. The instrumental NPs are quite constrained. This can be attributed by poor SM modeling within SR1 and SR2. However, almost all NPs are consistent with their nominal value within 1 standard deviation.

Since no significant excess is observed above the SM background, 95\% CL exclusion limits are set on the signal hypotheses.

\begin{figure}[h]
    \centering
    \includegraphics[width=0.89\linewidth]{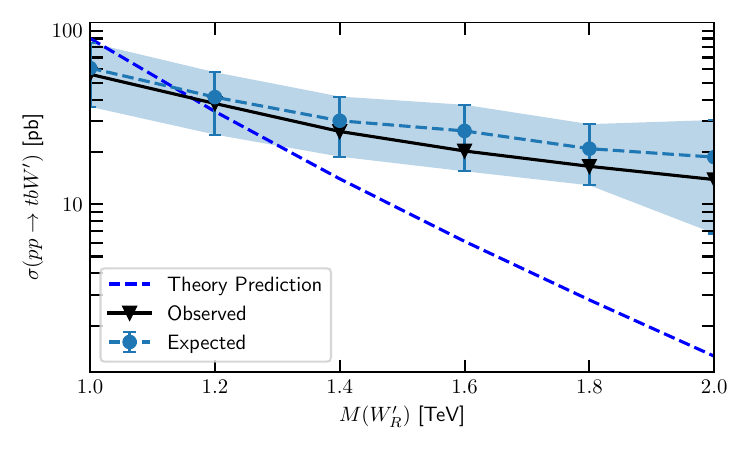} \\
    \includegraphics[width=0.89\linewidth]{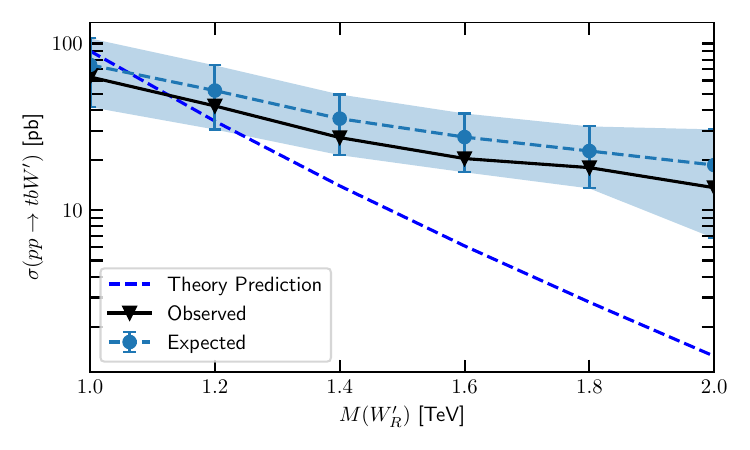}
    \caption{Limit plots derived from the 95\% CL exclusion upper limit calculation on expected background and data fits. An upper limit on the $W'$ boson mass is 1.15 TeV for the RH $W'$ boson  and 1.18 TeV for the LH $W'$ boson where $g'/g=2$.}
    \label{fig:limit_plot}
\end{figure}

Figure~\ref{fig:limit_plot} shows the limit plots for the $W'$ boson signal under the $g{\prime}/g = 2$ hypothesis. The expected limits are obtained using an Asimov dataset, while the observed limits are derived from fits to the actual data. The blue bands around the expected limit represent the $\pm1\sigma$ variations, obtained by recalculating exclusion limits using pseudo-datasets with event counts fluctuated according to Poisson statistics. The exclusion limits are evaluated only at the discrete $W'$ mass points simulated in the analysis, and the limit curves are interpolated between these points. These plots show that the observed upper limit falls within $1\sigma$ of the expected upper limit. The limit curves are quite smooth, with minor fluctuations. The exclusion curve drops as the $W'$ boson mass increases because the kinematic properties of a heavier resonance become easier to distinguish from $t\bar{t}$ events. There are also slight differences between the fits to the LH and RH $W'$ boson signal samples. As shown in Fig.~\ref{fig:LH_RH_nn}, differences in the neural network output distributions for the two cases lead to variations in the fit results.
\chapter{Conclusions}

A search for a heavy-philic $W'$ boson with couplings exclusive to the third generation of quarks is performed using ATLAS Run 2 data. This $W'$ boson is produced in association with a top and bottom quark, leading to a final state of $tbtb$. The search is conducted in the single lepton channel and targets mass ranges of $1000 < m_{W'} < 2000$~GeV. The kinematic phase space includes events where there are at least five jets and more than three of them are b-tagged.

Reconstructing the heavy-philic $W'$ presents a significant challenge due to the complex final state of two top quarks and two bottom quarks. Accurately assigning reconstructed jets to the correct decay products of the $W'$ is nontrivial, given the combinatorial ambiguity and overlapping kinematic properties. To address this, a machine learning strategy is developed to exploit subtle correlations in the event topology. The analysis uses a unique approach where two neural networks are trained. One neural network extracts high-level physics variables and focuses on fully reconstructing the $W'$. The other neural network focuses on signal and background discrimination. This architecture allows for physics-motivated interpretation at each stage in the analysis, avoiding the opacity often associated with fully end-to-end machine learning pipelines.

The output of this analysis yields binned distributions of a final neural network output score, with negative values representing SM-like events and positive values representing events that resemble heavy-philic $W'$ boson signal. A binned-maximum likelihood fit is performed on these distributions to test for the presence of the heavy-philic $W'$ boson in the ATLAS Run 2 dataset. The results of this fit show that there is no excess above the SM background.

Exclusion limits on the production cross section are set as a function of the mass of the $W'$ boson with its coupling constant set to twice the strength of the Standard Model $W$ boson ($g'/g=2$). Limits of $m_{W'} < 1.18$~TeV for a $W'$ boson with left-handed Standard-Model-like couplings and $m_{W'} < 1.12$~TeV for a $W'$ boson with right-handed couplings are excluded. These are the first limits on heavy-philic $W'$ boson production.

Further studies would benefit from the addition of jets defined with a larger radius. For a $W'$ boson with a mass of a few TeV, the resulting top quark is expected to be highly boosted, causing the jets from its decay products to overlap. Larger radius jets are therefore more suitable for top quarks with high momentum compared to the jets used in this study. These results would also benefit from the use of ongoing Run 3 data at the LHC, which provides not only a larger dataset but also a higher center-of-mass energy. The search could also be extended to lower mass ranges, where exclusion limits would be easier to establish. 

%
%
%
%
%
%
\printbibliography

%
%

\begin{appendices}
\chapter{PDF Uncertainty Studies}
With the upcoming high luminosity large hadron collider (HL-LHC) \cite{ApollinariG.:2017ojx}, the next generation of precision measurements will be made. These measurements will be extremely precise, and will require lower theoretical uncertainties. Parton distribution functions (PDFs) are becoming the more dominant theoretical uncertainty in measurements like top quark pair production. However, many other measurements will also require reduced PDF uncertainties \cite{Schwienhorst:2022yqu}. 

Colliders that will offer useful data that can significantly reduce PDF uncertainty, like the Electron Ion Collider \cite{khalek2022snowmass}, are far in the future. In the meantime, PDF uncertainty can be reduced using HL-LHC data. 

Machine learning techniques can be used to pre-process data and distill useful information to reduce uncertainty in specific regions of the parton distribution functions (PDFs) where uncertainties are currently large. Traditional approaches have incorporated one-, two-, or three-dimensional projections of the high-dimensional collider phase space into the global PDF fit~\cite{Yan_2023}. Variables such as the rapidity and longitudinal momentum ($p_Z$) of the top quark are typical examples, though these do not fully capture the available kinematic information. This appendix shows preliminary results from this ongoing effort.

A sample of $t\bar{t}$ plus one jet events with a center of mass energy of 14 TeV was generated using Madgraph at next-to-leading order (NLO) \cite{Alwall:2014hca}. A total of 7.5 million events were generated. The PDF set that was selected to study in detail was the CT18NLO PDF set \cite{Hou:2019efy}. The study looks at the truth level of $t\bar{t}j$ events without decaying the top quarks. The aim of this study is to constrain the high x region of the gluon PDF. $t\bar{t}j$ has been shown to be a process that has good potential to reduce this region of the PDFs \cite{gombas2022dependence}. 

\begin{figure}
    \centering
    \includegraphics[width=0.7\textwidth]{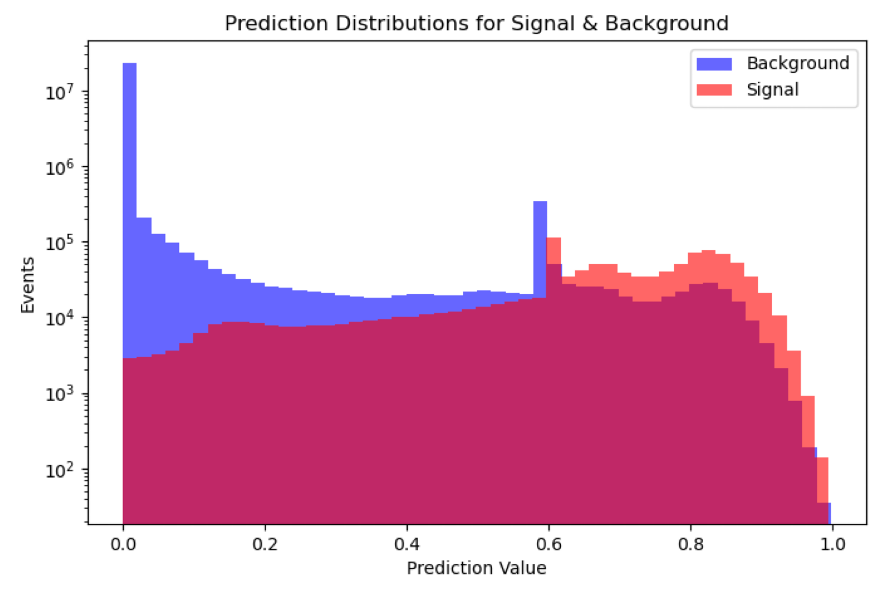}
    \caption{The MLP output scores for the trained MLP. Events that are closer to 1 are events that the MLP predicts to have an initial gluon parton whose initial momentum is greater than 2 TeV. Events that are closer to 0 are any other event. The inputs to this MLP are the kinematic 4-vectors of the final state $t\bar{t}j$. There are 3 fully connected hidden layers. The peak at about 0.6 is currently not well understood. }
    \label{fig:MLP_output}
\end{figure}

To test the idea of using machine learning to improve PDF fits, a MLP was developed to separate events with an initial gluon parton that had greater than 2 TeV longitudinal momentum. These $t\bar{t}j$ events were considered signal. Events with less than 2 TeV longitudinal momentum were considered background. The inputs to the MLP were the kinematic 4-vectors of the final state particles ($t\bar{t}j$). Decent separation was achieved which, not surprisingly, indicates that there is information about the initial colliding partons (flavor and initial momentum) in just the kinematics of the final state particles. If the MLP output score was higher than 0.7, it was considered signal and passed the MLP "filter". The MLP output scores can be seen in Fig.~\ref{fig:MLP_output}.

\begin{figure}[htp]
    \centering
    \includegraphics[width=0.3\textwidth]{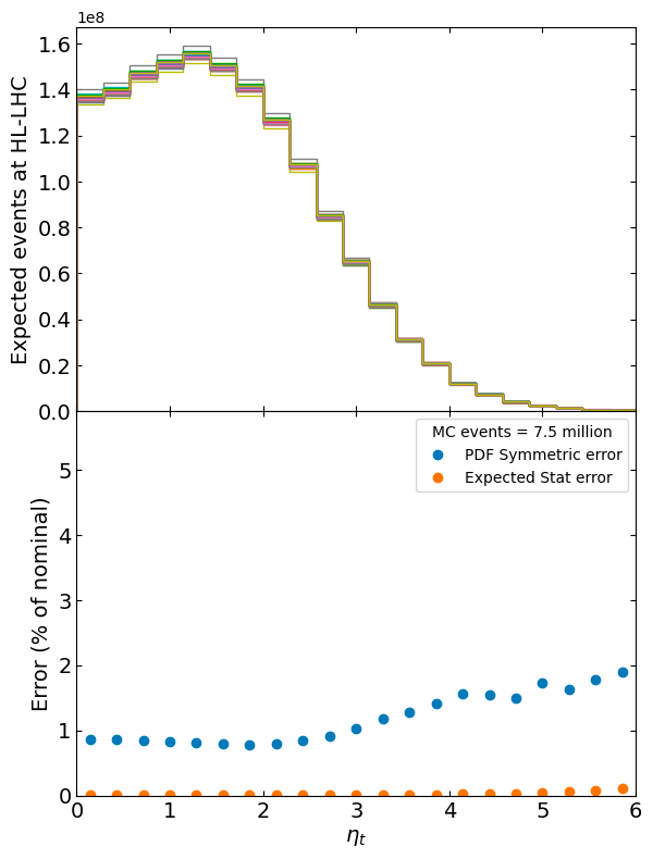}\hfill
    \includegraphics[width=0.3\textwidth]{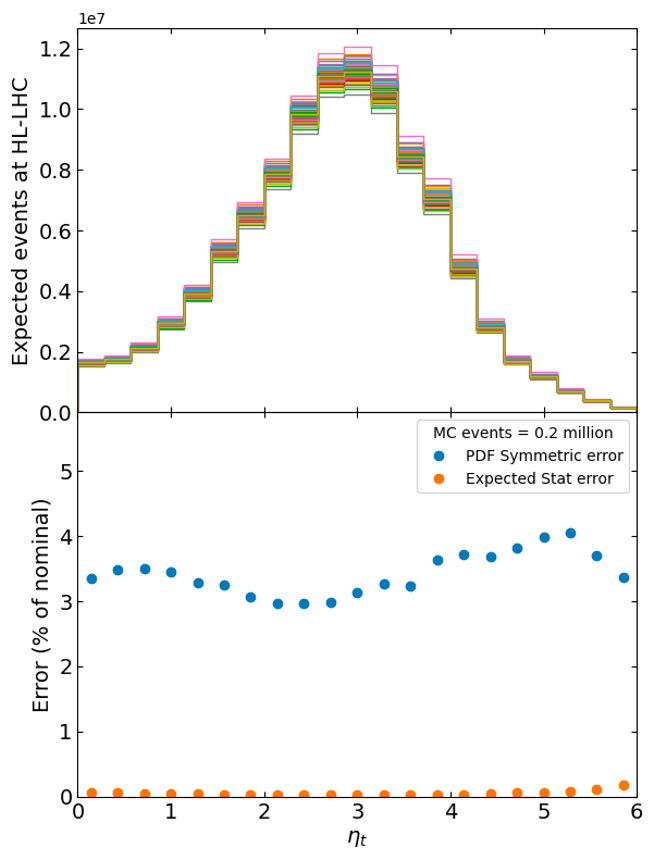}\hfill
    \includegraphics[width=0.3\textwidth]{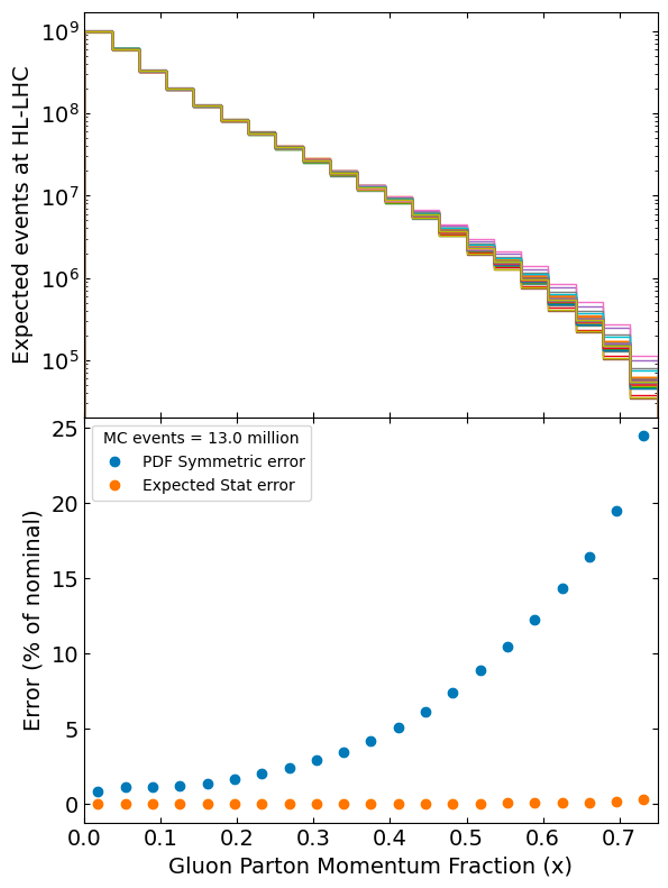}
    \caption{Differential, pseudo-data distributions that were input into ePump to update the PDF uncertainty.}
    \label{fig:epump_input}
\end{figure}

Two different differential distributions were created to compare how different methods can reduce PDF uncertainty. One histogram was filled if it passed the MLP filter and another which included every event. The rapidity of the top quark was chosen to be the kinematic variable. These differential distributions were then fed into ePump \cite{Schmidt:2018hvu} to see how much each could constrain the PDF uncertainty bands. They can be seen in Fig.~\ref{fig:epump_input} which only shows the rapidity.

A systematic uncertainty of 1\% was assigned to each bin in the distribution, correlated across all bins, approximating the expected uncertainty at the HL-LHC where the statistical uncertainty will be negligible. This can be confirmed in Fig.~\ref{fig:epump_input} which shows that the PDF uncertainty is larger than the expected statistical uncertainty.

As a best-case-scenario, the gluon PDF is directly fed into ePump with 1\% systematic uncertainties. This provides a useful upper limit to how much the gluon PDF uncertainty can be reduced with HL-LHC data. 

\begin{figure}[htp]
    \centering
    \includegraphics[width=0.3\textwidth]{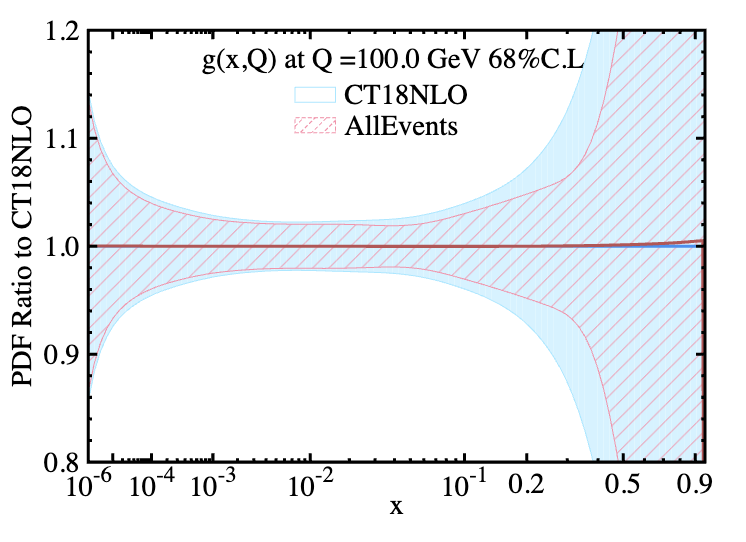}\hfill
    \includegraphics[width=0.3\textwidth]{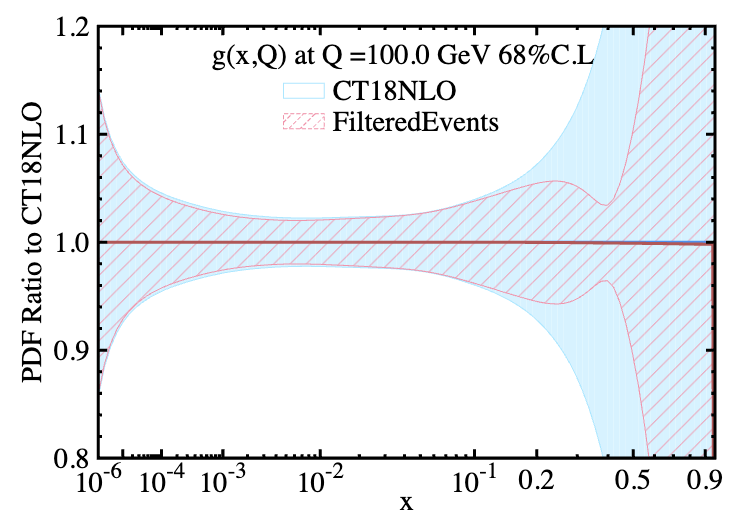}\hfill
    \includegraphics[width=0.3\textwidth]{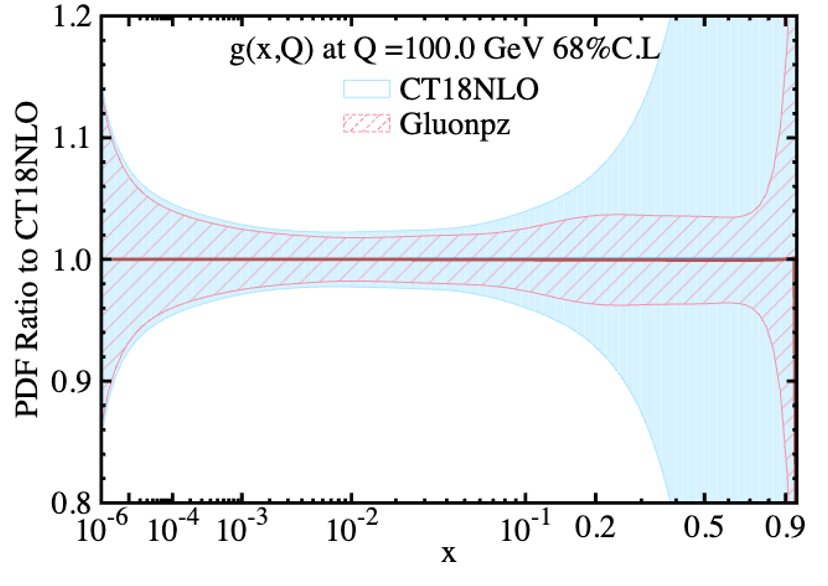}
    \caption{Updated CT18NLO PDF uncertainty bands before and after ePump updates. Left: Updating with the top rapidity distribution from every event. Center: Updating with the top rapidity distribution with events that pass the MLP filter. Right: Updating with the gluon PDF.}
    \label{fig:pdf_updates}
\end{figure}

It can be seen in Fig.~\ref{fig:pdf_updates} that the uncertainty in the high-${x}$ gluon region of the PDF set is heavily reduced when filtering  events.

The updated PDF uncertainty band of the best-case-scenario shows quite dramatic improvements in the gluon PDF uncertainty bands as expected. The uncertainty band can be narrowed up to x=0.9. This shows that there is an opportunity to improve the gluon PDF set with HL-LHC data using machine learning techniques up to a very high parton momentum fraction. 

This study shows that there is potential to reduce PDF uncertainties by forming variables with machine learning techniques because traditional techniques do not include the full information available for a given process.

\chapter{Asimov Fit Results}

This appendix includes results from all Asimov fits. Only the pre-fit and post-fit results from the fit to the 1 TeV $W'$ boson sample are shown. The pre-fit and post-fit plots for the other masses are very similar. 

\begin{figure}[h]
    \centering
    \includegraphics[width=0.8\linewidth]{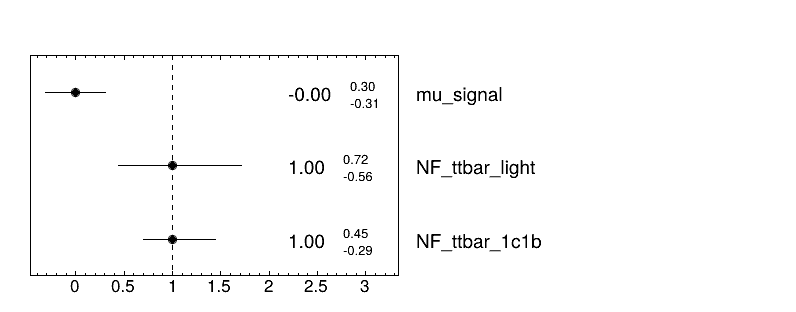}
    \caption{Normalization factors determined by the binned maximum likelihood fit for the LH 1 TeV $W'$ boson signal where g'/g=2.}
    \label{norms_LH_1tev_asimov}
\end{figure}

\begin{figure}[h]
    \centering
    \includegraphics[width=0.49\linewidth]{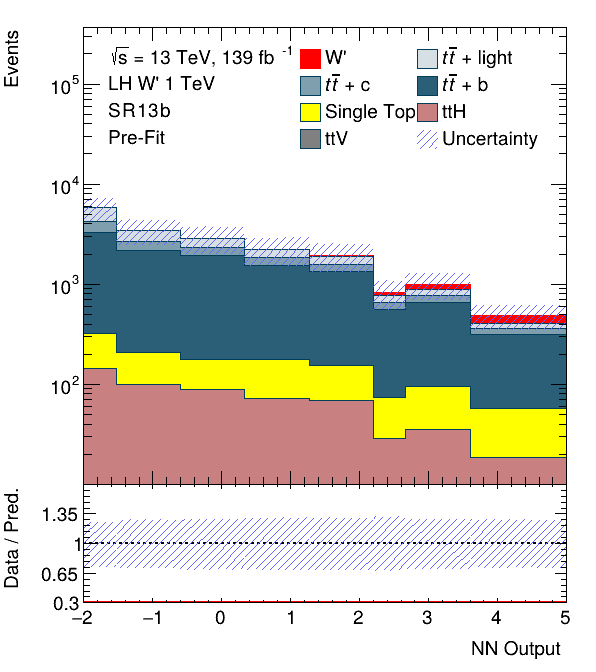}
    \includegraphics[width=0.49\linewidth]{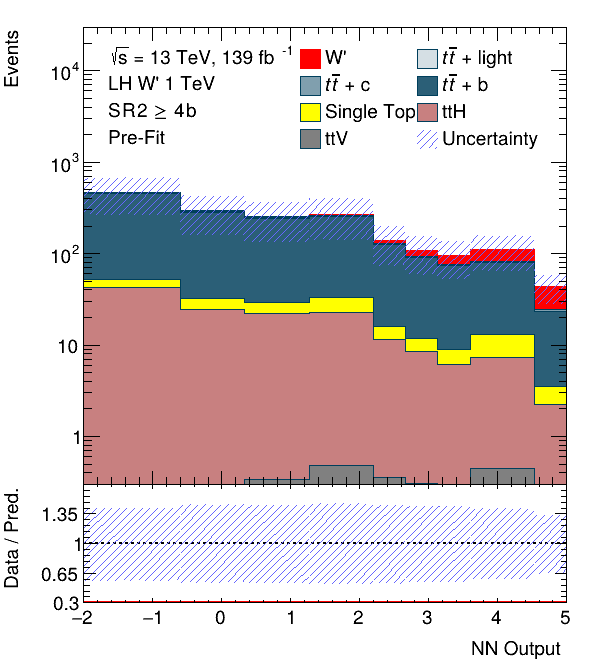}
    \caption{Asimov pre-fit plots for LH 1 TeV $W'$ boson signal sample where g'/g=2. All other mass points have similar distributions with the two signal regions, with very minor differences.}
    \label{expected_LH_1tev_prefit}
\end{figure}

\begin{figure}[h]
    \centering
    \includegraphics[width=0.49\linewidth]{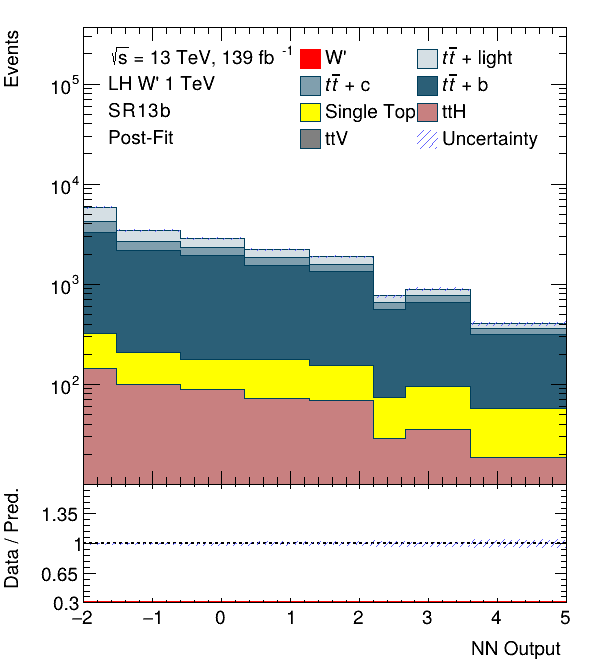}
    \includegraphics[width=0.49\linewidth]{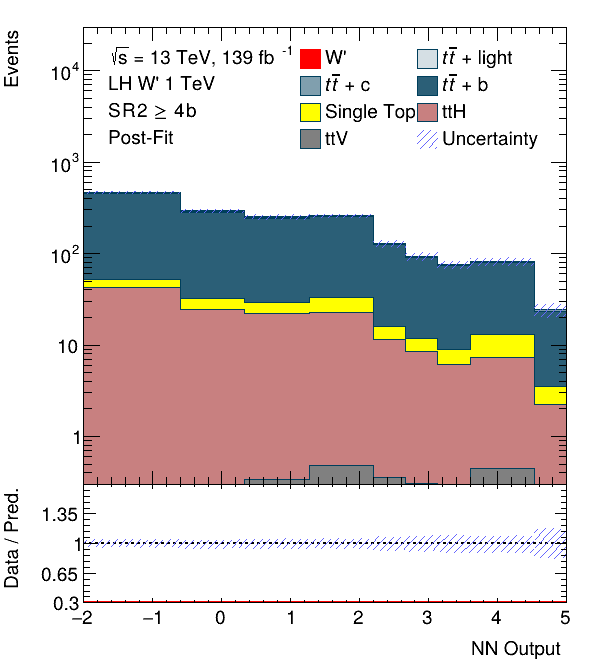}
    \caption{Asimov post-fit plots for LH 1 TeV $W'$ boson signal sample where g'/g=2. All other mass points have similar distributions with the two signal regions, with very minor differences.}
    \label{expected_LH_1tev_postfit}
\end{figure}

\begin{figure}[h]
    \centering
    \includegraphics[width=0.45\linewidth]{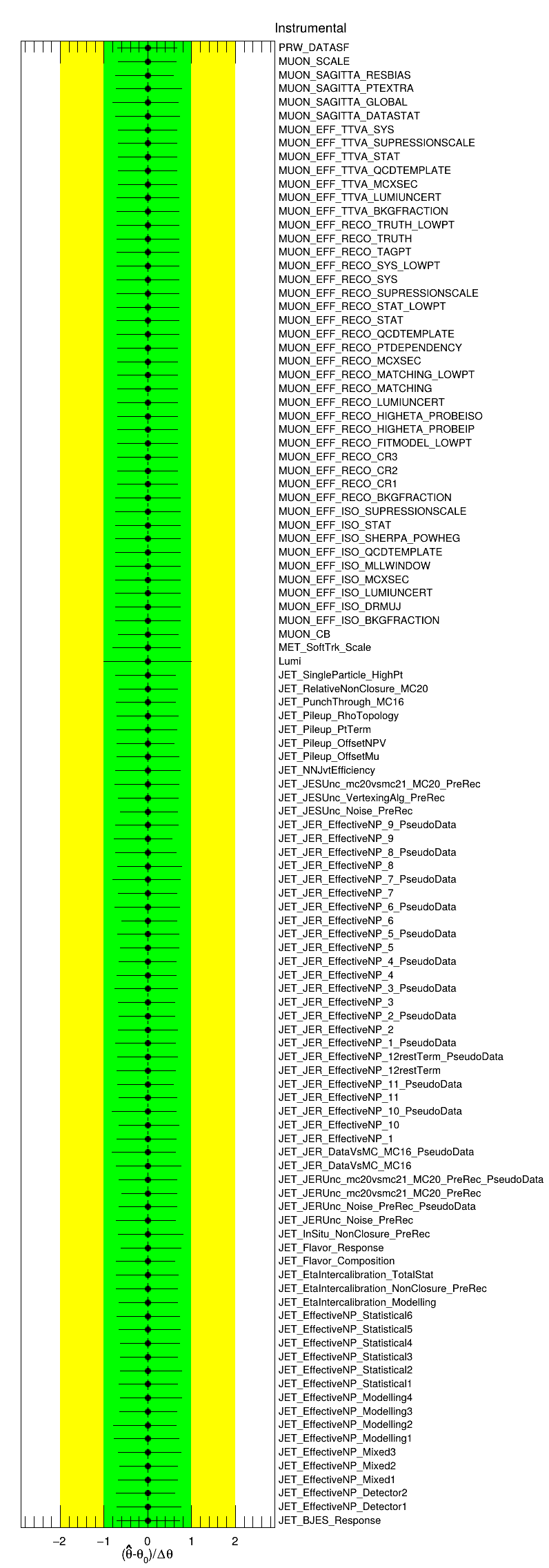}
    \caption{Detector systematic uncertainty pull plot for Asimov fit on LH 1 TeV $W'$ boson sample where g'/g=2.}
    \label{nuispar_inst_LH_1tev_asimov}
\end{figure}

\begin{figure}[h]
    \centering
    \includegraphics[width=0.49\linewidth]{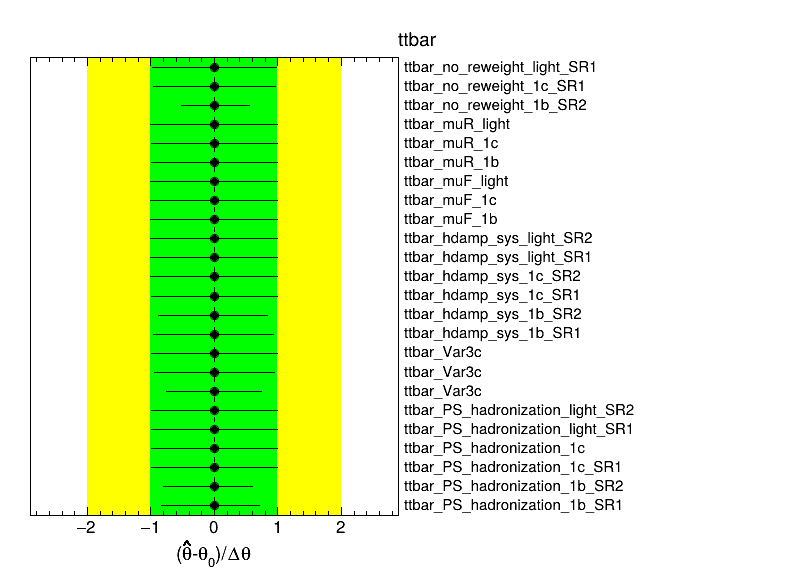}
    \includegraphics[width=0.49\linewidth]{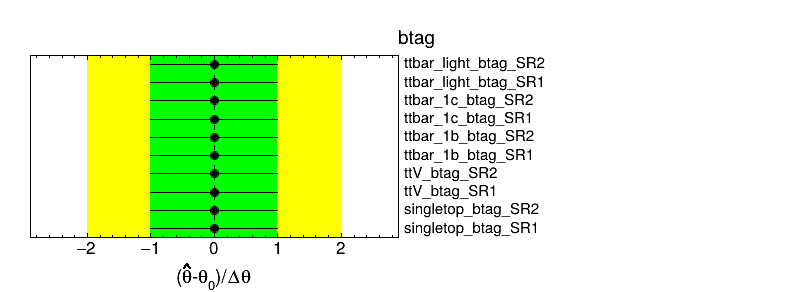}\\
    \includegraphics[width=0.49\linewidth]{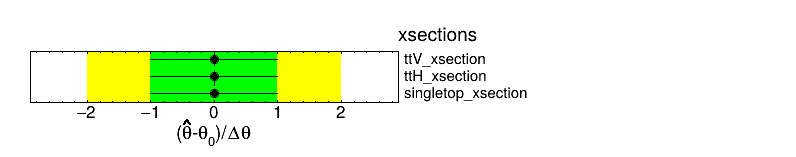}
    \caption{$t\bar{t}$, b-tagging, and cross-section systematic uncertainty pull plot for Asimov fit on LH 1 TeV $W'$ boson sample where g'/g=2.}
    \label{nuispar_all_LH_1tev_asimov}
\end{figure}

\begin{figure}[h]
    \centering
    \includegraphics[width=0.8\linewidth]{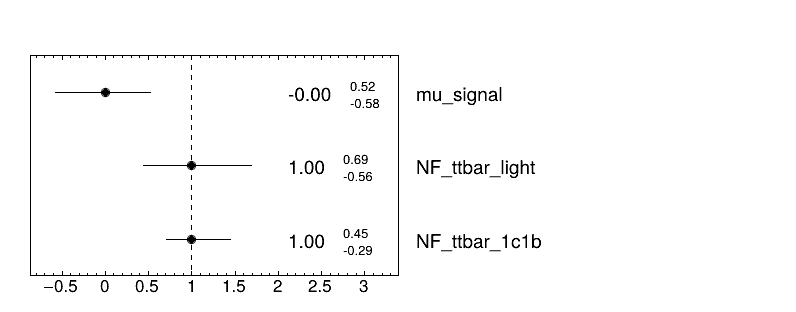}
    \caption{Normalization factors determined by the binned maximum likelihood fit for the LH 1.2 TeV $W'$ boson signal where $g'/g=2$.}
    \label{norms_LH__1p2tev}
\end{figure}

\begin{figure}[h]
    \centering
    \includegraphics[width=0.45\linewidth]{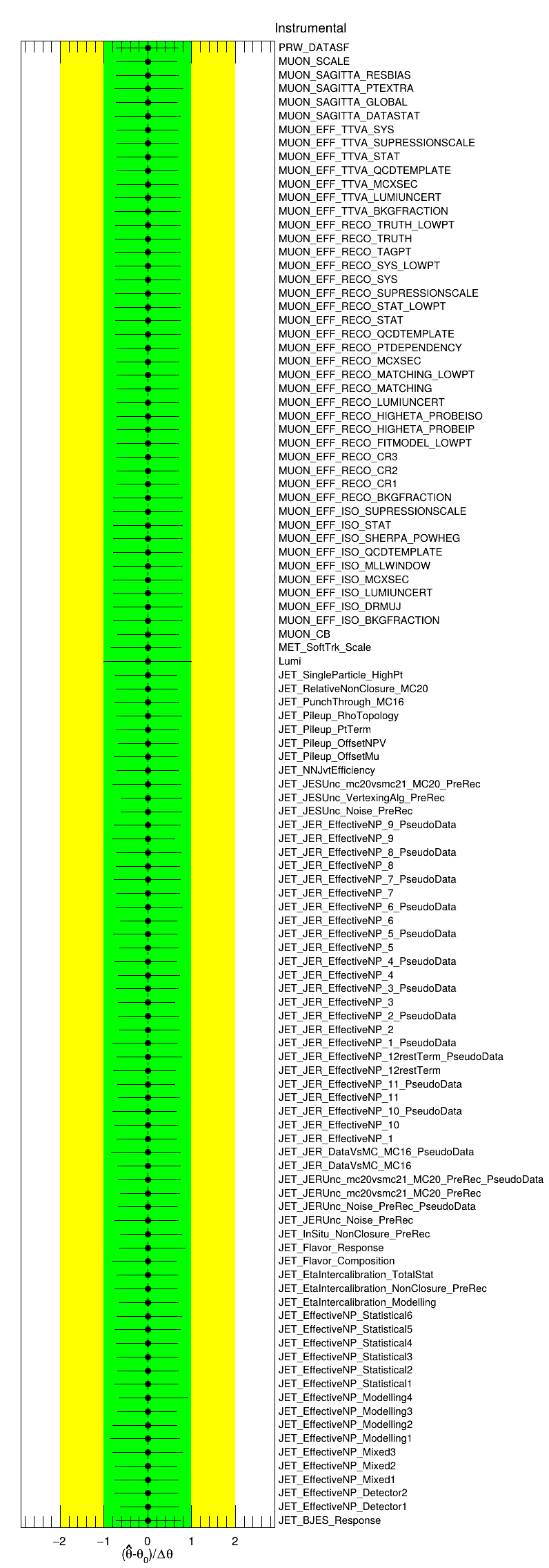}
    \caption{Detector systematic uncertainty pull plot for Asimov fit on LH 1.2 TeV $W'$ boson sample where $g'/g=2$.}
    \label{nuispar_inst_LH_1p2tev_asimov}
\end{figure}

\begin{figure}[h]
    \centering
    \includegraphics[width=0.49\linewidth]{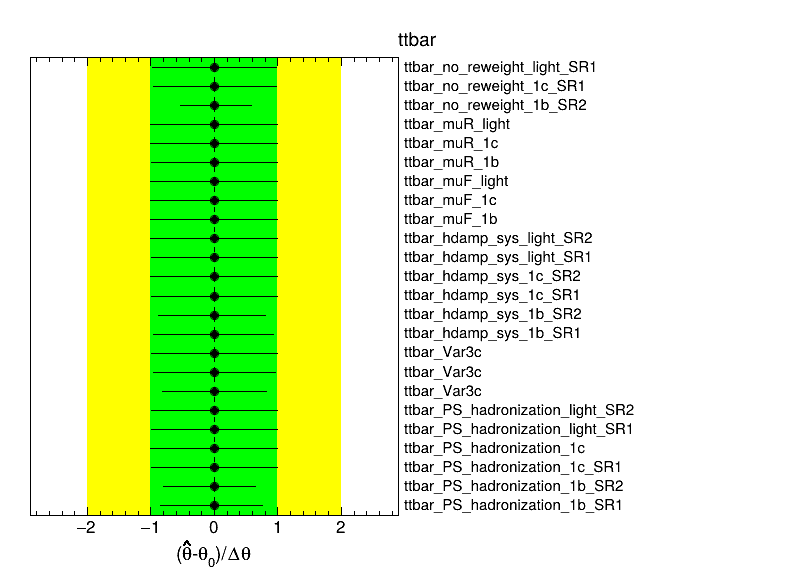}
    \includegraphics[width=0.49\linewidth]{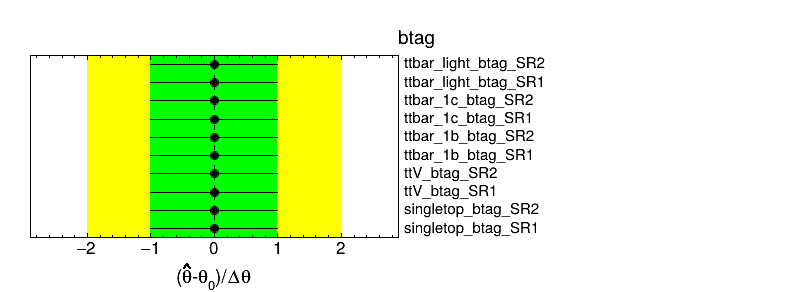}\\
    \includegraphics[width=0.49\linewidth]{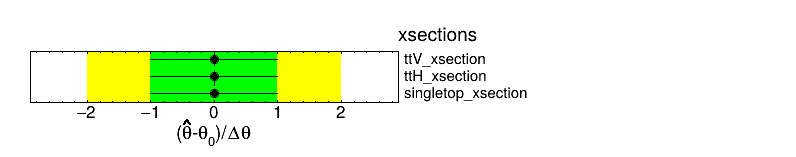}
    \caption{$t\bar{t}$, b-tagging, and cross-section systematic uncertainty pull plot for Asimov fit on LH 1.2 TeV $W'$ boson sample where $g'/g=2$.}
    \label{nuispar_all_LH_1p2tev_asimov}
\end{figure}

\begin{figure}[h]
    \centering
    \includegraphics[width=0.8\linewidth]{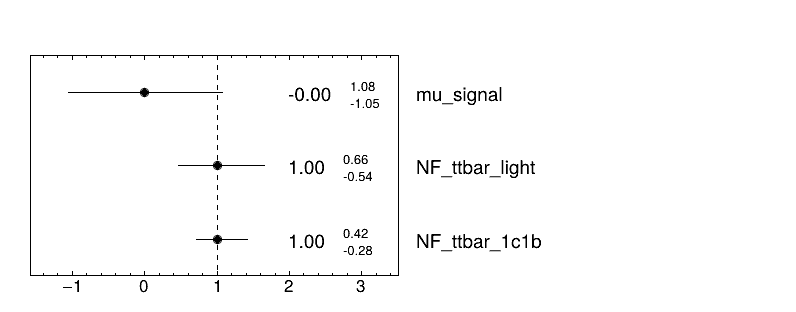}
    \caption{Normalization factors determined by the binned maximum likelihood fit for the LH 1.4 TeV $W'$ boson signal where $g'/g=2$.}
    \label{norms_LH__1p4tev}
\end{figure}

\begin{figure}[h]
    \centering
    \includegraphics[width=0.45\linewidth]{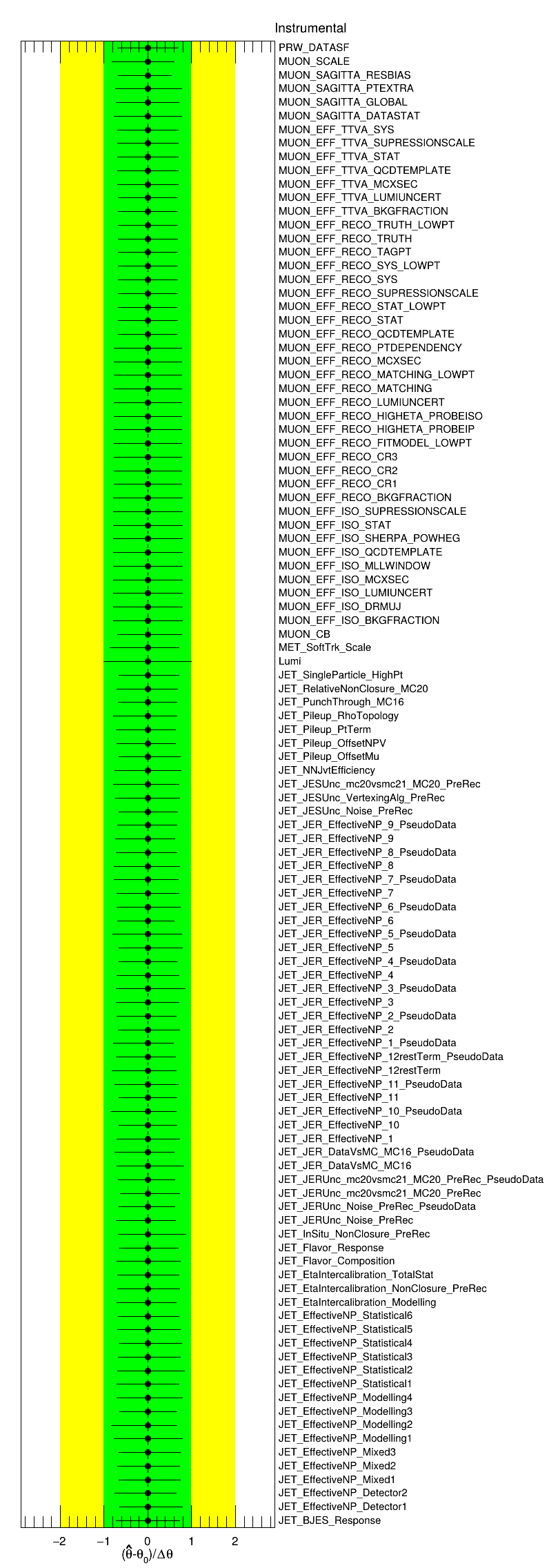}
    \caption{Detector systematic uncertainty pull plot for Asimov fit on LH 1.4 TeV $W'$ boson sample where $g'/g=2$.}
    \label{nuispar_inst_LH_1p4tev_asimov}
\end{figure}

\begin{figure}[h]
    \centering
    \includegraphics[width=0.49\linewidth]{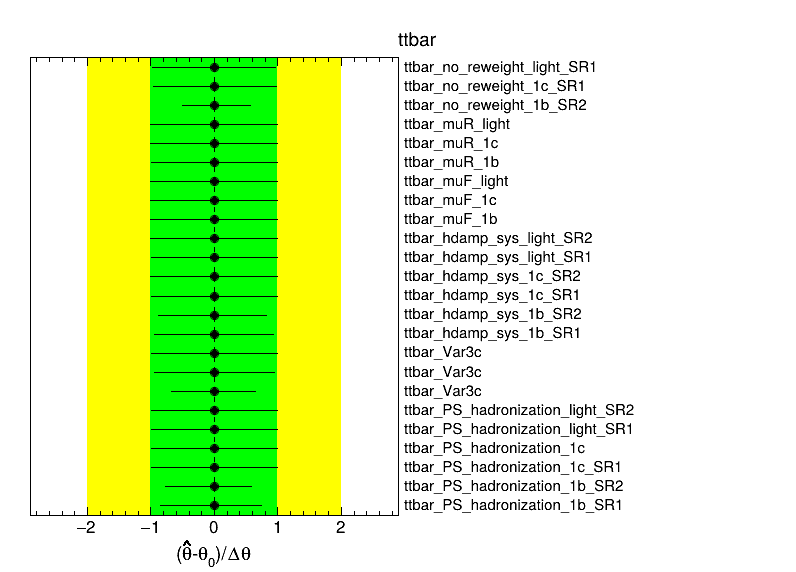}
    \includegraphics[width=0.49\linewidth]{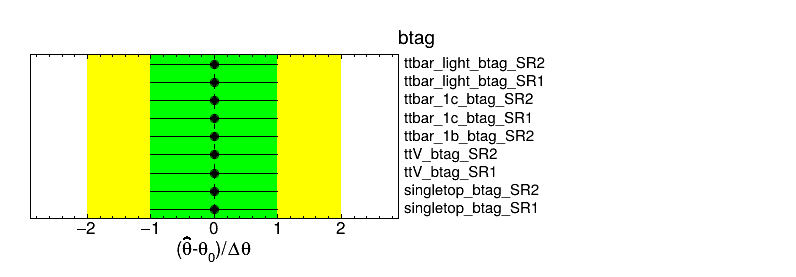}\\
    \includegraphics[width=0.49\linewidth]{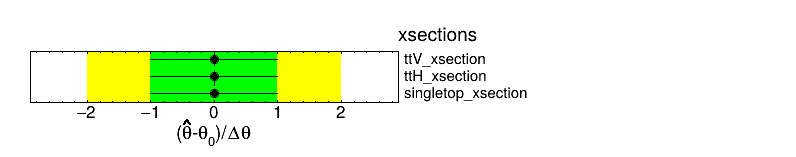}
    \caption{$t\bar{t}$, b-tagging, and cross-section systematic uncertainty pull plot for Asimov fit on LH 1.4 TeV $W'$ boson sample where $g'/g=2$.}
    \label{nuispar_all_LH_1p4tev_asimov}
\end{figure}

\begin{figure}[h]
    \centering
    \includegraphics[width=0.8\linewidth]{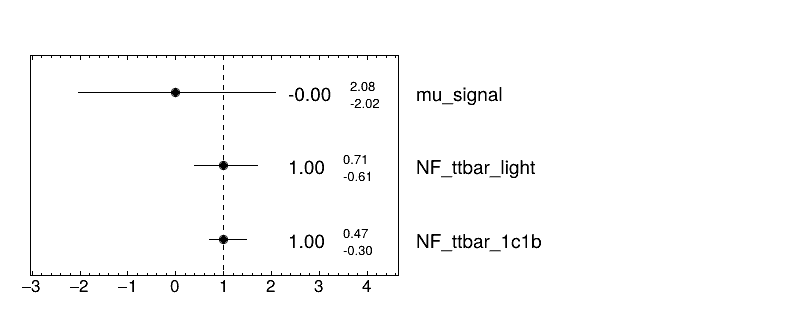}
    \caption{Normalization factors determined by the binned maximum likelihood fit for the LH 1.6 TeV $W'$ boson signal where $g'/g=2$.}
    \label{norms_LH__1p6tev}
\end{figure}

\begin{figure}[h]
    \centering
    \includegraphics[width=0.45\linewidth]{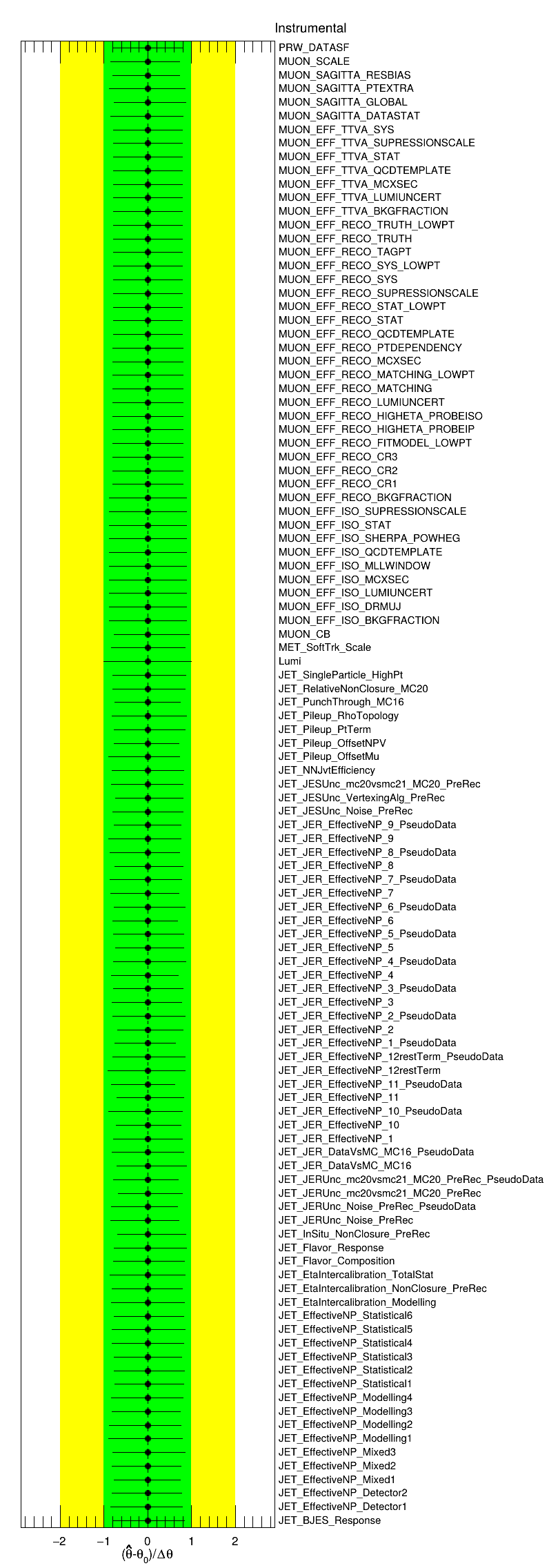}
    \caption{Detector systematic uncertainty pull plot for Asimov fit on LH 1.6 TeV $W'$ boson sample where $g'/g=2$.}
    \label{nuispar_inst_LH_1p6tev_asimov}
\end{figure}

\begin{figure}[h]
    \centering
    \includegraphics[width=0.49\linewidth]{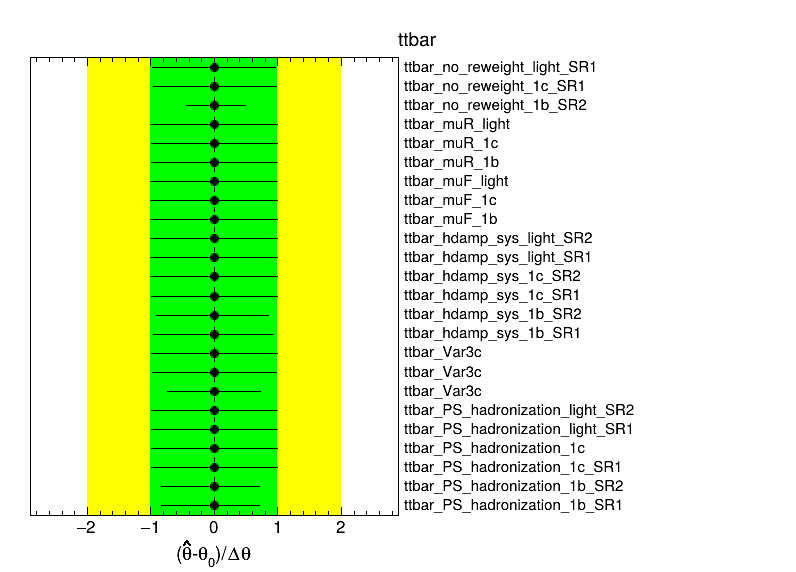}
    \includegraphics[width=0.49\linewidth]{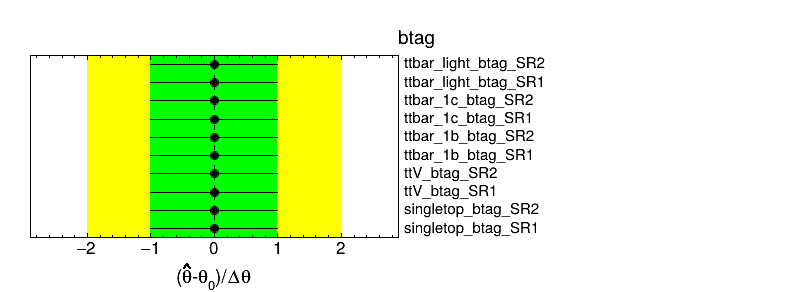}\\
    \includegraphics[width=0.49\linewidth]{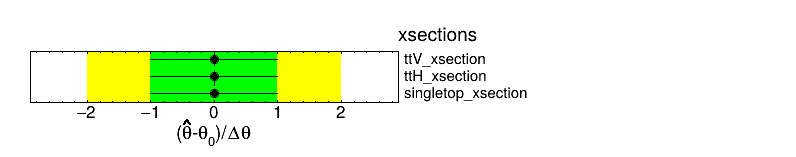}
    \caption{$t\bar{t}$, b-tagging, and cross-section systematic uncertainty pull plot for Asimov fit on LH 1.6 TeV $W'$ boson sample where $g'/g=2$.}
    \label{nuispar_all_LH_1p6tev_asimov}
\end{figure}

\begin{figure}[h]
    \centering
    \includegraphics[width=0.8\linewidth]{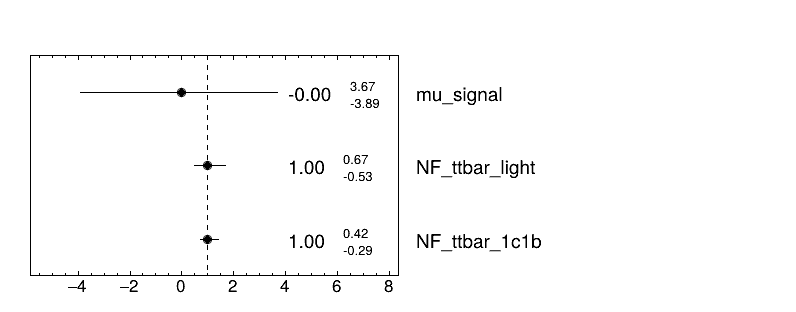}
    \caption{Normalization factors determined by the binned maximum likelihood fit for the LH 1.8 TeV $W'$ boson signal where $g'/g=2$.}
    \label{norms_LH__1p8tev}
\end{figure}

\begin{figure}[h]
    \centering
    \includegraphics[width=0.45\linewidth]{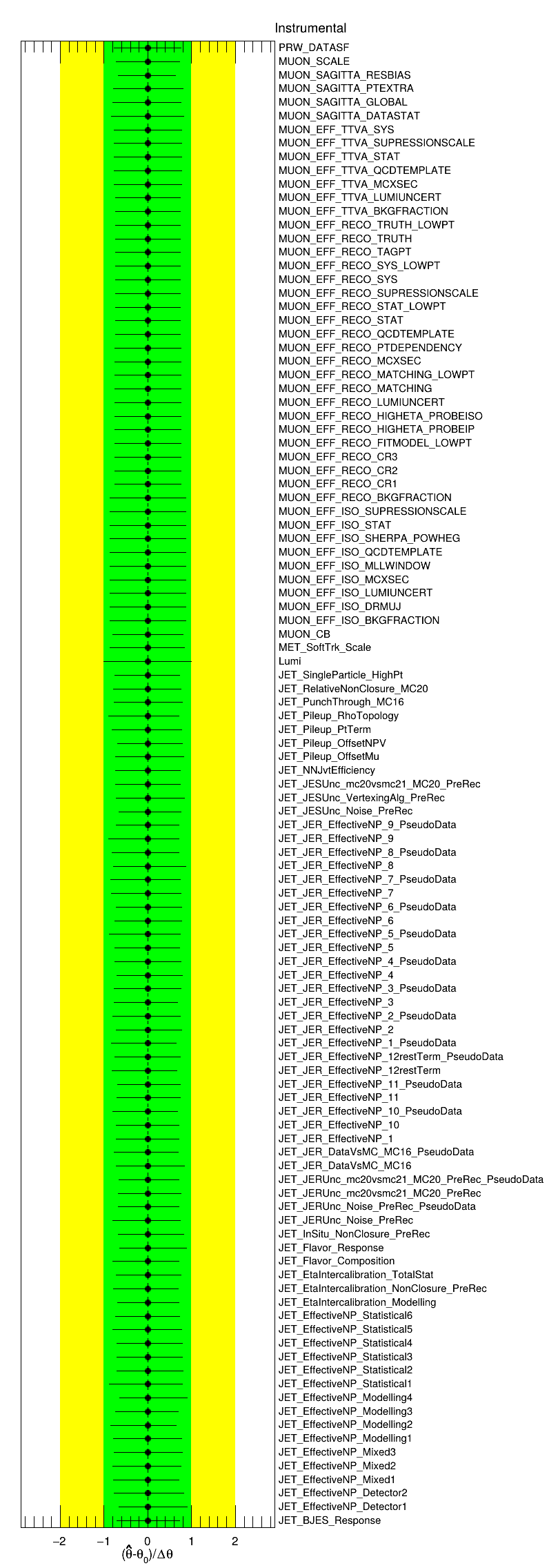}
    \caption{Detector systematic uncertainty pull plot for Asimov fit on LH 1.8 TeV $W'$ boson sample where $g'/g=2$.}
    \label{nuispar_inst_LH_1p8tev_asimov}
\end{figure}

\begin{figure}[h]
    \centering
    \includegraphics[width=0.49\linewidth]{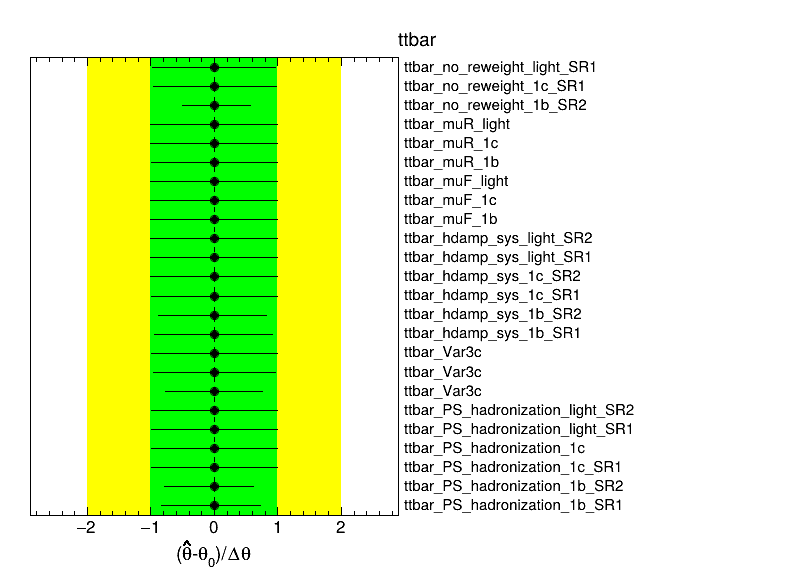}
    \includegraphics[width=0.49\linewidth]{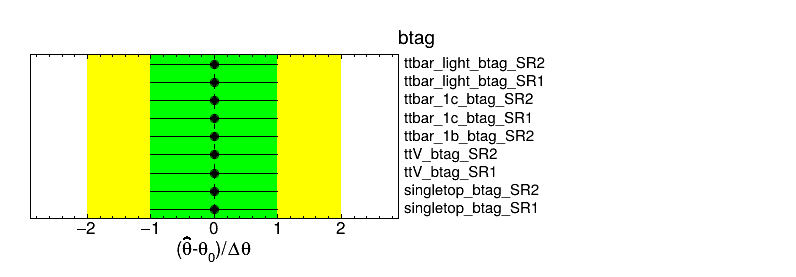}\\
    \includegraphics[width=0.49\linewidth]{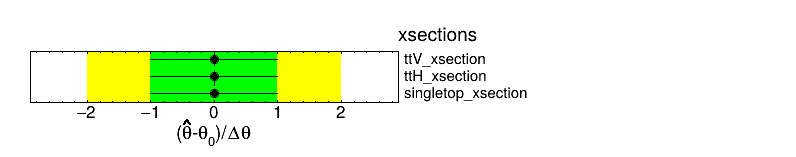}
    \caption{$t\bar{t}$, b-tagging, and cross-section systematic uncertainty pull plot for Asimov fit on LH 1.8 TeV $W'$ boson sample where $g'/g=2$.}
    \label{nuispar_all_LH_1p8tev_asimov}
\end{figure}

\begin{figure}[h]
    \centering
    \includegraphics[width=0.8\linewidth]{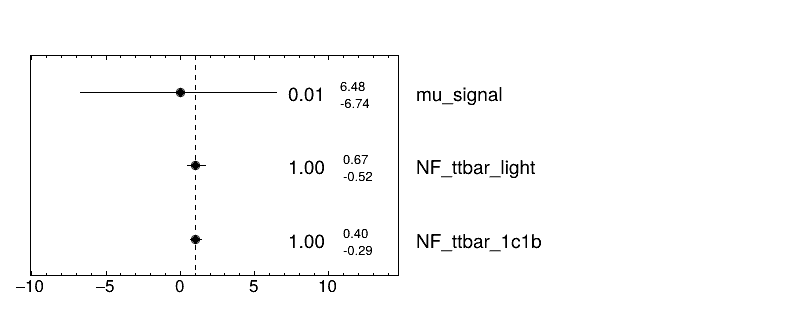}
    \caption{Normalization factors determined by the binned maximum likelihood fit for the LH 2 TeV $W'$ boson signal where $g'/g=2$.}
    \label{norms_LH__2tev}
\end{figure}

\begin{figure}[h]
    \centering
    \includegraphics[width=0.45\linewidth]{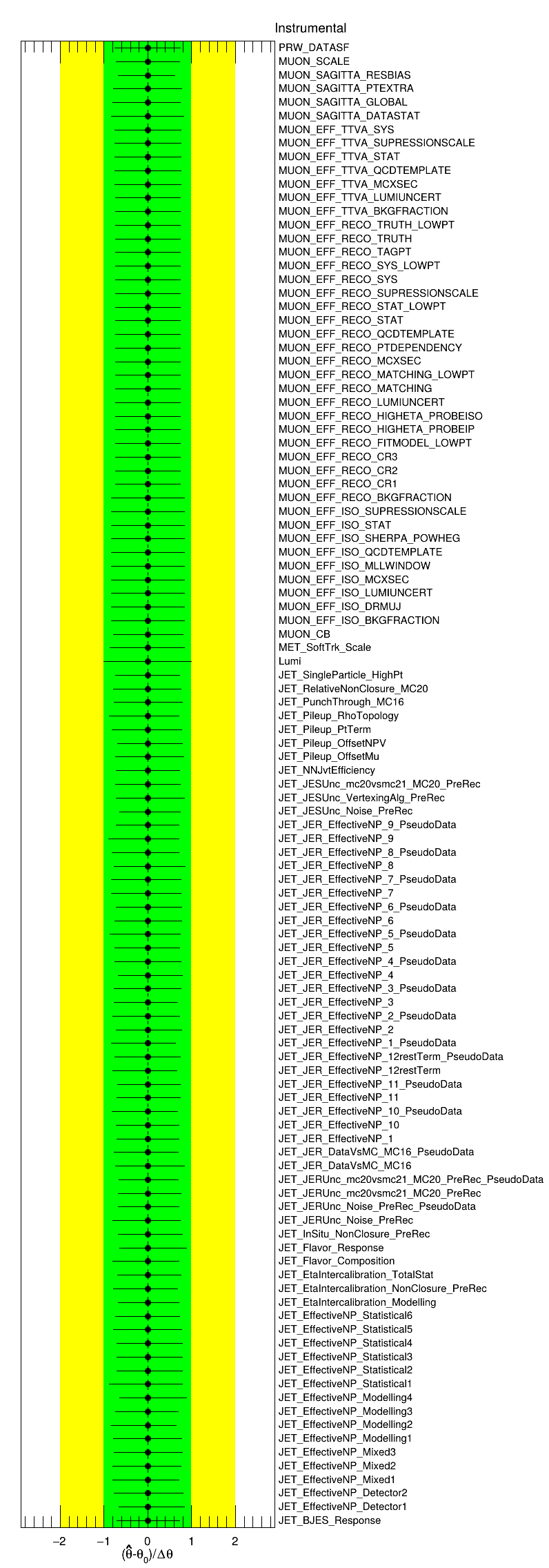}
    \caption{Detector systematic uncertainty pull plot for Asimov fit on LH 2 TeV $W'$ boson sample where $g'/g=2$.}
    \label{nuispar_inst_LH_2tev_asimov}
\end{figure}

\begin{figure}[h]
    \centering
    \includegraphics[width=0.49\linewidth]{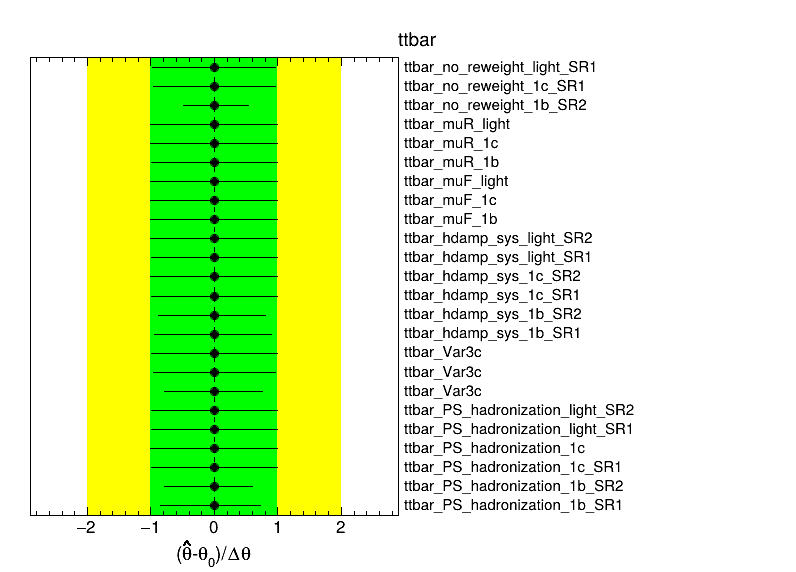}
    \includegraphics[width=0.49\linewidth]{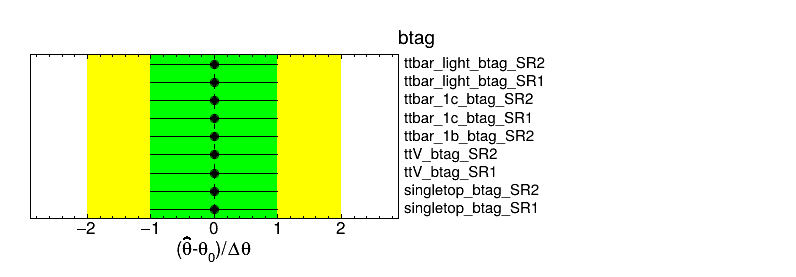}\\
    \includegraphics[width=0.49\linewidth]{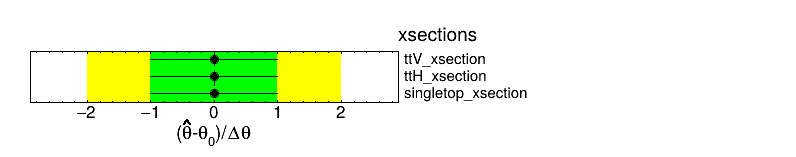}
    \caption{$t\bar{t}$, b-tagging, and cross-section systematic uncertainty pull plot for Asimov fit on LH 2 TeV $W'$ boson sample where $g'/g=2$.}
    \label{nuispar_all_LH_2tev_asimov}
\end{figure}

For the LH $W'$ boson, the pre-fit plot is shown in Fig.~\ref{expected_LH_1tev_prefit} and the post-fit plot is shown in Fig.~\ref{expected_LH_1tev_postfit}. Figures containing the resulting normalization factors can be found in Figs.~\ref{norms_LH_1tev_asimov},~\ref{norms_LH__1p2tev},~\ref{norms_LH__1p4tev},~\ref{norms_LH__1p6tev},~\ref{norms_LH__1p8tev}, and~\ref{norms_LH__2tev}. The normalization factors fluctuate slightly, but the uncertainty in the normalization parameter changes across different $W'$ boson mass points. The fitted results for the $t\bar{t}+\geq1b/1c$ and $t\bar{t}+\text{light}$ normalization parameters are nominal as expected. The pull plots on the instrument systematic uncertainties can be seen in Figs.~\ref{nuispar_inst_LH_1tev_asimov},~\ref{nuispar_inst_LH_1p2tev_asimov},~\ref{nuispar_inst_LH_1p4tev_asimov},~\ref{nuispar_inst_LH_1p6tev_asimov},~\ref{nuispar_inst_LH_1p8tev_asimov},~\ref{nuispar_inst_LH_2tev_asimov}. Each of these plots are fairly similar showing constraints on most of the systematic uncertainties, but not being pulled away from their nominal values. The pull plots on the systematic uncertainties for b-tagging, cross-sections, and $t\bar{t}$ modeling can be seen in Figs.~\ref{nuispar_all_LH_1tev_asimov},~\ref{nuispar_all_LH_1p2tev_asimov},~\ref{nuispar_all_LH_1p4tev_asimov},~\ref{nuispar_all_LH_1p6tev_asimov},~\ref{nuispar_all_LH_1p8tev_asimov},~\ref{nuispar_all_LH_2tev_asimov}. In general, the systematic uncertainties for the cross-sections and b-tagging are not constrained and do not deviate from their nominal values while some of the $t\bar{t}$ modeling systematic uncertainties are constrained slightly. Pull plots are very similar across each mass point.

\clearpage

\begin{figure}[h]
    \centering
    \includegraphics[width=0.8\linewidth]{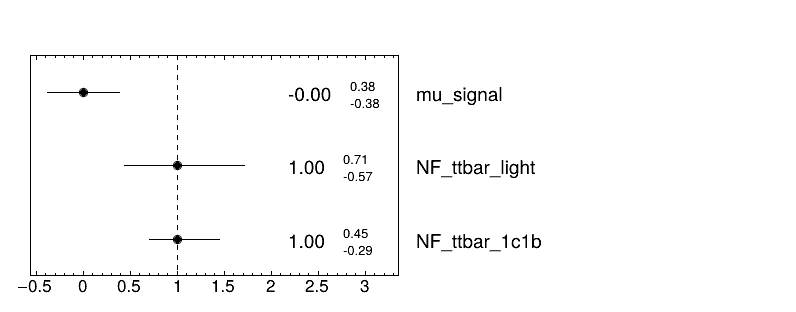}
    \caption{Normalization factors determined by the binned maximum likelihood fit for the RH 1 TeV $W'$ boson signal where g'/g=2.}
    \label{norms_RH_1tev_asimov}
\end{figure}

\begin{figure}[h]
    \centering
    \includegraphics[width=0.49\linewidth]{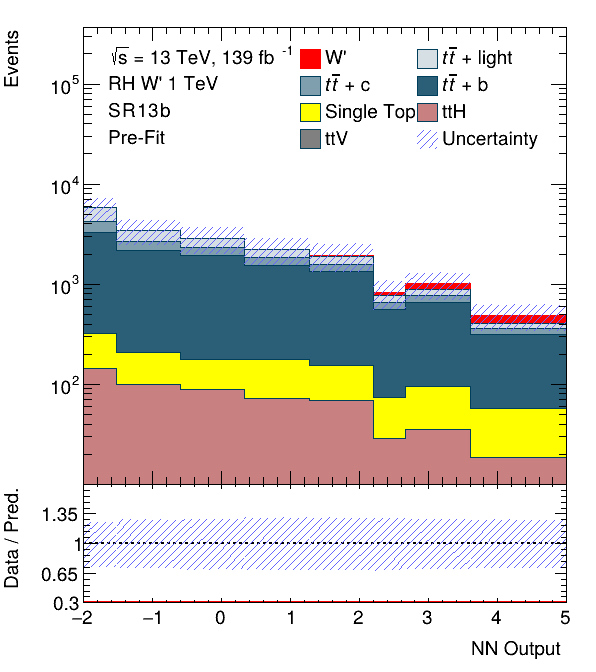}
    \includegraphics[width=0.49\linewidth]{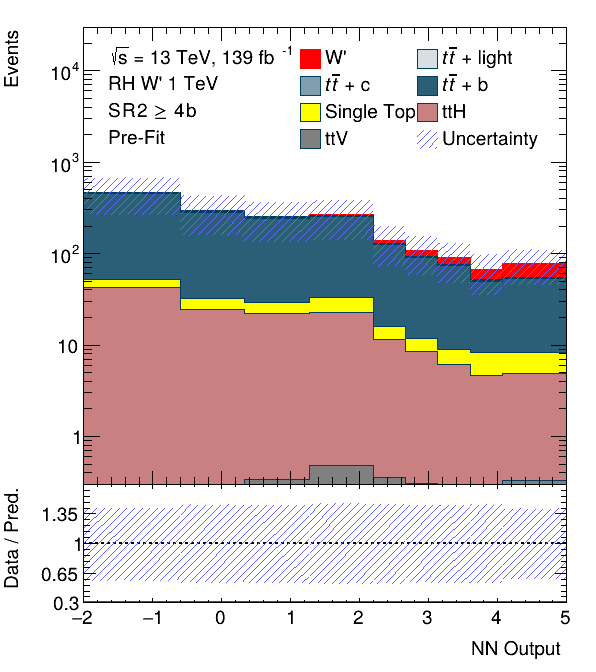}
    \caption{Asimov pre-fit plots for RH 1 TeV $W'$ boson signal sample where g'/g=2. All other mass points have similar distributions with the two signal regions, with very minor differences.}
    \label{expected_RH_1tev_prefit}
\end{figure}

\begin{figure}[h]
    \centering
    \includegraphics[width=0.49\linewidth]{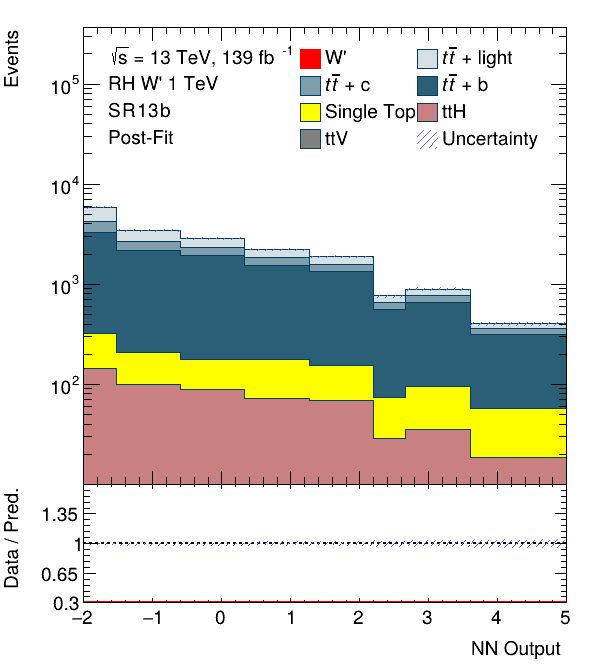}
    \includegraphics[width=0.49\linewidth]{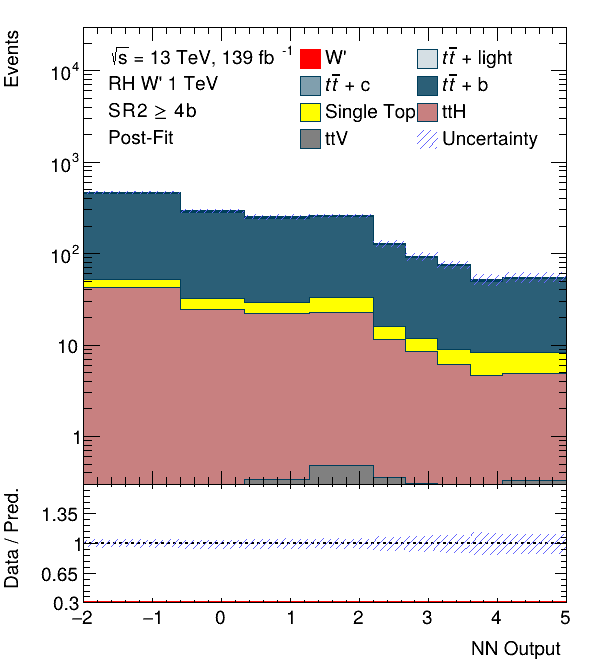}
    \caption{Asimov post-fit plots for RH 1 TeV $W'$ boson signal sample where g'/g=2. All other mass points have similar distributions with the two signal regions, with very minor differences.}
    \label{expected_RH_1tev_postfit}
\end{figure}

\begin{figure}[h]
    \centering
    \includegraphics[width=0.45\linewidth]{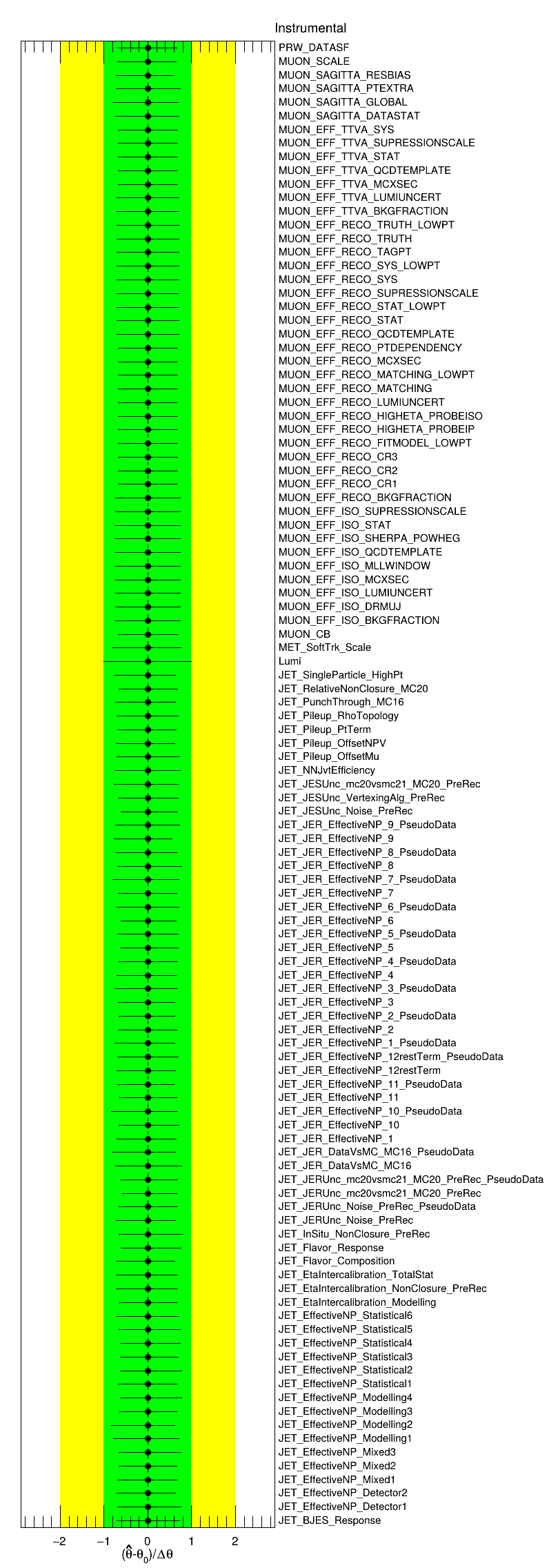}
    \caption{Detector systematic uncertainty pull plot for Asimov fit on RH 1 TeV $W'$ boson sample where g'/g=2.}
    \label{nuispar_inst_RH_1tev_asimov}
\end{figure}

\begin{figure}[h]
    \centering
    \includegraphics[width=0.49\linewidth]{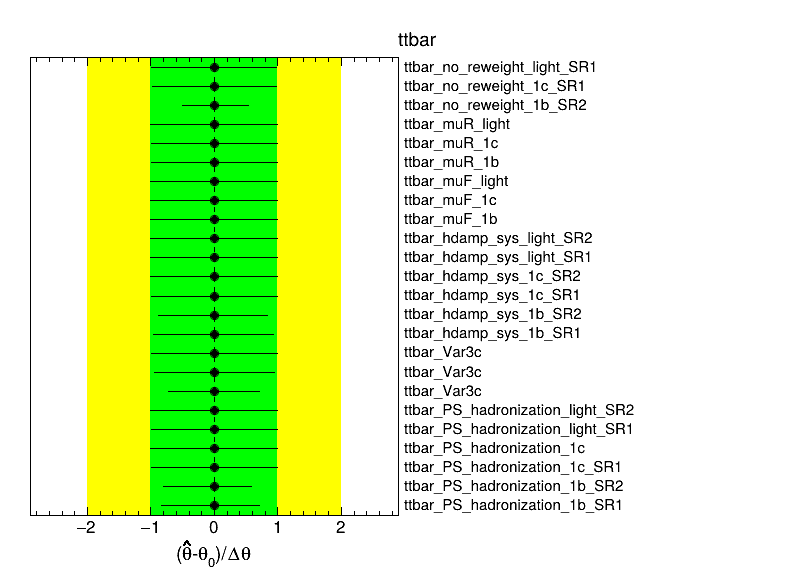}
    \includegraphics[width=0.49\linewidth]{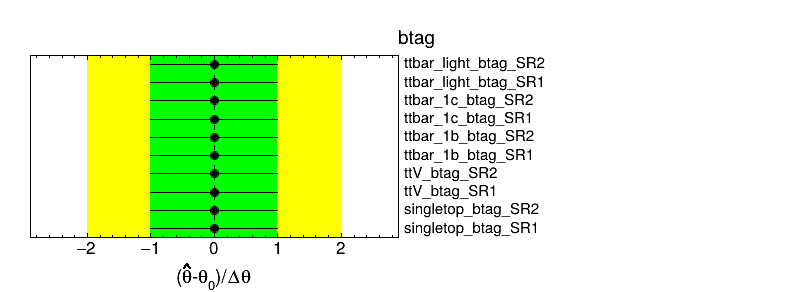}\\
    \includegraphics[width=0.49\linewidth]{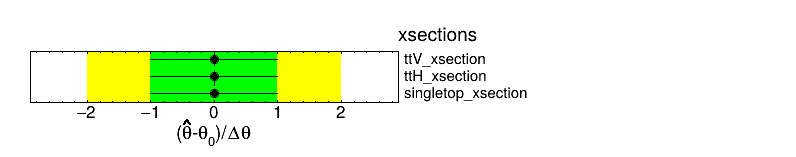}
    \caption{$t\bar{t}$, b-tagging, and cross-section systematic uncertainty pull plot for Asimov fit on RH 1 TeV $W'$ boson sample where g'/g=2.}
    \label{nuispar_all_RH_1tev_asimov}
\end{figure}

\begin{figure}[h]
    \centering
    \includegraphics[width=0.8\linewidth]{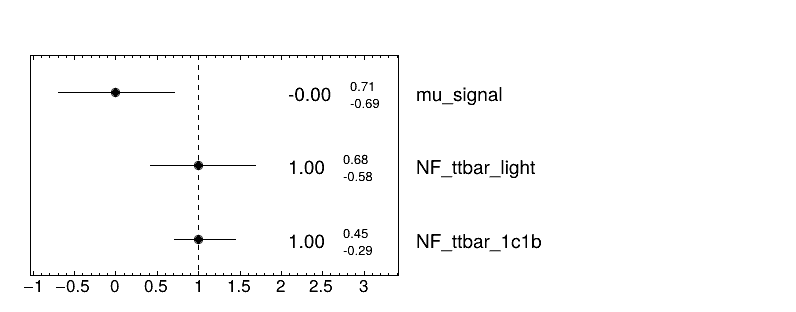}
    \caption{Normalization factors determined by the binned maximum likelihood fit for the RH 1.2 TeV $W'$ boson signal where $g'/g=2$.}
    \label{norms_RH_1p2tev_asimov}
\end{figure}

\begin{figure}[h]
    \centering
    \includegraphics[width=0.45\linewidth]{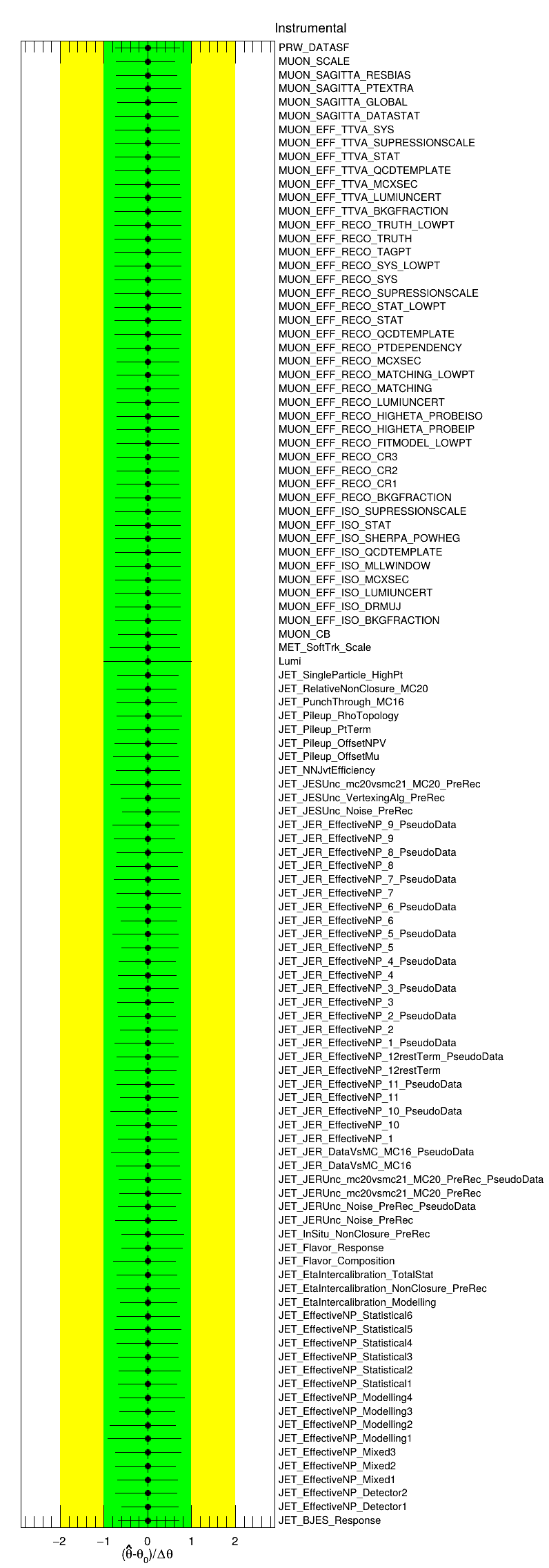}
    \caption{Detector systematic uncertainty pull plot for Asimov fit on RH 1.2 TeV $W'$ boson sample where $g'/g=2$.}
    \label{nuispar_inst_RH_1p2tev_asimov}
\end{figure}

\begin{figure}[h]
    \centering
    \includegraphics[width=0.49\linewidth]{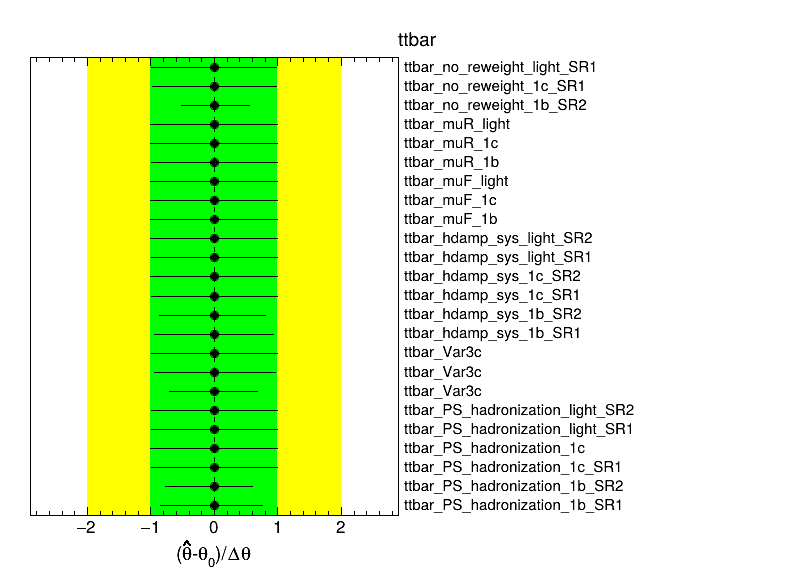}
    \includegraphics[width=0.49\linewidth]{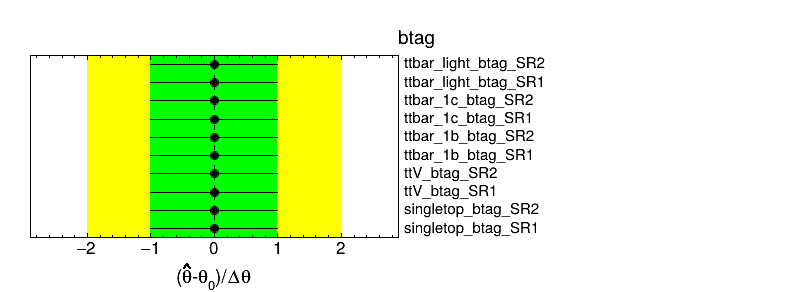}\\
    \includegraphics[width=0.49\linewidth]{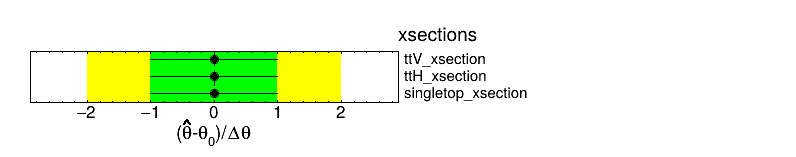}
    \caption{$t\bar{t}$, b-tagging, and cross-section systematic uncertainty pull plot for Asimov fit on RH 1.2 TeV $W'$ boson sample where $g'/g=2$.}
    \label{nuispar_all_RH_1p2tev_asimov}
\end{figure}

\begin{figure}[h]
    \centering
    \includegraphics[width=0.8\linewidth]{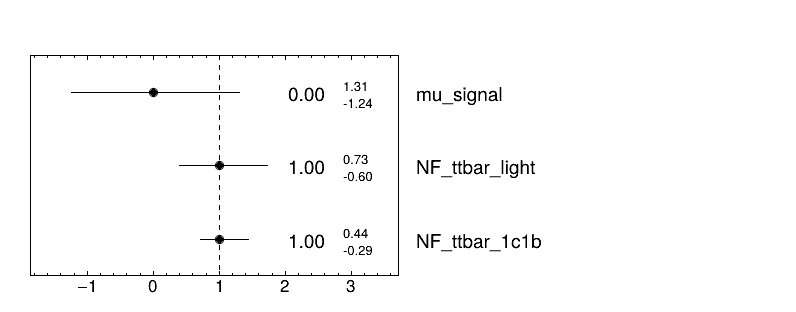}
    \caption{Normalization factors determined by the binned maximum likelihood fit for the RH 1.4 TeV $W'$ boson signal where $g'/g=2$.}
    \label{norms_RH_1p4tev_asimov}
\end{figure}

\begin{figure}[h]
    \centering
    \includegraphics[width=0.45\linewidth]{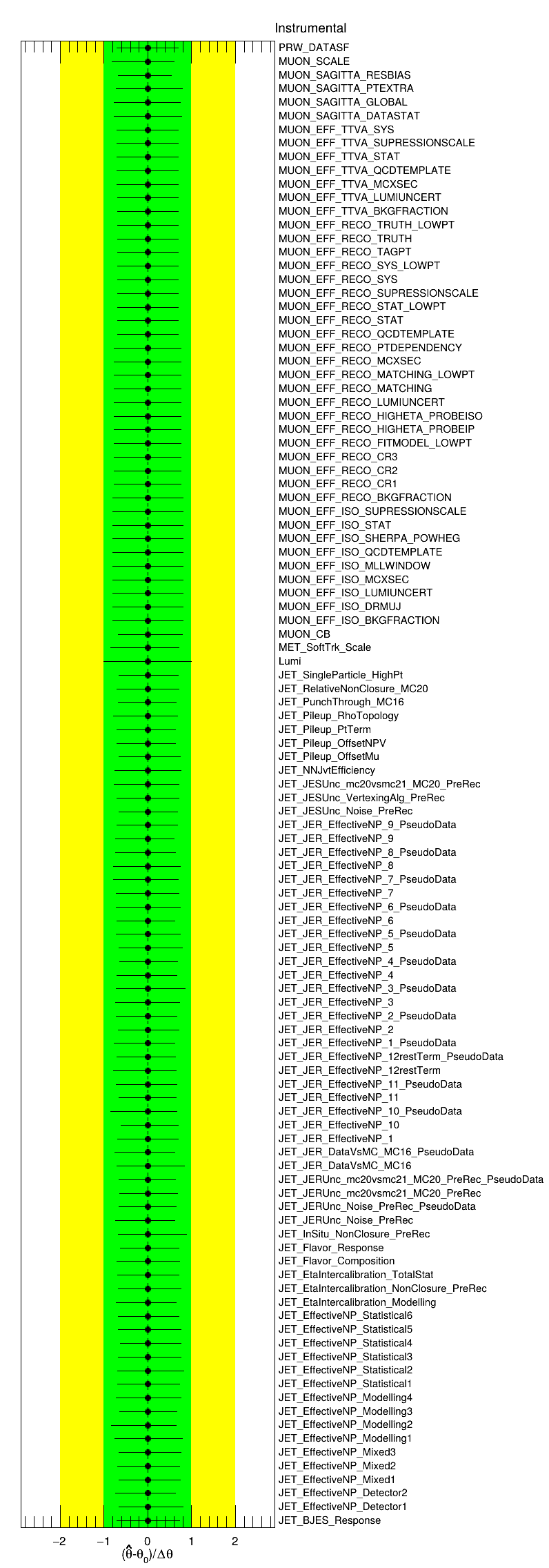}
    \caption{Detector systematic uncertainty pull plot for Asimov fit on RH 1.4 TeV $W'$ boson sample where $g'/g=2$.}
    \label{nuispar_inst_RH_1p4tev_asimov}
\end{figure}

\begin{figure}[h]
    \centering
    \includegraphics[width=0.49\linewidth]{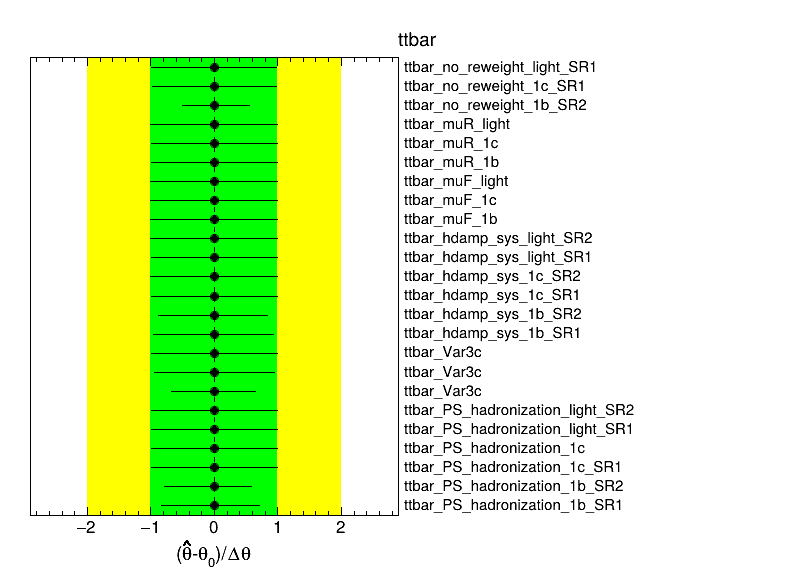}
    \includegraphics[width=0.49\linewidth]{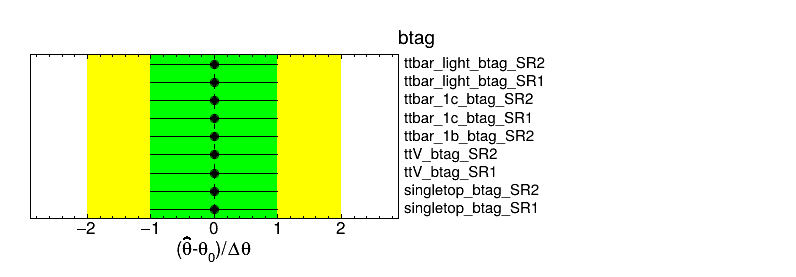}\\
    \includegraphics[width=0.49\linewidth]{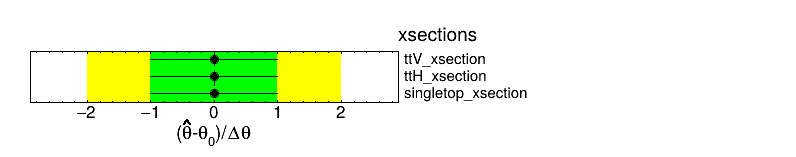}
    \caption{$t\bar{t}$, b-tagging, and cross-section systematic uncertainty pull plot for Asimov fit on RH 1.4 TeV $W'$ boson sample where $g'/g=2$.}
    \label{nuispar_all_RH_1p4tev_asimov}
\end{figure}

\begin{figure}[h]
    \centering
    \includegraphics[width=0.8\linewidth]{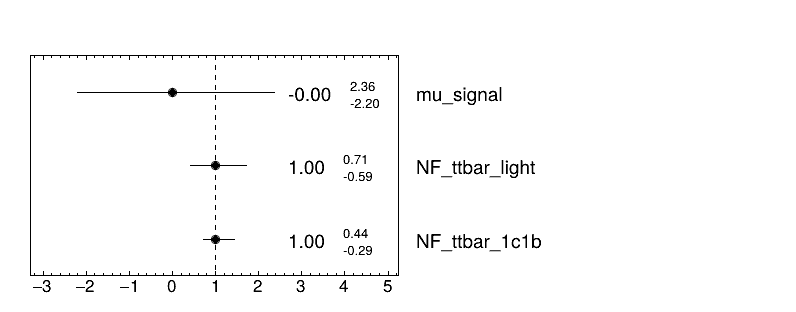}
    \caption{Normalization factors determined by the binned maximum likelihood fit for the RH 1.6 TeV $W'$ boson signal where $g'/g=2$.}
    \label{norms_RH_1p6tev_asimov}
\end{figure}

\begin{figure}[h]
    \centering
    \includegraphics[width=0.45\linewidth]{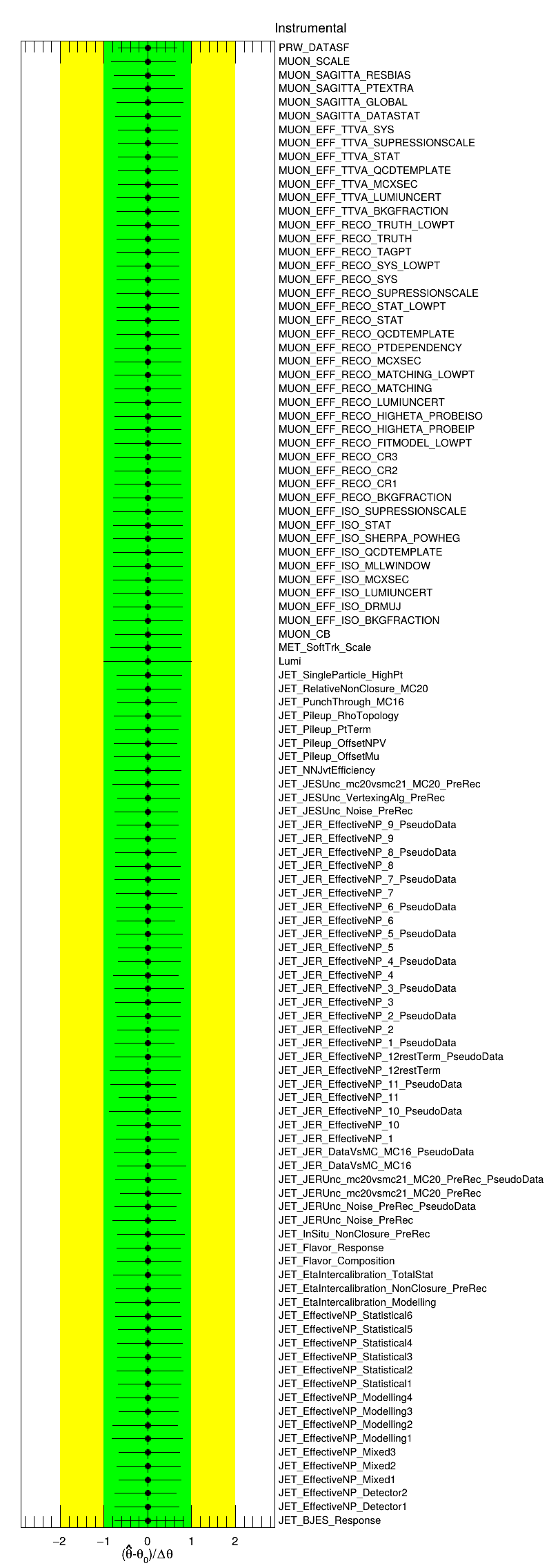}
    \caption{Detector systematic uncertainty pull plot for Asimov fit on RH 1.6 TeV $W'$ boson sample where $g'/g=2$.}
    \label{nuispar_inst_RH_1p6tev_asimov}
\end{figure}

\begin{figure}[h]
    \centering
    \includegraphics[width=0.49\linewidth]{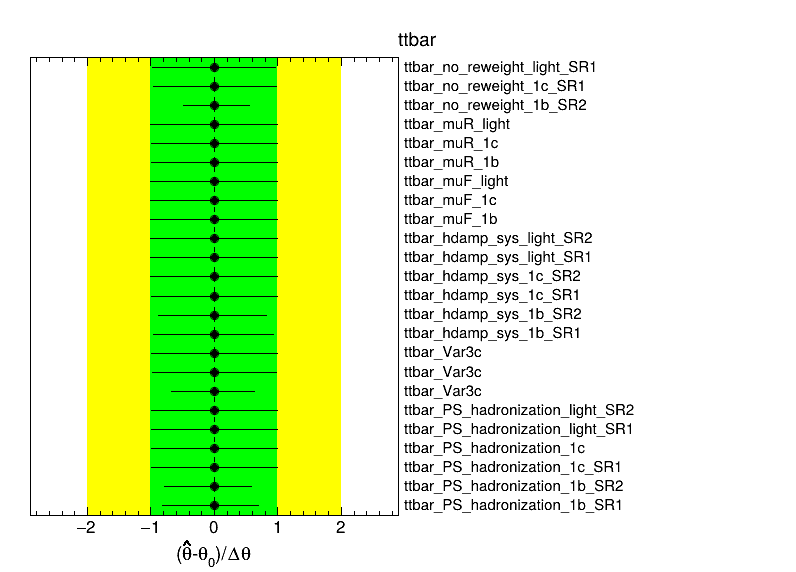}
    \includegraphics[width=0.49\linewidth]{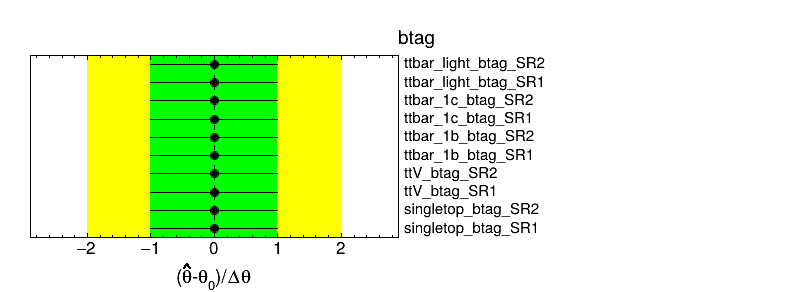}\\
    \includegraphics[width=0.49\linewidth]{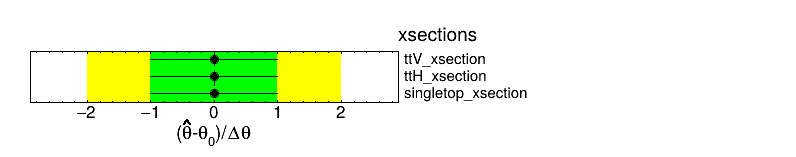}
    \caption{$t\bar{t}$, b-tagging, and cross-section systematic uncertainty pull plot for Asimov fit on RH 1.6 TeV $W'$ boson sample where $g'/g=2$.}
    \label{nuispar_all_RH_1p6tev_asimov}
\end{figure}

\begin{figure}[h]
    \centering
    \includegraphics[width=0.8\linewidth]{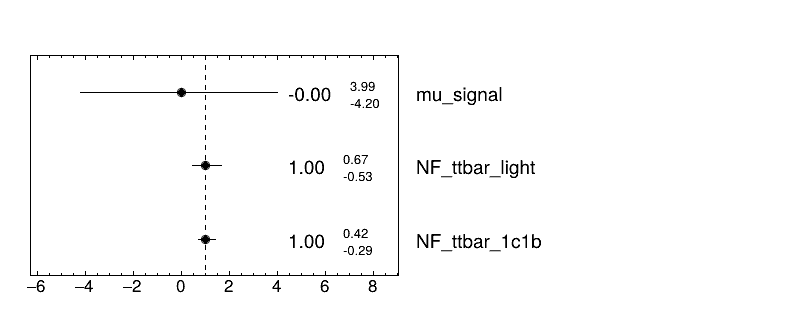}
    \caption{Normalization factors determined by the binned maximum likelihood fit for the RH 1.8 TeV $W'$ boson signal where $g'/g=2$.}
    \label{norms_RH_1p8tev_asimov}
\end{figure}

\begin{figure}[h]
    \centering
    \includegraphics[width=0.45\linewidth]{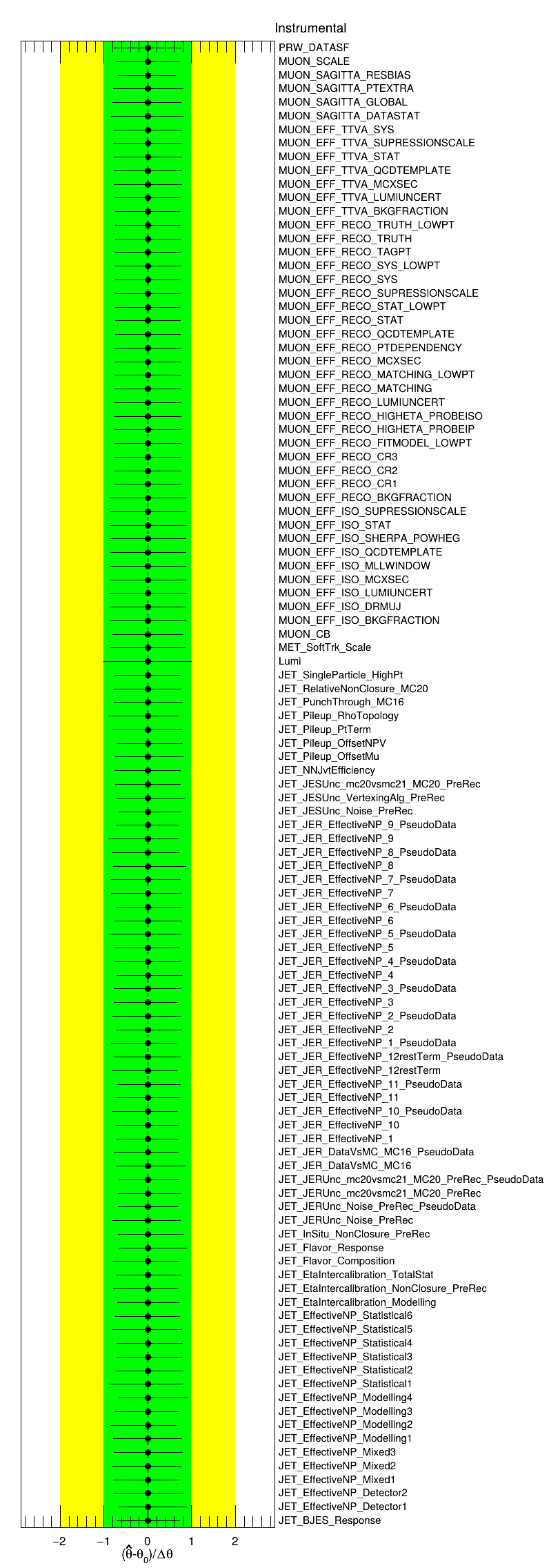}
    \caption{Detector systematic uncertainty pull plot for Asimov fit on RH 1.8 TeV $W'$ boson sample where $g'/g=2$.}
    \label{nuispar_inst_RH_1p8tev_asimov}
\end{figure}

\begin{figure}[h]
    \centering
    \includegraphics[width=0.49\linewidth]{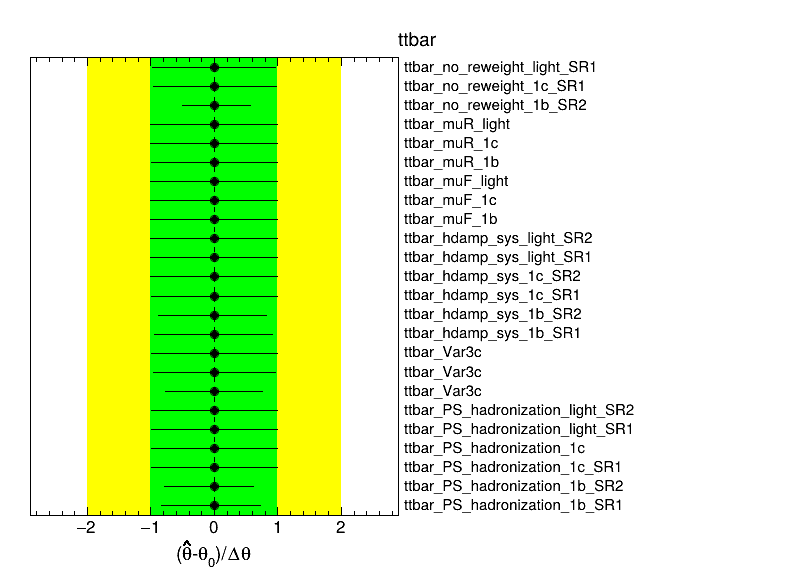}
    \includegraphics[width=0.49\linewidth]{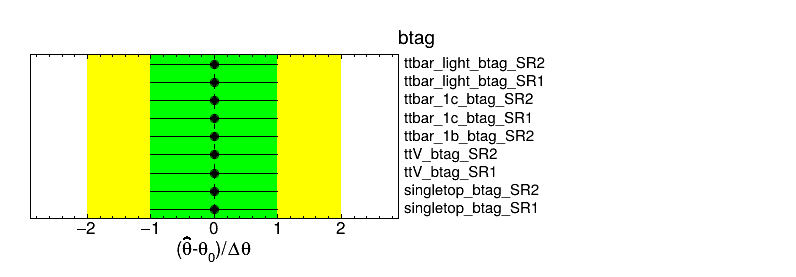}\\
    \includegraphics[width=0.49\linewidth]{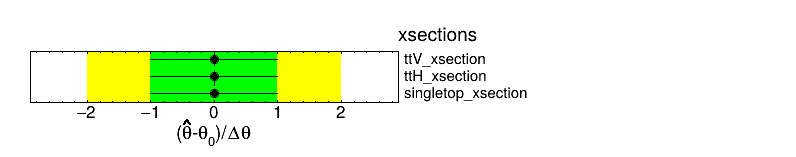}
    \caption{$t\bar{t}$, b-tagging, and cross-section systematic uncertainty pull plot for Asimov fit on RH 1.8 TeV $W'$ boson sample where $g'/g=2$.}
    \label{nuispar_all_RH_1p8tev_asimov}
\end{figure}

\begin{figure}[h]
    \centering
    \includegraphics[width=0.8\linewidth]{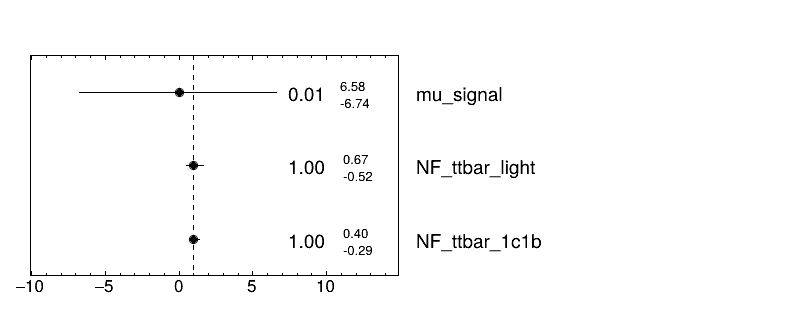}
    \caption{Normalization factors determined by the binned maximum likelihood fit for the RH 2 TeV $W'$ boson signal where $g'/g=2$.}
    \label{norms_RH_2tev_asimov}
\end{figure}

\begin{figure}[h]
    \centering
    \includegraphics[width=0.45\linewidth]{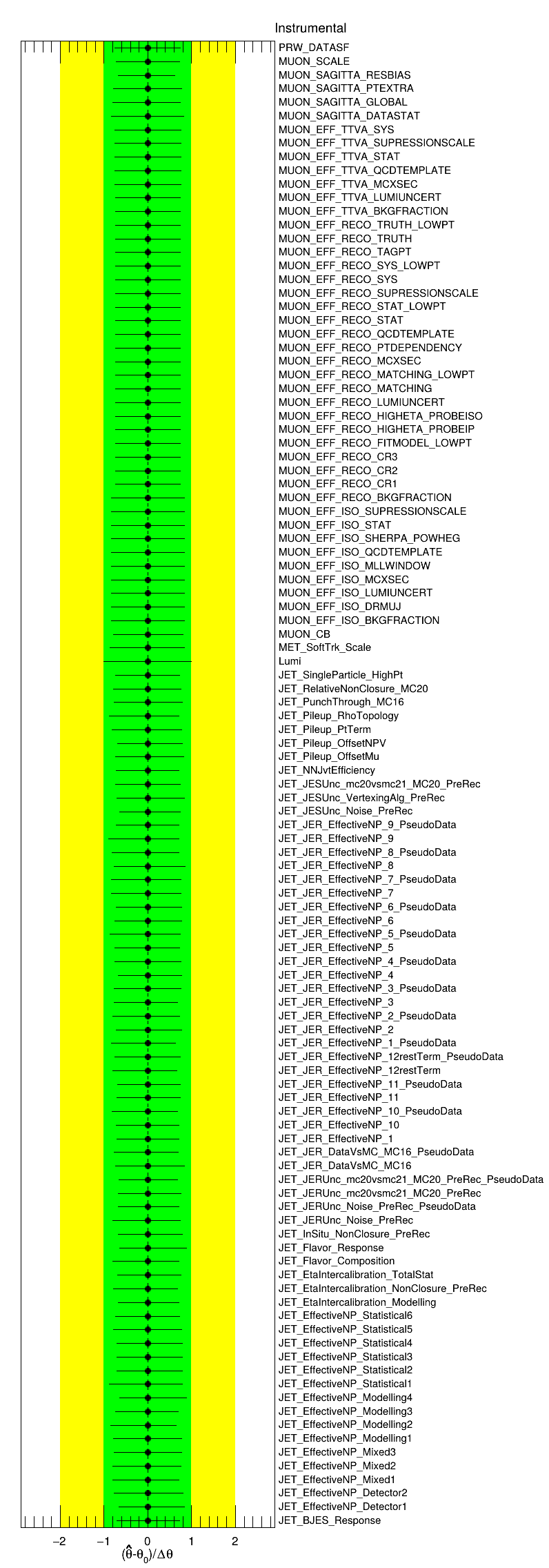}
    \caption{Detector systematic uncertainty pull plot for Asimov fit on RH 2 TeV $W'$ boson sample where $g'/g=2$.}
    \label{nuispar_inst_RH_2tev_asimov}
\end{figure}

\begin{figure}[h]
    \centering
    \includegraphics[width=0.49\linewidth]{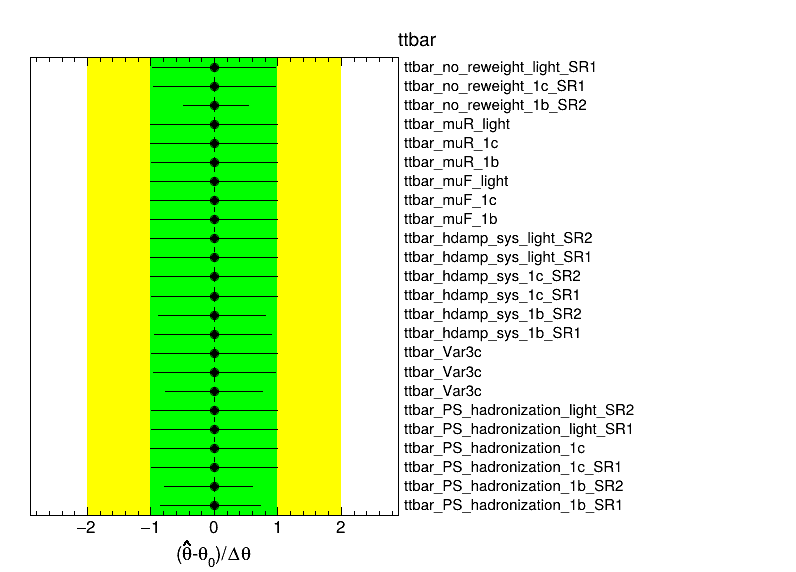}
    \includegraphics[width=0.49\linewidth]{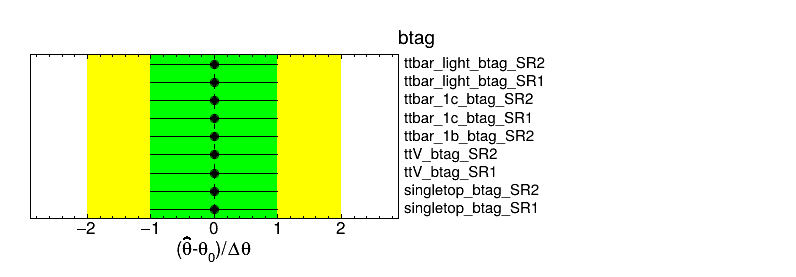}\\
    \includegraphics[width=0.49\linewidth]{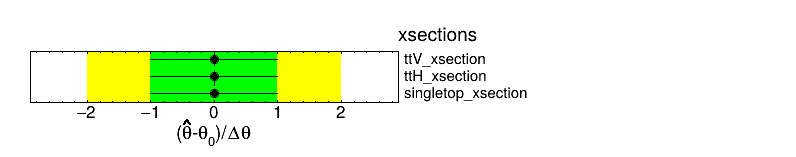}
    \caption{$t\bar{t}$, b-tagging, and cross-section systematic uncertainty pull plot for Asimov fit on RH 2 TeV $W'$ boson sample where $g'/g=2$.}
    \label{nuispar_all_RH_2tev_asimov}
\end{figure}

Similarly for the RH $W'$ boson, the pre-fit plot is shown in Fig.~\ref{expected_RH_1tev_prefit} and the post-fit plot is shown in Fig.~\ref{expected_RH_1tev_postfit}. Figures containing the resulting normalization factors can be found in Figs.~\ref{norms_RH_1tev_asimov},~\ref{norms_RH_1p2tev_asimov},~\ref{norms_RH_1p4tev_asimov},~\ref{norms_RH_1p6tev_asimov},~\ref{norms_RH_1p8tev_asimov}, and~\ref{norms_RH_2tev_asimov}. Differences between the LH and RH $W'$ boson fits are very small. The pull plots on the instrument systematic uncertainties can be seen in Figs.~\ref{nuispar_inst_RH_1tev_asimov},~\ref{nuispar_inst_RH_1p2tev_asimov},~\ref{nuispar_inst_RH_1p4tev_asimov},~\ref{nuispar_inst_RH_1p6tev_asimov},~\ref{nuispar_inst_RH_1p8tev_asimov},~\ref{nuispar_inst_RH_2tev_asimov}. The pull plots on the systematic uncertainties for b-tagging, cross-sections, and $t\bar{t}$ modeling can be seen in Figs.~\ref{nuispar_all_RH_1tev_asimov},~\ref{nuispar_all_RH_1p2tev_asimov},~\ref{nuispar_all_RH_1p4tev_asimov},~\ref{nuispar_all_RH_1p6tev_asimov},~\ref{nuispar_all_RH_1p8tev_asimov},~\ref{nuispar_all_RH_2tev_asimov}. There are no large differences between these pull plots and the ones from the LH $W'$ boson fits.

\chapter{All Data Fit Results}

\begin{figure}[h]
    \centering
    \includegraphics[width=0.8\linewidth]{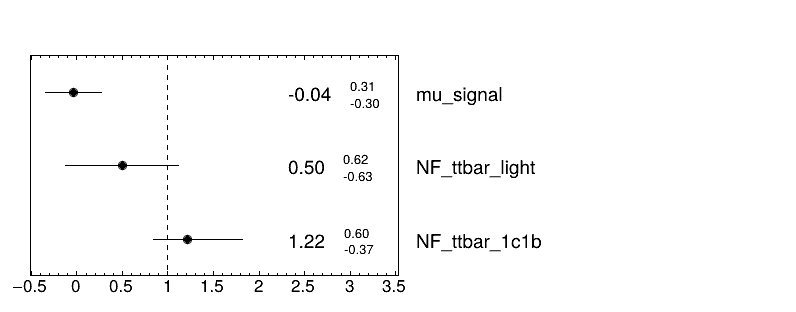}
    \caption{Normalization factors determined by the binned maximum likelihood fit for the LH 1 TeV $W'$ boson signal where g'/g=2.}
    \label{norms_LH_1tev}
\end{figure}

\begin{figure}[h]
    \centering
    \includegraphics[width=0.49\linewidth]{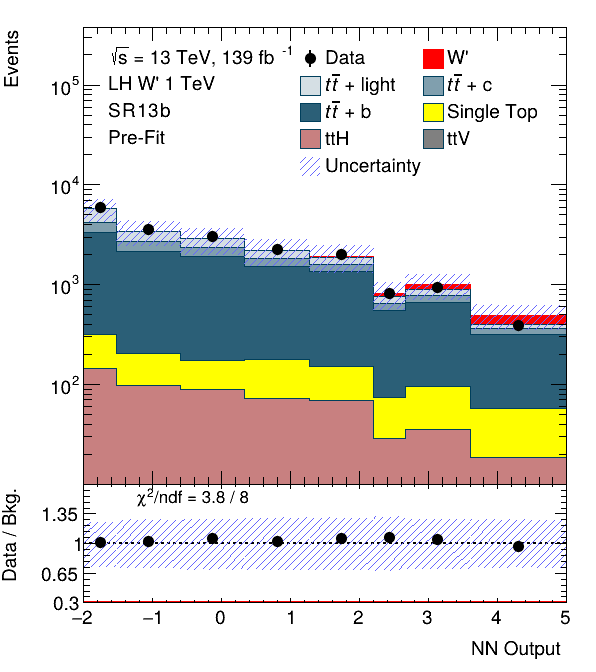}
    \includegraphics[width=0.49\linewidth]{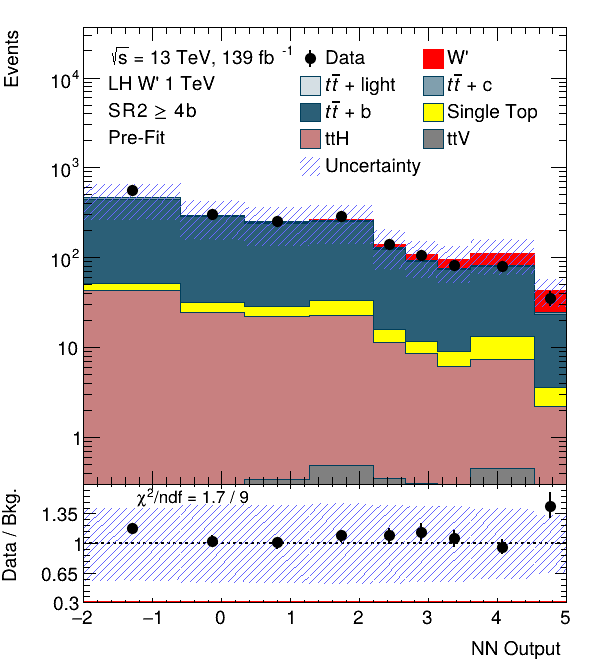}
    \caption{Asimov pre-fit plots for LH 1 TeV $W'$ boson signal sample where g'/g=2.}
    \label{data_LH_1tev_prefit}
\end{figure}

\begin{figure}[h]
    \centering
    \includegraphics[width=0.49\linewidth]{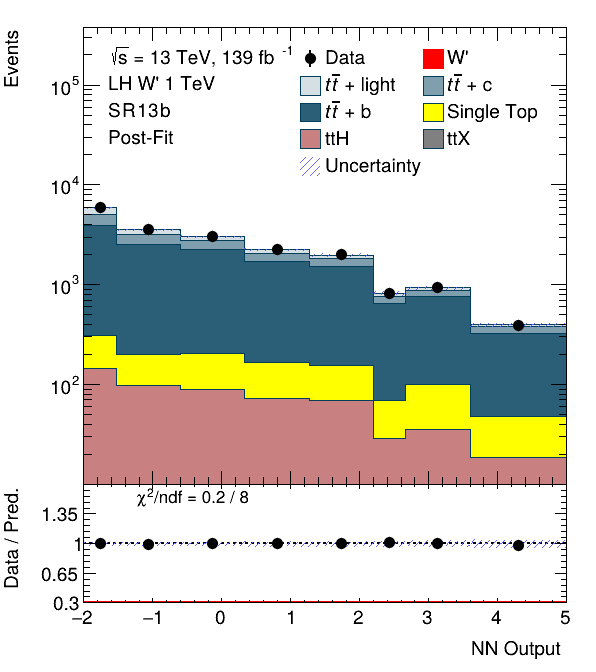}
    \includegraphics[width=0.49\linewidth]{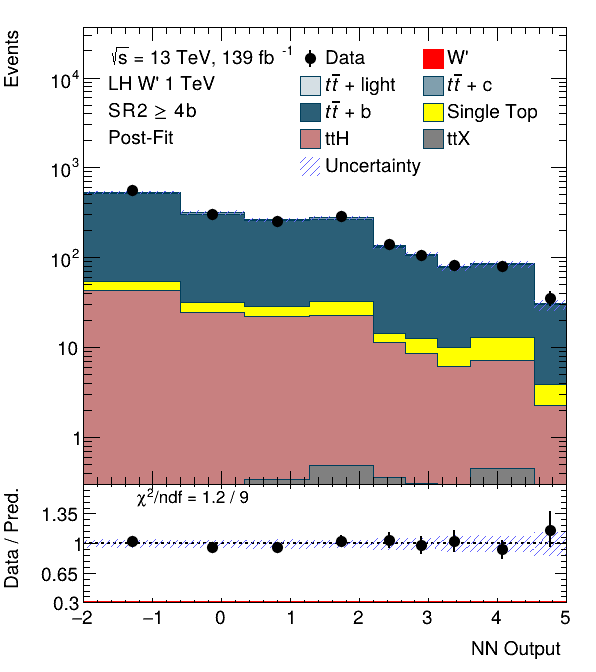}
    \caption{Asimov post-fit plots for LH 1 TeV $W'$ boson signal sample where g'/g=2.}
    \label{data_LH_1tev_postfit}
\end{figure}

\begin{figure}[h]
    \centering
    \includegraphics[width=0.8\linewidth]{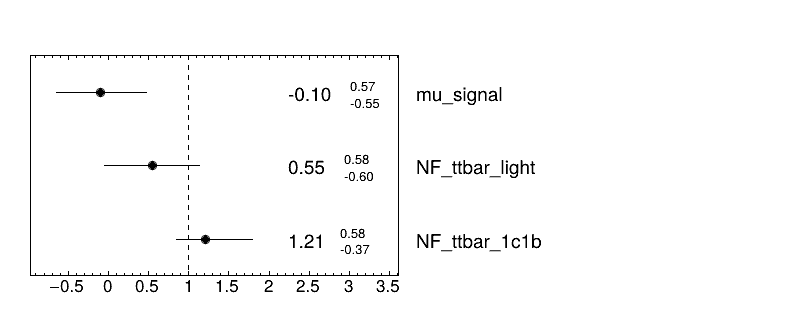}
    \caption{Normalization factors determined by the binned maximum likelihood fit for the LH 1.2 TeV $W'$ boson signal where $g'/g=2$.}
    \label{norms_LH_1p2tev}
\end{figure}

\begin{figure}[h]
    \centering
    \includegraphics[width=0.49\linewidth]{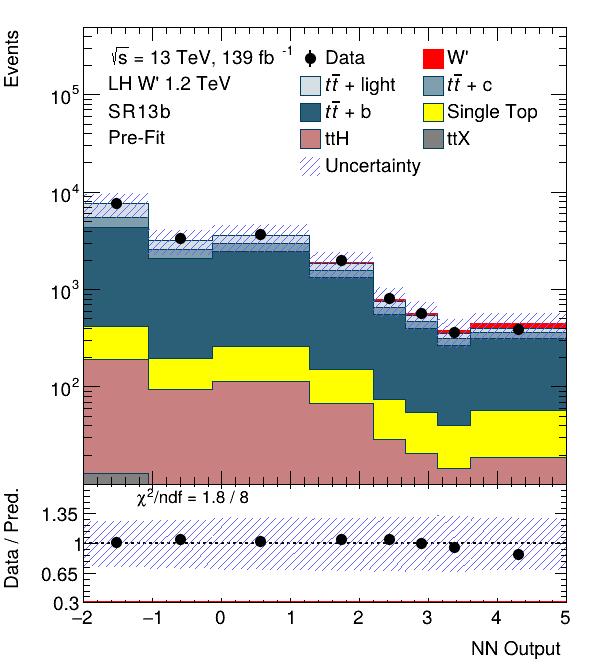}
    \includegraphics[width=0.49\linewidth]{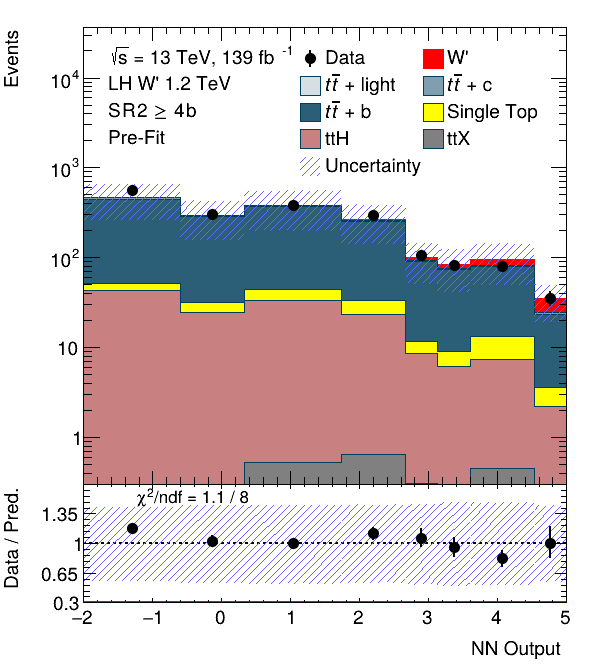}
    \caption{Asimov pre-fit plots for LH 1.2 TeV $W'$ boson signal sample where $g'/g=2$.}
    \label{data_LH_1p2tev_prefit}
\end{figure}

\begin{figure}[h]
    \centering
    \includegraphics[width=0.49\linewidth]{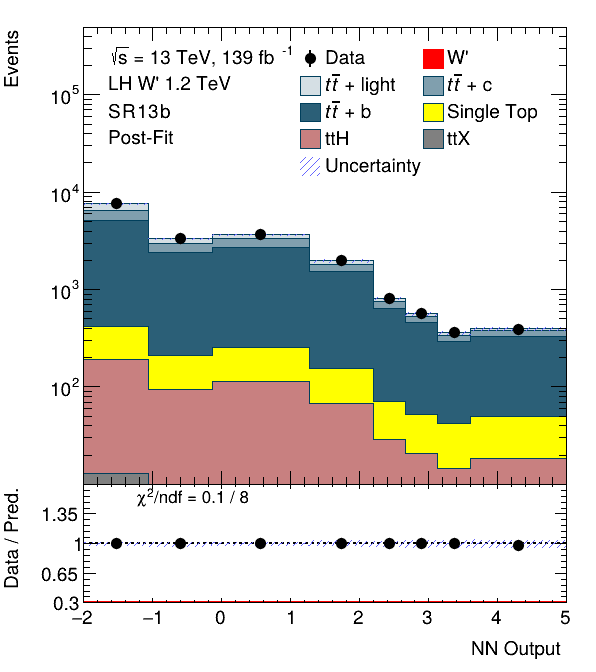}
    \includegraphics[width=0.49\linewidth]{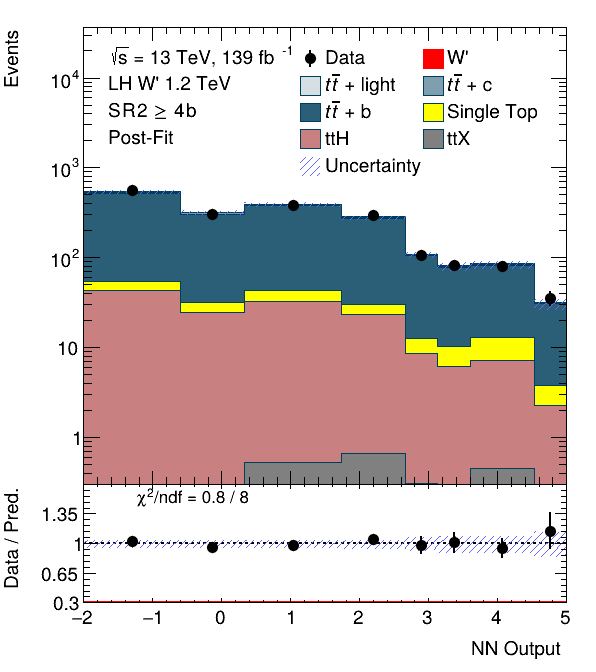}
    \caption{Asimov post-fit plots for LH 1.2 TeV $W'$ boson signal sample where $g'/g=2$.}
    \label{data_LH_1p2tev_postfit}
\end{figure}

\begin{figure}[h]
    \centering
    \includegraphics[width=0.8\linewidth]{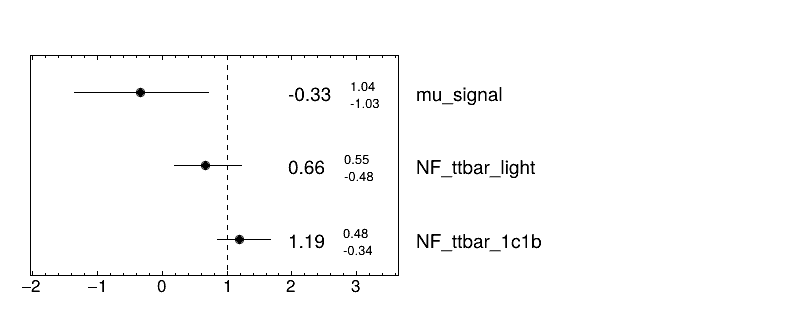}
    \caption{Normalization factors determined by the binned maximum likelihood fit for the LH 1.4 TeV $W'$ boson signal where $g'/g=2$.}
    \label{norms_LH_1p4tev}
\end{figure}

\begin{figure}[h]
    \centering
    \includegraphics[width=0.49\linewidth]{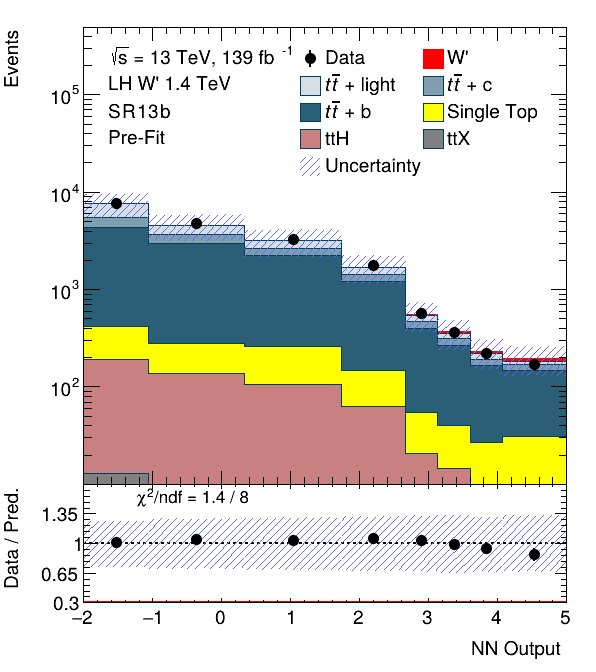}
    \includegraphics[width=0.49\linewidth]{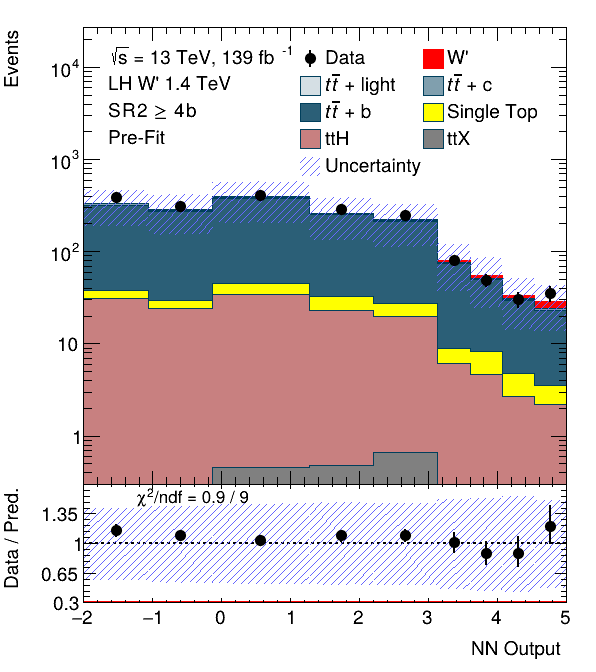}
    \caption{Asimov pre-fit plots for LH 1.4 TeV $W'$ boson signal sample where $g'/g=2$.}
    \label{data_LH_1p4tev_prefit}
\end{figure}

\begin{figure}[h]
    \centering
    \includegraphics[width=0.49\linewidth]{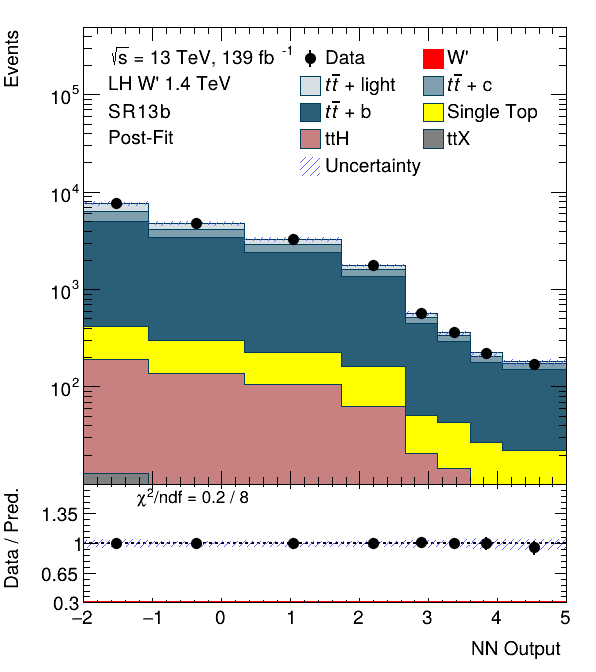}
    \includegraphics[width=0.49\linewidth]{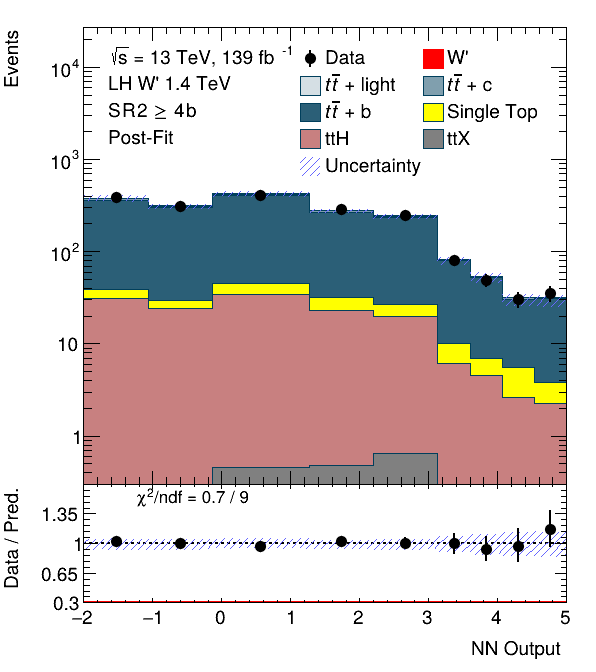}
    \caption{Asimov post-fit plots for LH 1.4 TeV $W'$ boson signal sample where $g'/g=2$.}
    \label{data_LH_1p4tev_postfit}
\end{figure}

\begin{figure}[h]
    \centering
    \includegraphics[width=0.8\linewidth]{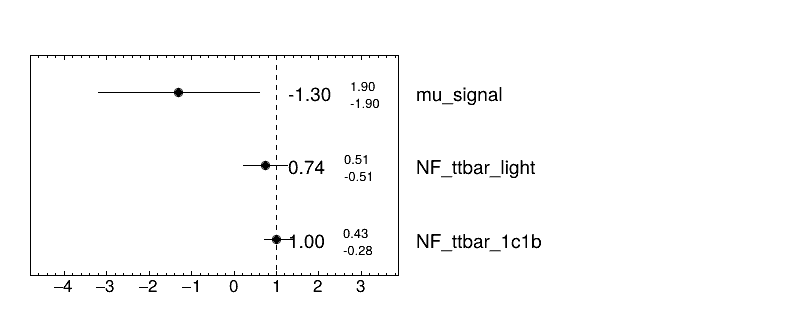}
    \caption{Normalization factors determined by the binned maximum likelihood fit for the LH 1.6 TeV $W'$ boson signal where $g'/g=2$.}
    \label{norms_LH_1p6tev}
\end{figure}

\begin{figure}[h]
    \centering
    \includegraphics[width=0.49\linewidth]{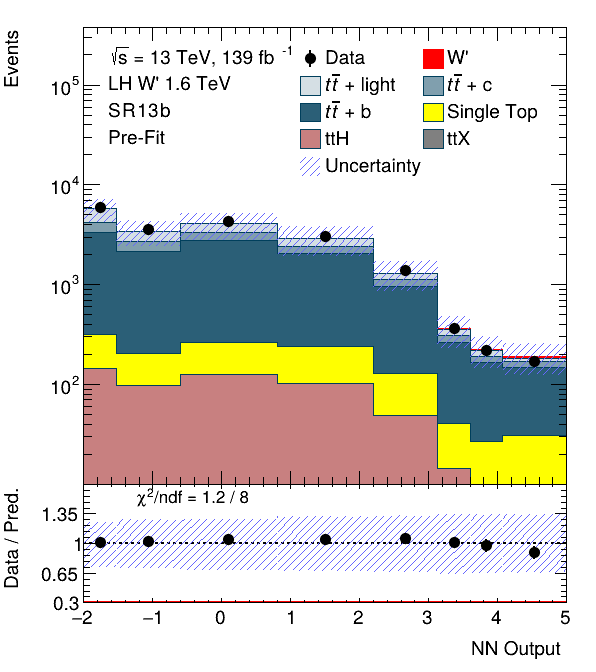}
    \includegraphics[width=0.49\linewidth]{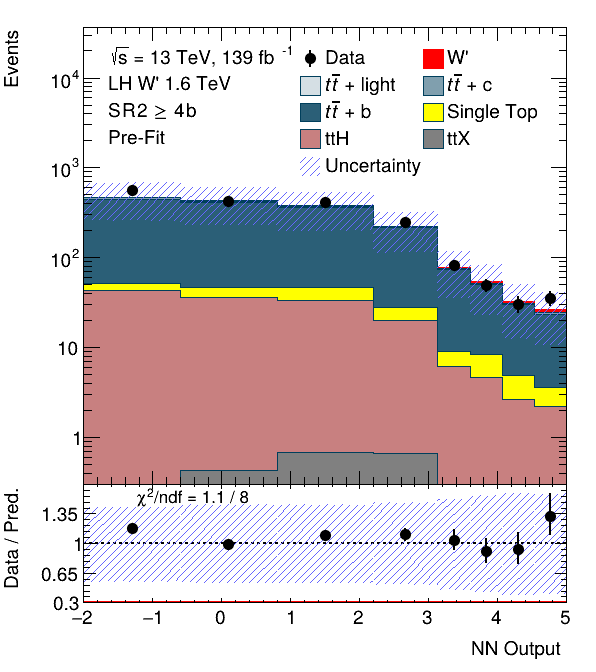}
    \caption{Asimov pre-fit plots for LH 1.6 TeV $W'$ boson signal sample where $g'/g=2$.}
    \label{data_LH_1p6tev_prefit}
\end{figure}

\begin{figure}[h]
    \centering
    \includegraphics[width=0.49\linewidth]{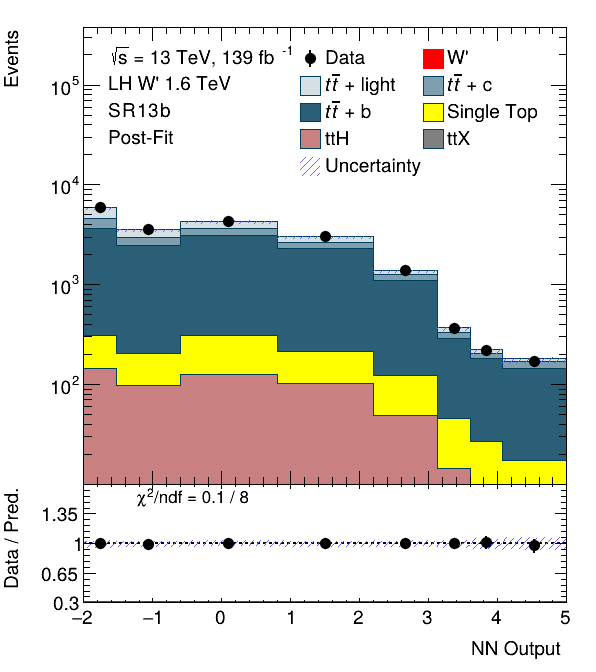}
    \includegraphics[width=0.49\linewidth]{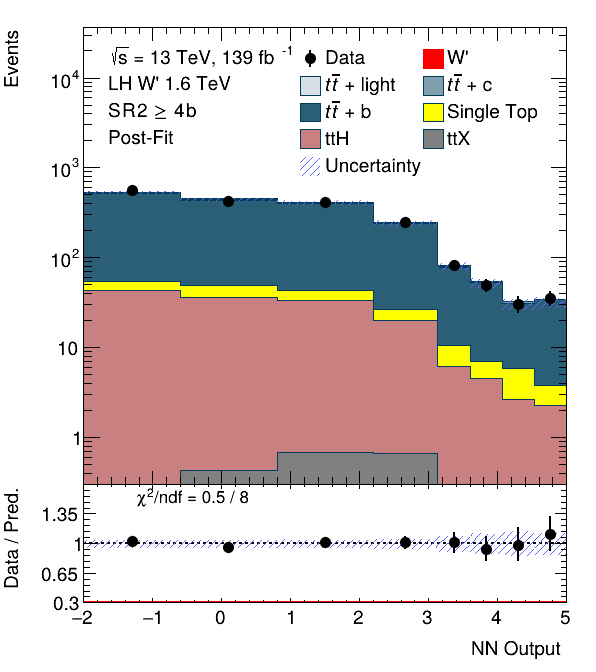}
    \caption{Asimov post-fit plots for LH 1.6 TeV $W'$ boson signal sample where $g'/g=2$.}
    \label{data_LH_1p6tev_postfit}
\end{figure}

\begin{figure}[h]
    \centering
    \includegraphics[width=0.8\linewidth]{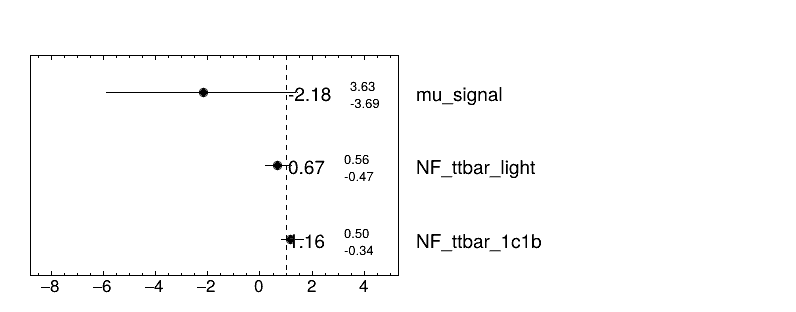}
    \caption{Normalization factors determined by the binned maximum likelihood fit for the LH 1.8 TeV $W'$ boson signal where $g'/g=2$.}
    \label{norms_LH_1p8tev}
\end{figure}

\begin{figure}[h]
    \centering
    \includegraphics[width=0.49\linewidth]{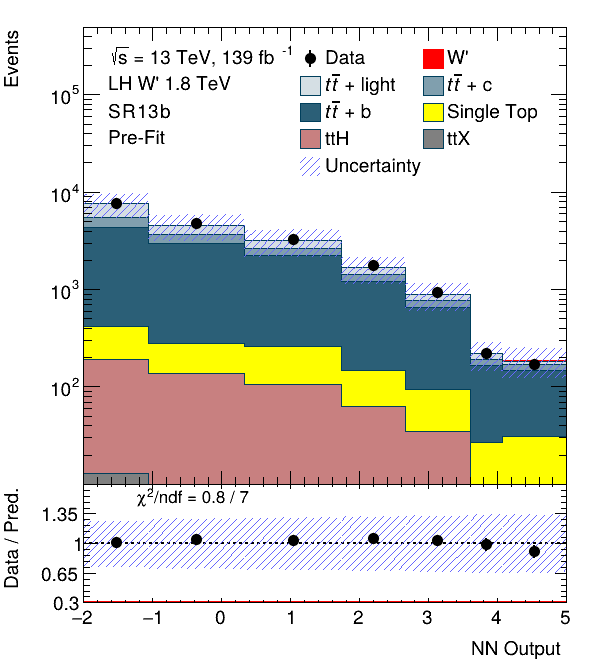}
    \includegraphics[width=0.49\linewidth]{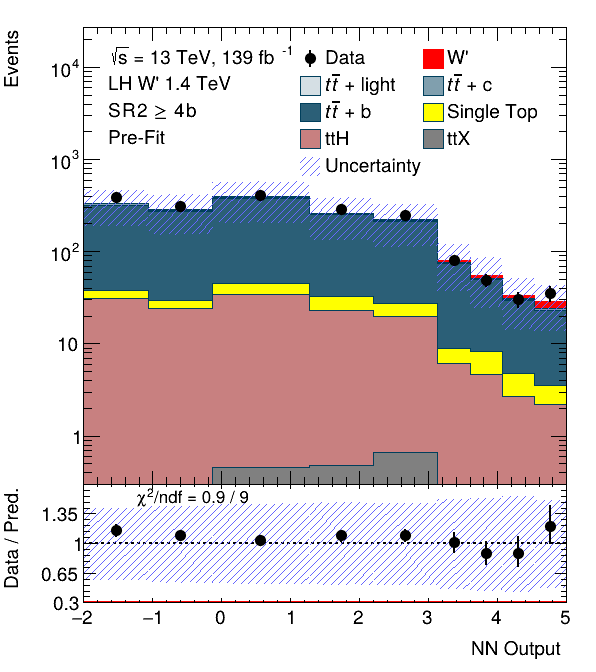}
    \caption{Asimov pre-fit plots for LH 1.8 TeV $W'$ boson signal sample where $g'/g=2$.}
    \label{data_LH_1p8tev_prefit}
\end{figure}

\begin{figure}[h]
    \centering
    \includegraphics[width=0.49\linewidth]{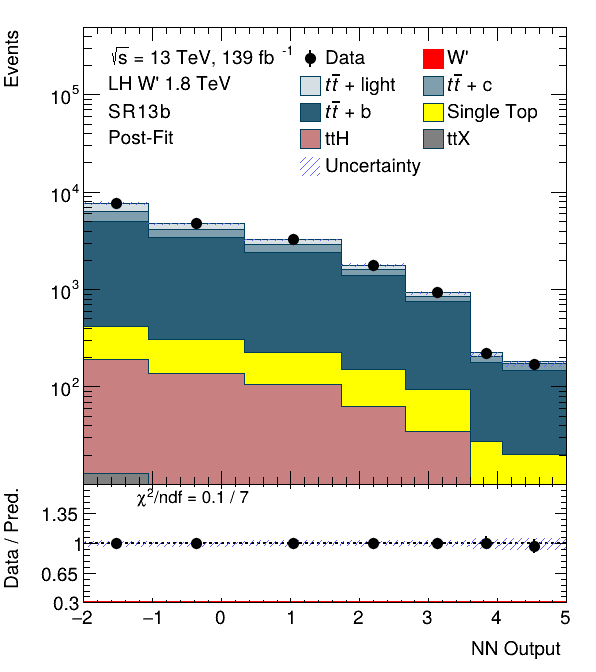}
    \includegraphics[width=0.49\linewidth]{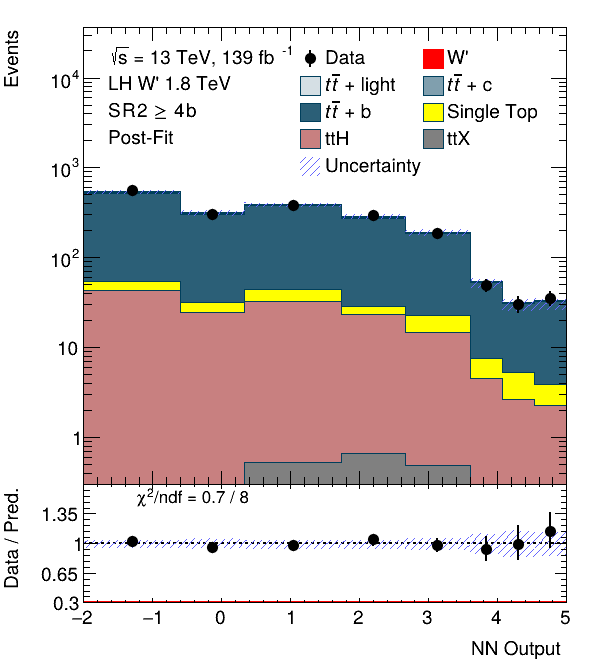}
    \caption{Asimov post-fit plots for LH 1.8 TeV $W'$ boson signal sample where $g'/g=2$.}
    \label{data_LH_1p8tev_postfit}
\end{figure}

\begin{figure}[h]
    \centering
    \includegraphics[width=0.8\linewidth]{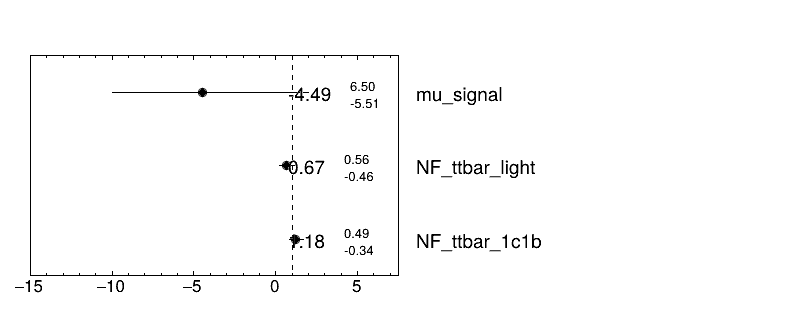}
    \caption{Normalization factors determined by the binned maximum likelihood fit for the LH 2 TeV $W'$ boson signal where $g'/g=2$.}
    \label{norms_LH_2tev}
\end{figure}

\begin{figure}[h]
    \centering
    \includegraphics[width=0.49\linewidth]{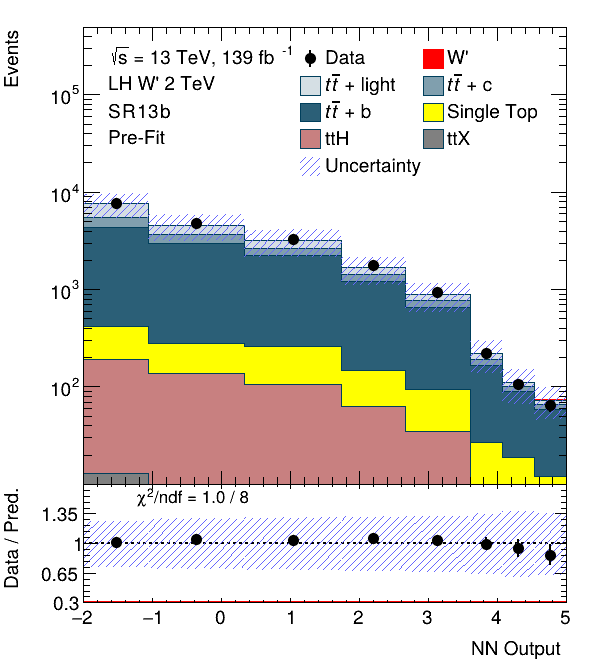}
    \includegraphics[width=0.49\linewidth]{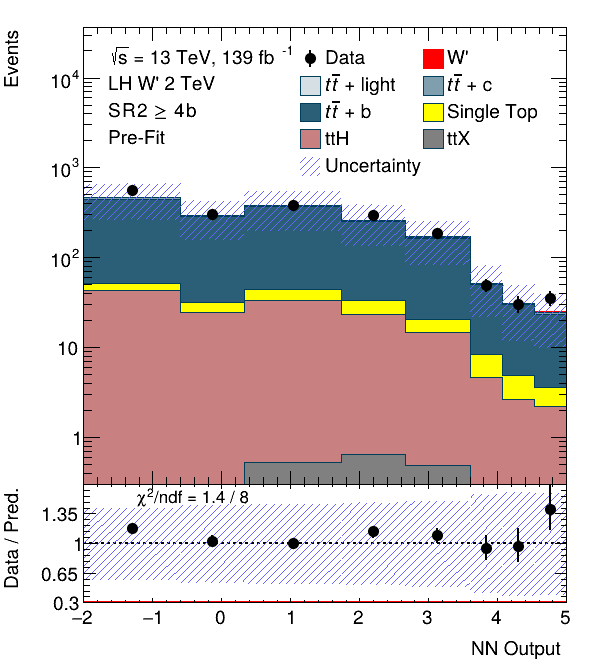}
    \caption{Asimov pre-fit plots for LH 2 TeV $W'$ boson signal sample where $g'/g=2$.}
    \label{data_LH_2tev_prefit}
\end{figure}

\begin{figure}[h]
    \centering
    \includegraphics[width=0.49\linewidth]{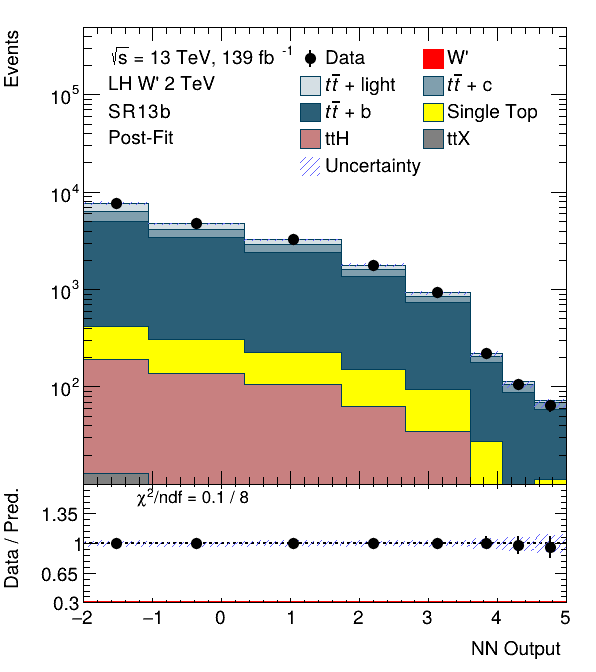}
    \includegraphics[width=0.49\linewidth]{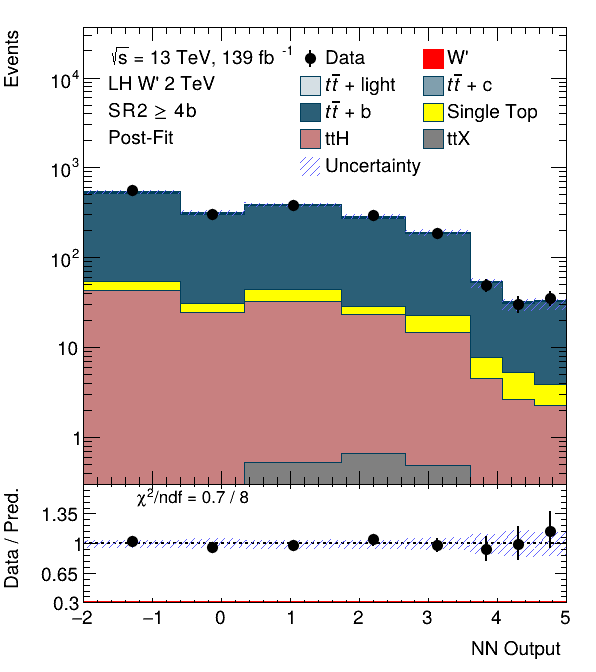}
    \caption{Asimov post-fit plots for LH 2 TeV $W'$ boson signal sample where $g'/g=2$.}
    \label{data_LH_2tev_postfit}
\end{figure}

This appendix includes results from all fits to data. For the LH $W'$ boson, the pre-fit plots are shown in Figs.~\ref{data_LH_1tev_prefit},~\ref{data_LH_1p2tev_prefit},~\ref{data_LH_1p4tev_prefit},~\ref{data_LH_1p6tev_prefit},~\ref{data_LH_1p8tev_prefit}, and~\ref{data_LH_2tev_prefit}. The pre-fit plots show overall good agreement between data and background. The major difference between the different mass points is the shrinking size of the signal. The post-fit plots are shown in Figs.~\ref{data_LH_1tev_postfit},~\ref{data_LH_1p2tev_postfit},~\ref{data_LH_1p4tev_postfit},~\ref{data_LH_1p6tev_postfit},~\ref{data_LH_1p8tev_postfit}, and~\ref{data_LH_2tev_postfit}. All post-fit plots show good convergence and the fit results in a large reduction of the uncertainty in each bin. Figures containing the resulting normalization factors can be found in Figs.~\ref{norms_LH_1tev},~\ref{norms_LH_1p2tev},~\ref{norms_LH_1p4tev},~\ref{norms_LH_1p6tev},~\ref{norms_LH_1p8tev}, and~\ref{norms_LH_2tev}. At large masses, the signal is very small and the POI, $\mu$, goes increasingly negative. A negative $\mu$ value is non-physical, which means that the fit is no longer sensitive to the $W'$ boson signal samples with high $W'$ boson mass.

\begin{figure}[h]
    \centering
    \includegraphics[width=0.45\linewidth]{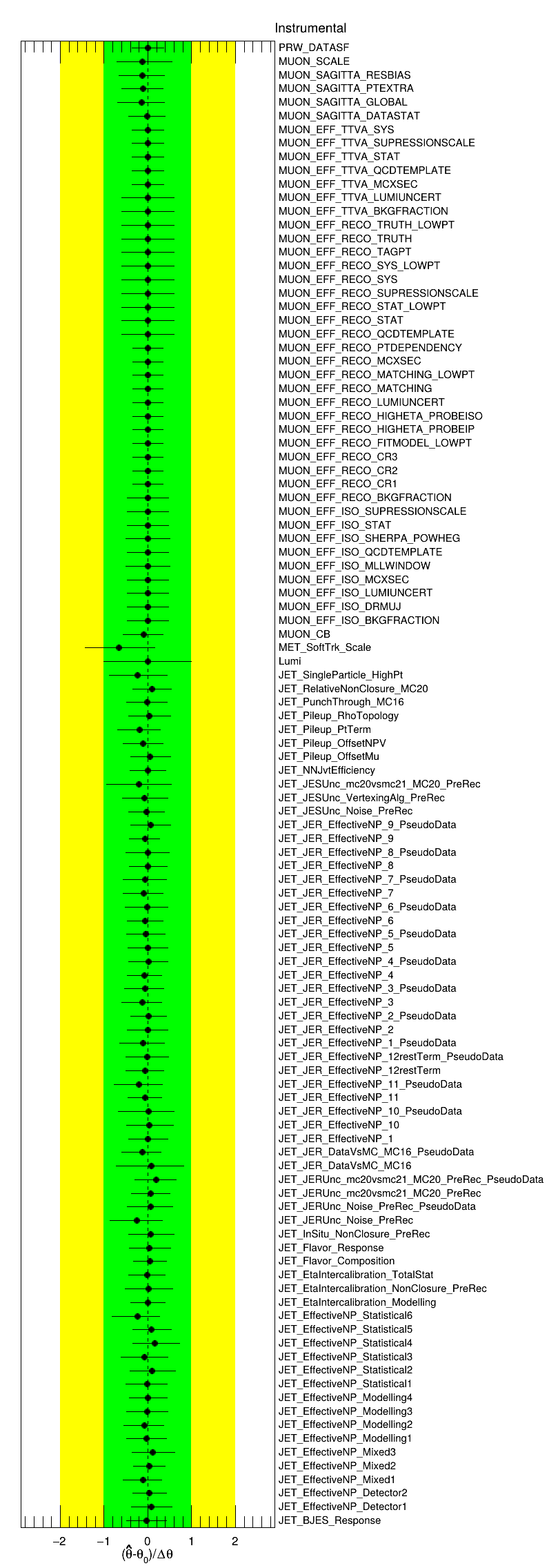}
    \caption{Detector systematic uncertainty pull plot for Asimov fit on LH 1 TeV $W'$ boson sample where g'/g=2.}
    \label{nuispar_inst_LH_1tev}
\end{figure}

\begin{figure}[h]
    \centering
    \includegraphics[width=0.49\linewidth]{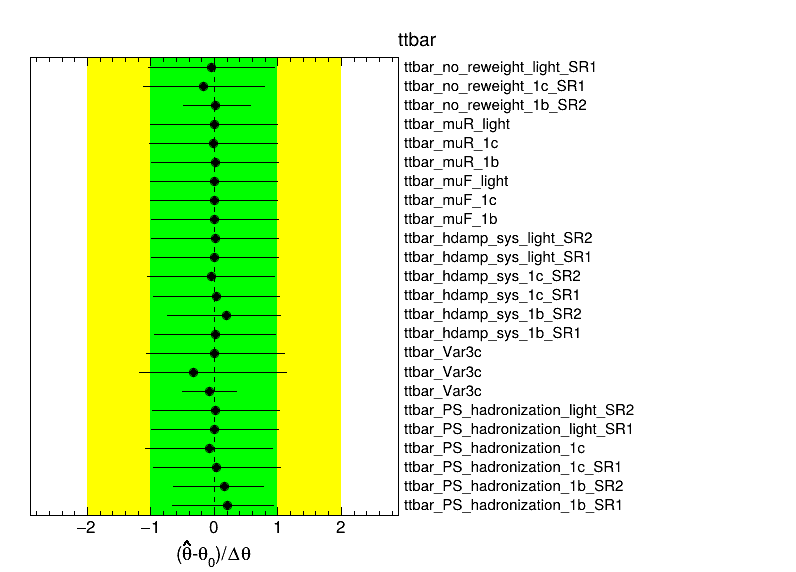}
    \includegraphics[width=0.49\linewidth]{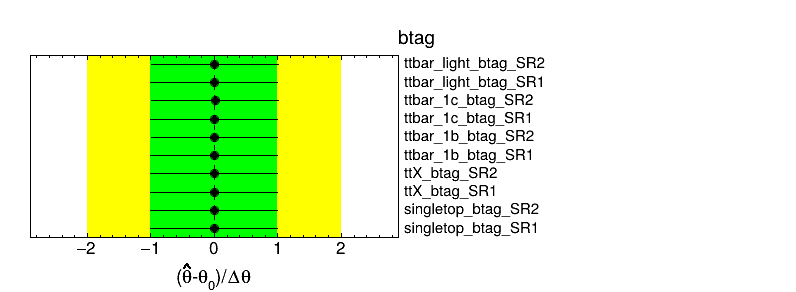}\\
    \includegraphics[width=0.49\linewidth]{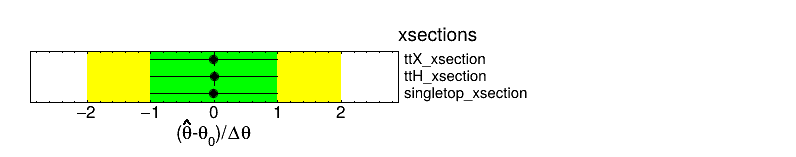}
    \caption{$t\bar{t}$, b-tagging, and cross-section systematic uncertainty pull plot for Asimov fit on LH 1 TeV $W'$ boson sample where g'/g=2.}
    \label{nuispar_all_LH_1tev}
\end{figure}

\begin{figure}[h]
    \centering
    \includegraphics[width=0.45\linewidth]{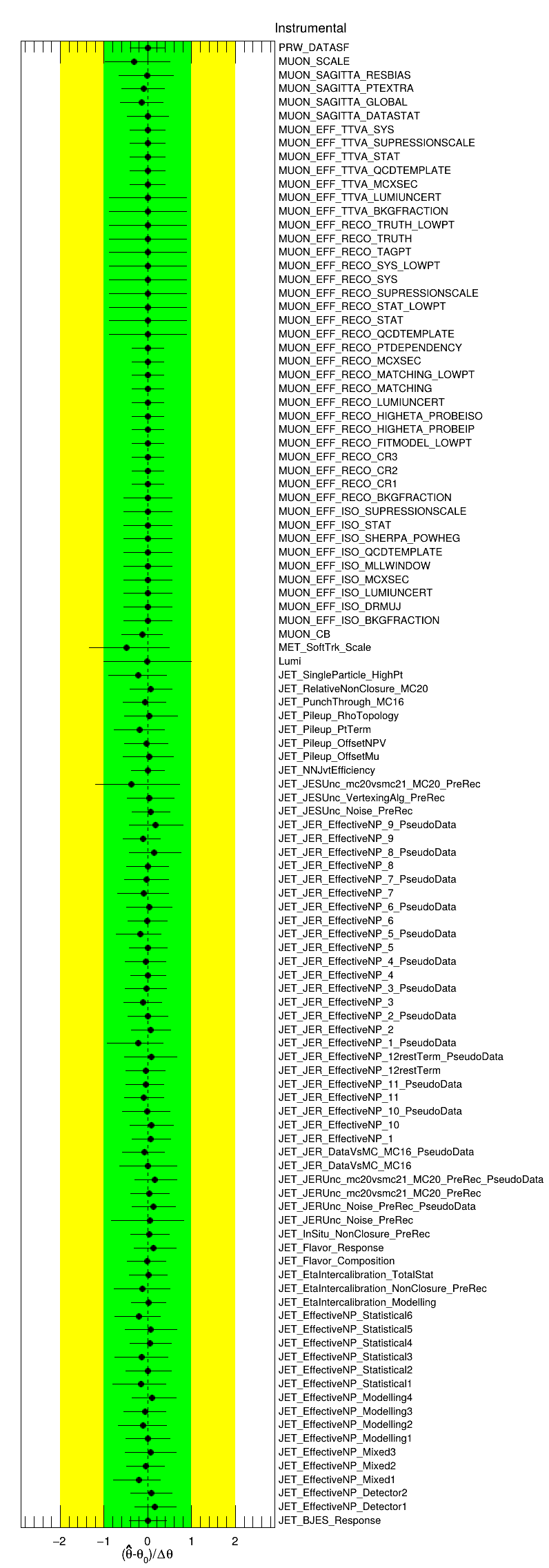}
    \caption{Detector systematic uncertainty pull plot for Asimov fit on LH 1.2 TeV $W'$ boson sample where $g'/g=2$.}
    \label{nuispar_inst_LH_1p2tev}
\end{figure}

\begin{figure}[h]
    \centering
    \includegraphics[width=0.49\linewidth]{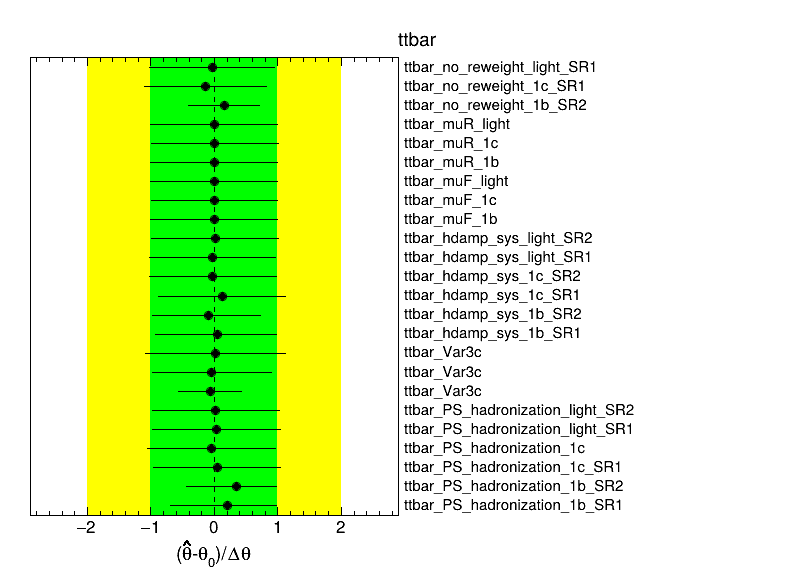}
    \includegraphics[width=0.49\linewidth]{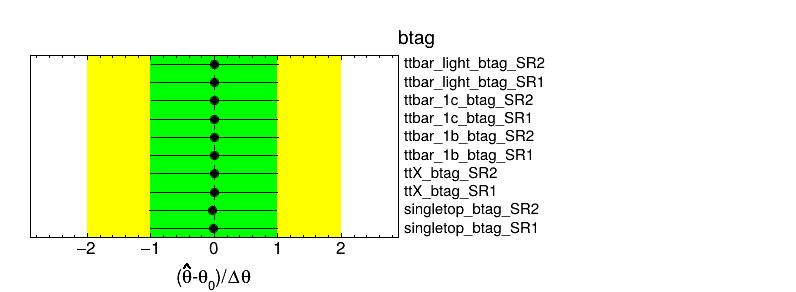}\\
    \includegraphics[width=0.49\linewidth]{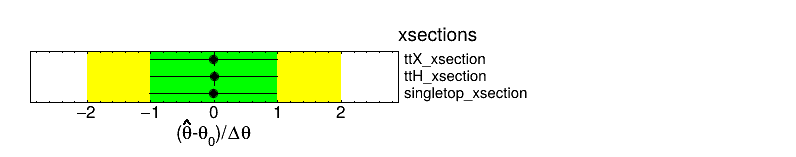}
    \caption{$t\bar{t}$, b-tagging, and cross-section systematic uncertainty pull plot for Asimov fit on LH 1.2 TeV $W'$ boson sample where $g'/g=2$.}
    \label{nuispar_all_LH_1p2tev}
\end{figure}

\begin{figure}[h]
    \centering
    \includegraphics[width=0.45\linewidth]{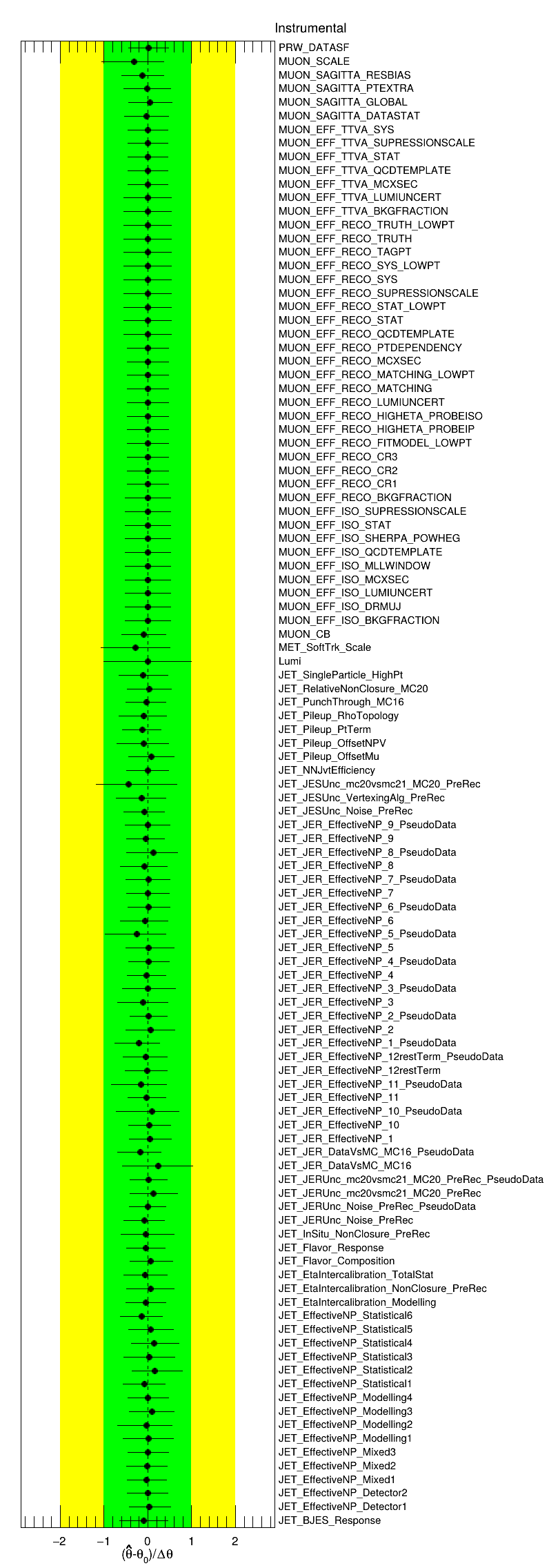}
    \caption{Detector systematic uncertainty pull plot for Asimov fit on LH 1.4 TeV $W'$ boson sample where $g'/g=2$.}
    \label{nuispar_inst_LH_1p4tev}
\end{figure}

\begin{figure}[h]
    \centering
    \includegraphics[width=0.49\linewidth]{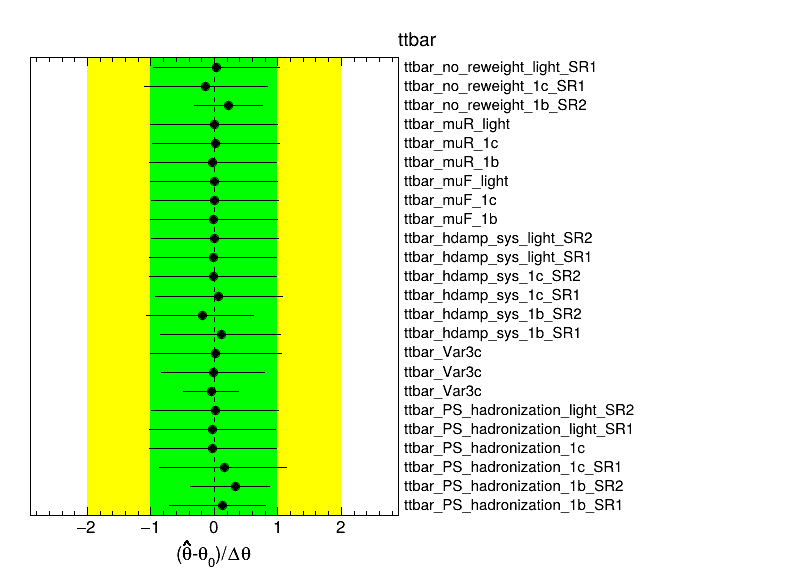}
    \includegraphics[width=0.49\linewidth]{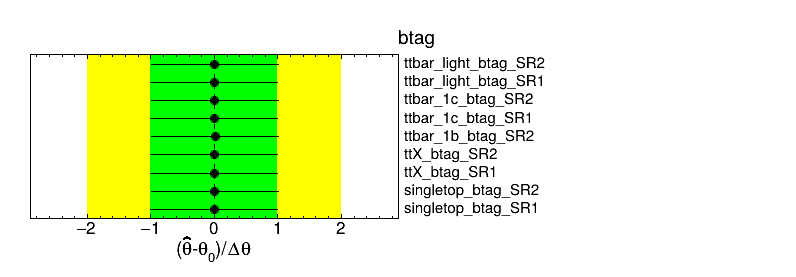}\\
    \includegraphics[width=0.49\linewidth]{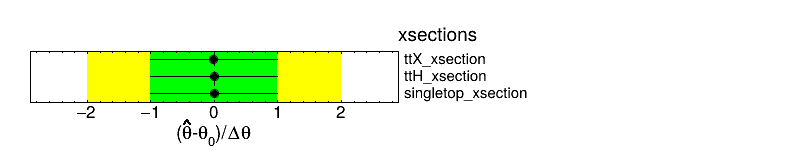}
    \caption{$t\bar{t}$, b-tagging, and cross-section systematic uncertainty pull plot for Asimov fit on LH 1.4 TeV $W'$ boson sample where $g'/g=2$.}
    \label{nuispar_all_LH_1p4tev}
\end{figure}

\begin{figure}[h]
    \centering
    \includegraphics[width=0.45\linewidth]{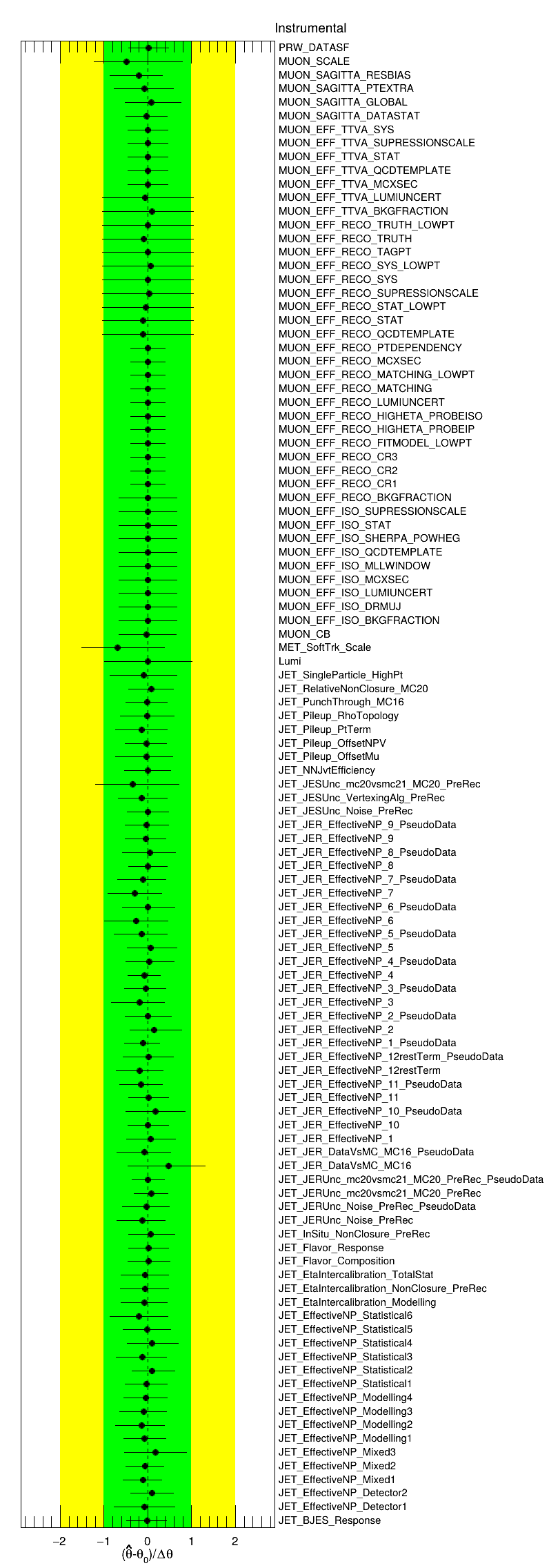}
    \caption{Detector systematic uncertainty pull plot for Asimov fit on LH 1.6 TeV $W'$ boson sample where $g'/g=2$.}
    \label{nuispar_inst_LH_1p6tev}
\end{figure}

\begin{figure}[h]
    \centering
    \includegraphics[width=0.49\linewidth]{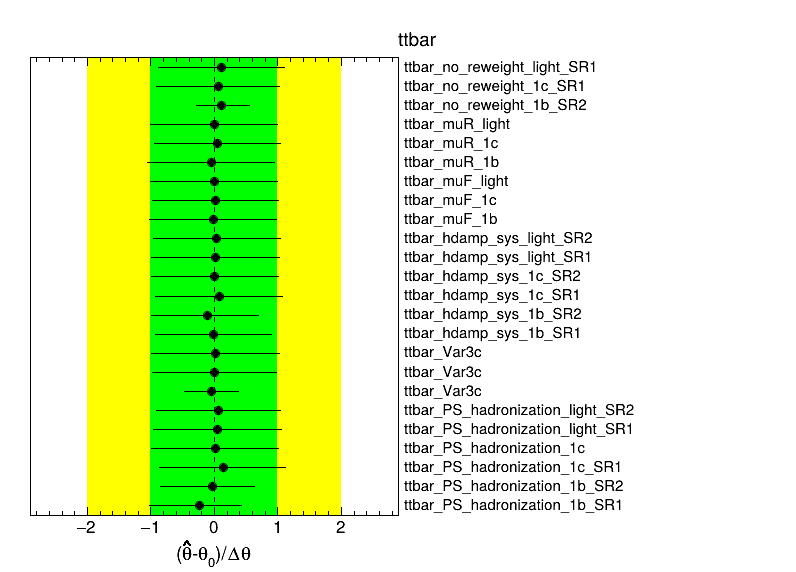}
    \includegraphics[width=0.49\linewidth]{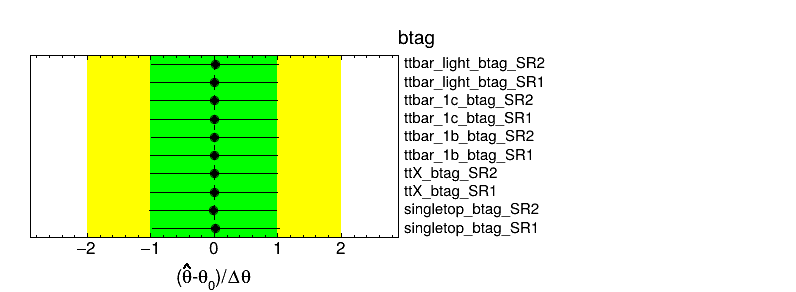}\\
    \includegraphics[width=0.49\linewidth]{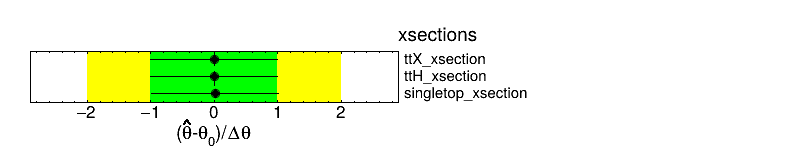}
    \caption{$t\bar{t}$, b-tagging, and cross-section systematic uncertainty pull plot for Asimov fit on LH 1.6 TeV $W'$ boson sample where $g'/g=2$.}
    \label{nuispar_all_LH_1p6tev}
\end{figure}

\begin{figure}[h]
    \centering
    \includegraphics[width=0.45\linewidth]{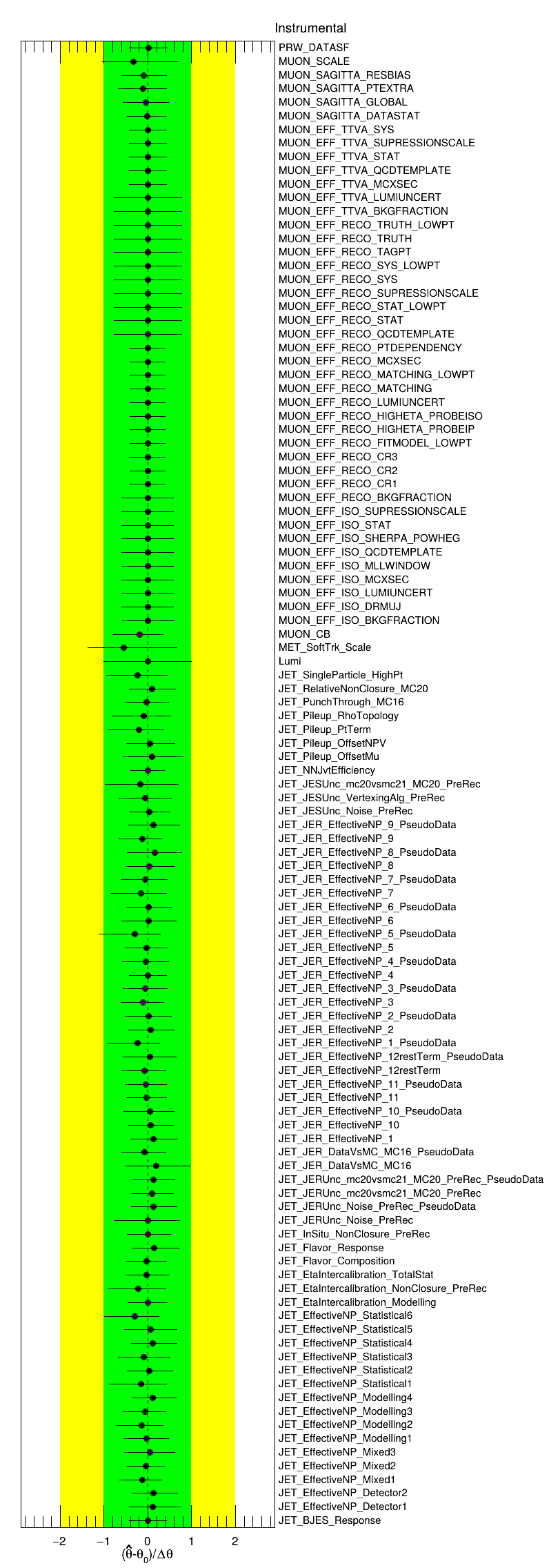}
    \caption{Detector systematic uncertainty pull plot for Asimov fit on LH 1.8 TeV $W'$ boson sample where $g'/g=2$.}
    \label{nuispar_inst_LH_1p8tev}
\end{figure}

\begin{figure}[h]
    \centering
    \includegraphics[width=0.49\linewidth]{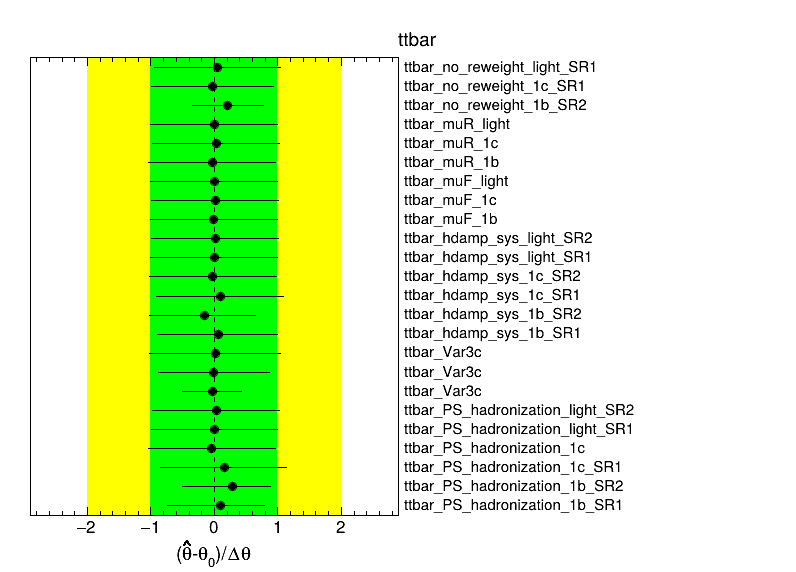}
    \includegraphics[width=0.49\linewidth]{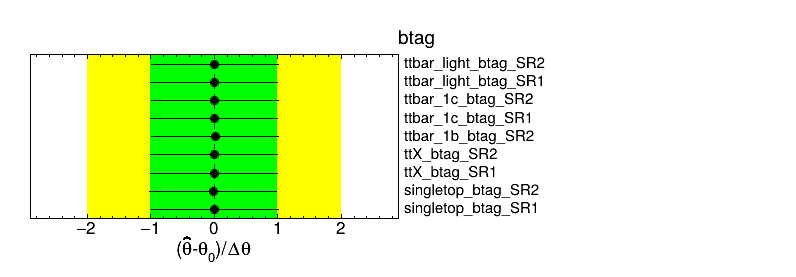}\\
    \includegraphics[width=0.49\linewidth]{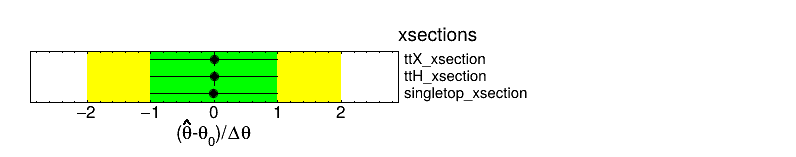}
    \caption{$t\bar{t}$, b-tagging, and cross-section systematic uncertainty pull plot for Asimov fit on LH 1.8 TeV $W'$ boson sample where $g'/g=2$.}
    \label{nuispar_all_LH_1p8tev}
\end{figure}

\begin{figure}[h]
    \centering
    \includegraphics[width=0.45\linewidth]{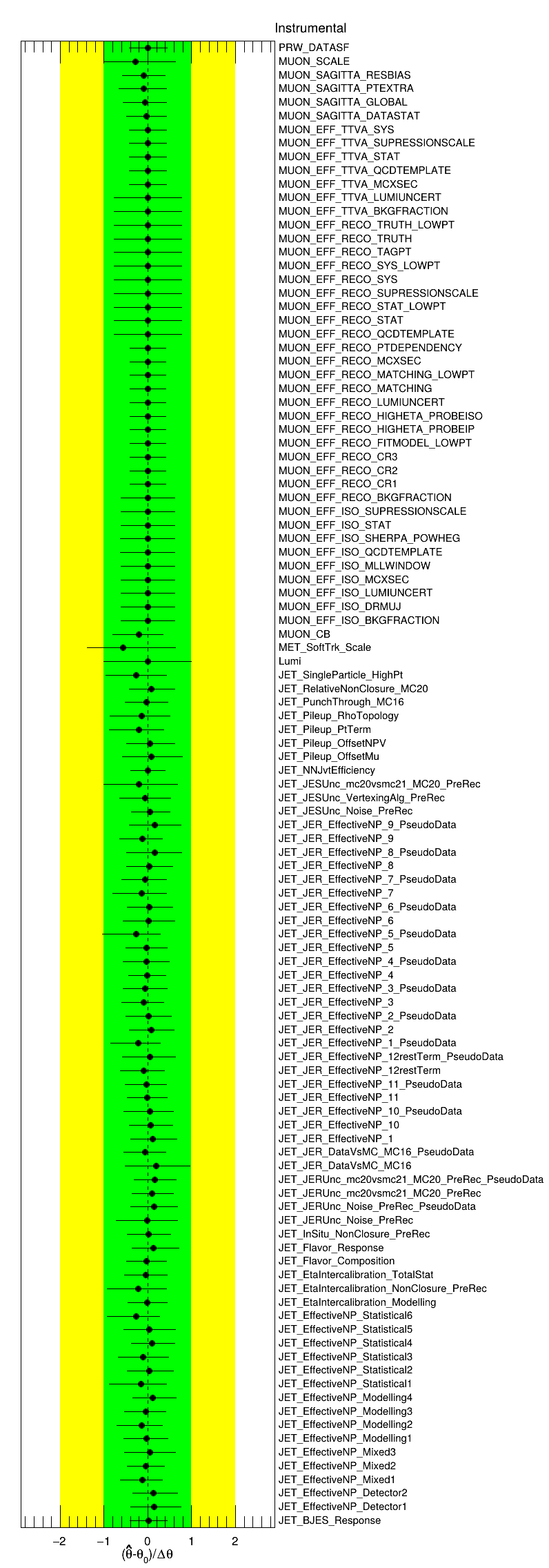}
    \caption{Detector systematic uncertainty pull plot for Asimov fit on LH 2 TeV $W'$ boson sample where $g'/g=2$.}
    \label{nuispar_inst_LH_2tev}
\end{figure}

\begin{figure}[h]
    \centering
    \includegraphics[width=0.49\linewidth]{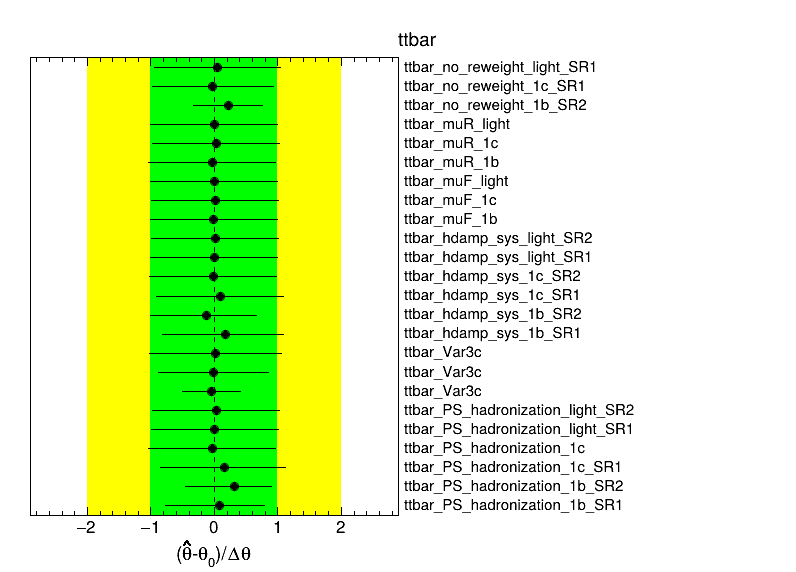}
    \includegraphics[width=0.49\linewidth]{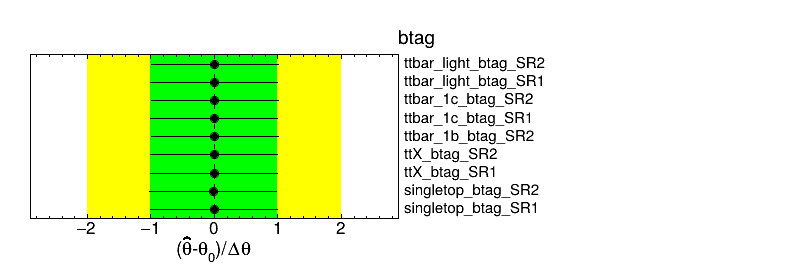}\\
    \includegraphics[width=0.49\linewidth]{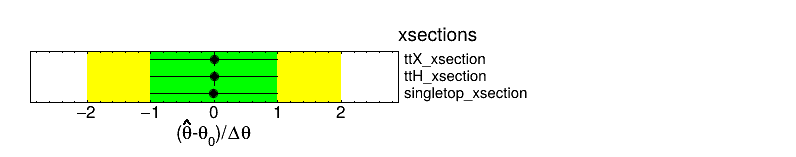}
    \caption{$t\bar{t}$, b-tagging, and cross-section systematic uncertainty pull plot for Asimov fit on LH 2 TeV $W'$ boson sample where $g'/g=2$.}
    \label{nuispar_all_LH_2tev}
\end{figure}

The pull plots on the instrument systematic uncertainties for the LH $W'$ boson can be seen in Figs.~\ref{nuispar_inst_LH_1tev}, ~\ref{nuispar_inst_LH_1p2tev}, ~\ref{nuispar_inst_LH_1p4tev}, ~\ref{nuispar_inst_LH_1p6tev}, ~\ref{nuispar_inst_LH_1p8tev}, ~\ref{nuispar_inst_LH_2tev}. Each of these plots are fairly similar showing constraints on most of the systematic uncertainties. There are a few systematic uncertainties that are pulled from their nominal values, but none that exceed a $1\sigma$ difference. The pull plots on the systematic uncertainties for b-tagging, cross-sections, and $t\bar{t}$ modeling can be seen in Figs.~\ref{nuispar_all_LH_1tev}, ~\ref{nuispar_all_LH_1p2tev}, ~\ref{nuispar_all_LH_1p4tev}, ~\ref{nuispar_all_LH_1p6tev}, ~\ref{nuispar_all_LH_1p8tev}, ~\ref{nuispar_all_LH_2tev}. In general, the systematic uncertainties for the cross-sections and b-tagging are not constrained and do not deviate from their nominal values. However, some of the $t\bar{t}$ modeling systematic uncertainties are constrained and pulled away from their nominal value. Pull plots are very similar across each mass point.

\begin{figure}[h]
    \centering
    \includegraphics[width=0.8\linewidth]{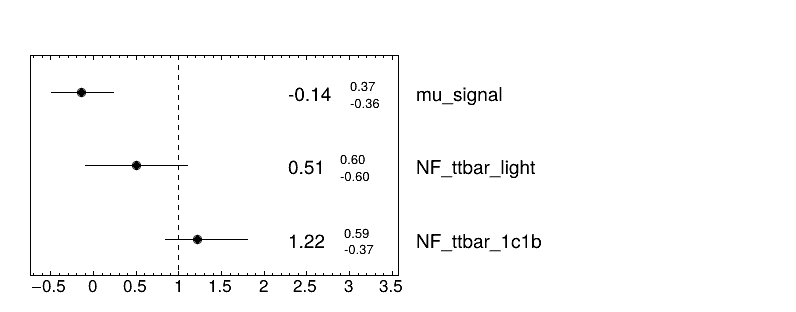}
    \caption{Normalization factors determined by the binned maximum likelihood fit for the RH 1 TeV $W'$ boson signal where g'/g=2.}
    \label{norms_RH_1tev}
\end{figure}

\begin{figure}[h]
    \centering
    \includegraphics[width=0.49\linewidth]{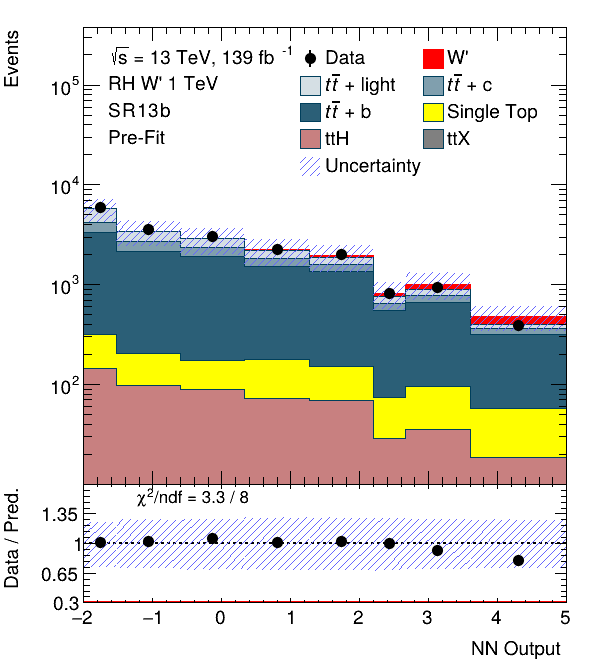}
    \includegraphics[width=0.49\linewidth]{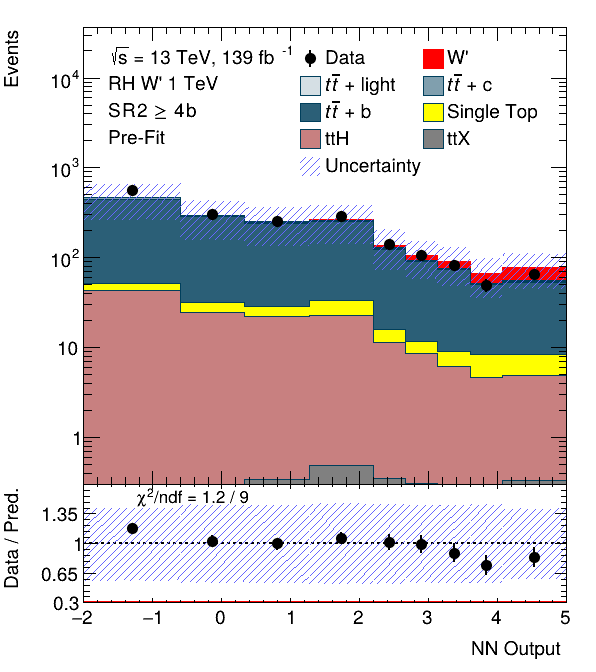}
    \caption{Asimov pre-fit plots for RH 1 TeV $W'$ boson signal sample where g'/g=2.}
    \label{data_RH_1tev_prefit}
\end{figure}

\begin{figure}[h]
    \centering
    \includegraphics[width=0.49\linewidth]{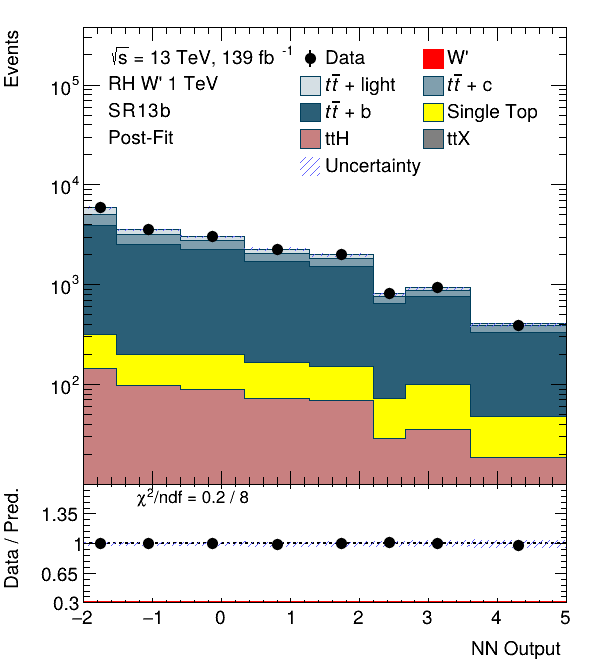}
    \includegraphics[width=0.49\linewidth]{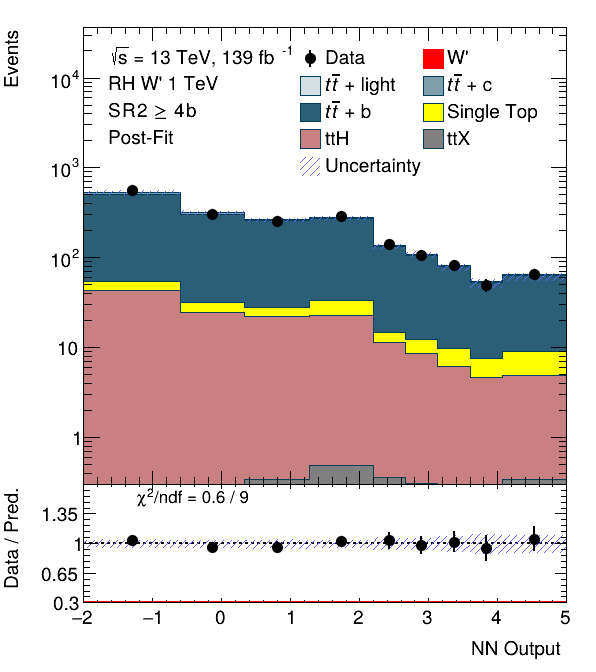}
    \caption{Asimov post-fit plots for RH 1 TeV $W'$ boson signal sample where g'/g=2.}
    \label{data_RH_1tev_postfit}
\end{figure}

\begin{figure}[h]
    \centering
    \includegraphics[width=0.8\linewidth]{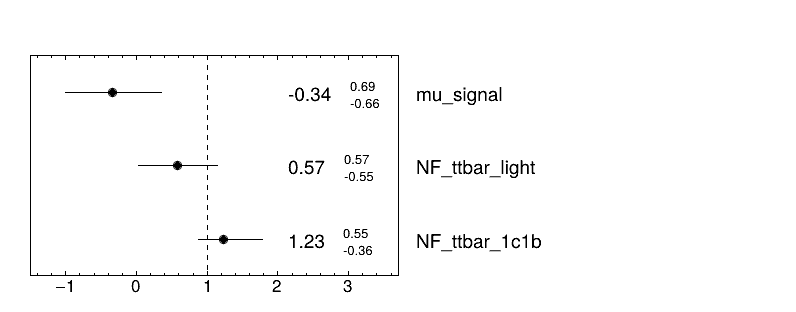}
    \caption{Normalization factors determined by the binned maximum likelihood fit for the RH 1.2 TeV $W'$ boson signal where $g'/g=2$.}
    \label{norms_RH_1p2tev}
\end{figure}

\begin{figure}[h]
    \centering
    \includegraphics[width=0.49\linewidth]{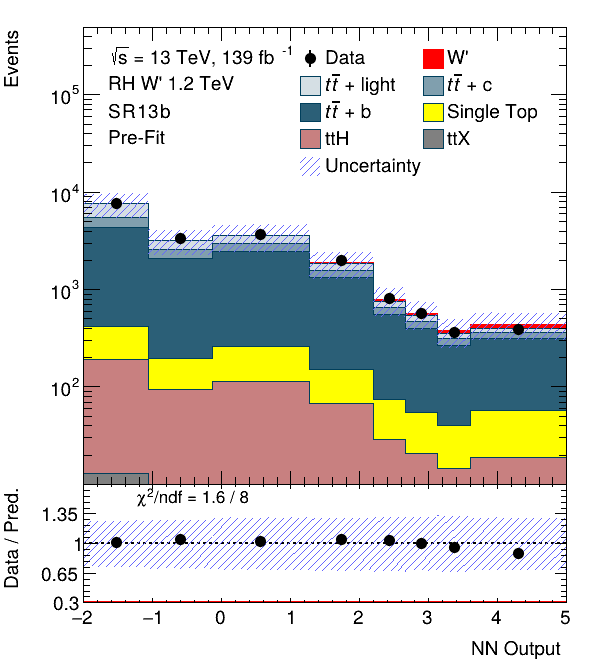}
    \includegraphics[width=0.49\linewidth]{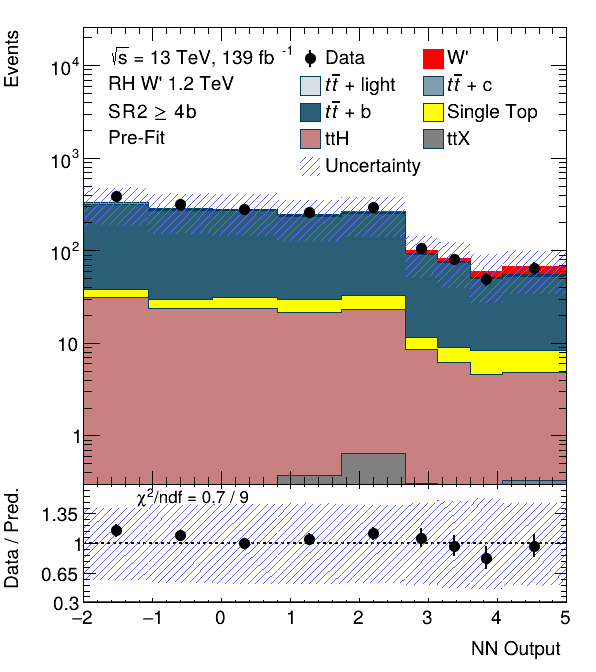}
    \caption{Asimov pre-fit plots for RH 1.2 TeV $W'$ boson signal sample where $g'/g=2$.}
    \label{data_RH_1p2tev_prefit}
\end{figure}

\begin{figure}[h]
    \centering
    \includegraphics[width=0.49\linewidth]{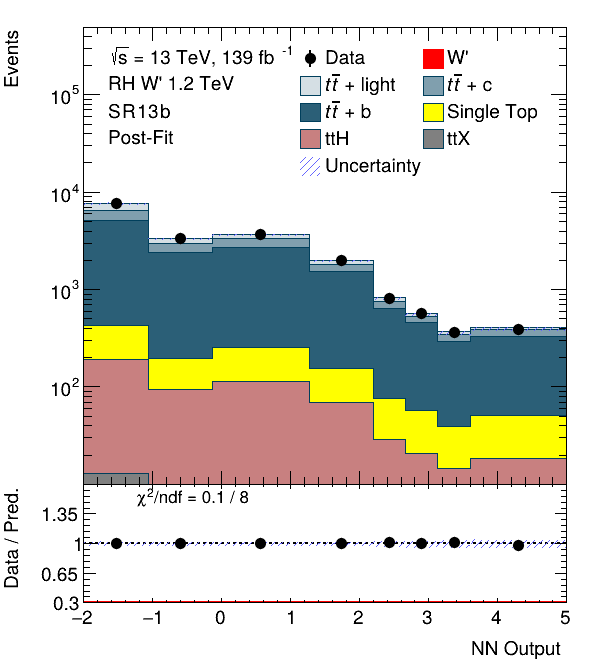}
    \includegraphics[width=0.49\linewidth]{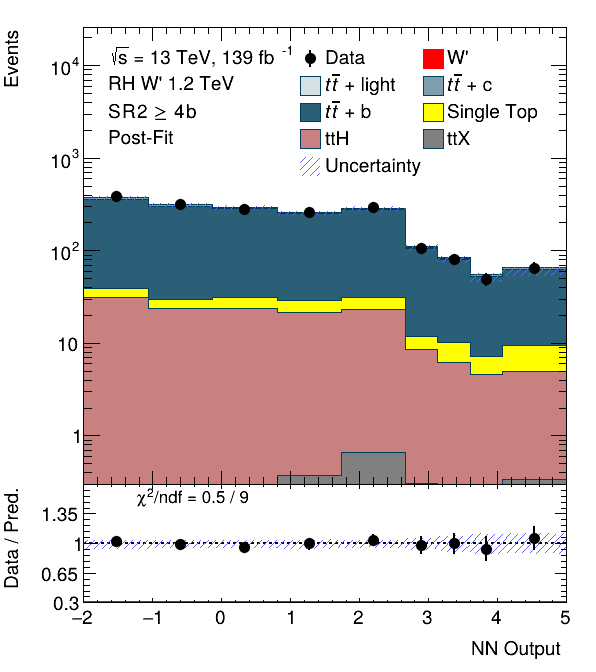}
    \caption{Asimov post-fit plots for RH 1.2 TeV $W'$ boson signal sample where $g'/g=2$.}
    \label{data_RH_1p2tev_postfit}
\end{figure}

\begin{figure}[h]
    \centering
    \includegraphics[width=0.8\linewidth]{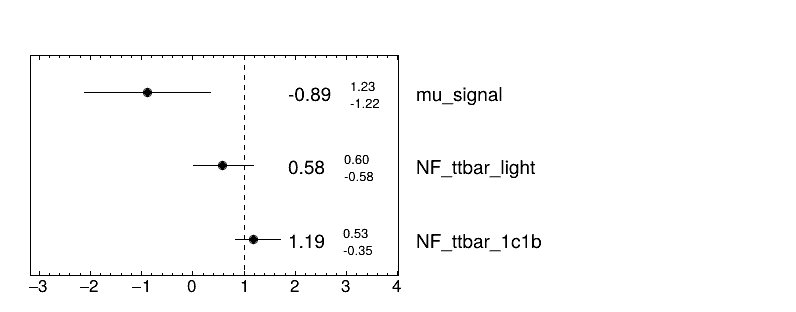}
    \caption{Normalization factors determined by the binned maximum likelihood fit for the RH 1.4 TeV $W'$ boson signal where $g'/g=2$.}
    \label{norms_RH_1p4tev}
\end{figure}

\begin{figure}[h]
    \centering
    \includegraphics[width=0.49\linewidth]{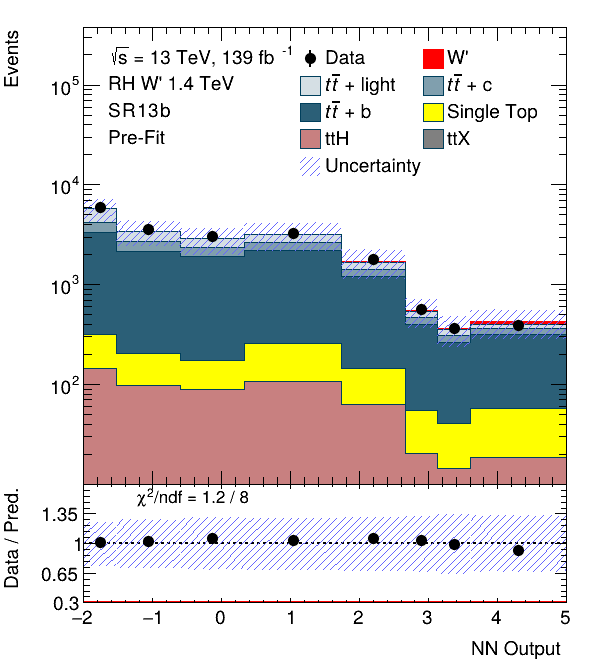}
    \includegraphics[width=0.49\linewidth]{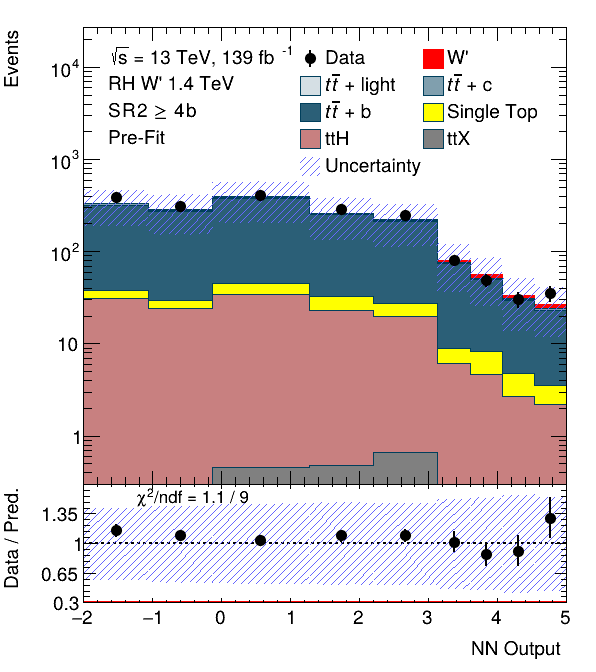}
    \caption{Asimov pre-fit plots for RH 1.4 TeV $W'$ boson signal sample where $g'/g=2$.}
    \label{data_RH_1p4tev_prefit}
\end{figure}

\begin{figure}[h]
    \centering
    \includegraphics[width=0.49\linewidth]{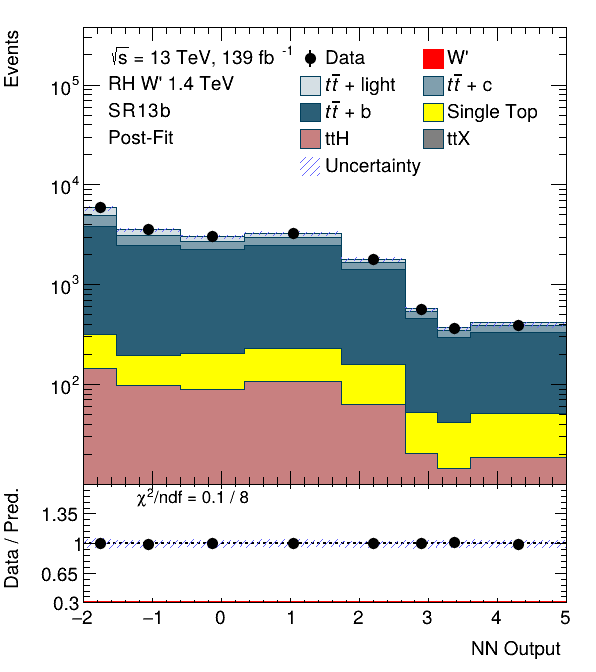}
    \includegraphics[width=0.49\linewidth]{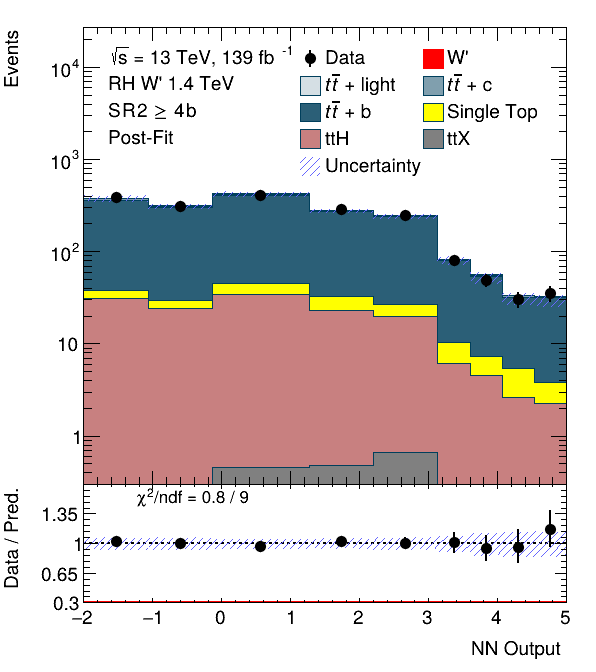}
    \caption{Asimov post-fit plots for RH 1.4 TeV $W'$ boson signal sample where $g'/g=2$.}
    \label{data_RH_1p4tev_postfit}
\end{figure}

\begin{figure}[h]
    \centering
    \includegraphics[width=0.8\linewidth]{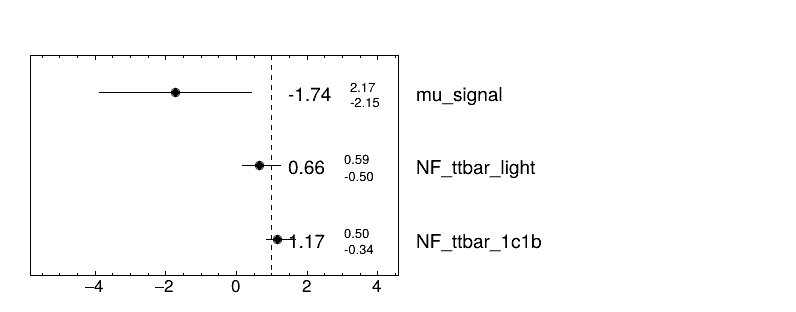}
    \caption{Normalization factors determined by the binned maximum likelihood fit for the RH 1.6 TeV $W'$ boson signal where $g'/g=2$.}
    \label{norms_RH_1p6tev}
\end{figure}

\begin{figure}[h]
    \centering
    \includegraphics[width=0.49\linewidth]{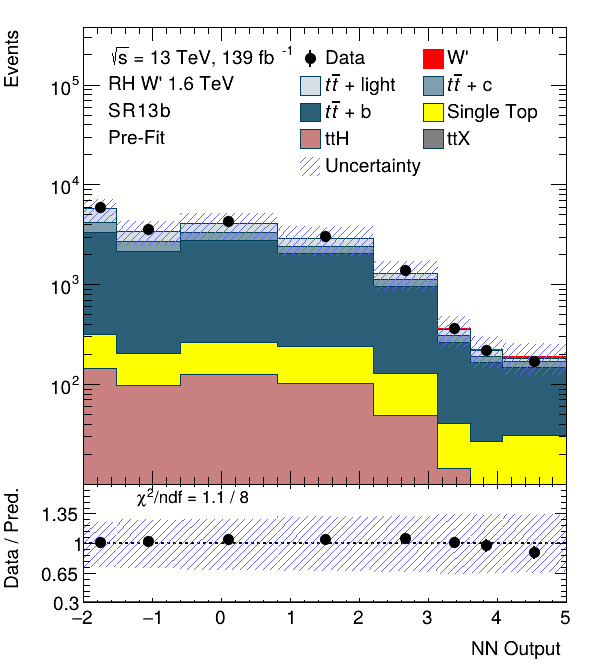}
    \includegraphics[width=0.49\linewidth]{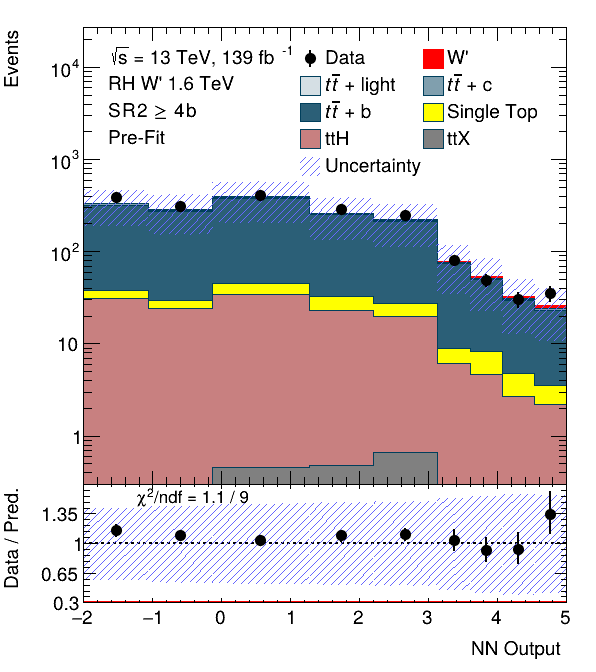}
    \caption{Asimov pre-fit plots for RH 1.6 TeV $W'$ boson signal sample where $g'/g=2$.}
    \label{data_RH_1p6tev_prefit}
\end{figure}

\begin{figure}[h]
    \centering
    \includegraphics[width=0.49\linewidth]{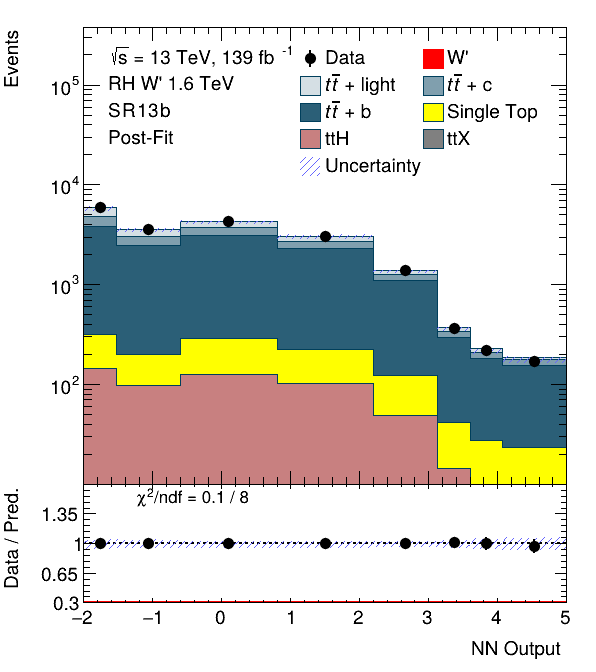}
    \includegraphics[width=0.49\linewidth]{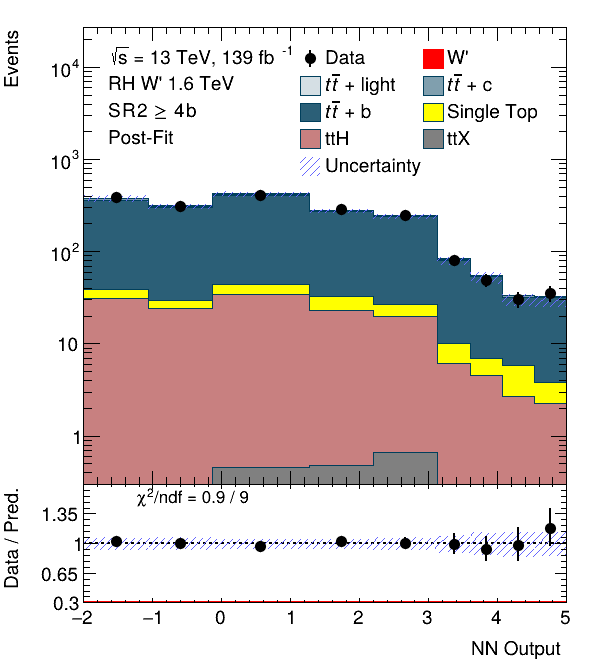}
    \caption{Asimov post-fit plots for RH 1.6 TeV $W'$ boson signal sample where $g'/g=2$.}
    \label{data_RH_1p6tev_postfit}
\end{figure}

\begin{figure}[h]
    \centering
    \includegraphics[width=0.8\linewidth]{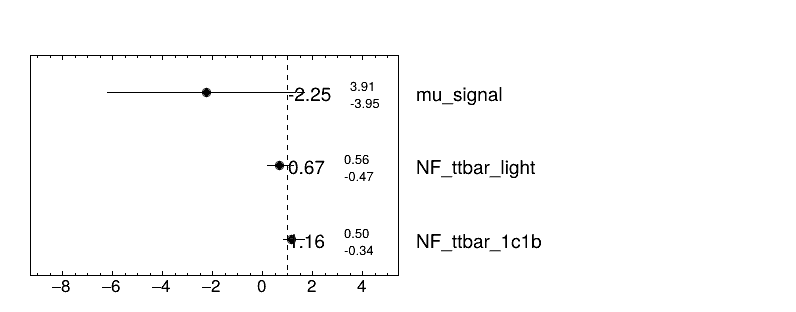}
    \caption{Normalization factors determined by the binned maximum likelihood fit for the RH 1.8 TeV $W'$ boson signal where $g'/g=2$.}
    \label{norms_RH_1p8tev}
\end{figure}

\begin{figure}[h]
    \centering
    \includegraphics[width=0.49\linewidth]{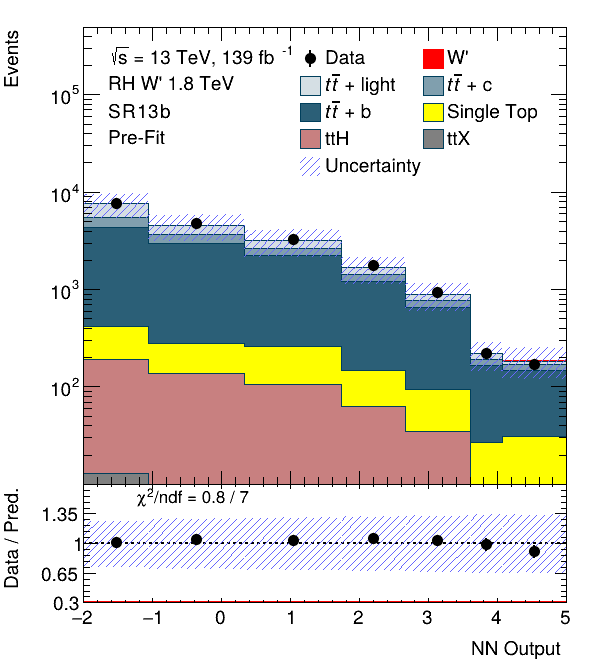}
    \includegraphics[width=0.49\linewidth]{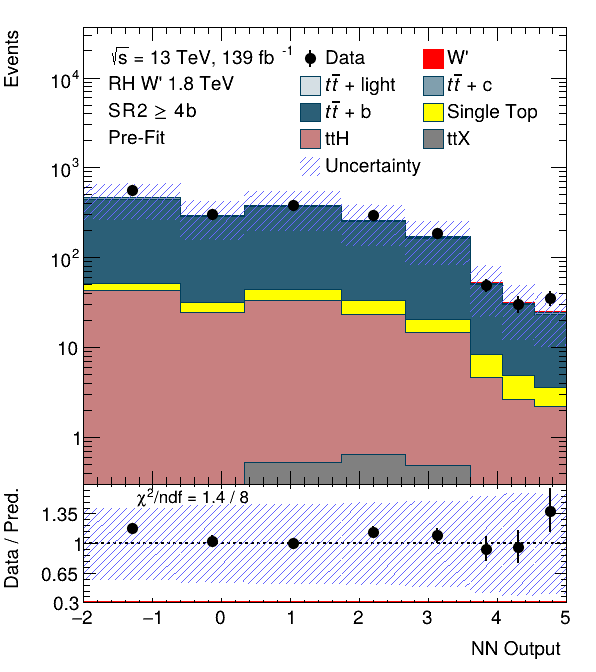}
    \caption{Asimov pre-fit plots for RH 1.8 TeV $W'$ boson signal sample where $g'/g=2$.}
    \label{data_RH_1p8tev_prefit}
\end{figure}

\begin{figure}[h]
    \centering
    \includegraphics[width=0.49\linewidth]{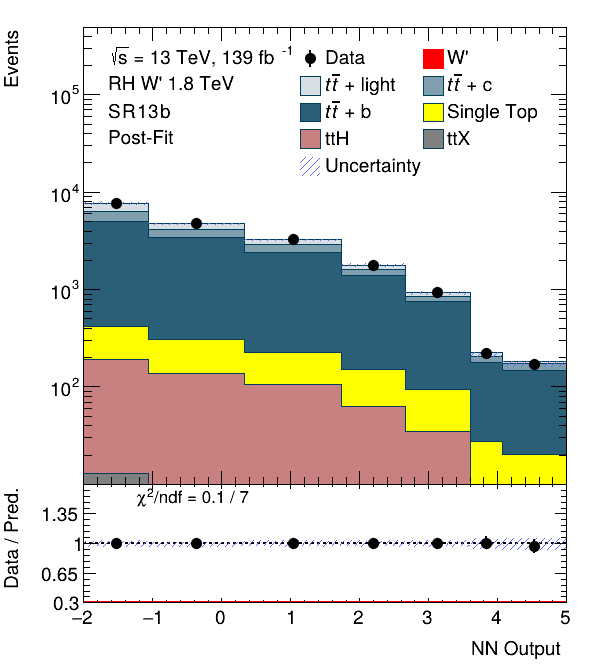}
    \includegraphics[width=0.49\linewidth]{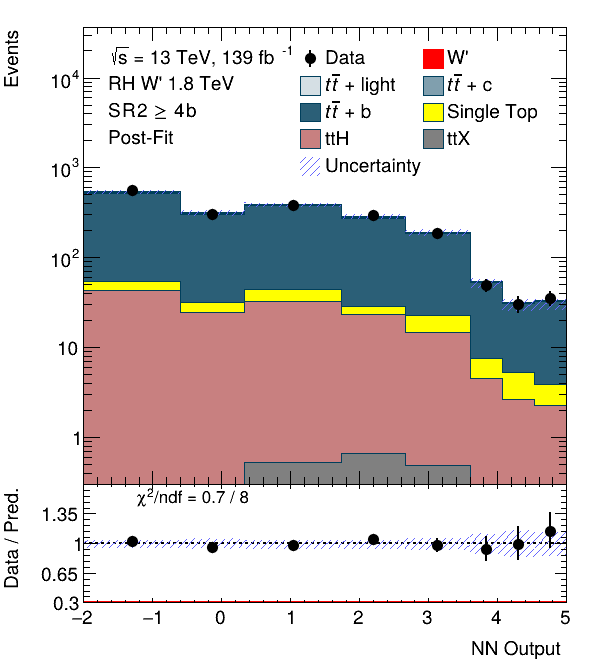}
    \caption{Asimov post-fit plots for RH 1.8 TeV $W'$ boson signal sample where $g'/g=2$.}
    \label{data_RH_1p8tev_postfit}
\end{figure}

\begin{figure}[h]
    \centering
    \includegraphics[width=0.8\linewidth]{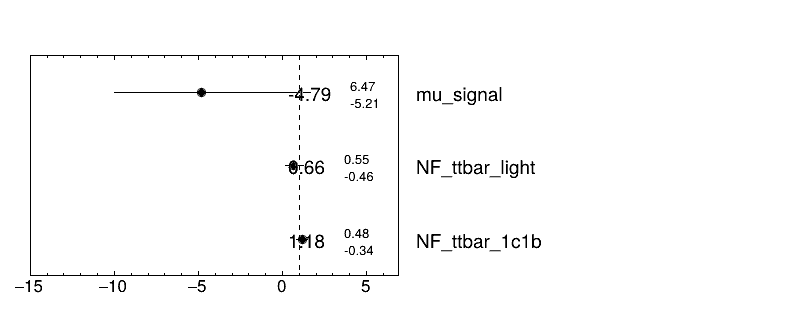}
    \caption{Normalization factors determined by the binned maximum likelihood fit for the RH 2 TeV $W'$ boson signal where $g'/g=2$.}
    \label{norms_RH_2tev}
\end{figure}

\begin{figure}[h]
    \centering
    \includegraphics[width=0.49\linewidth]{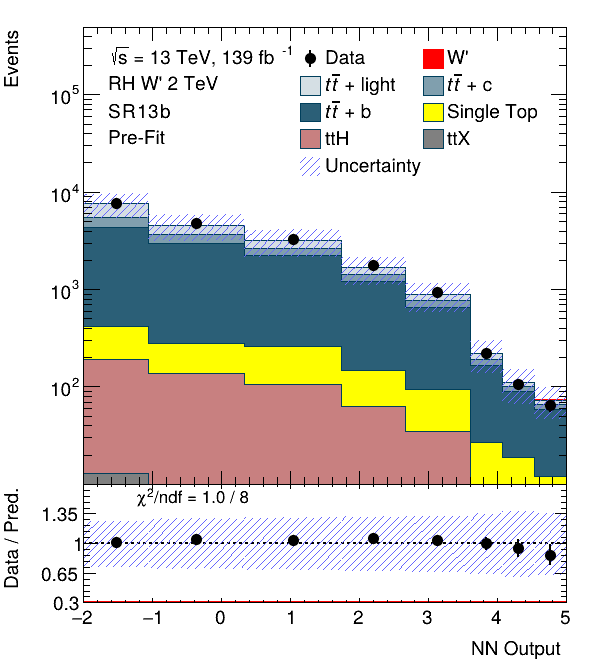}
    \includegraphics[width=0.49\linewidth]{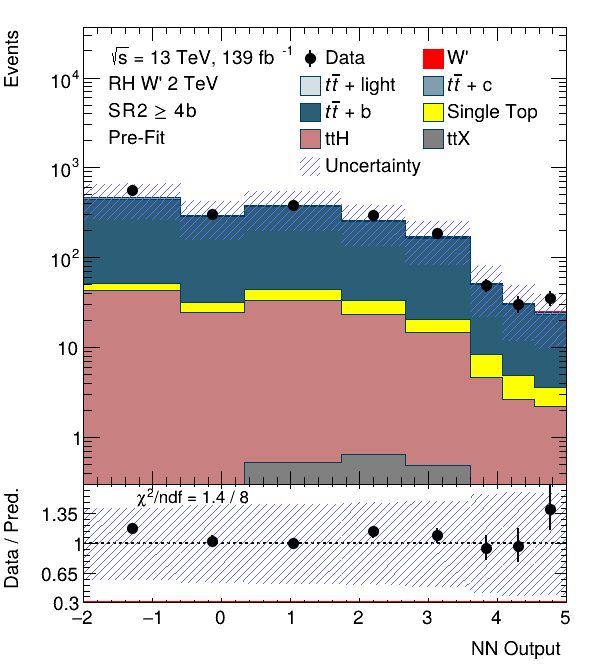}
    \caption{Asimov pre-fit plots for RH 2 TeV $W'$ boson signal sample where $g'/g=2$.}
    \label{data_RH_2tev_prefit}
\end{figure}

\begin{figure}[h]
    \centering
    \includegraphics[width=0.49\linewidth]{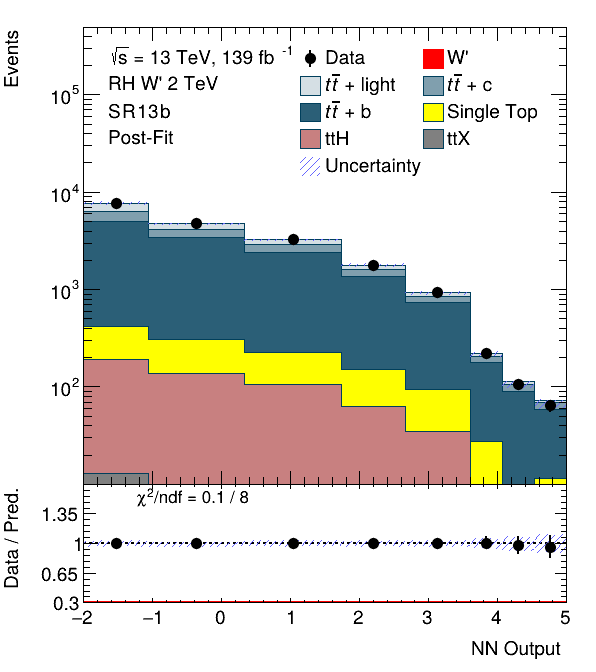}
    \includegraphics[width=0.49\linewidth]{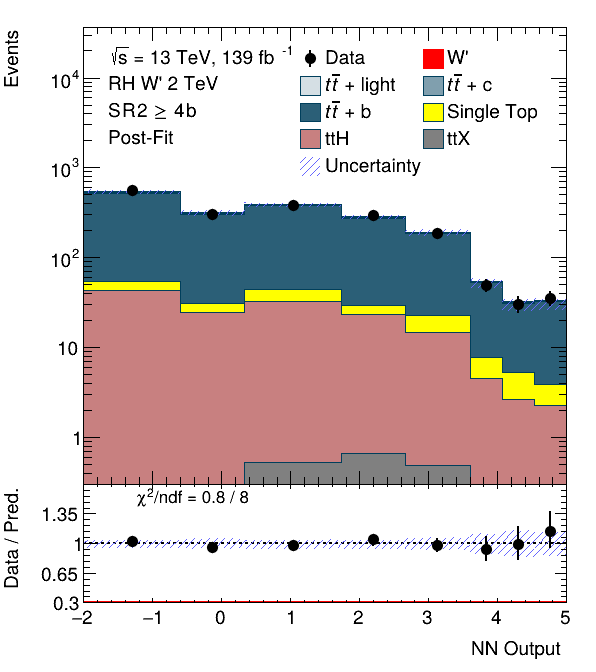}
    \caption{Asimov post-fit plots for RH 2 TeV $W'$ boson signal sample where $g'/g=2$.}
    \label{data_RH_2tev_postfit}
\end{figure}

\clearpage
For the RH $W'$ boson, the pre-fit plots are shown in Figs. ~\ref{data_RH_1tev_prefit}, ~\ref{data_RH_1p2tev_prefit}, ~\ref{data_RH_1p4tev_prefit}, ~\ref{data_RH_1p6tev_prefit}, ~\ref{data_RH_1p8tev_prefit}, and ~\ref{data_RH_2tev_prefit}, and the post-fit plots are shown in Figs.~\ref{data_RH_1tev_postfit}, ~\ref{data_RH_1p2tev_postfit}, ~\ref{data_RH_1p4tev_postfit}, ~\ref{data_RH_1p6tev_postfit}, ~\ref{data_RH_1p8tev_postfit}, and ~\ref{data_RH_2tev_postfit}. Figures containing the resulting normalization factors can be found in Figs.~\ref{norms_RH_1tev}, ~\ref{norms_RH_1p2tev}, ~\ref{norms_RH_1p4tev}, ~\ref{norms_RH_1p6tev}, ~\ref{norms_RH_1p8tev}, and ~\ref{norms_RH_2tev}. There are no major differences between the LH and RH $W'$ boson fits. 

\begin{figure}[h]
    \centering
    \includegraphics[width=0.45\linewidth]{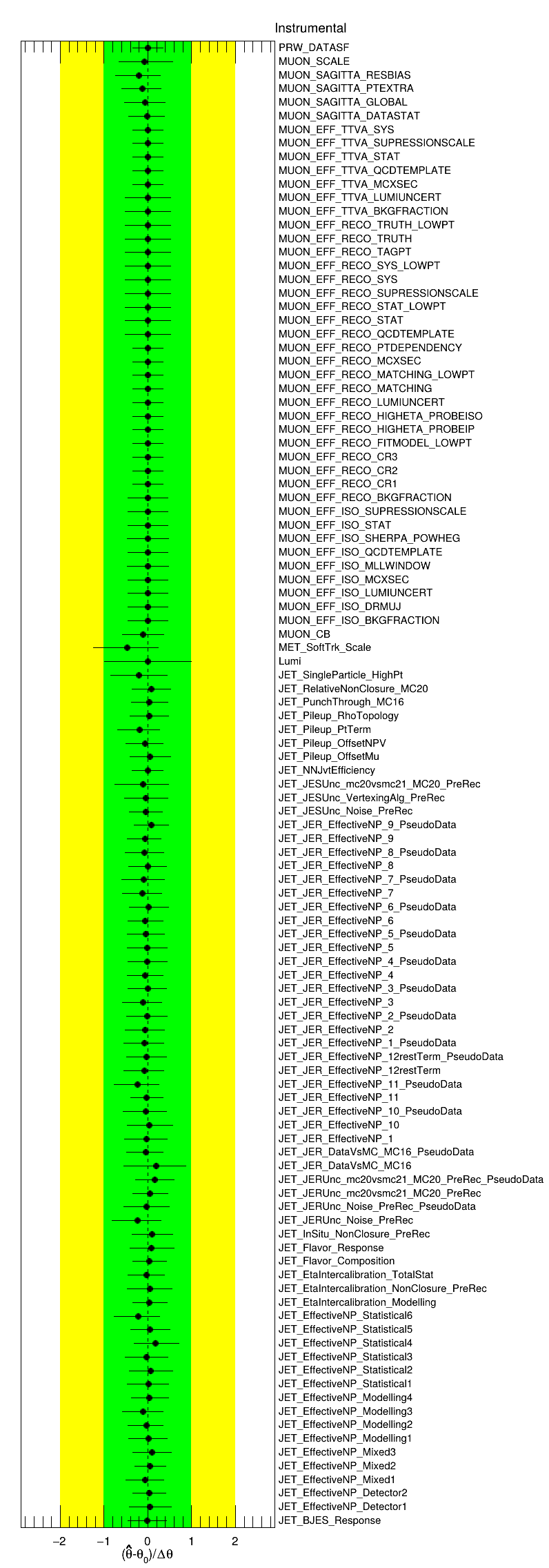}
    \caption{Detector systematic uncertainty pull plot for Asimov fit on RH 1 TeV $W'$ boson sample where g'/g=2.}
    \label{nuispar_inst_RH_1tev}
\end{figure}

\begin{figure}[h]
    \centering
    \includegraphics[width=0.49\linewidth]{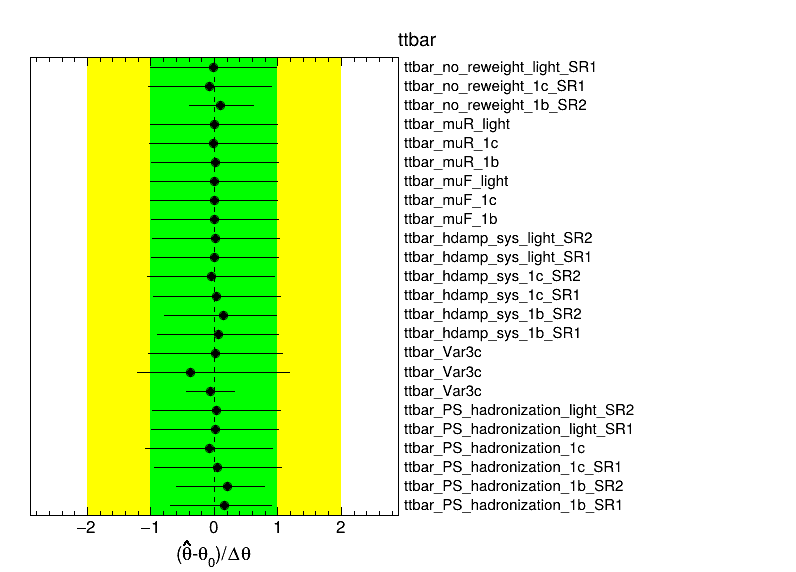}
    \includegraphics[width=0.49\linewidth]{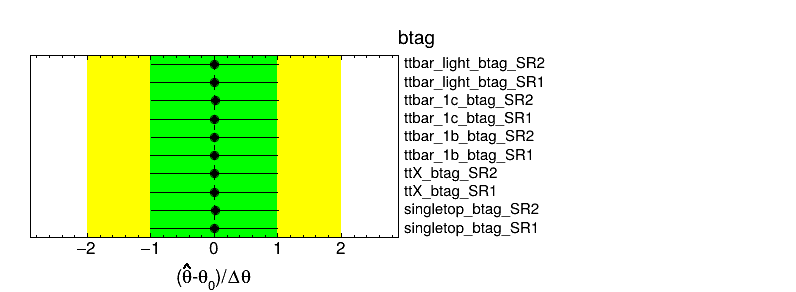}\\
    \includegraphics[width=0.49\linewidth]{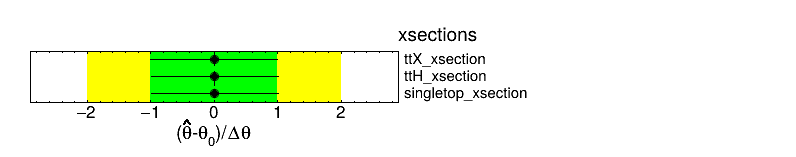}
    \caption{$t\bar{t}$, b-tagging, and cross-section systematic uncertainty pull plot for Asimov fit on RH 1 TeV $W'$ boson sample where g'/g=2.}
    \label{nuispar_all_RH_1tev}
\end{figure}

\begin{figure}[h]
    \centering
    \includegraphics[width=0.45\linewidth]{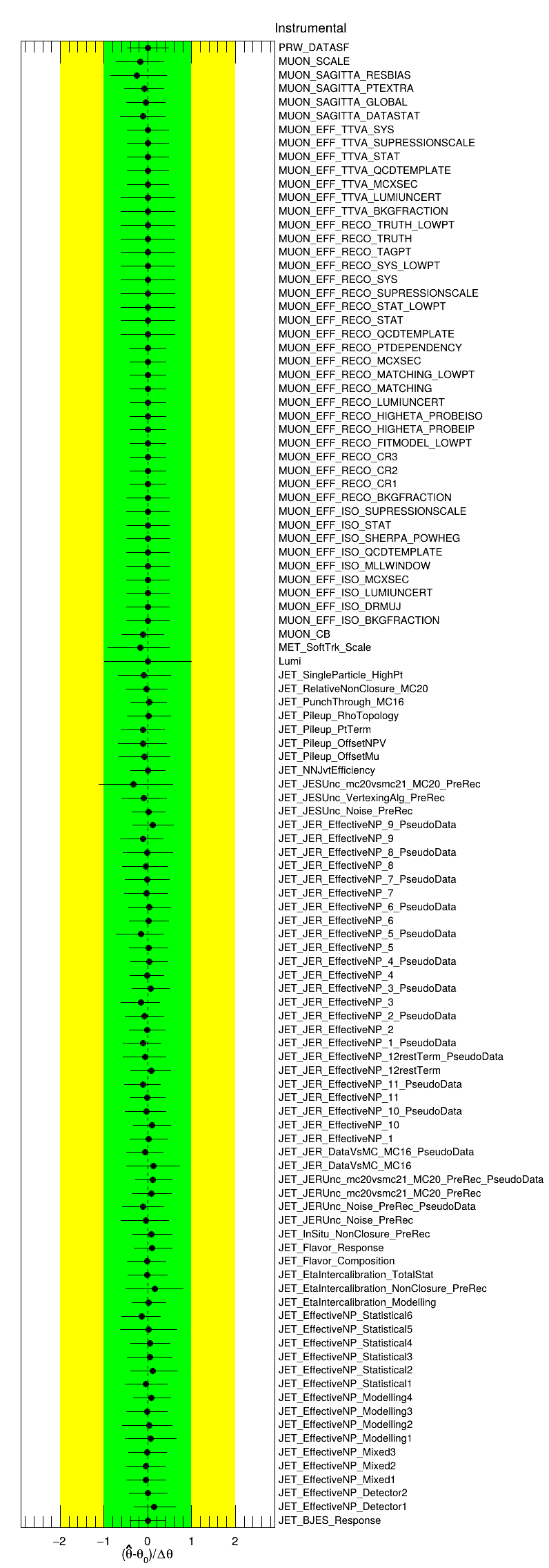}
    \caption{Detector systematic uncertainty pull plot for Asimov fit on RH 1.2 TeV $W'$ boson sample where $g'/g=2$.}
    \label{nuispar_inst_RH_1p2tev}
\end{figure}

\begin{figure}[h]
    \centering
    \includegraphics[width=0.49\linewidth]{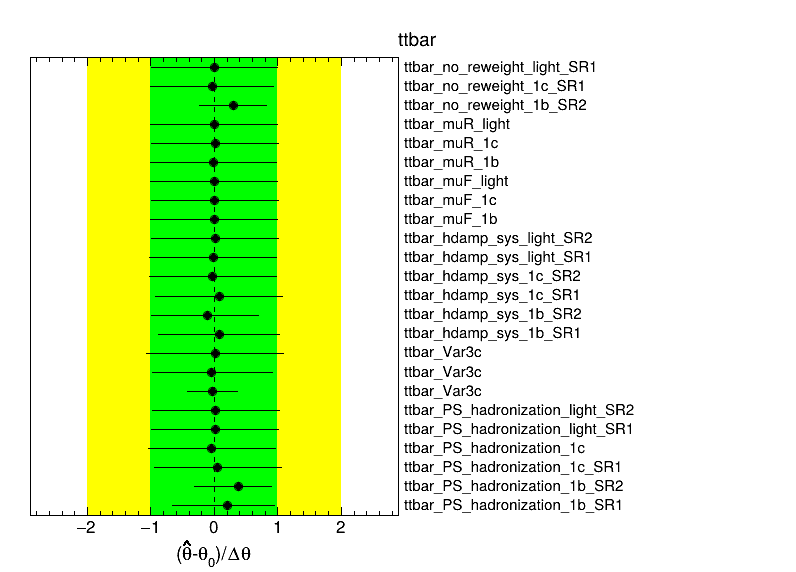}
    \includegraphics[width=0.49\linewidth]{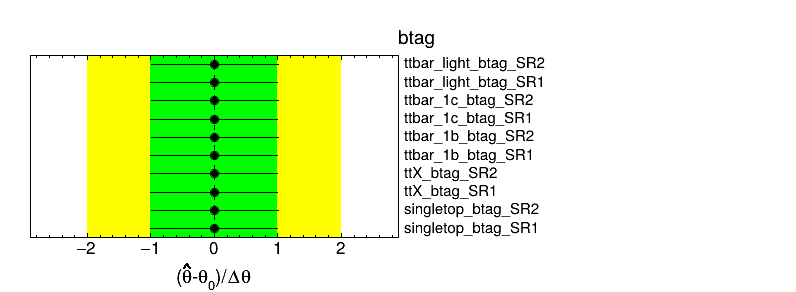}\\
    \includegraphics[width=0.49\linewidth]{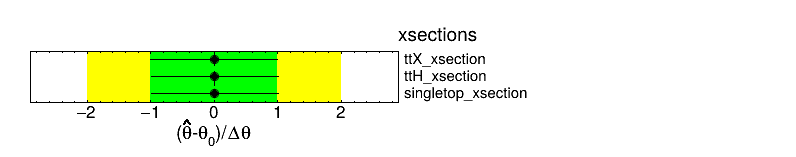}
    \caption{$t\bar{t}$, b-tagging, and cross-section systematic uncertainty pull plot for Asimov fit on RH 1.2 TeV $W'$ boson sample where $g'/g=2$.}
    \label{nuispar_all_RH_1p2tev}
\end{figure}

\begin{figure}[h]
    \centering
    \includegraphics[width=0.45\linewidth]{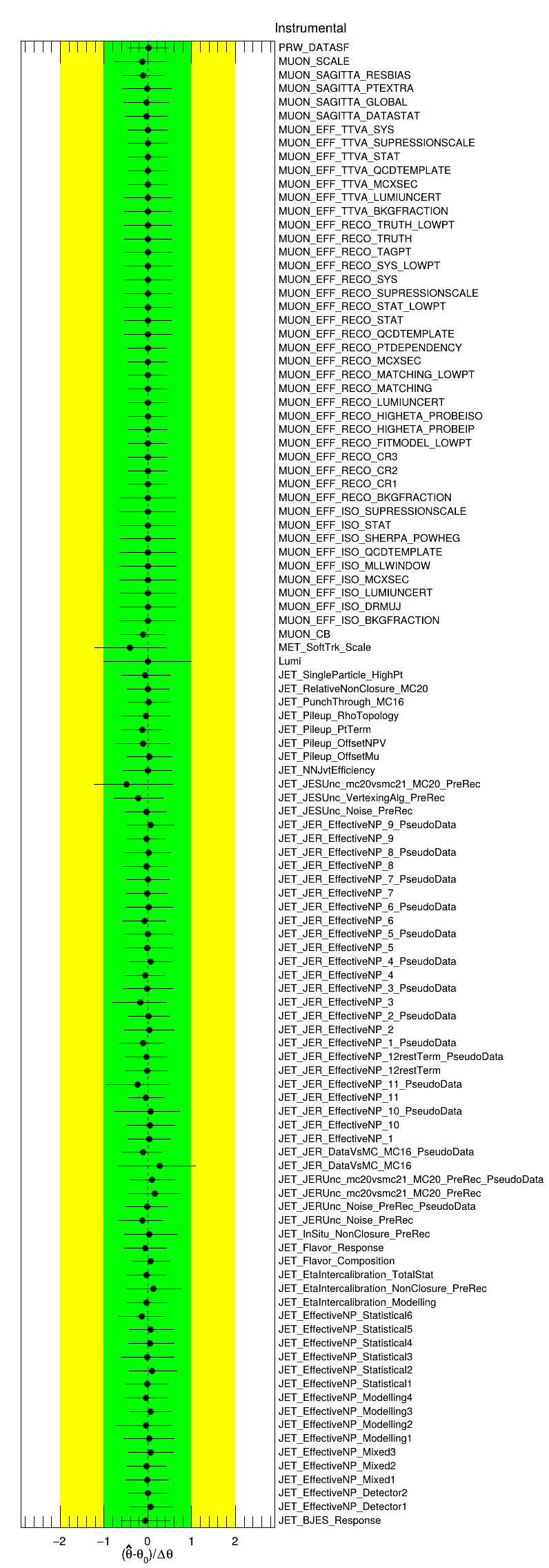}
    \caption{Detector systematic uncertainty pull plot for Asimov fit on RH 1.4 TeV $W'$ boson sample where $g'/g=2$.}
    \label{nuispar_inst_RH_1p4tev}
\end{figure}

\begin{figure}[h]
    \centering
    \includegraphics[width=0.49\linewidth]{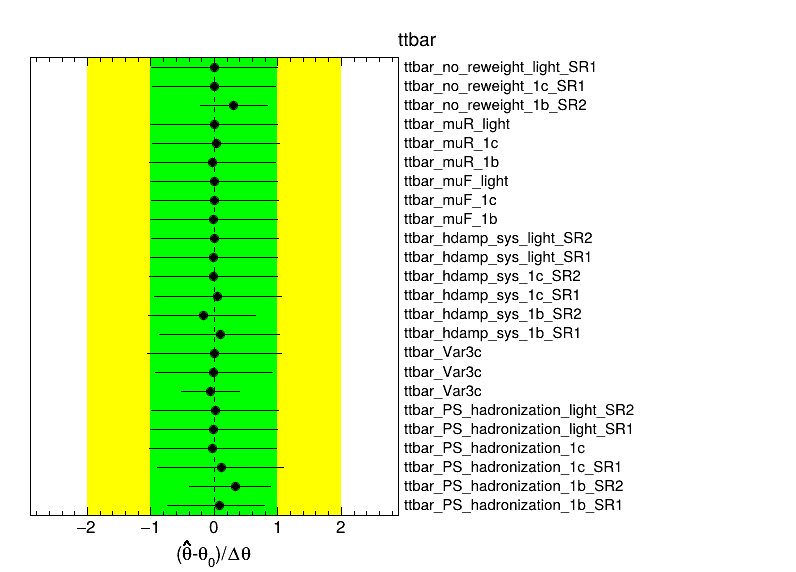}
    \includegraphics[width=0.49\linewidth]{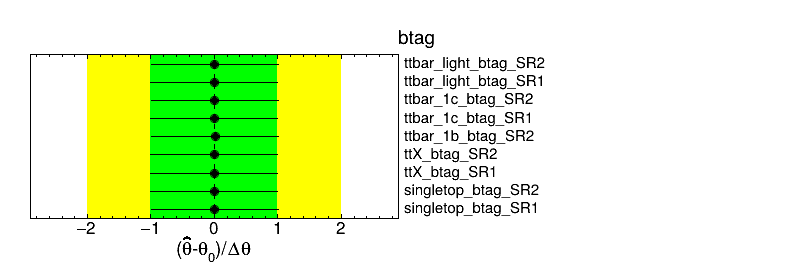}\\
    \includegraphics[width=0.49\linewidth]{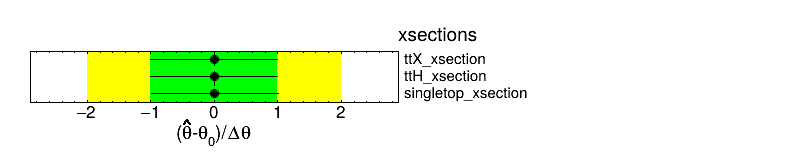}
    \caption{$t\bar{t}$, b-tagging, and cross-section systematic uncertainty pull plot for Asimov fit on RH 1.4 TeV $W'$ boson sample where $g'/g=2$.}
    \label{nuispar_all_RH_1p4tev}
\end{figure}

\begin{figure}[h]
    \centering
    \includegraphics[width=0.45\linewidth]{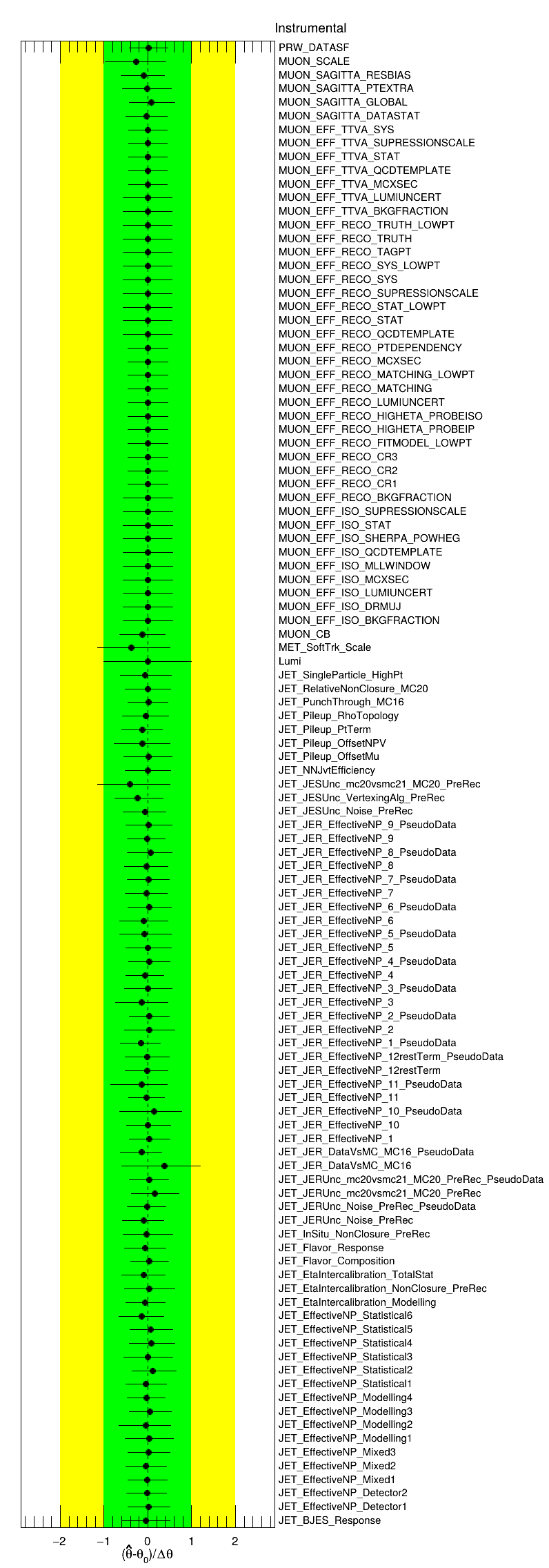}
    \caption{Detector systematic uncertainty pull plot for Asimov fit on RH 1.6 TeV $W'$ boson sample where $g'/g=2$.}
    \label{nuispar_inst_RH_1p6tev}
\end{figure}

\begin{figure}[h]
    \centering
    \includegraphics[width=0.49\linewidth]{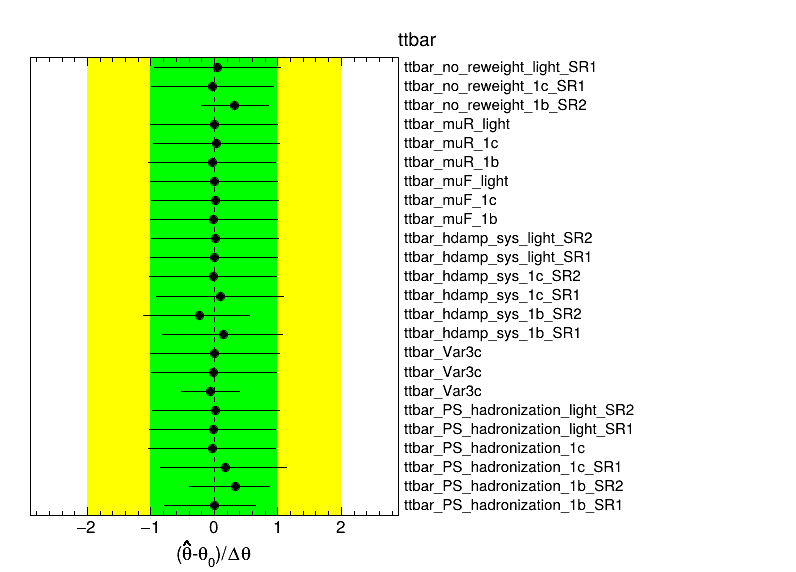}
    \includegraphics[width=0.49\linewidth]{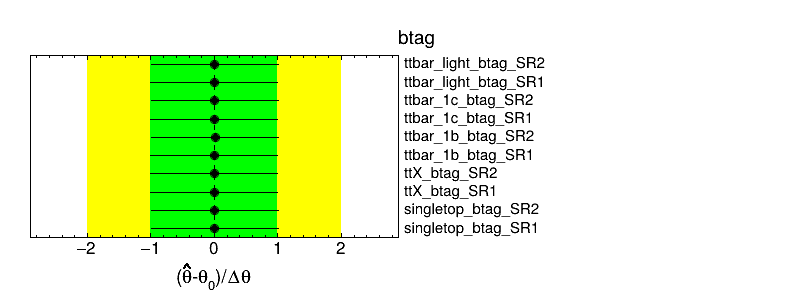}\\
    \includegraphics[width=0.49\linewidth]{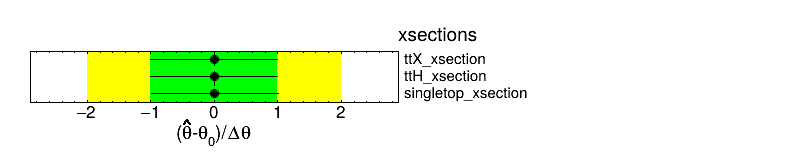}
    \caption{$t\bar{t}$, b-tagging, and cross-section systematic uncertainty pull plot for Asimov fit on RH 1.6 TeV $W'$ boson sample where $g'/g=2$.}
    \label{nuispar_all_RH_1p6tev}
\end{figure}

\begin{figure}[h]
    \centering
    \includegraphics[width=0.45\linewidth]{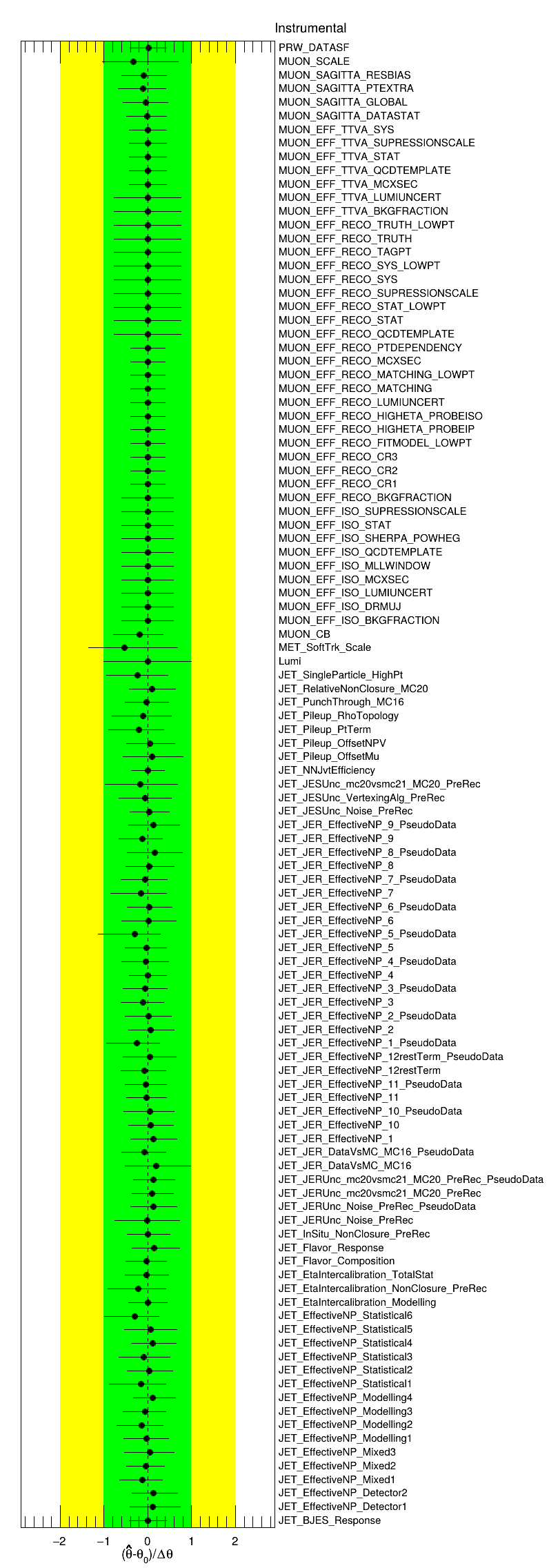}
    \caption{Detector systematic uncertainty pull plot for Asimov fit on RH 1.8 TeV $W'$ boson sample where $g'/g=2$.}
    \label{nuispar_inst_RH_1p8tev}
\end{figure}

\begin{figure}[h]
    \centering
    \includegraphics[width=0.49\linewidth]{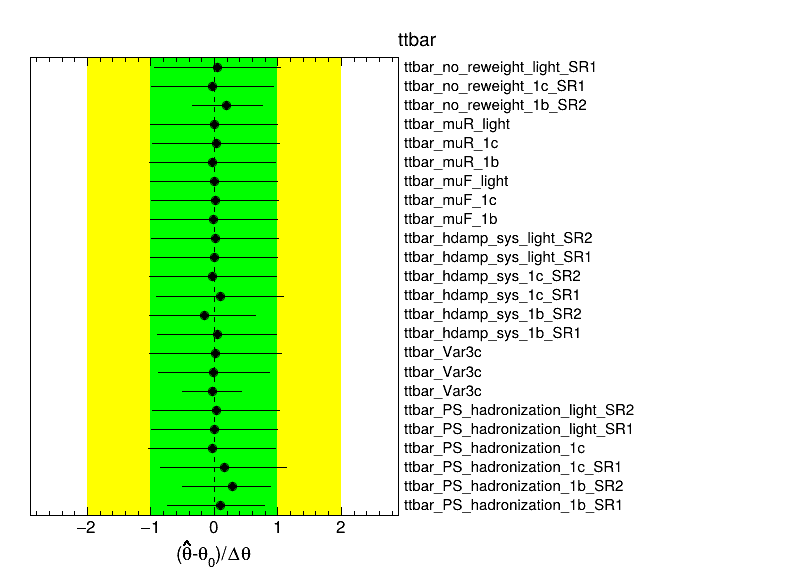}
    \includegraphics[width=0.49\linewidth]{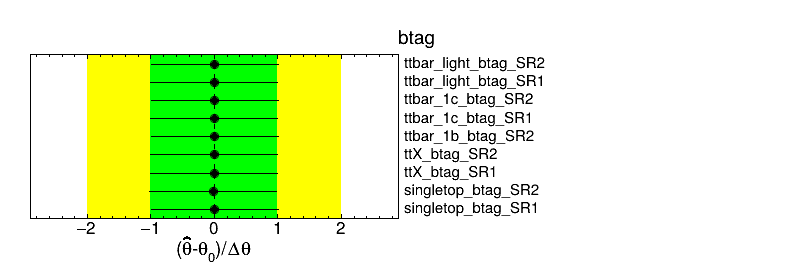}\\
    \includegraphics[width=0.49\linewidth]{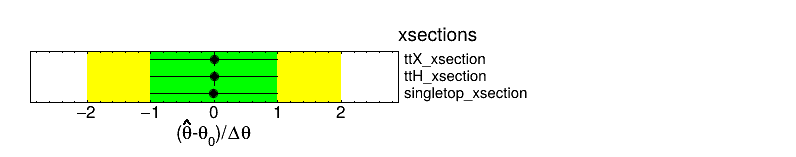}
    \caption{$t\bar{t}$, b-tagging, and cross-section systematic uncertainty pull plot for Asimov fit on RH 1.8 TeV $W'$ boson sample where $g'/g=2$.}
    \label{nuispar_all_RH_1p8tev}
\end{figure}

\begin{figure}[h]
    \centering
    \includegraphics[width=0.45\linewidth]{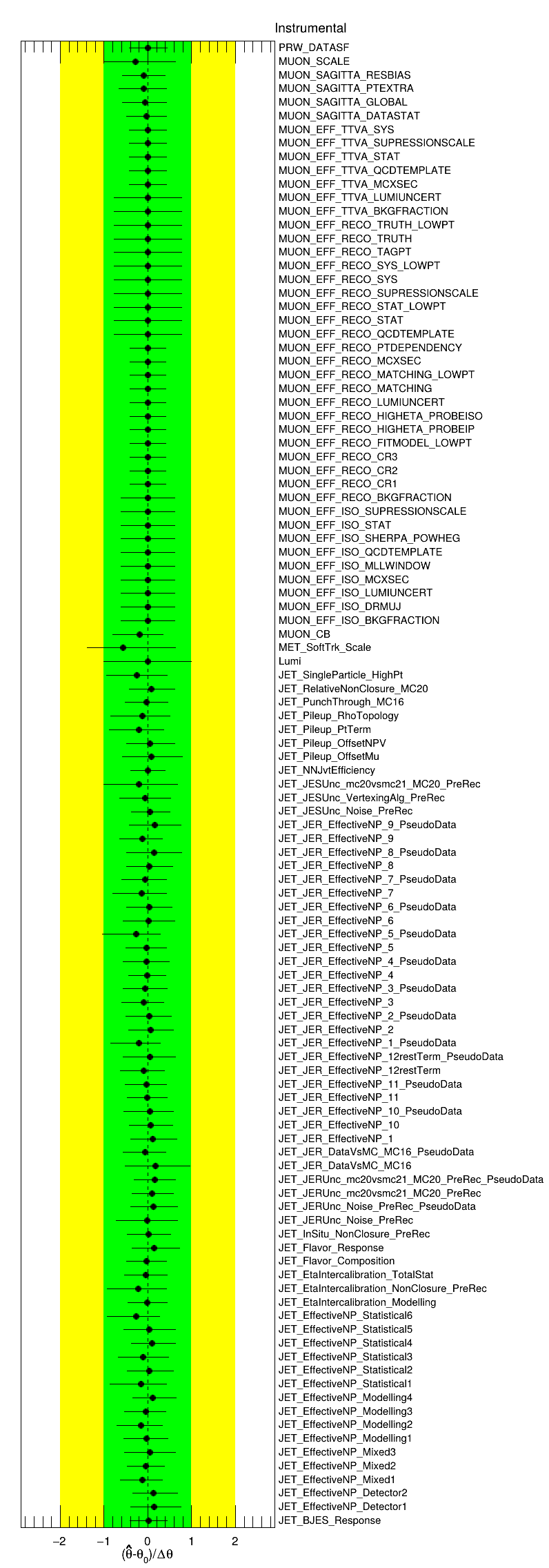}
    \caption{Detector systematic uncertainty pull plot for Asimov fit on RH 2 TeV $W'$ boson sample where $g'/g=2$.}
    \label{nuispar_inst_RH_2tev}
\end{figure}

\begin{figure}[h]
    \centering
    \includegraphics[width=0.49\linewidth]{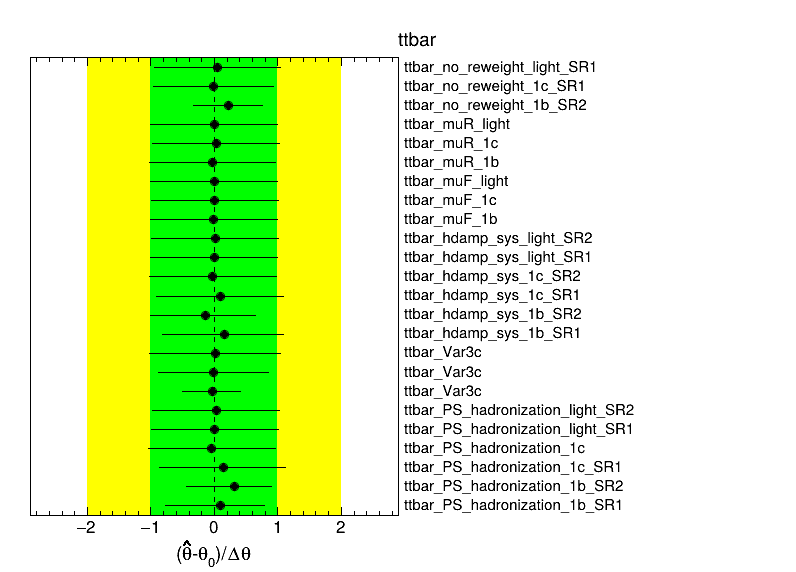}
    \includegraphics[width=0.49\linewidth]{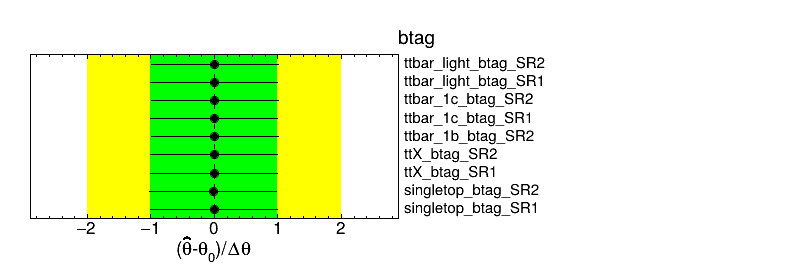}\\
    \includegraphics[width=0.49\linewidth]{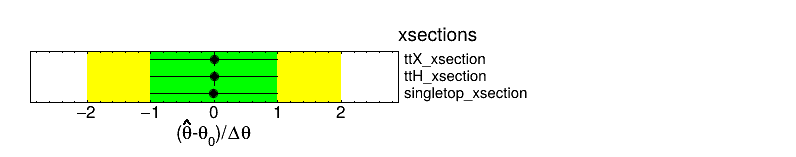}
    \caption{$t\bar{t}$, b-tagging, and cross-section systematic uncertainty pull plot for Asimov fit on RH 2 TeV $W'$ boson sample where $g'/g=2$.}
    \label{nuispar_all_RH_2tev}
\end{figure}

The pull plots on the instrument systematic uncertainties can be seen in Figs.~\ref{nuispar_inst_RH_1tev}, ~\ref{nuispar_inst_RH_1p2tev}, ~\ref{nuispar_inst_RH_1p4tev}, ~\ref{nuispar_inst_RH_1p6tev}, ~\ref{nuispar_inst_RH_1p8tev}, ~\ref{nuispar_inst_RH_2tev}. The pull plots on the systematic uncertainties for b-tagging, cross-sections, and $t\bar{t}$ modeling can be seen in Figs.~\ref{nuispar_all_RH_1tev}, ~\ref{nuispar_all_RH_1p2tev}, ~\ref{nuispar_all_RH_1p4tev}, ~\ref{nuispar_all_RH_1p6tev}, ~\ref{nuispar_all_RH_1p8tev}, ~\ref{nuispar_all_RH_2tev}. In general, pull plots are very similar between LH and RH $W'$ boson fits.

\chapter{Heavy-philic $W'$ Samples}
Table~\ref{tab:wp_dataset_info} summarizes the $W’$ boson signal samples used in this analysis. It lists key details such as the $W’$ mass, chirality, production cross section, generator filter efficiency (GenFiltEff), number of generated Monte Carlo (MC) events, and the corresponding dataset ID numbers (DSID).

\begin{table}[ht]
    \centering
    \caption{List of the generated W' LH and RH samples. All samples are simulated with FullSim and available in the appropriate proportions of MC20a, MC20d, and MC20e.}
    \begin{tabular}{cccccc}
        \hline
        \hline
        DSID & Mass (GeV) & Chirality & Cross Section (pb) & GenFiltEff & MC Events \\
        \hline
        510889 & 1000  & LH & 0.022541  & 5.556070E-01 & 0.5M \\
        510890 & 1200  & LH & 0.0085582 & 5.568781E-01 & 0.5M \\
        510891 & 1400  & LH & 0.0035013 & 5.572652E-01 & 0.5M \\
        510892 & 1600  & LH & 0.0015285 & 5.591470E-01 & 0.5M \\
        510893 & 1800  & LH & 0.00070273 & 5.565144E-01 & 0.5M \\
        510894 & 2000  & LH & 0.00033325 & 5.538775E-01 & 0.5M \\
        510895 & 2500  & LH & 5.9788E-05 & 5.528410E-01 & 0.5M \\
        510896 & 3000  & LH & 1.1918E-05 & 5.563817E-01 & 0.5M \\
        510897 & 4000  & LH & 5.5018E-07 & 5.535567E-01 & 0.5M \\
        510898 & 1000  & RH & 0.02266   & 5.500035E-01 & 0.5M \\
        510899 & 1200  & RH & 0.0085155 & 5.578584E-01 & 0.5M \\
        510900 & 1400  & RH & 0.0035008 & 5.551680E-01 & 0.5M \\
        510901 & 1600  & RH & 0.0015201 & 5.565374E-01 & 0.5M \\
        510902 & 1800  & RH & 0.00069754 & 5.530268E-01 & 0.5M \\
        510903 & 2000  & RH & 0.00033302 & 5.533983E-01 & 0.5M \\
        510904 & 2500  & RH & 5.9378E-05 & 5.537889E-01 & 0.5M \\
        510905 & 3000  & RH & 1.1875E-05 & 5.542597E-01 & 0.5M \\
        510906 & 4000  & RH & 5.4839E-07 & 5.560649E-01 & 0.5M \\
        \hline
        \hline
    \end{tabular}
    \label{tab:wp_dataset_info}
\end{table}
\end{appendices}
%
%
%
%
%
%
\end{document}